\documentclass[12pt]{yalephd}

\pdfoutput=1

\usepackage{units}
\usepackage{amsmath}
\usepackage{amssymb}
\usepackage{graphicx}
\usepackage{bm}
\usepackage{multirow}
\usepackage{color}
\usepackage{relsize}
\usepackage[normalem]{ulem}
\usepackage{microtype}
\usepackage{mathtools}
\usepackage[hyperref=true,
	sorting=nyt, %none
        url=false,
        isbn=false,
        backref=true,
        style=numeric-comp,
        citereset=chapter,
        maxcitenames=3,
        maxbibnames=100,
        block=none]{biblatex}
\usepackage{comment} % for easy commenting of large text blocks
\usepackage{braket} % for bra-ket notation
\usepackage{makecell}
\usepackage{booktabs} % for prettier tables
\usepackage{soul} % for \hl and \st
\usepackage{varioref} % for page numbers
\usepackage{hyperref} % for hyperlinking refs
\usepackage[capitalise]{cleveref} % for ranges of equation references
\usepackage{cancel} % cancelto
\usepackage{mathrsfs} % for mathscr
\usepackage{pifont} % for dingbat xmark
\usepackage{diagbox}
\renewbibmacro{in:}{%
  \ifboolexpr{%
     test {\ifentrytype{article}}%
     or
     test {\ifentrytype{inproceedings}}%
  }{}{\printtext{\bibstring{in}\intitlepunct}}%
}

\newcommand{\be}{\begin{equation}}
\newcommand{\ee}{\end{equation}}

\newcommand{\ba}{\begin{align}}
\newcommand{\ea}{\end{align}}

\newcommand{\bg}{\begin{gather}}
\newcommand{\eg}{\end{gather}}

\newcommand{\re}[1]{\text{Re}\{#1\}}
\newcommand{\im}[1]{\text{Im}\{#1\}}

\newcommand{\Nh}{Non-hermitian}
\newcommand{\nh}{non-hermitian}

\newcommand{\heff}{$H_{\rm CMT}$}

\newcommand{\G}{\Gamma}
\newcommand{\g}{\gamma}
\newcommand{\e}{\varepsilon}
\newcommand{\z}{\zeta}
\newcommand{\gp}{\gamma_\perp}

\newcommand{\m}{\mu}
\newcommand{\w}{\omega}
\newcommand{\vph}{\varphi}
\newcommand{\dph}{\delta\varphi}
\newcommand{\Q}{$Q$}

\newcommand{\pr}{^\prime}
\newcommand{\prpr}{^{\prime\prime}}
\newcommand{\bx}{\bm{x}}

\newcommand{\cis}[2][]{e^{#1i#2}}

\newcommand{\fbar}{\bar{F}}

\newcommand{\wn}{\omega_0}

\newcommand{\defn}{\stackrel{\textrm{\scriptsize def}}{=}}

\newcommand{\bs}{\boldsymbol}

\newcommand{\scSPA}{self-consistent SPA}
\newcommand{\iSPA}{inverse SPA}

\newcommand{\cmark}{\ding{51}}%
\newcommand{\xmark}{\ding{55}}%

\newcommand{\Hcmt}{H_{\rm CMT}} % my (custom) shortcut commands

\author{William R. Sweeney}
\title{Electromagnetic Eigenvalue Problems and Nonhermitian Effects in Linear and Saturable Scattering}
\date{May 2020}
\advisor{A. Douglas Stone}

\begin{document}

\frontmatter

\begin{abstract}
In this thesis we address a series of new problems in non-hermitian optical scattering -- linear and nonlinear -- with increasing degrees of complexity.

We develop the theory of reflectionless scattering modes (RSMs), introducing a novel and broad class of impedance-matched eigenproblems: for a given structure, find the incident wavefronts and frequencies which are not partially reflected at all, but are instead transmitted through the scatterer, or dissipated within it.
The RSM framework includes critical coupling and coherent perfect absorption (CPA) as specific instances, and does not necessarily require intrinsic gain or loss to realize in most cases.
We analyze the symmetry properties of the RSMs, and find that they support exceptional points (EPs) which can be directly probed with steady-state excitation, and are accompanied by a quartic flattening of the reflection lineshape. 
These are distinct from the often-studied resonant and scattering EPs, and can be engineered in hermitian cavities with no gain or loss, in which case the transmission lineshape is also flattened.

We study degenerate coherent perfect absorption (CPA EP), which is a specific example of the new kind of RSM EP mentioned above.
Here, two perfectly absorbed states are brought together by tuning parameters of the scatterer.
In the case of a ring-resonator, it is known that resonant EPs take on a chiral flavor; we show that the same holds true for the CPA EP, and use this fact to design a patterned ring resonator which either predominantly absorbs or reflects light, depending on the direction of incidence.

We extend CPA from cavities with a linear dielectric response to include the saturating nonlinearity and dispersion of a two-level absorbing medium. %, which for some materials is a more realistic model of an absorber than a constant imaginary part of the refractive index.
By using the CPA theorem, which relates CPA in a lossy cavity to lasing in an amplifying cavity, augmented to account for both saturation and dispersion, we show that the SALT algorithm in the single-mode regime can also be used to find the saturable CPA modes through a simple mapping.
This demonstrates that between a lower and upper threshold for loss one can maintain CPA by continuously adjusting the pump strength.
We also clarify the bad-cavity limit of dispersive but linear CPA, identifying new modes that are hybrids of the cavity and atomic degrees of freedom, with a strongly dispersive response to changes in the pump.

We present and solve the general problem of scattering from an arbitrary cavity with a saturable two-level amplifying or absorbing medium, generalizing the known phenomenology of bistability that has mainly been studied in structures with little or no spatial complexity.
Unlike with lasing or CPA, these solutions have both incoming and outgoing flux.
We carefully analyze the validity of approximations used for isolated resonances, and find that the previously used inverse single-pole approximation requires some modification for the case of scattering.

\end{abstract}

\maketitle

\makecopyright

\setcounter{tocdepth}{3}
\tableofcontents

\listoffigures

\listoftables

\newpage
\
\vfil
\hfil {\it for} \ {\scshape Susan} \hfil

\hfil {\scshape and healthcare workers everywhere} \hfil
\vfil
\newpage

\chapter{Acknowledgments}
There are so many who have supported me and deserve my deepest thanks during my time at Yale.
First and foremost, I owe a special thanks to my thesis advisor A. Douglas Stone, who through the years has been both a source of steadfast encouragement and support, as well as a model of enthusiasm for physics and rigorous scholarship.
His door was always open to me whenever I had a question, and he was always willing to discuss and develop new ideas, many of which now populate the document you are reading. 
I am continually learning from his insightful thinking about physics, and am greatly indebted to him for giving me the opportunity to join his group, despite the unusual path I took to get there.
A special thanks also to my committee members, Hui Cao, Jack Harris, and Owen Miller, for many enlightening conversations, meetings, and lectures over the years.

I am grateful to Chia Wei Hsu for many years of successful collaboration; I consider myself lucky to have been able to work with and learn from him, and also to have shared many a delicious lunch at the carts with him.
Thanks as well to Alex Cerjan, who was invaluable in helping me to get started in the group.
I have learned a great deal talking to him over the years.

Thank you to my collaborators Stefan Rotter, Lan Yang, and Changqing Wang for many fruitful discussions about research, and for having the patience to discuss some of the ideas in this thesis with me.

A special thanks to Steven Johnson for agreeing to be my outside reader, and for being a wonderful host for the Stone group on several visits to MIT, as well as for introducing me to the Julia computing language many years ago.

To my parents, John and Gladys: I would never have made it this far in graduate school or in life without your continuous love, support, and guidance.
Thank you!
You are perhaps the only people in the world happier than I am that I'm finally done!

My years in New Haven have been some of the most meaningful of my life, and that is in no small part thanks to the many great people I have met here, too numerous to list.

To James Mulligan and Fangzhou Zhu, thank you for the many stimulating discussions at all hours of the day and night about math and physics, and especially Fangzhou for sharing in so many musical experiences with me.

To my old roommates and dear, dear friends, Ben Brubaker, Zack Lasner, Emily Brown, Dylan Mattingly: thank you for making apartment-life more enriching and fulfilling than it had any right to be.
Your ongoing friendship means the world to me.

Susan, words could never express the gratitude I have for you and everything you do.
Thank you for keeping me grounded, thank you for helping me soar.

\mainmatter

\chapter{Introduction \label{sec:chpt0: intro}}
%The quest to understand the interaction between light and matter has played a central role in physics for at least a century, with related Nobel Prizes awarded as early as 1921, and as recently as 2018~\cite{nobel_1921,nobel_2018}.
The spatial variation of the electromagnetic field is often neglected when the size of the material it interacts with is many orders of magnitude smaller than the wavelength of the light.
For example, an atomic radius is ${\cal O}(0.1\ {\rm nm})$, while optical wavelengths are ${\cal O}(100\ {\rm nm})$, a thousand times bigger.
Many important phenomena can be well-explained using this assumption, such as the population dynamics of an ensemble of irradiated atoms.
However, for some light-matter phenomena, the spatial complexity of the light field cannot be so handily factored out.
Steady-state lasing is one such example.

Lasing is the self-organized, coherent free-oscillation of the electromagnetic field in an open cavity, and is fundamentally important to modern science and technology.
Here, and throughout the rest of this thesis, ``open'' means that the light is imperfectly confined, and can leak out through the boundaries of the cavity in the form of far-field radiation.

``Laser'' is actually an acronym: {\bf L}ight {\bf A}mplification by {\bf S}timulated {\bf E}mission of {\bf R}adiation.
Read literally, it describes a mechanism for optical amplification.
If an optical cavity has a sufficiently high degree of amplification, it can spontaneously self-organize and oscillate.
Today, the word ``laser'' is used almost exclusively as synecdoche to refer to such a self-oscillating cavity, and not just to its amplifying mechanism.
It is more appropriate to call this device a {\it laser oscillator}.
When the amplifying power of the cavity is too small to sustain self-oscillation, then the device is a {\it laser amplifier} (a case we discuss in \Cref{chp:chpt5: saturable scattering}).

Lasers employed in technology and research range in size from hundreds of wavelengths to hundreds of thousands, and the space-dependence of the light field inside the laser cannot be neglected the way it can for an atomic scatterer in free space, except in the simplest of cases.
Lasers exhibit a rich variety of spatio-temporal dynamics, ranging from steady-state (also called ``continuous wave'' or CW), to regularly pulsed, to fully chaotic operation, all of which are described by the Maxwell-Bloch equations (MBE)~\cite{haken_sauermann_1963, lamb_1964, haken_1985}.
These are coupled, nonlinear, partial differential equations that describe Maxwell's equations coupled to a gain medium, which is modeled as an ensemble of independent two-level atoms. 
%More generally, semiclassical laser theory refers o Maxwell's equations coupled to quantum matter.
%In this thesis I will focus on generalizations and applications of the 
Steady-state {\it ab initio} laser theory (SALT)~\cite{tureci_2006, tureci_2007, tureci_2008, ge_2010, esterhazy_2014, cerjan_2015} describes the space of CW solutions to the MBE without making any prior assumptions about the space-dependence of the light field.
Within the scope of photonics, it is appropriately called {\it ab initio}, because the inputs to the theory are phenomenological parameters whose origins lie outside the scope of photonics, belonging instead to the domain of many-body and solid state physics. These inputs are a linear dielectric function $\e_c(\bx,\w)$, and a set of parameters describing the gain medium.
Nothing further is assumed about the geometry or dimensionality of the problem, except for steady-state operation with a discrete harmonic spectrum.
For laser sources, the number of lasing modes, their frequencies, their intracavity profiles, and their far-field radiation patterns are all determined by SALT, under the assumption of no phase correlation between the lasing modes.
Because of this generality, SALT describes many kinds of CW lasing cavities, such as a photonic crystal defect cavity~\cite{1999_painter_science, 2001_lin_ool, JJ_book}, a chaotic cavity~\cite{Cao:2015fv, 1996_nockel_ol, 1998_gmachl_science, 1986_qian_science, 1994_nockel_ol, 1997_nockel_nature, 2018_bittner_science, 2001_stone_ps}, or a disordered cavity (a ``random laser''~\cite{1996_pre_wiersma, 1999_cao_prl, 2003_cao_waves, tureci_2006, tureci_2008}), which are largely beyond the scope of other theories.

Though laser oscillation is not the focus of this thesis, it will serve as a familiar guidepost throughout.
It is a relevant starting point, since it is commonly encountered and shares many features with the novel results presented in this work on amplifiers, absorbers, and reflectionless scattering.
Additionally, SALT was the historical starting point in the development of essentially all of the ideas in this thesis.
The features of the CW lasing problem addressed by SALT that will recur throughout are
\begin{enumerate}
    
    \item Semiclassical: the general theory of light-matter interaction is so broad and complex that it is the subject of very many textbooks, and is still evolving as new physical regimes are reached.
    A complete theory treats light and matter on an equal quantum footing, and leads to many surprising predictions, such as vacuum Rabi oscillations~\cite{brune_1996, wallraff_2005, bochmann_2008}.
    However, for lasers it is typically sufficient to treat only the matter as quantum, while treating the electromagnetic fields as classical, i.e.,~as deterministic mean fields described by Maxwell's equations, hence {\it semi}-classical.
    This is because the intensity of light in lasers is large, and the average number of photons is high enough that it can be treated as a continuous quantity (neglecting fluctuations), as we do in the rest of this thesis.
    
    \item Openness: optical structures generally do not perfectly trap light. Instead the light irreversibly leaks out and is lost to the universe in the form of radiation.
    This is in contrast to the typical scenario for quantum matter, for example the way an electron can be bound indefinitely in an external Coulomb potential.
    Because of this, the role played by ``modes'' in optics is somewhat different from its quantum counterpart, as will be discussed in \Cref{chp:chpt1: theory of scattering}.
    
    \item Non-hermiticity: in addition to the power radiated away due to openness, the optical structures that we consider do not generally conserve energy flux, as photons are created or destroyed in the light-matter interaction.
    By specifying the material properties in different regions, one can locally amplify or locally attenuate the field, in which case the scattering region is said to be \nh.
    The consequences of non-hermiticity are reviewed in the next section.
    
     \item Saturable nonlinearity: the gain medium assumed in the MBE and in SALT is an ensemble of independent two-level quantum systems with a partially inverted population.
     Where the light intensity is high due to stimulated emission, the population of the lower and upper states is pushed closer to equality, so that the medium is just as likely to absorb a photon as it is to undergo stimulated emission.
    In those regions the medium becomes ineffective as an amplifier.
    When the field is uniform, this is known as {\it gain-saturation}; when non-uniform, the saturation happens unequally at different points in space, an effect known as {\it spatial-hole burning}, which will be important in this work.
    Spatial hole-burning in a saturating medium leads to non-trivial spatial-dependence that cannot be neglected unless some further assumption is made about the geometry.
    The MBE, SALT, and spatial-hole burning are briefly reviewed in \Cref{sec:chpt0: saturable two-level media}.
      
\end{enumerate}

It is a goal of this thesis to address certain properties of Maxwell's equations for \nh\ cavities in the spirit of SALT, without making any simplifying assumptions about their geometry.
Some chapters (\ref{chp:chpt2: rsms}, \ref{chp:chpt3: cpa ep}) are devoted to the study of special scattering solutions of the linear Maxwell equations, motivated by the SALT approach, and exploiting features particular to linear \nh\ systems.
Other chapters (\ref{chp:chpt4: saturable cpa},\ref{chp:chpt5: saturable scattering}) focus on novel nonlinear extensions of scattering theory, taking into account the saturating nonlinearity for both amplifiers and absorbers.

 	\section{\Nh\ Photonics \label{sec:chpt0: nh physics}}

Generally, linear scattering from an open wave system with steady-state, time-harmonic illumination has both incoming and outgoing propagating waves.
In the typical problem, the incoming component is fixed, while the outgoing component is unknown {\it a priori}, and must be solved for.
The linear scattering problem is guaranteed to have a solution on both physical and mathematical grounds.
If the atoms in the cavity absorb and emit photons with equal rates, so that the photon flux is conserved, then the total power radiated outward from the cavity is equal to the incident power, for any choice of input wave.
Such an energy-flux conserving scatterer is said to be {\it hermitian}; in contrast, a {\it \nh} scatterer, which has some intrinsic material absorption or amplification, does not conserve energy flux.

%The study of \nh\ physics in the context of optics is very relevant, as optical amplification is easily achieved, for example through the addition of ionic dopants, while loss is inevitable, whether by material absorption or free-space radiation due to imperfect confinement.
A variety of interesting and sometimes counterintuitive phenomena have been observed in a broad class of \nh\ photonic systems: enhancing transmission through the addition of loss in coupled waveguides~\cite{2009_Guo_PRL, 2010_Ruter_NatPhys}, suppressing lasing through the addition of gain in coupled cavities~\cite{2012_Liertzer_PRL, 2014_Brandstetter_NatComm}, the simultaneous vanishing of reflection together with unit transmission but only from one side of a scatterer~\cite{2011_Lin_PRL, 2011_Hernandez-Coronado_PLA, 2011_Longhi_JPA, 2012_Ge_PRA, 2012_Jones_JPA, 2013_Feng_nmat, 2014_Ramezani_PRL_2, 2014_Midya_PRA, 2015_Fleury_ncomms, 2016_Rivolta_PRA, 2016_Jin_SR, 2016_Chen_OE, 2016_Yang_OE, 2017_Sarisaman_PRA, 2018_Sarisaman_PRB}, power oscillations~\cite{2008_makris_PRL}, topological state-transfer~\cite{2016_Xu_Nat}, asymmetric mode-switching~\cite{2016_Doppler_Nat}, constant-intensity waves~\cite{2017_Makris_LSA, 2019_Brandstotter_PRB}, and chiral lasing from disk resonators~\cite{Wiersig:2011hs, Wiersig:2014bq, 2016_Peng_PNAS}.
This list of novel \nh\ effects in optics is certainly not exhaustive.

 	\subsection{Laser Oscillation, Coherent Perfect Absorption \label{sec:chpt0: lasers and cpa}}

\begin{figure}[t]
   \centering
   \includegraphics[width=\textwidth]{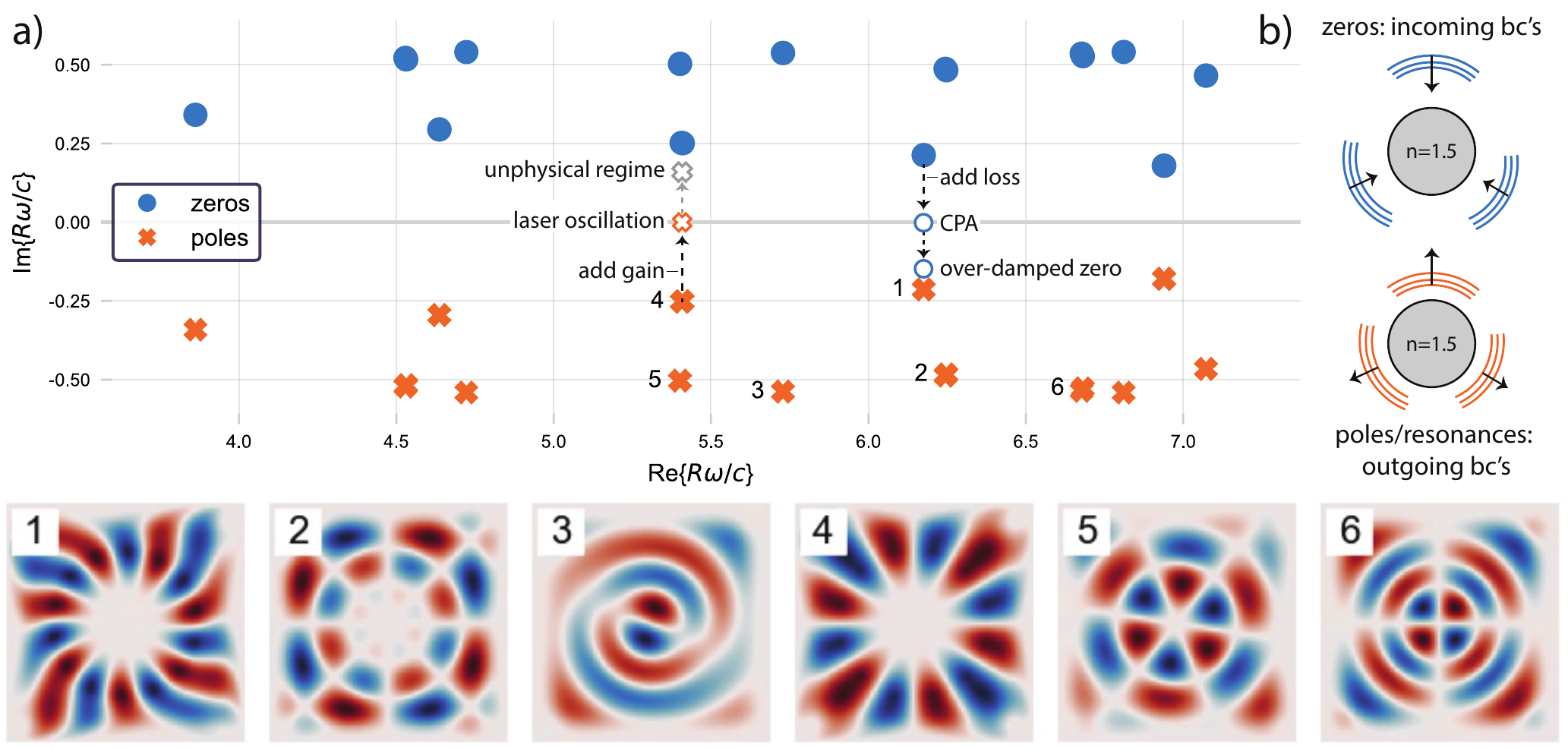}
   \caption[$S$-matrix poles and zeros as purely outgoing and incoming solutions]{$S$-matrix poles and zeros as purely outgoing and incoming solutions.
   The cavity is a two-dimensional disk of radius $R$ and refractive index $n=1.5$.
   {\bf (Top, a)}:~Locations of scattering poles (orange x's) and zeros (blue circles) in the complex frequency-plane for the hermitian cavity shown in (b).
   The poles and zeros are complex conjugates of each other.
   Black dashed arrows represent the flow of poles and zeros when gain and loss is added to the cavity, respectively.
   When a pole reaches the real axis, the cavity becomes unstable and self-oscillates at that frequency; further addition of gain does not move the pole into the upper half-plane.
   In contrast, the zero will flow into the lower half-plane as the amount of loss is increased past the CPA value.
   Each pole (and zero) is associated with a particular mode; the numbered poles have theirs shown in the correspondingly numbered panels at the bottom.
   {\bf (Top, b)}:~Schematic representing poles (orange) and zeros (blue) as satisfying purely outgoing and incoming boundary conditions.
   {\bf (Lower)}:~Examples of purely outgoing modes.
   Number labels correspond to the associated scattering poles shown in (a).
   The distortion of the modes at the edge of the computational cells is an artifact of the perfectly matched layers used to implement the outgoing boundary (see \Cref{chp:chpt2: rsms}, \Cref{sec:chpt2: PMLs}). 
   }
    \label{fig:chpt0: poles-zeros}
\end{figure}

Perhaps the most important \nh\ effects in a cavity are laser amplification and oscillation.
If the atoms within the cavity emit more photons than they absorb on average, then the cavity is an amplifier, radiating more power outwards than is being sent in.
As we increase the difference between the rates of photon emission and absorption, say by creating an inverted population via some external pump, this excess of radiated power also increases.
At some point, the net rate of photon emission from the atoms just equals the rate at which photons leak out of the cavity in the form of far-field radiation, which defines the threshold for laser oscillation.
Infinitesimally above threshold, the lasing problem corresponds to solving Maxwell's equations subject to purely outgoing boundary conditions, with vanishingly small intensity.
This is an over-determined problem that does not have a solution except at certain discrete frequencies, called {\it scattering poles} (for reasons that will become evident in \Cref{chp:chpt1: theory of scattering}), unlike the typical scattering problem mentioned above with both incoming outgoing components that always has a solution.
The purely-outgoing problem can also be solved when the degree of amplification is below the threshold for laser oscillation, in which case the discrete solutions will be complex and with negative imaginary parts, i.e.,~in the {\it lower half-plane}.
Each discrete scattering pole is associated with its own outgoing wavefront (see \cref{fig:chpt0: poles-zeros}).
Threshold lasing can thus be characterized by the condition that a scattering pole flows upward and just reaches the real axis as the amplification within the cavity approaches the threshold value~\cite{1973_Lang_PRA, esterhazy_2014}.

%In addition to EPs, unusual scattering states are also an important feature of \nh\ open systems.
%These are superpositions of incident and outgoing wavefronts that would otherwise be forbidden by flux conservation.
%Lasing is such a scattering state, albeit a special one in which a \nh\ cavity radiates in the absence of any incident flux, which is clearly forbidden in Hermitian systems.

Coherent perfect absorption (CPA) is another important \nh\ scattering state that can only be realized at discrete frequencies, similar to lasing.
CPA is a scattering phenomenon in which an incident wavefront impinges on an absorbing cavity and leads to no outgoing radiation at all, since the energy is entirely dissipated within the cavity, hence {\it perfect absorption}.
It is related to threshold laser oscillation through a simple time-reversal argument~\cite{Chong:2010ft, Wan:2011bz, Noh:2012wx, Piper:2014js, 2016_Zhou_Optica, Zhao:2016cd, Baranov:2017jv, 2019_Pichler_Nature},
which is codified in the CPA theorem:
\begin{description}
    \item {\bf Linear CPA Theorem:} The specific input state which is totally absorbed is the time-reverse of the threshold lasing mode for a similar cavity, but with gain replacing loss [$n(\bx) \to n^*(\bx)$].
\end{description}
Any other incident wavefront will be partially absorbed and partially transmitted or reflected.

The time-reverse of threshold lasing, which is CPA, corresponds to solving Maxwell's equations with purely {\it incoming} boundary conditions.
Generally, the purely incoming problem has complex-valued discrete solutions that we will call {\it scattering zeros}; achieving CPA requires tuning the degree of absorption in order to cause a scattering zero to flow downward to the real axis.
This will happen when the rate of absorption within the scatterer balances the in-coupling rate between the scatterer and the exciting field; in this sense CPA is a variant and multichannel generalization of the concept of critical coupling to a resonator~\cite{Yariv:2000dz}.

We will show in \Cref{chp:chpt1: theory of scattering} that for hermitian, flux-conserving cavities, the scattering zeros and poles, which are associated with purely incoming and outgoing states, occur in conjugate complex frequency pairs.
The addition of material gain or loss is necessary to move either of them to the real axis to achieve lasing or CPA.
Hence lasing and CPA are inherently \nh\ scattering phenomena.

One important difference between threshold lasing and CPA is their behavior when the gain/loss is increased beyond its threshold/critical value.
For laser-oscillation above threshold, the scattering/lasing-pole remains on the real axis, and does not pass into the upper half-plane; it is stabilized there by the nonlinearity of the gain medium, which is discussed later in this chapter, \Cref{sec:chpt0: saturable two-level media}.
At the same time, the output power increases from zero as the gain is increased past the lasing-threshold value.
For CPA, on the other hand, as the degree of loss in the system is made super-critical, for a fixed input power, the scattering-zero simply flows into the lower half-plane, so that CPA is no longer possible.
This is summarized in \cref{fig:chpt0: poles-zeros}a.

            \subsection{Exceptional Points \label{sec:chpt0: eps}}

\begin{figure}[t]
   \centering
   \includegraphics[width=\textwidth]{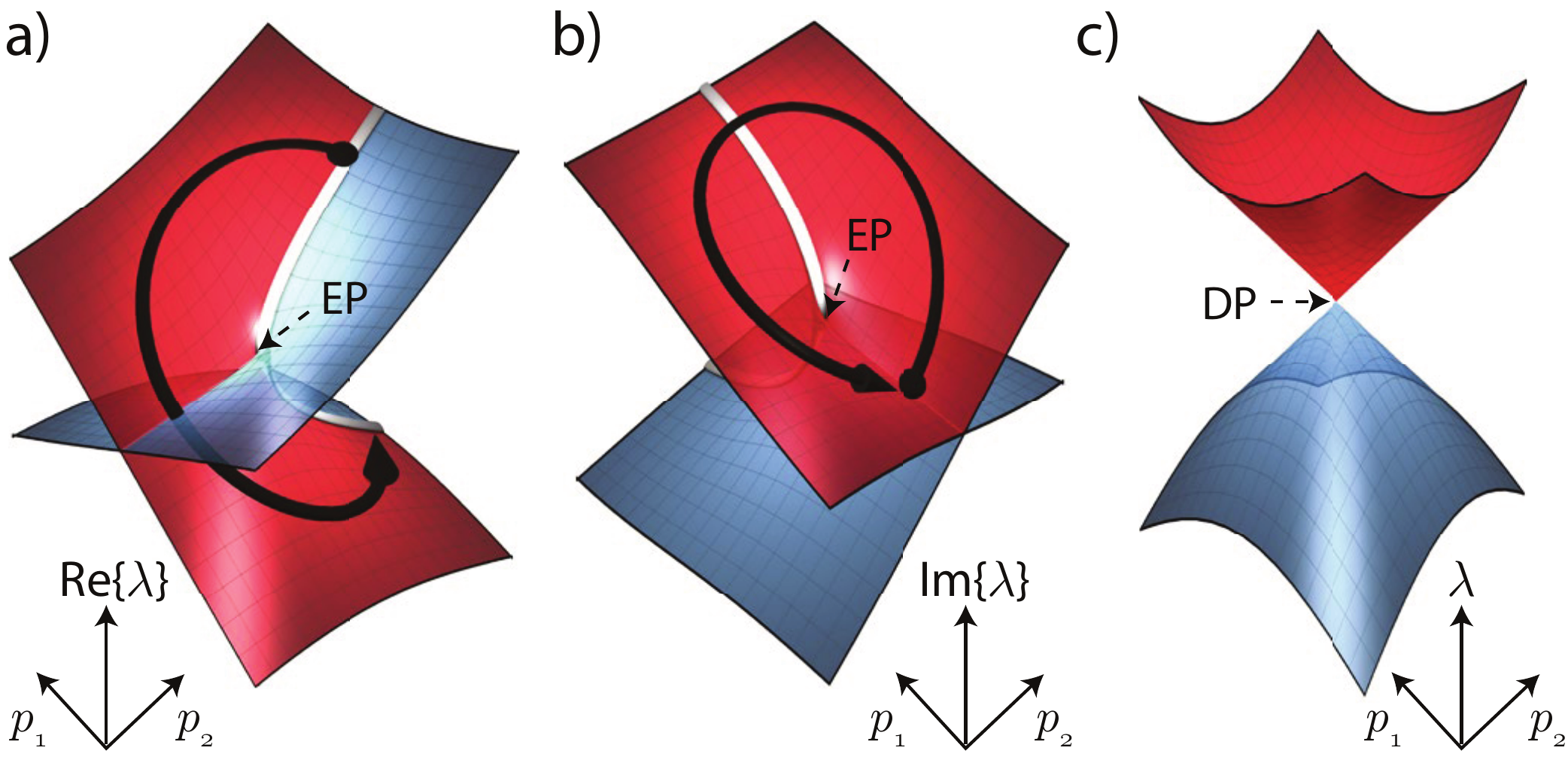}
   \caption[Eigenvalue manifold near an exceptional point, and hermitian degeneracy]{Illustration of the complex square-root manifold of eigenvalues ($\lambda$) in the vicinity of an exceptional point [EP] (a--b), and the conical manifold of an ordinary hermitian degeneracy [DP] (c).
   The horizontal plane is the two-dimensional parameter space (generic parameters $p_1$ and $p_2$) in which the degenerate point lives.
   Reproduced from Ref.~\cite{2019_Ozdemir_nmat}.
   {\bf(a)}:~Real part of $\lambda(p_1,p_2)$.
   The EP is the branch point of the square root, with the branch cut going along the negative $p_1$-axis.
   Red shading corresponds to sheet with positive imaginary value, blue for negative.
   {\bf(b)}:~Imaginary part of $\lambda(p_1,p_2)$.
   Black arrow follows one eigenvalue along a hypothetical loop in parameter space, encircling the EP.
   The same arrow in (a) does not close on itself.
   {\bf(c)}:~Hermitian case, with $\lambda(p_1,p_2)\in{\mathbb R}$.
   Near the degeneracy, known as a ``diabolic point'' (DP), $\lambda$ has a double cone structure, with two unambiguously defined sheets.
    It is named after its resemblance to the similarly shaped toy known as a diabolo.
   }
    \label{fig:chpt0: ep-manifold}
\end{figure}

A hallmark of \nh\ physics is the existence of exceptional points (EPs).
These are generic degeneracies of \nh\ systems, that is, the coalescence of eigenvalues {\it and} eigenvectors of a linear operator.
At an EP, the operator becomes defective in the sense that the dimension of the space spanned by its eigenbasis is reduced~\cite{Kato:1995bm, 2011_Moiseyev_book, 2012_Heiss_JPA, 2017_Feng_nphoton, 2018_ElGanainy_nphys, 2019_Miri_Science, 2019_Ozdemir_nmat}.
This is in contrast to the degeneracy of a hermitian operator, the typical case encountered in physics, which is constrained by symmetry to have the eigenvectors span a space whose dimension is equal to the number of degenerate eigenvalues.
EPs exist in a parameter space of the relevant operator, as illustrated in \cref{fig:chpt0: ep-manifold}. %for example in the original ${\cal PT}$ paper the operator was the Hamiltonian, and the parameter was the exponent defining the shape of the complex polynomial potential.
A universal feature of EPs is that the manifold of the eigenvalues in the space of parameters is non-analytic, and in the simplest case of a two-fold degeneracy is that of a complex square root, depicted in \cref{fig:chpt0: ep-manifold}.
Because of this, near an EP a resonant system shows enhanced frequency splitting under small perturbations that may lead to improved sensing, albeit with some caveats~\cite{Chen:2017jk, 2017_Hodaei_Nat, 2018_Zhang_EP_PRL}.
The same feature also makes asymmetric state transfer possible~\cite{2016_Doppler_Nat, 2016_Xu_Nat} because the topology of the eigenvalue manifold leads to chiral behavior under cyclic variation of the parameters.
Another example: chiral lasing from a disk resonator can be traced to the existence of an EP whose sole eigenvector has a definite chirality; a similar example of chiral behavior at an EP will play an important role in \Cref{chp:chpt3: cpa ep}.
In this thesis we will focus exclusively on EPs with a two-fold degeneracy, which is the simplest case, though higher-order degeneracies are possible and exhibit interesting physics~\cite{2016_ding_prx, 2017_Achilleos_PRB, 2018_hoeller_arxiv}.

Two types of EPs have been extensively studied in physics: resonant and scattering.
In resonant EPs, which were the first to be studied, the linear operator that becomes defective is an effective wave operator, $\hat A_{\rm eff}$, describing an open system.
It is worth noting that in many studies, an approximate effective Hamiltonian \heff\ is introduced as the starting point instead; we will address this in \Cref{chp:chpt1: theory of scattering}.
%We will show that in most contexts in photonics it is correctly defined as an operator describing the resonant structure of the scattering matrix.
As the controlling parameters of $\hat A_{\rm eff}$ are swept through an EP, the scattering spectrum exhibits the merging of two resonance peaks that previously had distinct resonance frequencies but similar linewidths.
This is followed by a width-bifurcation, where the resonance lineshape becomes a superposition of two Lorentzians with different linewidths.
In unitary systems (e.g.,~no imaginary part of the refractive index), resonant EPs can only occur at complex frequencies below the real axis, and do not correspond to physical steady-state solutions, although they can still strongly influence the scattering properties for real frequencies~\cite{Peng:2014kl, Zhen:bl, Zhou:2018dy}.
By adding gain to an electromagnetic cavity one may bring the resonant EP to a real frequency, corresponding to exceptional lasing at threshold.
But an amplifying system is not ideal for the study of EPs, due to the large amplified spontaneous emission noise at threshold, and the necessity of accounting for the non-linearity of the medium, which stabilizes lasing above threshold.

In scattering EPs, the defective linear operator is the scattering matrix itself, $S(\w)$, which is the operator that connects the far-field scattered wavefront to the incident one.
Scattering EPs cannot occur for real frequencies in a system without gain or loss, but arise naturally in systems with joint parity and time-reversal (${\cal PT}$) symmetry, and have mainly been studied in that context.
In the ${\cal PT}$-symmetric case, the scattering eigenchannels make a transition from flux-conserving to amplifying or attenuating propagation at an EP~\cite{Chong:2011ev, 2012_Ge_PRA, Ambichl:2013gq}.
The eigenstates of $S(\w)$, which typically have both incoming and outgoing components, whether at an EP or not, are distinct from the resonances, which is to say $\hat A_{\rm eff}$ is not generally defective simultaneously with $S$.

The prototypical example from elementary physics of a system with an EP is an undriven damped harmonic oscillator, with a single degree of freedom, $x(t)$, obeying
\be
\label{eq:chpt0: damped harmonic oscillator}
    \ddot x(t) + 2 \g \dot x(t) + \w_0^2 x(t) = 0.
\ee
The angular frequency of the undamped oscillator is $\w_0$, and $\g$ is the damping rate.
As we will see, the EP corresponds to the parameter values at which two typically linearly-independent harmonic solutions become degenerate.
This system is known to have three kinds of behavior:
\begin{enumerate}

    \item $\g<\w_0$, called ``underdamped''.
    In this regime the system oscillates harmonically in time, with an exponentially decaying envelope.
    The time-dependence can be generally written as the sum of two terms, each of the exponential form $\cis[-]{(\pm \w - i\g)t}$, where $\w$ and $\g$ are real numbers, and $\w=\sqrt{\w_0^2-\g^2}$.

    \item $\g>\w_0$, called ``overdamped''.
    Here the system decays exponentially (more accurately as a sum of decaying exponentials), with no oscillation at all.
    As with the underdamped case, the time dependence can be written as the sum of two exponentials: $e^{-\g_\pm t}$, where $\g_\pm=\g\pm\sqrt{\g^2-\w_0^2}$ are two real decay rates.

    \item $\g=\w_0$, called ``critically damped''. At this one point in parameter space, which we will see is actually an EP, the system does not oscillate, but also does not decay as a sum of pure exponentials.
    Instead, one term is of the form $e^{-\g t}$, and the other is of the form $t e^{-\g t}$.
\end{enumerate}

Except for critical damping, the two solutions are each of the form $\cis[-]{\alpha t}$, for some complex frequency $\alpha$, and together they span the two-dimensional space of initial conditions $x(0)$, $\dot x(0)$.
If we constrain the solutions to have a complex exponential form, then at critical damping, when the solutions coalesce, we appear to lose one, though in fact it has simply moved out of the subspace that we are considering.
The remaining solution $x_2(t)$ can be constructed by $x_2(t) = \partial x_1(t)/\partial\w|_{\w=\g}$.
The spectral representation of the impulse response $h(t)$ of the system [satisfying $\hat L h(t) = \delta(t)$] is
\be
    h(\w) = \frac{-1}{\w_+-\w_-} \left( \frac{1}{\w-\w_+} - \frac{1}{\w-\w_-} \right),
\ee
where $\hat L \defn d^2/dt^2 + 2\g d/dt + \w_0^2$, and $\w_\pm \defn -i\g \pm \sqrt{\w_0^2-\g^2}$.
So long as $\w_\pm$ are not the same, i.e.,~away from critical damping, the response has two simple poles, and its lineshape is the sum of two distinct Lorentzians.
As critical damping is approached, $\w_+ \rightarrow \w_-$, and $h(\w)$ becomes (using L'Hopital's rule)
\be
    h_{\rm crit}(\w) = \frac{\partial h(\w)}{\partial \w_+} \bigg|_{\w_+ = \g} = \frac{-1}{(\w+i\g)^2},
\ee
which has a second-order pole and is a squared-Lorentzian.

This simple system exhibits all the essential properties of an EP.
As a set of parameters is varied, the eigenfrequencies come together in a square-root singularity.
At the point of degeneracy, a complex time-harmonic solution becomes modified by a factor which is linear in time, and is therefore lost to the subspace of such solutions.
Simultaneously, the response function associated with the operator modifies its spectral lineshape quadratically.
The same mathematical structure as this simple example applies to solutions of Maxwell's equations, where again the existence of EPs denotes the absence of two linearly-independent, time-harmonic solutions at particular parameter values, and the necessity of looking for an additional solution outside of this space.
We will see other examples of Maxwellian EPs in \Cref{chp:chpt2: rsms,chp:chpt3: cpa ep}, which are not degeneracies of scattering resonances.
These EPs will have interesting consequences for the spatial behavior of the system as well.
%these results generalize beyond the effective Hamiltonian that defines resonances, and have notable consequences when spatial degrees of freedom are included.

EPs and ordinary hermitian degeneracies differ in more ways than just the nature of eigenvalue splitting in their vicinity.
They are each characterized by a different number of parameters that must be tuned to engineer one.
For hermitian linear operators, the condition for degeneracy imposes one constraint: that the difference between the real eigenvalues vanishes.\footnote{Another consequence of hermiticity is that this condition cannot be met unless two different positive quantities independently vanish. The result is that a hermitian degeneracy still requires two-parameter tuning (see \cref{fig:chpt0: ep-manifold}c).}
EPs, which exist only for \nh\ operators, have two constraints: that the real {\it and imaginary} differences between eigenvalues vanish, which requires the tuning of at least two parameters in the generic case.
Any further ``design'' requirements on the EP, such as having the degenerate eigenvalue be real, will increase the number of necessary tuning parameters.
For example, an EP resonance which is also real (i.e.,~exceptional point laser) will generally require the simultaneous tuning of three parameters, and if we further require the degenerate frequency to have a prescribed value, then four parameters must be tuned.

The numerology of parameters changes when the operator possesses symmetries, and is therefore not generic.
If two states transform under different irreducible representations of the symmetry, then they are protected from coalescing at an EP, even with parameter tuning.
For example, a parity-symmetric, one-dimensional cavity has an even and an odd set of resonances, and no amount of tuning can bring two states of opposite parity (even/odd) to an EP, so long as the tuning respects the symmetry.
On the other hand, symmetries can also {\it reduce} the number of necessary tuning parameters, assuming that the symmetry is preserved under their variation.
For example, a system that is invariant under a simultaneous parity and time-reversal operation (${\cal PT}$) has eigenvalues that are either real or paired with another eigenvalue that is its complex conjugate.
The ${\cal PT}$ symmetry constraint reduces the required number of tuning parameters from two to one~\cite{2012_Heiss_JPA, 2017_Feng_nphoton, 2018_ElGanainy_nphys}, which will be relevant in \Cref{chp:chpt2: rsms} on reflectionless EPs.

    \subsubsection{${\cal PT}$ Symmetry and Exceptional Points}

In a pioneering study~\cite{1998_Bender_PRL}, Bender and Boettcher analyzed the behavior of a quantum-mechanical particle subject to a nonconservative Hamiltonian which does not satisfy the usual hermitian symmetry, but instead has ${\cal PT}$-symmetry.
They found that a portion --- sometimes all --- of its eigenstates are nevertheless associated with {\it real} eigenvalues, which is remarkable given the absence of hermitian symmetry.
Their original paper was understood to be relevant in optics by Christodoulides and coworkers~\cite{2005_Ruschhaupt_JPA, 2007_ElGanainy_OL, 2008_Klaiman_PRL, 2008_makris_PRL, 2008_Musslimani_PRL, 2009_Guo_PRL, 2010_Ruter_NatPhys}, and launched the very active field of \nh\ photonics, a major part of which focuses on ${\cal PT}$-symmetric optics.

%That an operator with ${\cal PT}$-symmetry can have a spectrum with an entirely real sector generalizes to optics, where \nh\ time-evolution means that flux is not conserved, and there is some amplification or attenuation of the field in the medium in which it propagates.

There is a special connection between EPs and ${\cal PT}$-symmetry, which is that the EP is a critical point in parameter space beyond which the system exhibits spontaneous symmetry-breaking.
A {\it system} is said to have ${\cal PT}$ symmetry if its governing equation is unchanged by the ${\cal PT}$ operator, so that the equation transforms under the trivial representation of ${\cal PT}$.
As with the damped harmonic oscillator \cref{eq:chpt0: damped harmonic oscillator}, a ${\cal PT}$-symmetric system also has three kinds of behavior: unbroken, broken, and exceptional.
In the unbroken phase, eigenvalues are real and separated by some finite distance, and the corresponding eigenvectors each map back to themselves under ${\cal PT}$, just like their defining linear operator.
In contrast, in the broken phase, the eigenvalues are complex and come as complex-conjugate pairs, so that if $\w$ is an eigenvalue then so is $\w^*$.
Each corresponding eigenstate transforms not into itself but into its conjugate partner under ${\cal PT}$, different from the linear operator.
Since symmetry prevents the eigenvalues from leaving the real axis unless paired with another eigenvalue which is its complex conjugate, any ${\cal PT}$-preserving perturbation which is not sufficiently large to close the gap between distant real eigenvalues cannot move them off of the real axis.
Therefore, there are two symmetry phases in parameter space, with the boundary between them being the EP.

    	\section{Saturable Two-Level Media \label{sec:chpt0: saturable two-level media}}

Up to this point we have treated \nh\ effects in {\it linear} absorbing or amplifying media.
Linear absorption or amplification is phenomenologically captured by a positive or negative imaginary part of the index of refraction, respectively.
However, this simple macroscopic prescription is not built on a microscopic model, which is needed for a realistic treatment of \nh\ media.
For many gain media, an appropriate microscopic model is an ensemble of two-level quantum systems, each consisting of a ground state and an excited metastable state.
We will refer to this as a {\it two-level} or {\it atomic medium}.
An external pump delivers energy to the system and partially populates the excited state, which can radiatively decay through stimulated emission, or through another channel which is unobserved and irreversible.
If the two-level medium is pumped to a non-equilibrium inverted state, i.e.,~with more ``atoms'' in the excited state than in the ground state, then stimulated emission tends to amplify any light propagating within the medium.
Conversely, in the absence of inversion, the two-level system is a net sink of photon flux.
Many gain media have substantially more complicated spectra, yet are well described by an effective two-level structure~\cite{ge_thesis_2010,cerjan_2015_oe}.

There are three fields needed to describe the two-level medium: the vector fields ${\bf E}$, ${\bf P}_{\rm in}$, and the scalar field $D$.
${\bf E}(\bx,t)$ is the electric field, which will be treated classically as explained previously.
The polarization density ${\bf P}_{\rm nl}(\bx,t)$ and the inversion density $D(\bx,t)$ describe the atomic medium, which must still be treated quantum mechanically.
${\bf P}_{\rm nl}(\bx,t)$ describes the dipolar response of the atoms, and the $D(\bx,t)$ is the average number of excited atoms less ground state atoms per unit volume.
The subscript ``${\rm nl}$", for ``nonlinear'', is appended to the atomic polarization density field to distinguish it from the that part of the material polarization which is not due to the gain atoms, which is assumed to be linear in the electric field ${\bf E}$, and is captured in the cavity dielectric function $\e_c(\bx)$.
The equations of motion of the combined light-matter system are the Maxwell-Bloch equations (MBE)~~\cite{haken_sauermann_1963,lamb_1964,haken_1985,tureci_2006}
\begin{gather}
    \nabla \times \nabla \times {\bf E} + c^{-2}\e_c \partial_t^2{\bf E} = c^{-2}\partial_t^2{\bf P}_{\rm nl} \label{eq:chpt0: maxwell}\\
    \partial_t {\bf P}^+_{\rm nl} = -i(\w_a - i\gp){\bf P}^+_{\rm nl} - \frac{i}{\hbar}D({\bf g}\cdot {\bf E}){\bf g}^* \label{eq:chpt0: polarization}\\
    \partial_t D = -\g_\parallel (D - D_0(\bx)) - \frac{4}{\hbar}\im{{\bf P}^+_{\rm nl}}\cdot{\bf E} \label{eq:chpt0: inversion}\\
    {\bf P}_{\rm nl}(t,\bx) = 2\re{{\bf P}^+_{\rm nl}(t,\bx)}. \label{eq:chpt0: polarization positive}
\end{gather}

\Cref{eq:chpt0: maxwell} is Maxwell's equation for the electric field in matter assuming that there are no free charges, driven by a fluctuating current source which is the oscillating atomic polarization.
The first term on the right in \cref{eq:chpt0: polarization} gives the free oscillation $\sim\cis[-]{\w_at}$ of the atomic polarization with a decaying envelope $e^{-\gp t}$, where $\gp$ is the polarization dephasing rate.
The second term is nonlinear in $D$ and ${\bf E}$, and modifies this precession rate of the polarization.
In \cref{eq:chpt0: inversion} for the inversion density field, the first term governs the relaxation at a rate $\g_\parallel$ of the field to some given spatially-varying value $D_0(\bx)$, which we refer to as the ``pump profile", while the second term causes the inversion to decrease or increase, depending on whether the polarization and the electric field are in- or out-of phase, respectively.
The fields ${\bf E}$, ${\bf P}_{\rm nl}$, and  $D$ are real, while the field ${\bf P}_{\rm nl}^+$ is the positive-frequency part of ${\bf P}_{\rm nl}$ and is complex.
\Cref{eq:chpt0: maxwell,eq:chpt0: polarization,eq:chpt0: inversion,eq:chpt0: polarization positive} are nonlinear dynamical equations for four real spatio-temporal fields, which without further constraint are generally chaotic~\cite{haken_1985,lugiato_1992,sunada_2005_pre,arecchi_1999}, and support a rich variety of behaviors.

Steady state {\it ab initio} laser theory (SALT) aims to describe the set of un-driven solutions which are multi-harmonic ($\sim\sum_\mu\cis[-]{\w_\mu t}$) at asymptotically long times, and are therefore relevant for the stable operation of CW lasers~\cite{tureci_2006, tureci_2007, tureci_2008, ge_2010, esterhazy_2014, cerjan_2015}.
The solutions are un-driven in the sense that there is no coherent input wave at or near the laser frequencies; however, there must be some incoherent pump to create the inverted population, which is included in $D_0(\bx)$ from the inversion \cref{eq:chpt0: inversion}.
Below, we will consider both multi-frequency (multimode lasing) and single-mode lasing.
For the single-mode case, the nonlinear term in \cref{eq:chpt0: inversion} is time-independent in the rotating wave approximation, and solutions will exist for which the inversion is also time-independent (stationary inversion).
For the multimode case, SALT assumes that the fast oscillations in the same nonlinear term self-average to a small value over the natural response time of the slow field $D$, which is $\g_\parallel^{-1}$, so that in steady-state, $D$ is constant in time:
\be
    \label{eq:chpt0: stationary inversion}
    \partial_t D(t,\bx) = 0.
\ee
\Cref{eq:chpt0: stationary inversion} is the stationary inversion approximation (SIA).
This approximation makes the multi-harmonic solutions of the MBE tractable, and reduces them to a set of coupled, nonlinear, frequency-domain equations for the electric field:
\begin{gather}
    {\bf E}(t,\bx) = \sum_{\mu=1}^{N}{\bf E}_\mu(\bx) \cis[-]{\w_\mu t} + {\rm c.c.}\\
    \label{eq:chpt0: salt}
    \nabla \times \nabla \times {\bf E}_\mu(\bx) - \left[ \e_c(\bx) + \chi_{\rm nl}(\w_\mu,\bx) \right]\left(\w_\mu/c\right)^2 {\bf E}_\mu(\bx) = 0 \\
    \chi_{\rm nl}(\w,\bx) \defn \frac{\g(\w) D_0(\bx)}{1+ \sum_\nu \G(\w_\nu)|{\bf E}_\nu(\bx)|^2},
\end{gather}
where we have rescaled the fields ${\bf E} \to (\hbar \sqrt{\gp \g_\parallel} /2g) {\bf E}$ and $D \to (\hbar \gp / g^2) D$~\cite{ge_2010}.
The ${\bf E}_\mu(\bx)$'s are the space-dependent lasing modes associated with frequencies $\w_\mu$.
The lorentzian gain-curve of the atomic medium is 
\begin{gather}
    \g(\w) \defn \frac{\gp}{\w-(\w_a-i\gp)},\qquad \G(\w) \defn |\g(\w)|^2.
\end{gather}
\Cref{eq:chpt0: salt} is $N$ coupled nonlinear equations in the ${\bf E}_\mu$'s, where $N$ is itself determined by maximizing the number of simultaneous solutions that  are mutually consistent with these equations.
The second term in the square brackets of \cref{eq:chpt0: salt}, $\chi_{\rm nl}$, is a {\it saturating nonlinearity}, with the distinctive feature that in regions where the intensity is high, the atomic medium is more transparent.
SALT has been shown to be an excellent approximation for the steady-state, multi-mode solutions of the MBE over its expected region of validity~\cite{ge_2008_oe, cerjan_2015_thesis, pick_2015_pra}.

\Cref{eq:chpt0: salt}, subject to the purely outgoing boundary conditions appropriate for lasing, is the core of SALT.
However, when it is subjected to more general scattering boundary conditions, with both incident and outgoing radiation, it instead describes the steady-state electromagnetic response of a cavity $\e_c$ with a saturating two-level susceptibility, whether amplifying or absorbing.
The stationary inversion condition for the validity of SALT is also satisfied by \cref{eq:chpt0: salt} for monochromatic scattering, since there is only one frequency component and therefore no beating terms that might spoil the SIA.
Therefore a SALT-like theory of scattering from a saturable medium is valid under rather general conditions, so long as $\g_\parallel$ is the slowest time scale in the system.

We will leave further discussion of saturable two-level media until \Cref{chp:chpt4: saturable cpa,chp:chpt5: saturable scattering}.
For now, the key takeaway is that \cref{eq:chpt0: salt}, which describes steady-state lasing in two-level media, can be extended to describe arbitrary scattering from such media, and therefore includes saturable absorption and amplification in addition to lasing.
To our knowledge, a SALT-based generalization of this type constitutes the first general framework for treating saturable amplifiers and absorbers for systems with arbitrary spatial complexity and quality factor.

	\section{Overview of This Thesis \label{sec:chpt0: thesis overview} }

In this thesis I address a series of new problems in non-hermitian optical scattering, with increasing degrees of complexity.
But first, in \Cref{chp:chpt1: theory of scattering}, I give a brief overview of the properties of linear scattering, developing the framework on which the rest of the thesis is built.
I will cursorily review the theory of resonances and their connection to scattering.
I will also present the effective Hamiltonian approach~\cite{1969_Mahaux_book, 1997_Beenakker_RMP, 2003_Viviescas_PRA, 2017_Rotter_RMP}, which will play an important role in \Cref{chp:chpt2: rsms,chp:chpt3: cpa ep}.
Finally, I will present several equivalent formulations of the scattering problem in terms of sources and boundary conditions that are used in various derivations and numerical calculations throughout.%, with special focus on the perfectly matched layer (PML) method.

In \Cref{chp:chpt2: rsms}, I develop the theory of reflectionless scattering modes (RSMs).
This novel work is an application of the effective-Hamiltonian approach to a broad class of scattering problems: for a given structure, find the incident wavefronts and frequencies which are not partially reflected at all, but are instead  transmitted through the scatterer, or dissipated within it.
The RSM framework includes critical coupling and CPA as specific instances, and in a sense threshold lasing too.
However, unlike with CPA or lasing, the generic RSM problem does not necessarily require intrinsic gain or loss to realize.
I analyze the symmetry properties of the RSMs, and find that they support exceptional points which can be directly probed, and are distinct from the often-studied resonant EPs.

In \Cref{chp:chpt3: cpa ep}, I discuss degenerate coherent perfect absorption (CPA EP), which is a specific example of the new kind of exceptional point introduced in \Cref{chp:chpt2: rsms}.
Here, two perfectly absorbed states are brought together by tuning parameters of the scatterer.
In the case of a ring-resonator, it is known that resonant EPs take on a chiral flavor in that the exceptional state can be made mostly clockwise or mostly counterclockwise.
I show that the same holds true for the CPA EP, and use this fact to design a patterned ring resonator which either predominantly absorbs or reflects light, depending on the direction of incidence.
I also discuss the important distinction between an exceptional point of the underlying CPA wave operator and the scattering matrix.

\Cref{chp:chpt4: saturable cpa} generalizes CPA to include the saturating nonlinearity and dispersion of a two-level absorbing medium, which for some materials is a more realistic model of an absorber than a constant imaginary part of the refractive index.
%For this absorbing case, the two-level medium has a negative pump parameter $D_0<0$.
While the problem of treating CPA with a saturable two-level system was partially addressed in Longhi~\cite{longhi_2011_pra}, that work assumes a simple ring-laser geometry, and incorrectly identifies the behavior in the bad-cavity limit, when the rate at which energy leaks out of the cavity is comparable to the dephasing rate of the medium, $\gp$.
I extend saturable CPA to arbitrary geometry by connecting it to lasing through a version of the time-reversal argument used for linear CPA, but augmented to account for both saturation and dispersion.
Moreover, I show that the SALT algorithm in the single-mode regime can also be used to find the saturable CPA modes through a simple mapping, which demonstrates that between a lower and upper threshold for loss one can maintain CPA by continuously adjusting the pump strength.
I also clarify the bad-cavity limit of dispersive CPA, showing that the modes are hybrids of the cavity and atomic degrees of freedom, with a strongly dispersive response to changes in the pump.

Finally, in \Cref{chp:chpt5: saturable scattering}, I will present and solve the general problem of scattering from a saturable two-level medium.
Unlike with lasing or CPA, these solutions have both incoming and outgoing flux, and unlike the linear case described in \Cref{chp:chpt1: theory of scattering}, they must be solved for self-consistently.
Nevertheless, solutions have been previously found for the simplest configurations, such as ring resonators that support purely traveling waves.
Here I address the saturable scattering problem for arbitrary geometry, and show that the results known from ring-resonators about the bistability of solutions and the nature of the transition between mono- and bistability hold even for arbitrary geometry.
I also carefully analyze the validity of approximations used for isolated resonances, and find that the previously used single-pole approximation~\cite{ge_2010} requires some modification for the case of scattering.

The results presented in Chapters \ref{chp:chpt2: rsms}~--~\ref{chp:chpt5: saturable scattering} are my original contributions, and are also presented in Refs.~\cite{sweeney_rsm_2019} and~\cite{2019_Sweeney_PRL}.
I have also contributed to Ref.~\cite{2018_Zhang_EP_PRL} on the theory of quantum noise in a narrow-band amplifying EP sensor, and Ref.~\cite{wang_2020} on a novel, polarization-sensitive unidirectional transparency in indirectly coupled optical resonators, but these lie outside the scope of this thesis.

\chapter{Theory of Scattering \label{chp:chpt1: theory of scattering}}
The theory of scattering aims to describe the outcome of sending a specified waveform towards an object, letting it interact, and observing what has come back out a long time later.
Scattering is relevant for any open wave system, ranging from ocean acoustics to elementary particles, including classical electromagnetism, which is the focus of this thesis.
The most general theory of scattering is cumbersome and often intractable, since it must accommodate a very broad set of phenomena, such as long-range, nonlinear interactions and time-varying potentials.
The literature on the topic is consequently dense and often irrelevant for classical electromagnetism.
Our goal in this chapter is to summarize in one place those aspects of the theory that are important for much of classical electromagnetic scattering from a restricted class of dielectric objects.
These are dielectric scatterers that are finite, time-invariant, and local in the spatial degrees of freedom.
We will also require that the scatterers are linear in the electric field amplitude in this chapter and in \Cref{chp:chpt3: cpa ep}; we will address nonlinear susceptibilities in \Cref{chp:chpt4: saturable cpa,chp:chpt5: saturable scattering}.

In much of the literature on scattering, the input waveform is taken to be a plane wave, which is a valid starting point for scatterers much smaller than the wavelength of the probe beam.
However, in this thesis it is crucial that we consider arbitrarily-shaped wavefronts, as most of our focus is on the resonant response of scatterers or structures larger than the wavelength of light.

We will formalize the scattering problem as an initial value problem, in terms of the time-evolution of an initial wave packet into a final scattered state at asymptotically late times.
By passing from the time-domain to the frequency-domain, the initial value problem becomes a boundary value problem (BVP), either with scattering boundary conditions, or with purely outgoing boundary conditions and a source term.
For the TM polarization of light in one- and two-dimensional scatterers, the vectorial Maxwell equations in the frequency domain, generally relevant for all of classical electromagnetism, reduce to the scalar Helmholtz equation.  % for one component of the electric field, which describes  
This greatly simplifies the analyses in this chapter and in the remainder of the thesis.
Nevertheless, the fundamental concepts in this work, such as coherent perfect absorption, reflectionless scattering modes, and exceptional point physics, will also apply to the full vectorial Maxwell solutions, once the polarization degrees of freedom are taken into account.

\begin{figure}
    \centering
    \includegraphics[width=\textwidth]{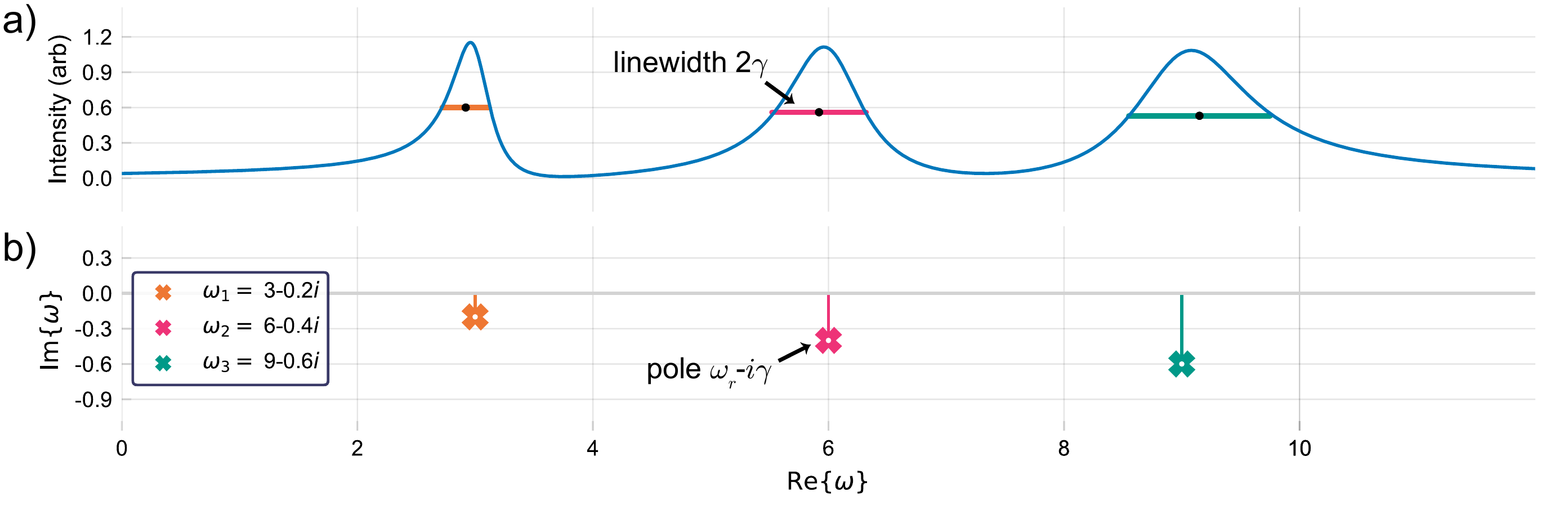}
    \caption[Representation of the relationship between resonant features of $S(\w)$ and its complex-valued poles]{Representation of the relationship between resonant features of $S(\w)$ and its complex-valued poles.
    {\bf(a)}:~Hypothetical intracavity intensity as a function of frequency $\w$.
    It is generated from the sum of three Lorentzians with unit amplitude and parameters given in the legend of (b), it is not the result of an actual scattering calculation.
    Horizontal lines represent FWHM linewidth, bisected by black dot, for visual comparison with the imaginary parts shown in (b).
    {\bf(b)}:~Locations of the complex-valued poles of $S(\w)$.
    Were this an actual calculation, there would be a quasi-normal mode associated with each.
    In this schematic we are imagining fairly isolated resonances with approximately Lorentzian lineshapes.
    In that case, the central frequency of a resonant peak is $\re{\w_n}$, and the FWHM is $2\im{\w_n}$, for some $n$.
    However, when multiple resonances overlap and interfere, the lineshapes can be more complicated, featuring both peaks and dips (Fano lineshapes), and the real and imaginary parts of the $\w_n$'s will not be obviously visible in the scattering lineshapes.
    }
    \label{fig:chpt1: s-resonance schematic}
\end{figure}

The scattering matrix $S(\w)$ ($S$-matrix) describes the deterministic and linear relationship between the incident and outgoing wave amplitudes.
Near a certain {\it resonant} frequency, $\w_0$, $S(\w)$ varies rapidly, large field amplitudes are attained within the scatterer, and the interaction time grows long in the time-domain picture of scattering.
A resonance can be characterized by $\w_0$ and its FWHM\footnote{
    The full width at half max (FWHM) is the spectral width of a resonant feature, such as the intracavity intensity, measured at half of its peak value.
}  {\it linewidth}, $2\g_0$, as shown in \cref{fig:chpt1: s-resonance schematic}a.
For isolated resonances, these two features are often combined into a single figure of merit $Q_0$, which is a standard measure of how well a cavity traps light at $\w_0$:
\begin{equation}
    \label{eq:chpt1: definition of Q}
    Q_0 \defn \w_0/2\g_0.
\end{equation}
Throughout this thesis we will assume approximately steady-state experiments for which $\w$ may only take on real values.
However, complex-valued frequencies play a central role in the analysis of scattering and resonance effects.

There is a deep connection, given by the effective Hamiltonian formalism~\cite{1969_Mahaux_book,1997_Beenakker_RMP,2003_Viviescas_PRA,2017_Rotter_RMP}, between the frequencies and linewidths of resonances and the {\it quasi-normal modes}, also called Gamow states, of the wave operator with purely outgoing boundary conditions.~\cite{1928_Gamow_ZFP,1981_Bohm_JMP,1998_Ching_RMP,2018_Lalanne_LPR}
The quasi-normal modes of open systems, which, as noted above, are generally associated with complex eigenfrequencies $\{\w_n\}$, are the analog of normal modes in energy-conserving, closed systems.
We will see that the quasi-normal mode eigenfrequencies are the poles of the analytically continued $S$-matrix (see \cref{fig:chpt1: s-resonance schematic}b), which is simply $S(\w)$ with $\w$ promoted from a real to a complex variable.
In simple cases, the real part of $\w_n$ is the center frequency of a resonant peak as seen in scattering, and the width of the peak is given by twice the imaginary part of $\w_n$.
Because of this close connection, we will often refer to both the real-frequency $S$-matrix phenomenon and the quasi-normal eigenmodes and complex eigenfrequencies as ``resonances'', but only when the context makes it clear which one we are talking about; when more clarity is needed we will call the complex eigenfrequencies {\it $S$-matrix poles}.
In general, the quasi-normal modes are not physically realizable steady-states due to their exponential growth at infinity.
However in electromagnetic scattering with gain, a quasi-normal mode {\it can} be realized physically, which corresponds to the onset of laser emission~\cite{1973_Lang_PRA, esterhazy_2014, tureci_2006, tureci_2008, ge_2010}.

The effective Hamiltonian formalism relates two electromagnetic problems: the scattering problem, where the incident wave is specified at any frequency, and the outgoing wave can always be found, and a source-free problem subject to purely outgoing boundary conditions.
The latter is overdetermined, so that solutions can only be found at certain complex frequencies.
We will refer to such overdetermined scattering problems as {\it electromagnetic eigenvalue problems}, where the complex frequencies are the eigenvalues.
The connection between these eigenproblems and conventional scattering is central to understanding reflectionless scattering modes (RSMs) and coherent perfect absorption (CPA), which are prominent in \Cref{chp:chpt2: rsms,chp:chpt3: cpa ep,chp:chpt4: saturable cpa}.
Because of this, we will spend some time describing the implementation of the boundary conditions and sources needed for solving both the scattering and eigenvalue problems.

We will also describe temporal coupled-mode theory (TCMT)~\cite{Haus_book, 2003_Fan_JOSAA, 2004_Suh_JQE, 2018_Wang_OL, 2019_Zhao_PRA, 2017_Alpeggiani_PRX}, formally similar to the effective Hamiltonian method, which gives an approximate connection between the $S$-matrix and the eigenfrequencies and modes of a \nh\ Hamiltonian.
TCMT is sufficiently accurate for many photonics applications, particularly those involving high-\Q\ resonances.
It is a standard tool of photonics because it is simple, yet widely applicable, and is often used for designing and analyzing optical devices~\cite{2012_Verslegers_PRL, Peng:2014kl, 2014_Hsu_NL, Zhen:bl}, as we do in \Cref{chp:chpt2: rsms,chp:chpt3: cpa ep}.
However, we prefer the effective Hamiltonian formalism when we need to rigorously justify statements about scattering, such as the existence of RSMs and their symmetry properties.

    \section{Preliminaries \label{sec:chpt1: general description}}

\begin{figure}
    \centering
    \includegraphics[width=\textwidth]{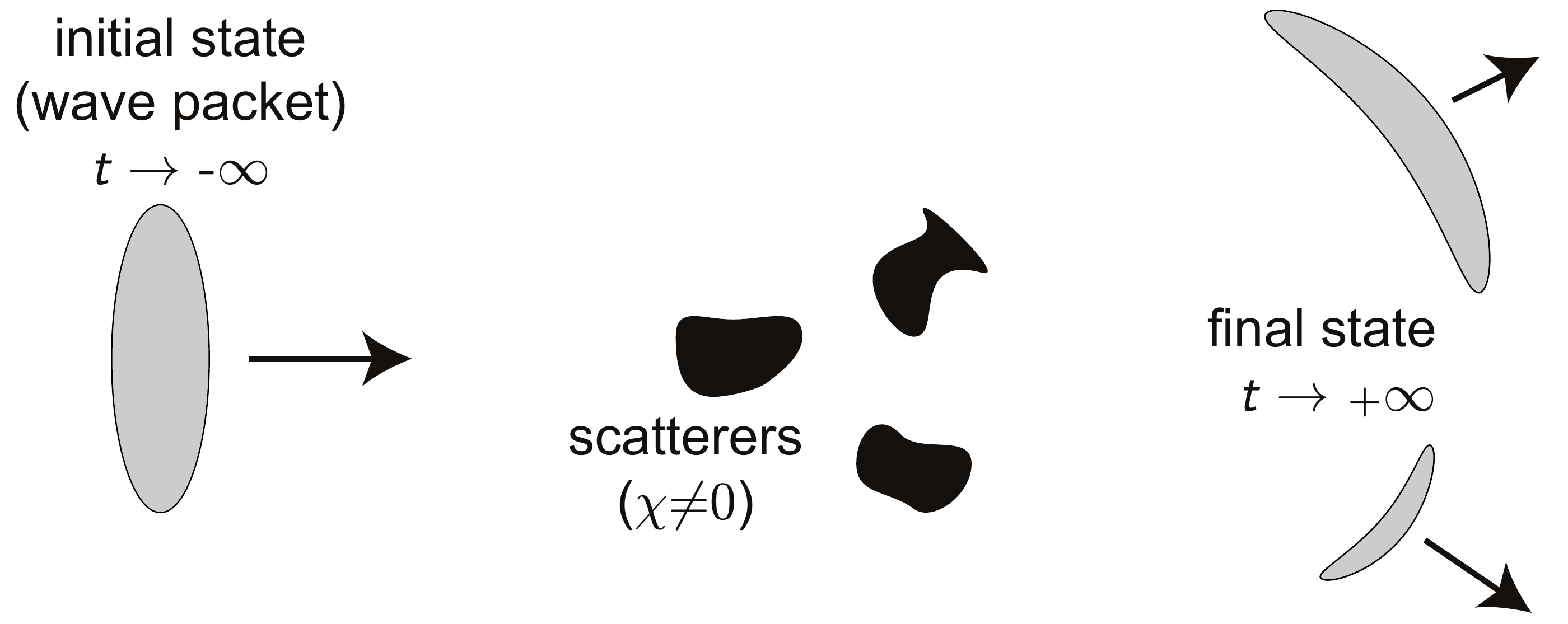}
    \caption[Schematic of a general scattering process]{Schematic description of the scattering problem, read from left-to-right.
    An initial wave packet is specified far in the past.
    It propagates forward in time and inward towards the scatterer.
    A long time later, after interacting with the scatterer, the wave can be observed propagating outwards.
    The goal of scattering theory is to predict the final form of the wave from its initial state, given some arbitrary scattering region, which is either a region with non-vanishing susceptibility ($\chi\neq0$), or, in the case of metallic waveguide, a boundary deformation.
    }
    \label{fig:chpt1: ivp schematic}
\end{figure}

In a wave scattering problem, an initial waveform is specified far away and in the distant past.
Over time, it propagates inward, interacts with some object ({\it the scatterer}), and a long time later propagates away from the scatterer and travels a great distance before it is detected, as depicted in \cref{fig:chpt1: ivp schematic}.
The formal solution of the scattering problem uses the wave operator, including all the effects due to the scatterer, to time-evolve the initial state from the distant past to the distant future.
In the absence of the scatterer, the initial wave freely propagates for all times; a scatterer disturbs the free propagation and changes the spatiotemporal distribution of the final state.
If the scattering is weak, then the final state has a high overlap with the hypothetical case of no scatterer at all.
The opposite limit of strong scattering has a final state which is very different from the result of free propagation, for example the angular distribution of a scattered wave might be very large compared to that of the incident wave.

A monochromatic wave (CW) with frequency $\w$, which is relevant for steady-state excitation, is not localized in time, and therefore does not easily fit into the scattering paradigm just described.
Instead it is useful to think of an initial {\it wave packet}, which is a sinusoidal wave whose amplitude-modulation is slow compared to its frequency $\w$.
The wave packet contains a great many oscillations and can be roughly localized in both the frequency and time domains.
When the packet impinges on the scatterer, it will spend some time partially confined within it, while also leaking out as radiation.
The leakage rate defines an interaction time $\tau(\w)$ for the scattering process.
A striking phenomenon emerges when $\w$ is near a scatterer-specific resonant frequency $\w_0$: a dramatic increase of the scattering strength, accompanied by a long interaction time and large peak amplitude within the scatterer.
The linewidth of the resonance, $2\g_0$, is related to the interaction time by $\g_0=1/2\tau(\w_0)$.
When $\w$ is within an isolated resonance linewidth, the field amplitude within the scatterer behaves like a driven damped harmonic oscillator with quality factor $Q_0=\w_0/2\g_0$, with the wave packet acting as the external drive.
Resonances do not occur when the frequency $\w$ is too small, such that the wavelength within the scatterer is much larger than its characteristic size.
We will assume in all cases discussed in this thesis that we are probing the scattering region at a frequency high enough for the system to support resonances.

The scattering and resonance problems that we consider in this work are monochromatic, that is, they do not mix frequencies.
In some contexts, scatterers can change the energy of the photons and shift their frequency away from $\w$, as happens with Raman and Brillouin scattering.
We only consider scatterers which absorb and emit photons with the same energy as the incident photons, although we do not require that photon flux, and hence energy flux, is conserved.
This also excludes cases where the dielectric function has explicit time-dependence, such as happens when the refractive index is dynamically-modulated through, say, an electro-optic effect~\cite{1964_Yariv_IEEE, 2004_Almeida_nat, 2008_Dong_PRL, 2009_Yu_natph}, i.e.,
\begin{equation}
    \e(\bx,\bx\pr;t,t\pr) \equiv \e(\bx,\bx\pr;t-t\pr).
\end{equation}
Such dielectrics are called {\it time-invariant}.

We also limit our analysis to spatially local susceptibilities.
Nonlocal susceptibilities are usually not relevant in electromagnetic scattering from dielectrics, though they are known to be important in certain nano-scale structures~\cite{2008_Garcia_JPC, 2010_McMahon_PRB, 2011_Raza_PRB, 2020_Gubbin_PRX, 2020_Yang_PRB}.
The scattering problem is greatly simplified for {\it local} dielectric responses:
\begin{equation}
    \e(\bx,\bx\pr;t-t\pr) \equiv \delta(\bx-\bx\pr)\e(\bx,t-t\pr).
\end{equation}

Initially, we also neglect any nonlinear field interactions.
These effects will be discussed extensively in \Cref{chp:chpt4: saturable cpa,chp:chpt5: saturable scattering}, where we address the local interaction mediated by the saturation of a two-level medium.

By requiring that the dielectric is linear and time-invariant, the scattering of a wave packet can be analyzed into the scattering of each of its frequency components, from which the final outcome can be synthesized.
Because of this, we never need to consider an explicit form for the initial wave-packet; we can work in the idealized limit of a monochromatic initial state $\psi^{\rm in}(\bx,\w) = \delta(\w-\w_0)\psi^{\rm in}(\bx)$.
Nevertheless, thinking in terms of temporally-localized packets is very helpful in clarifying ambiguities that arise when analyzing the scattering problem in the frequency domain, where past and future are not meaningfully distinct.

    \subsection{From Maxwell to Helmholtz \label{sec:chpt1: maxwell to helmholtz}}

In this section we will show that a significant simplification can be made to Maxwell's equations for one polarization in one- and two-dimensional structures.

The equations governing electromagnetism are Maxwell's equation (given in the SI convention, see, e.g., Ref~\cite[818]{1975_Jackson_book}):
\begin{gather}
    \label{eq:chpt1: maxwell 1}
    \nabla \cdot {\bf D} = \rho_{\rm free} \\
    \nabla \cdot {\bf B} = 0 \\
    \nabla \times {\bf E}  = -\partial_t {\bf B} \\
    \nabla \times {\bf H} = {\bf J}_{\rm free} + \partial_t {\bf D}.
\end{gather}
In this thesis, we are only treating scattering from neutral dielectrics with no free charge or current densities [$\rho_{\rm free}(\bx) \equiv 0$; ${\bf J}_{\rm free}(\bx)\equiv{\bf 0}$].
Also, as discussed before, the dielectric materials are time-invariant and local, so that 
\begin{equation}
    {\bf D}(\bx,t) = \e_0 \int dt\pr\,\e_r(\bx,t-t\pr) {\bf E}(\bx,t\pr).
\end{equation}
$\e_0$ is the free-space permittivity, and $\e_r$ is the relative permittivity of the scatterer.
Meanwhile the magnetic permeability is spatially uniform, taking the free-space value $\mu_0$.

With these constraints, the last two Maxwell equations in the frequency domain are
\begin{gather}
%    \nabla \cdot {\bs(}\e_{\rm r}(\bx,\w) {\bf E}{\bs)} = 0 \\
%    \nabla \cdot {\bf B} = 0 \\
        \label{eq:chpt1: maxwell 2}
    \nabla \times {\bf E}  = i \w {\bf B} \\
        \label{eq:chpt1: maxwell 3}
    \nabla \times {\bf B} = -i \w c^{-2} \e_r(\bx,\w) {\bf E}.
\end{gather}
For the sake of conciseness we will take $c = 1$ for the rest of this chapter, and drop the subscript from the relative permittivity.
Plugging \cref{eq:chpt1: maxwell 3} into the curl of \cref{eq:chpt1: maxwell 2} gives a second-order partial differential equation (PDE) for ${\bf E}$ alone:
\begin{equation}
    \nabla \times \nabla \times {\bf E} - \w^2 \e(\bx,\w) {\bf E} = {\bf 0}.
\end{equation}
This can be written in terms of the components of ${\bf E}$:
\begin{equation}
    \label{eq:chpt1: matrix maxwell}
    \begin{pmatrix}
        -\partial_y^2 - \partial_z^2 - \w^2 \e & \partial_x\partial_y & \partial_x\partial_z \\
        \partial_x\partial_y &  -\partial_x^2 - \partial_z^2 - \w^2 \e & \partial_y\partial_z \\
        \partial_x\partial_z & \partial_y\partial_z &  -\partial_x^2 - \partial_y^2 - \w^2 \e \\
    \end{pmatrix}
    \begin{pmatrix}
    E_x \\ 
    E_y \\
    E_z
    \end{pmatrix} = {\bf 0}.
\end{equation}
We have assumed that $\e$ is isotropic, i.e.,~rotationally invariant, so that the dielectric tensor is proportional to the identity.
In this thesis we will only consider one- or two-dimensional structures, for the sake of simplicity.
These can be modeled as three-dimensional structures that have a translation invariance in one direction, say, along $z$.
Therefore ${\bf E}(x,y,z) = {\bf E}(x,y)\cis[-]{k_z z}$, and $\partial_z \to -ik_z$.
For in-plane illumination we have $k_z \equiv 0$, so that \cref{eq:chpt1: matrix maxwell} reduces to
\begin{equation}
    \begin{pmatrix}
        -\partial_y^2 - \w^2 \e & \partial_x \partial_y & 0 \\
        \partial_x\partial_y &  -\partial_x^2 - \w^2 \e & 0 \\
        0 & 0 &  -\partial_x^2 - \partial_y^2 - \w^2 \e \\
    \end{pmatrix}
    \begin{pmatrix}
    E_x(x,y) \\ 
    E_y(x,y) \\
    E_z(x,y)
    \end{pmatrix} = {\bf 0}.
\end{equation}
In this form it is obvious that, for one polarization, $E_x=E_y\equiv0$, since the $E_z$-component completely decouples from them.
This polarization is transverse-magnetic (TM) since, by \cref{eq:chpt1: maxwell 2}, ${\bf B}$ must be transverse to the one remaining field component, i.e., ${\bf B}$ must be in the transverse $x-y$ plane.

In the rest of the thesis we will consider only the in-plane propagation of the TM polarization for the planar geometry just described.
The remaining field component $E_z$ satisfies a scalar PDE
\begin{equation}
    \label{eq:chpt1: helmholtz}
    \{\nabla^2 + \w^2 \e(\w,\bx)\}E_z(\bx) = 0,
\end{equation}
which is the Helmholtz equation.
With only one field component to keep track of from now on, there is little benefit to calling it $E_z$, and instead we will often use a generic field variable such as $\psi$ or $\phi$.

A one-dimensional electromagnetic structure can be similarly reduced to a scalar Helmholtz equation, by considering it as the $k_y=k_z=0$ limit of a structure with both $y$ and $z$-invariance.
%In both one-dimensional and two-dimensional cases we will write the Helmholtz equation as it appears in \cref{eq:chpt1: helmholtz}.

There are other ways to derive the Helmholtz equation for a TM-like polarization of Maxwell's equations~\cite{2005_prog_in_optics}, though this is perhaps the simplest.

    \subsection{Scattering as a Boundary Value Problem \label{sec:chpt1: scattering as BVP} }

In this section we will show that the initial-value formulation of scattering in the time-domain, as described in \cref{sec:chpt1: general description}, is equivalent to an inhomogeneous boundary-value problem (BVP) in the frequency-domain.
In both the time- and frequency-domains, the scattering problem is not overdetermined, and a solution exists for any choice of incoming wavefront and frequency.

The particular scattering problem that is a major focus of this thesis is
\begin{equation}
    \label{eq:chpt1: scattering pde}
    \{\nabla^2 + \w^2 \e(\bx,\w)\}\psi(\bx) = 0,
\end{equation}
where 
\begin{equation}
    \label{eq:chpt1: scattering bc}
    \psi(|\bx|\to \infty) \sim \psi^{\rm in}(\bx) + \psi^{\rm out}(\bx).
\end{equation}
\Cref{eq:chpt1: scattering bc} is a {\it scattering boundary condition}, and it specifies the asymptotic behavior of the field $\psi$ far away from the scatterer.
The incoming part of the boundary condition $\psi^{\rm in}$ is specified {\it a priori}; $\psi^{\rm out}$ is purely outgoing and must be solved for.
The in/out decomposition of \cref{eq:chpt1: scattering bc} is not well-defined for an arbitrary function $f$; we will justify its validity for the scattering solution $\psi$ in the next section. %but follows from $\psi$ being a solution to the Helmholtz equation in free space asymptotically far from the scatterer.
%For now we will rely on the intuitive notion that we can unambiguously decompose $\psi(\bx)$ into an incoming and outgoing component sufficiently far from the cavity.

\Cref{eq:chpt1: scattering pde} is an elliptic PDE~\cite{1970_Mathews_mm}.
As such, its solution is uniquely determined within some simply-connected domain $\Omega$ by specifying the boundary conditions on $\partial\Omega$, which must be of the form
\begin{equation}
    \label{eq:chpt1: general boundary conditions}
    \left[\ \psi(\bx) + \oint_{\partial\Omega} \ \beta(\bx,\bx\pr) \nabla \psi(\bx\pr) \cdot d{\bf S}\pr\ \right]_{\bx \in \partial\Omega} = 0,
\end{equation}
where $\partial\Omega$ is the boundary of $\Omega$.\footnote{
    In principle, the RHS of \cref{eq:chpt1: general boundary conditions} can be an arbitrary function of position, but this will never be the case in this work.
}
The vectorial area element $d{\bf S}\pr$ is normal to the surface $\partial\Omega$.

The scattering boundary condition \cref{eq:chpt1: scattering bc} is not apparently of this form, but a simple field transformation fixes this.
The Helmholtz equation for 
\begin{equation}
    \label{eq:chpt1: definition of phi}
    \phi(\bx) \defn \psi(\bx) - \psi^{\rm in}(\bx),\quad \forall \bx
\end{equation}
is
\begin{gather}
    \label{eq:chpt1: scattering pde inhomogeneous}
    \{\nabla^2 + \w^2 \e(\bx,\w)\}\phi(\bx) = j(\bx), \\
    \label{eq:chpt1: scattering source init}
    j(\bx) \defn -\{\nabla^2 + \w^2\e(\bx,\w)\}\psi^{\rm in}(\bx).
\end{gather}
By definition, $\lim_{|\bx|\to\infty}\phi(\bx)\sim\psi^{\rm out}(\bx)$, though within the scattering region $\phi(\bx) \neq \psi^{\rm out}(\bx)$. 
%Note that $\phi$ is not specified in the asymptotic region only, so that $\phi(\bx) \neq \psi^{\rm out}(\bx)$, though $\lim_{|\bx|\to\infty}\phi(\bx)\sim\psi^{\rm out}(\bx)$.
\Cref{eq:chpt1: scattering pde inhomogeneous} for the transformed field differs from \cref{eq:chpt1: scattering pde} for the original field $\psi$ by having a source $j$ on its RHS. 
%The inhomogeneous RHS $j(\bx)$ is known because $\psi^{\rm in}$ was specified as part of the definition of the scattering problem.
The source [\cref{eq:chpt1: scattering source init}] can be simplified because, as we will see in the next section, $\psi^{\rm in}(\bx)$ satisfies the free-space Helmholtz equation, with $\e(\bx) \equiv 1$.
Then \cref{eq:chpt1: scattering source init} simplifies to
\begin{gather}
    \label{eq:chpt1: scattering source}
    j(\bx) = - \w^2 \chi(\bx,\w) \psi^{\rm in}(\bx), \\
    \chi(\bx,\w) \defn \e(\bx,\w) - 1.
\end{gather}
%The function $\chi(\bx,\w)$ is the usual relative dielectric susceptibility of electromagnetism.
All the scatterers that we will consider are finite in size, so that $\chi(\bx,\w)$ has compact support and vanishes everywhere outside some finite region, as does $j(\bx)$.

The boundary condition for the transformed field $\phi$ is
\begin{equation}
    \label{eq:chpt1: scattering bc outgoing}
    \phi(|\bx|\to \infty) \sim \psi^{\rm out}(\bx).
\end{equation}
We will see later in \cref{sec:chpt2: boundary matching} that the incoming and outgoing asymptotic fields $\psi^{\rm in/out}$ generally satisfy a boundary condition relating the field to its derivative at points on the boundary, exactly in the form of \cref{eq:chpt1: general boundary conditions}.
Therefore the transformed field $\phi$ satisfies the inhomogeneous Helmholtz equation \ref{eq:chpt1: scattering pde inhomogeneous}, with a boundary condition consistent with the usual elliptic PDE boundary value problem (BVP).

We emphasize again that scattering solutions exist for any real-valued $\w$.
This will be contrasted later with related eigenvalue problems that only have solutions at discrete, complex-valued $\w$.
Mathematically, the existence is verified in the time-domain by providing an explicit construction of the final scattered state: the initial solution propagated forward in time from $t=-\infty$ to $t=+\infty$.
In this section we have mapped the initial-value problem in the time-domain to an inhomogeneous BVP in the frequency-domain, which is also known to always have a solution~\cite{1970_Mathews_mm}.
The eigenvalue problems that we will encounter later amount to solving \cref{eq:chpt1: scattering pde inhomogeneous} with a vanishing RHS.

There are two major components to finding a scattering solution which have become evident in the course of this analysis.
The first is the current source on the RHS of \cref{eq:chpt1: scattering pde inhomogeneous}, and the second is the boundary condition \cref{eq:chpt1: general boundary conditions}.
There are multiple ways to construct each of these; we will discuss them later in this chapter, in \cref{sec:chpt2: boundary conditions,sec:chpt2: equivalent sources}.

    \subsection{Last Scattering Surface (LSS), Asymptotic Regions \label{sec:chpt1: LSS and asymptotics}}

The geometry of the scattering problem consists of a finite scattering region, outside of which are asymptotic regions that extend to infinity.
Any surface which has no scattering defects in its exterior, so that the waves propagate freely there, is a {\it last scattering surface} (LSS).
It is generally convenient to choose a LSS such that entirety of the scatterer is contained within the volume that it encloses, and no scattering happens on the LSS itself.
For example, a convenient LSS for a two-dimensional scatterer in free-space is a sufficiently large circle, regardless of the shape or material composition of the scatterer within it.
%The LSS so defined is fictitious, and the final result of the scattering analysis will not depend on it, nevertheless it is a useful construction in what follows.

Within the LSS, a linear and time-invariant photonic structure is described by its dielectric function $\e(\bx,\w)$.
In \Cref{chp:chpt4: saturable cpa,chp:chpt5: saturable scattering} we will relax the linearity condition, as a consequence the scattering solutions will not necessarily be unique there.
If $\e$ is complex-valued, then its imaginary part describes absorption and/or gain from matter reservoirs connected to the scattering system; otherwise, it is energy-flux conserving.

\begin{figure}
    \centering
    \includegraphics[width=\textwidth]{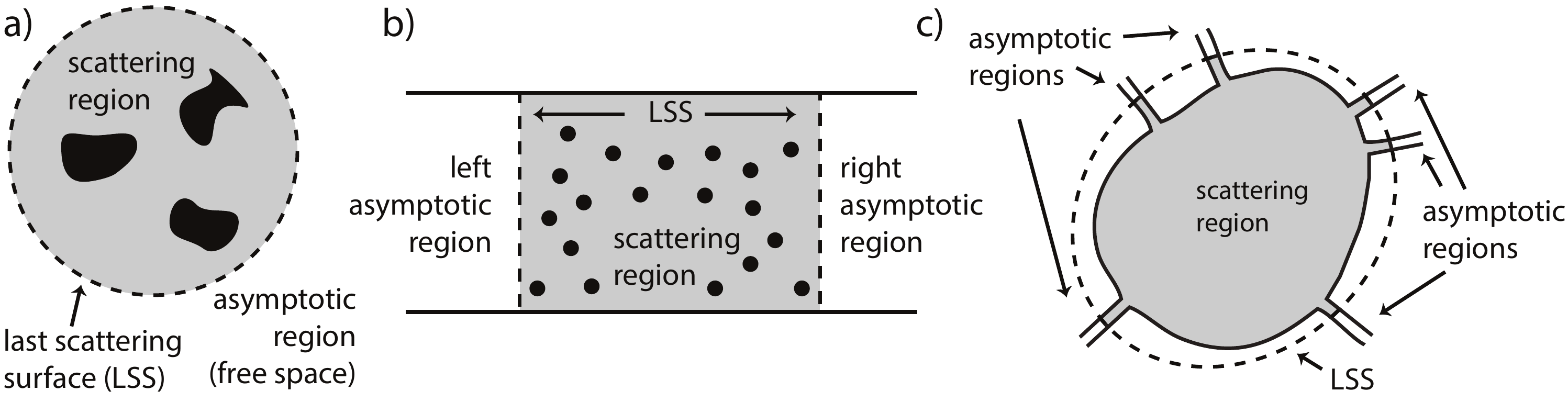}
    \caption[Examples of asymptotic regions and last scattering surface in different geometries]{Examples of scattering geometries: all have an interior scattering region and exterior asymptotic region, separated by a last scattering surface (LSS).
    {\bf(a)}:~Two-dimensional free-space geometry, where the LSS is a circle containing all of the scatterers.
    There is a single asymptotic region.
    {\bf(b)}:~Two-dimensional strip waveguide geometry.
    In this case there are two asymptotic regions, one to the left, another to the right.
    This reduces to the one-dimensional scattering case in two different limits.
    First, when the transverse width of the waveguide goes to infinity and supports plane-wave propagation at any incident angle (slab-geometry), then scattering at normal-incidence ($k_y=0$) is one-dimensional.
    Second, when the transverse width of the waveguide is sufficiently small to support only single-mode propagation, then the one propagating mode can be effectively described as one-dimensional.
    {\bf(c)}:~A multi-waveguide chaotic junction, similar to \vref{fig:chpt2: octopus}.
    The LSS extends a little into each waveguide.
    There are six asymptotic regions, one per waveguide.
    In principle, each waveguide can support either a single mode or multiple modes, depending on the input frequency.
    }
    \label{fig:chpt1: lss}
\end{figure}

Outside the LSS are the asymptotic regions, which can take many forms.
\Cref{fig:chpt1: lss} summarizes the geometry for a few different cases.
We assume that they are time-reversal invariant, and also have some form of translational invariance.
Examples include the vacuum or a uniform dielectric, a finite set of disjoint metallic waveguides, and an infinite photonic crystal waveguide.
This last has a discrete translational symmetry, while all the others have a continuous symmetry.
There is a definite ``inward'' direction toward the LSS for each distinct region.
The wavefunctions within each asymptotic region that propagate towards or away from the scattering region are the {\it channel functions}.
A consequence of translational invariance is that the propagation within the asymptotic region can be characterized by some constant $\beta(\w)$, while time-reversal invariance means that $-\beta(\w)$ is another constant associated with the same frequency, corresponding to propagation in the opposite direction.
Since any freely-propagating solution at $\w$ is a sum of these two channel functions, it follows that the scattering solution in the asymptotic regions can be decomposed in the in/out channel basis, as mentioned at the beginning of \cref{sec:chpt1: scattering as BVP}.

A two-dimensional free-space asymptotic region (\cref{fig:chpt1: lss}a) has an additional rotational symmetry, so that the channel functions can also be characterized by an angular momentum number $m$:
\begin{equation}
    \phi_m(r,\theta,\w) = H^\pm_m(kr) \cis{m\theta},
\end{equation}
where wavenumber $k=\w$, since $c=1$, and $H^+_m(x)$ is a Hankel function of the first kind [alternately $H^{(1)}_m(x)$], and $H^-_m(x)$ of the second [$H^{(2)}_m(x)$], both of which satisfy
\begin{equation}
    f\prpr(r) + \frac{1}{r}f\pr(r) + \left[ k^2 - \left(\frac{m}{r}\right)^2 \right] f = 0.
\end{equation}
The differential equation for $g(r) = r^{1/2}f(r)$ has a radial translational symmetry in the asymptotic limit $r \to \infty$, so that its solutions can have a definite inward and outward sense of propagation (corresponding to $H^-$ and $H^+$, respectively), connected through time-reversal.

When the asymptotic regions are waveguides, there is a finite number $2N$ of propagating channels for a given $\w$.
Based on the direction of their fluxes, the channels can be unambiguously grouped into $N$ incoming and $N$ outgoing states.
On the other hand, a two-dimensional finite scatterer in free space has a countably infinite set of propagating angular-momentum channels.
However, a finite scatterer of linear scale $R$ will interact with only a finite number of angular momentum states, with $m<L_{\max}$, with $L_{\rm max} \sim \sqrt{ \bar{\e} }R \w$, where $\bar{\e}$ is the spatially-averaged dielectric function in the scattering region.
Hence we can reasonably truncate the infinite-dimensional channel-space to a finite, $N$-dimensional subspace of relevant channels for each $\w$, similar to the waveguide case.

    \section{The Scattering Matrix \label{sec:chpt1: S-matrix} }

\begin{figure}
    \centering
    \includegraphics[width=.6\textwidth]{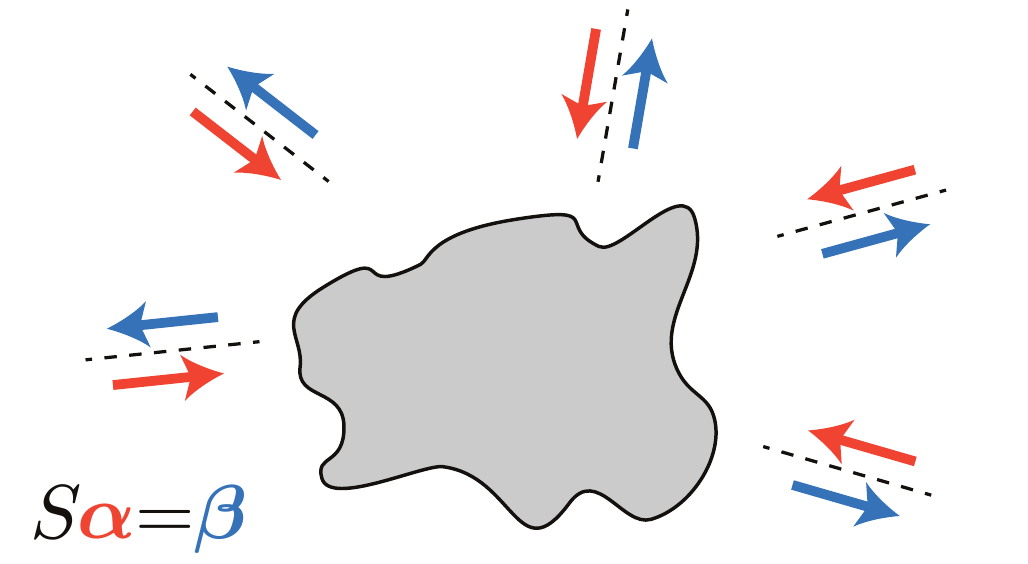}
    \caption[Depiction of scattering matrix]{Schematic depicting a general scattering process.
    A finite scatterer interacts with a finite set of asymptotic incoming and outgoing channels (dashed lines), indicated by the red and blue arrows, respectively.
    Channels may be localized in position space (e.g., waveguide channels) or in momentum space (e.g., angular-momentum channels).
    }
    \label{fig:chpt2: s-matrix-schematic}
\end{figure}

The general scattering process consists of the inward propagation of light along the $N$ incoming channels, followed by an interaction with the scatterer, and then outward propagation towards infinity along the $N$ outgoing channels. % as illustrated in \cref{fig:chpt2: s-matrix-schematic}.
The partial scattering of a single incoming channel into the outgoing channel which is its time-reverse defines a reflection coefficient, and scattering into each other channel defines the transmission coefficients.
In two- and three-dimensions, the terminology of reflection and transmission is not normally used, and one simply speaks of backscattering into the same channel and inter-channel scattering.
In the channel basis, the wavefronts of the incoming and outgoing fields are given by length-$N$ vectors ${\bs \alpha}$ and ${\bs \beta}$, normalized such that ${\bs \alpha}^\dagger {\bs \alpha}$ and ${\bs \beta}^\dagger {\bs \beta}$ are equal in magnitude to the total incoming and total outgoing energy flux, respectively.
This is illustrated in \cref{fig:chpt2: s-matrix-schematic}.

The $N$-by-$N$ {\it scattering matrix} $S(\w)$, which relates ${\bs \alpha}$ and ${\bs \beta}$ at frequency $\w$, is defined by
\begin{equation}
    \label{eq:chp1: S}
    S(\w) {\bs \alpha} = {\bs \beta},
\end{equation}
and comprises all of the reflection and transmission coefficients.
If the scatterer is lossless, i.e.,~$\e(\bx,\w)$ is real everywhere, then each incoming state leads to a flux-conserved output, and the $S$-matrix is unitary:
\begin{equation}
    \label{eq:chpt1: S unitary}
    S(\w)S(\w)^\dagger = S(\w)^\dagger S(\w) = I_N,
\end{equation}
where $I_N$ is the $N$-by-$N$ identity matrix.
When the material susceptibilities are isotropic, linear in the electric field, and time-invariant --- all of which we have already assumed --- then Lorentz reciprocity holds, and the $S$-matrix is symmetric~\cite{2013_Jalas_nphoton}:
\begin{equation}
    \label{eq:chpt1: S reciprocal}
    S(\w)=S(\w)^T.
\end{equation}

The $S$-matrix, being well-defined at all real frequencies, absent self-oscillating solutions (laser oscillation), can be extended to complex frequencies via analytic continuation.
When we do this, it is important to remember that the time-domain behavior that we wish to describe is the evolution of an initial state from $t=-\infty$ to $t=+\infty$.
This constrains the analytic behavior of $S(\w)$ in the complex plane, which we now show.

The initial time-domain wave packet $\psi^{\rm in}(\bx,t)$ can be expanded in the channel basis $\phi^{\rm in/out}_m$ as
\begin{equation}
    \lim_{t\to-\infty} \psi^{\rm in}(\bx,t) \sim \sum_m \int d\w\ a^{\phantom{\rm in}}_m(\w) \phi^{\rm in}_m(\bx,\w)\cis[-]{\w t}.
\end{equation} 
According to the definition of the scattering matrix from \cref{eq:chp1: S}, the final configuration of the wave packet after scattering is
\begin{equation}
    \lim_{t\to+\infty} \psi^{\rm out}(\bx,t) \sim \sum_{m,m\pr} \int d\w\ S(\w)^{\phantom{\rm in}}_{m,m\pr} a^{\phantom{\rm in}}_{m\pr}(\w) \phi^{\rm out}_{m\pr}(\bx,\w)\cis[-]{\w t}.
\end{equation}
The frequency integral can be done by contour integration in the complex $\w$ plane.
At very late times ($t \to +\infty$), the exponential behaves as $\cis[-]{\re{\w}t}e^{\im{\w}|t|}$, so that the contour may be closed along a large semicircle in the {\it lower} half-plane.
Therefore the integral can be evaluated in terms of the residues at the poles of $S(\w)$ in the lower half plane.
Any poles of $S(\w)$ that are in the upper half-plane do not contribute to scattering, and correspond to instabilities in the system; they are not consistent with scattering formulated as an initial value problem at asymptotically early times ($t\to-\infty$).
This is similar to why the WKB connection formulas are directional in nature~\cite{1970_Mathews_mm, 1981_morse_book, 2010_bender_book}, familiar from the study of the one-dimensional Schr{\"o}dinger equation in quantum mechanics~\cite{1994_Sakurai_qm,1994_Shankar_book}.
%, and therefore must represent bound-states, which are known not to exist for finite scatterers in electromagnetism~[WHAT SHOULD I CITE?].

Another analytic property of $S(\w)$ can be seen by considering the time-reverse of a scattering process.
For a scatterer defined by refractive index $n(\bx,\w)$, which we will temporarily denote by $S(n,\w)$, time-reversing \cref{eq:chp1: S} gives~\cite{Chong:2010ft, Wan:2011bz, Noh:2012wx}
\begin{equation}
    \label{eq:chp1: S time reverse}
    S^*(n^*,\w^*) = [S(n,\w)]^{-1}.
\end{equation}
If $S(n,\w)$ has a pole at the complex $\w^p$, then $S(n^*,\w)$ has a zero at $(\w^{p})^*$.
If the scatterer is flux-conserving ($n\in{\mathbb R}$), then the zeros and poles of the analytically continued $S$-matrix come in complex conjugate pairs.
If absorption is added to the scatterer, then the complex-valued poles and zeros generally move downward.
Therefore \cref{eq:chp1: S time reverse} implies that when gain is added, the poles and zeros move upward.

When enough gain is added to the system, a pole that started in the lower half-plane reaches the real axis, which corresponds to the onset of the lasing instability.
In the linear theory, more gain cannot be added to the system while preserving the causal structure of the scattering problem.
In contrast, zeros may safely flow into the lower half-plane by the addition of sufficient absorption within the scatterer, without breaking causality.
In the context of electromagnetic scattering, this corresponds to overdamping a resonance.

    \subsection[Effective Hamiltonian Representation of the $S$-matrix]{Effective Hamiltonian Representation of the $S$-matrix\label{sec:chpt1: effective hamiltonian} }

In this section we introduce a wave-operator representation of $S(\w)$ and $\det S(\w)$, which establishes the connection, mentioned in the introduction to this chapter, among scattering resonances ($\w\in{\mathbb R}$), $S$-matrix poles ($\w\in{\mathbb C}$), and open-boundary electromagnetic eigenvalue problems.
This is done by analyzing scattering from an open system in terms of an equivalent closed one that is then coupled to the channel continuum.

Choose a LSS that contains all of the scattering inhomogeneities, which divides the interior scattering region $\Omega$ from the exterior infinite asymptotic regions $\bar{\Omega}$, such that the LSS is $\partial\Omega$ (see \vref{fig:chpt1: lss}).
The details of the coupling between the interior and exterior will depend on the specific choice of LSS, however, the $S$-matrix will not, as must be the case on physical grounds.
Nevertheless, different choices will be more or less suitable for the solution of a given problem~\cite{1950_Teichmann_PR,1958_Lane_RMP}.

We recast the Helmholtz \cref{eq:chpt1: scattering pde} as a linear differential operator which is a function of the frequency $\w$.
The scattering solution at that frequency is the state that is annihilated by it:
\begin{equation}
    \hat A(\w)\ket{\w} = 0.
\end{equation}
The matrix elements of this operator are
\begin{equation}
    \label{eq:chp1: elements of A}
    \braket{\bx\pr|\hat A(\w)|\bx}=\delta(\bx-\bx\pr) \left\{ \w^2 - \frac{1}{\e(\bx)} \nabla^2 \right\}.
\end{equation}
We separate the Helmholtz $\hat A(\w)$ into three pieces, one for each domain and the LSS:
\begin{equation}
    \hat A(\w) = \hat A_0(\w)+\hat A_c(\w)+\hat V(\w).
\end{equation}
The closed-cavity Helmholtz operator $\hat A_0(\w)$ is identical to the original operator $\hat A(\w)$ within region $\Omega$, but is zero elsewhere.
Likewise, the channel operator $\hat A_c(\w)$ is identical to $\hat A(\w)$ in region $\bar{\Omega}$ and zero elsewhere, and the coupling operator $\hat V(\w)$ is zero everywhere except at the LSS.
This partition is advantageous because we know how to invert $\hat A_0$ on $\Omega$, since it is a closed system, and $\hat A_c$ on $\bar{\Omega}$, since it has translational invariance.
The coupling $\hat V$ will act as a perturbation on the operator $\hat A_0 + \hat A_c$.

The closed-cavity wave operator $\hat A_0(\w)$ on $\Omega$, which we do not assume is hermitian, admits a discrete spectrum of the form $\hat A_0(\w_\mu)\ket{\mu}=0$ with eigenvalues $\{\w_\mu\}$.
The {\it matrix} $A_0(\w)$ (without a hat) is defined by the matrix elements of the operator $\hat A_0$:
\begin{equation}
    \label{eq:chpt1: A0}
    A_0(\w)_{\mu \nu} \defn \braket{\mu|\hat A_0(\w)|\nu}.
\end{equation}

The channel-space wave operator $\hat A_c(\w)$ on $\bar\Omega$ has a continuous spectrum.
It generates the propagating channel modes via $\hat A_c(\w)\ket{\w,n}=0$,  where $n$ is an integer or set of integers which uniquely specify the asymptotic channels for each real frequency $\w$.

The only non-vanishing matrix elements of $\hat V$ are those between closed and continuum states.
The matrix $W(\w)$, not necessarily square, is the off-diagonal submatrix of $\hat V$:
\begin{equation}
    W(\w)_{\mu n} \defn \braket{\mu|\hat V(\w)|n,\w}.
\end{equation}
It contains all the information about $\hat V(\omega)$, since 
\begin{equation}
    \hat V(\w) = \sum_{\mu,n,\w} W(\w)_{\mu n}\ket{\mu}\bra{n,\w} + \text{h.c.}
\end{equation}

A general relation~\cite{1969_Mahaux_book, 1997_Beenakker_RMP, 2003_Viviescas_PRA, 2009_Rotter_JPA, 2017_Rotter_RMP} between the matrices $S$, $A_0$, and $W$, originally developed in the study of nuclear reactions as the continuum-shell model or shell-model approach
~\cite{1957_Bloch, 1958_Feshbach_ARNS, 1958_Feshbach_AP, 1961_Fano_PR, 1966_Lane_PR, 1969_Mahaux_book, 1985_Nishioka_PLB, 2000_Dittes_PR, 2003_Sadreev_JPA}, is
\begin{equation}
    \label{eq:chpt1: Heidelberg}
    S(\omega) = I_N - 2 \pi i W_p^\dagger(\w) G_{\rm eff}(\w) W_p(\w).
\end{equation}
This is the first central result of this section, which we refer to as the {\it effective Hamiltonian}, or {\it wave-operator representation} of the $S$-matrix. 
%$I_N$ is the $N$-by-$N$ identity matrix in the space of propagating channel amplitudes.
$W_p$ is the submatrix of $W$ that is restricted to the propagating channels.
In one- and two-dimensional free-space geometries $W_p \equiv W$, while for quasi one-dimensional waveguides that can support non-propagating transverse modes within the scattering region, $W_p\neq W$.
The effective Green function $G_{\rm eff}(\w)$ is the inverse of an effective wave operator $A_{\rm eff}(\w)$:
\begin{gather}
\begin{split}
    \label{eq:chpt1: Aeff_and_Sigma}
    G_{\rm eff}(\w) \defn A_{\rm eff}(\w)^{-1}, \qquad
    A_{\rm eff}(\w) \defn A_0\pr(\w) - \Sigma^R(\w),\\
    \Sigma^R(\w) \defn \Delta(\w) - i \Gamma(\w), \qquad
    \Gamma(\w) \defn \pi W_p(\w) W_p(\w)^\dagger,\\
    \Delta(\w) \defn {\rm p.v.} \int d\w\pr \frac{W_p(\w\pr) W_p^\dagger(\w\pr)}{\w\pr-\w}. \qquad
\end{split}
\end{gather}
$A_0\pr(\w)$ is the closed-cavity wave operator $A_0$ plus a hermitian modification that includes the effect of evanescent channel states what were previously excluded from $W_p$.
The matrix $\Sigma^R$ is the {\it retarded self-energy}, and acts only at the LSS; $\Delta(\w)$ and $-i\Gamma(\w)$ are its hermitian and anti-hermitian components, respectively.
The integral in $\Delta(\w)$ is the Cauchy principle value integral~\cite{1970_Mathews_mm}.

\Cref{eq:chpt1: Heidelberg} has a useful intuitive interpretation.
It says that the product of scattering is a superposition of the illuminating field (that's the first term), with a scattered field, given by the second term.
The scattered field must couple into and out of the resonant structure (the two $W$'s); meanwhile between the coupling events it will evolve according to the cavity Green function, modified to account for leakage into the outgoing channels.

Rarely is it useful or even possible to explicitly compute the various matrices defined in \cref{eq:chpt1: Aeff_and_Sigma}.
Their primary value is that they provide an exact and analytic connection between the $S$-matrix and the underlying operators $\hat A_0$, $\hat A_c$, and $\hat V$, each of which can be easily constructed and characterized.

The derivation of \cref{eq:chpt1: Heidelberg} is rather involved, and lies far outside the scope of this thesis.
Nevertheless, a concise description of it is hard to come by, so we will outline its basic steps here:
\begin{quote}
    Start with the standard relation describing scattering from some general potential $\hat V_{\rm sc}$~\cite[284]{2015_Weinberg_QM}\footnote{
        There is a sign-difference in the second term between \cref{eq:chpt1: weinberg scattering} and the usual form of its quantum counterpart.
        This is because the quantum analog of $\hat A(\w)$ in \cref{eq:chp1: elements of A} is $E-H_0-V$, which differs by an overall sign from the convention for quantum mechanics.
    }:
    \begin{equation}
        \label{eq:chpt1: weinberg scattering}
        S_{mn}(\w,\w\pr) = \delta_{mn}\delta(\w-\w\pr) + 2\pi i \delta(\w-\w\pr) \braket{\phi_m|\hat V_{\rm sc}|\psi^{\rm out}_n},
    \end{equation}
    where $\{\ket{\phi_m}\}$ are the known solutions in the absence of $\hat V_{\rm sc}$, and $\{\ket{\psi^{\rm out}_n}\}$ are the outgoing solutions (at late times) of the full scattering operator, which includes $\hat V_{\rm sc}$.
    We can use the relation\footnote{
    %    This is not the Lippmann-Schwinger equation~\cite{1994_Sakurai_qm}, which implicitly solves for $\ket{\psi_n^{\rm out}}$ and does not have $\hat V$ in the operator inverse.
    %    In contrast, \cref{eq:chpt1: lippmann-schwinger-like} solves for $\ket{\psi_n^{\rm out}}$ explicitly in terms of an inverse operator containing $\hat V$.
           This is not the Lippmann-Schwinger equation~\cite{1994_Sakurai_qm}.
           It is verified by operating on both sides with $\hat A(\w+i0^-)$, noting that by definition $(\hat A-\hat V_{\rm sc})\ket{\phi_n}=0$.
    } between $\ket{\psi^{\rm out}_n}$ and $\ket{\phi_n}$,
    \begin{equation}
        \label{eq:chpt1: lippmann-schwinger-like}
        \ket{\psi^{\rm out}_n} = \ket{\phi_n} - [\hat A(\w-i0^-)]^{-1} \hat V_{\rm sc} \ket{\phi_n},
    \end{equation}
    to evaluate the matrix element on the RHS of \cref{eq:chpt1: weinberg scattering}:
    \begin{equation}
        \label{eq:chpt1: heidelberg middle derivation}
       \braket{\phi_m|\hat V|\psi^{\rm out}_n} = \cancelto{0}{\braket{\phi_m|\hat V|\phi_n}} - \braket{\phi_m| \hat V [\hat A(\w-i0^-)]^{-1} \hat V |\phi_n}.
    \end{equation}
    The first term on the RHS is zero because $\hat V$ has vanishing matrix elements between channel states and other channel states.
    The addition of a small negative imaginary part $i0^-$ to the frequency in the inverse is necessary to select the ``out'' states; had we instead used $i0^+$ we would have selected the ``in'' states.
    %Finally, the operator $[\hat A(\w+i0^-)]^{-1}$ in the basis of bound states and channel states is
    %\begin{equation}
    %    \begin{pmatrix}
    %        A_0(\w+i0^-) & W_{\mu n}(\w+i0^-) \\
    %        W^\dagger_{\mu n}(\w+i0^-) & A_c(\w+i0^-)
    %    \end{pmatrix}.
    %\end{equation}

    The remaining matrix element on the RHS of \cref{eq:chpt1: heidelberg middle derivation} is between channel states and other channel states, so that in block-matrix form, we need to evaluate
    \begin{equation}
        \label{eq:chpt1: block matrix inverse}
        \begin{matrix}
            \Big( W^\dagger & 0 \Big) \\
        \mbox{}
        \end{matrix}
        \begin{pmatrix}
            A_0 & W \\
            W^\dagger & A_c
        \end{pmatrix}^{-1}
        \begin{pmatrix}
            W \\
            0
        \end{pmatrix} 
        =
        W^\dagger\left[A_0 - WA_c^{-1}(\w-i0^-)W^\dagger\right]^{-1}W,
    \end{equation}
    which follows from a standard relation of linear algebra involving the Schur Complement~\cite{Handbook_of_LA}.

    The final step is to reduce $\hat V \hat A_c^{-1}(\w-i0^-)\hat V$, which is the operator-equivalent of the second term in brackets on the RHS of \cref{eq:chpt1: block matrix inverse}.
    This can be done by inserting the complete channel states on either side of $\hat A_c^{-1}$ and evaluating the resulting integrals, accounting for the $i0^-$ pole.
    Plugging this back into \cref{eq:chpt1: weinberg scattering} gives the effective Hamiltonian relation \cref{eq:chpt1: Heidelberg}.
\end{quote}

The quasi-normal modes of the system are the eigenmodes of the \nh, nonlinear eigenvalue problem 
\begin{equation}
    \label{eq:chpt1: effective wave equation}
    \hat{A}_{\rm eff}(\omega^p_n) \ket{\omega^p_n} = 0,
\end{equation}
with eigenvalues $\{\omega^p_n\}$ that are the complex-valued resonance frequencies, which generally form a countably infinite set.
The frequency dependence of the self-energy $\Sigma^R(\w)$ makes the eigenproblem nonlinear in its eigenvalue\footnote{
    Without self-energy there is still a nonlinearity in eigenvalue $\w$, since the Helmholtz operator depends on $\w^2$.
    This is not the nonlinearity that we are referring to when we call \cref{eq:chpt1: effective wave equation} a nonlinear eigenvalue problem, since it is trivially accounted for by defining $\lambda = \w^2$, with eigenvalue $\lambda$.
}.
Since $\Gamma(\w)$ is positive semidefinite, it generally contributes a negative imaginary part to the resonance frequencies, pushing the poles of $G_{\rm eff}(\w)$ and $S(\w)$ into the lower half-plane, exactly capturing the effect of openness.

A series of algebraic manipulations applied to \cref{eq:chpt1: Heidelberg} yields a powerful identity for $\det S(\w)$.
Using the ``push-through identity'' of linear algebra~\cite{Bernstein_Matrix_book} (see \Cref{chp:app: linalg}), we move the interaction matrix $W_p$ across $G_{\rm eff}$ to obtain
\begin{equation}
    \label{eq:chpt1: k-matrix rep}
    S(\w) = {\bs(} I + i K(\w){\bs)}/{\bs(} I - i K(\w){\bs)},
\end{equation}
where $K(\w)$ is the reactance matrix~\cite{1967_MacDonald_PR, 1982_Newton_book}
\begin{equation}
    K(\w) \defn -\pi W_p(\w)^\dagger [A_0\pr(\w) - \Delta(\w)]^{-1} W_p(\w).
\end{equation}
Taking the determinant of both sides and multiplying numerator and denominator by $\det (A_0\pr - \Delta)$, we arrive at the second central result of this section:
\begin{equation}
\label{eq:chpt1: det(S)}
    \det S(\w) = \frac{\det {\bs(} A_0\pr(\w) - \Delta(\w) - i\Gamma(\w){\bs)}}{\det {\bs(} A_0\pr(\w) - \Delta(\w) + i \Gamma(\w){\bs)} }.
\end{equation}
\Cref{eq:chpt1: det(S)} is not a simple identity of linear algebra: the LHS is the determinant of an $N$-by-$N$ square matrix, while the right-hand side is a ratio of functional determinants of differential operators on an infinite-dimensional Hilbert space.
See \Cref{sec:app: detailed_derivation_S} for more details on the derivation of \cref{eq:chpt1: det(S)} from \cref{eq:chpt1: Heidelberg}.
%A related expression was given in Refs.~\cite{2013_Grigoriev_PRA, 2019_Krasnok_arXiv}.

\Cref{eq:chpt1: det(S)} demonstrates the analytic connection that we had previously mentioned between the resonant poles of the $S$-matrix and the eigenvalues of the operator $\hat A_{\rm eff}(\w)$, which appears in the denominator of this expression.
It also shows that $S(\w)$ is zero at the countably infinite set of complex frequencies $\{\w^z_n\}$ that are the eigenvalues of the operator defined in the numerator of \cref{eq:chpt1: det(S)}:
\begin{gather}
    \hat A_z(\w) \defn \hat A_0\pr(\w) - \hat \Sigma^A(\w), \\
    \hat \Sigma^A(\w) \defn \hat \Delta(\w) + i \hat \Gamma(\w).
\end{gather}
$\hat \Sigma^A(\w)$ is the {\it advanced self-energy}, which differs from the retarded self-energy of \cref{eq:chpt1: Aeff_and_Sigma} in the sign of $i\hat\Gamma$.
We call the set $\{\w^z_n\}$ the {\it zeros of the $S$-matrix}.
Coherent perfect absorption (CPA) arises when, by tuning the degree of absorption in the system, one member of this set reaches a real frequency and becomes a steady-state solution.
Conversely, the threshold of laser oscillation occurs when sufficient gain is added to the system such that a member of $\{\w^p_n\}$ first reaches the real axis, corresponding to a purely outgoing solution at a real frequency.
The two are linked by time-reversal, as mentioned in \Cref{sec:chpt0: lasers and cpa} of the previous chapter.
A lossless scatterer has a real-valued $\e(\bx)$, and therefore $\hat A_0\pr(\w)$ is hermitian when $\w$ is real.
In this case, time-reversal symmetry implies that $\w^z_n = \w^{p*}_n$.
This is consistent with \cref{eq:chpt1: det(S)} since $[ \hat \Sigma^A(\w)]^\dagger = \hat \Sigma^R(\w)$ when $A_0\pr(\w)$ is hermitian, which follows from $[\hat{\cal O}(\w^*)]^\dagger = \hat{\cal O}(\w)$ for $\hat {\cal O} \in \{\hat A_0\pr, \Delta, \Gamma\}$, by a generalization of the Schwartz reflection principle~\cite{1970_Mathews_mm}.

While an eigenvalue of $\hat A_0\pr(\w) - \hat \Sigma^R(\w)$ corresponds to a pole of $\det S(\w)$ and an eigenvalue of $\hat A_0\pr(\w) - \hat \Sigma^A(\w)$ to a zero, there is one important exception.
\Cref{eq:chpt1: det(S)} shows that if $\w^p_n = \w^z_n$ is the simultaneous eigenvalue of both, then $\det S(\w)$ may be neither zero nor infinite.
Such an exception happens at a bound state in the continuum (BIC)~\cite{2016_Hsu_NRM}, which contains neither incoming nor outgoing radiation, exists at a real frequency, and does not affect $S(\w)$ since it is decoupled from far-field radiation; it is neither a pole nor a zero of $S(\w)$.
The other case, where $S(\w)$ has both a diverging and vanishing eigenvalue but $\det S(\w)$ is generic, can also happen, for example in ${\cal PT}$-symmetric cavities that exhibit CPA and lasing at the same frequency~\cite{2010_longhi_pra, Chong:2011ev, Feng:2014gp, 2016_wong_natphot}.
%In the topological-defect picture, a BIC implies that a $+1$ charge ($S$-matrix zero) annihilates with an $-1$ charge (resonance) on the real axis.

    \subsection{Temporal Coupled Mode Theory (TCMT) \label{sec:chpt1: tcmt} }

Temporal coupled mode theory (TCMT)~\cite{Haus_book, 2003_Fan_JOSAA, 2004_Suh_JQE, 2018_Wang_OL, 2019_Zhao_PRA, 2017_Alpeggiani_PRX} is formally similar to the effective Hamiltonian approach described above.
It can be derived from symmetry constraints~\cite{Haus_book, 2003_Fan_JOSAA, 2004_Suh_JQE, 2018_Wang_OL, 2019_Zhao_PRA} rather than from first principles.
The main difference between it and the effective Hamiltonian approach is that TCMT is not an exact construction, based on a wave equation, but is instead an approximate analytic model that faithfully captures most of the phenomenology of resonances. 
%Because it is simple and widely applicable, it is a standard tool of photonics, used widely in the design and analysis of optical devices~\cite{2012_Verslegers_PRL, 2014_Peng_nphys, 2014_Hsu_NL, 2015_Zhen_Nature}; we will use it for this purpose in \Cref{chp:chpt2: rsms,chp:chpt3: cpa ep}.
In this section, we outline the phenomenological theory, and highlight the similarities to and differences from the effective Hamiltonian formalism.

TCMT typically assumes harmonic time dependence in a reciprocal scattering region that supports $M$ internal resonances, each coupled to $N$ external propagating channels. 
The length-$N$ vectors ${\bs \alpha}$ and ${\bs \beta}$ contain incoming and outgoing field amplitudes, respectively.
The field within the cavity is described by a superposition of the $M$ resonances, whose amplitudes are contained in the length-$M$ vector ${\bf a}$.
The amplitudes of the internal resonances evolve in time under some $M$-by-$M$ Hamiltonian $H_{\rm 0}$, whose eigenvalues are the normal mode frequencies of the closed system.
TCMT relates all of these quantities by~\cite{2004_Suh_JQE}
\begin{gather}
    \label{eq:TCMT_1}
    -i \w {\bf a}  =  -i \Hcmt {\bf a}  + D^T {\bs \alpha} , \\
    \label{eq:TCMT_2}
    {\boldsymbol \beta} = S_0 {\bs \alpha} + D {\bf a}.
\end{gather}
The $M$-by-$M$ Hamiltonian of the open system in TCMT is
\begin{equation}
    \label{eq:chpt1: H_cmt}
    \Hcmt \defn H_0 - i \frac{D^\dagger D}{2}.
\end{equation}
The $N$-by-$M$ matrix $D$ contains the coupling coefficients between the resonances and the channels; the $m^{\rm th}$ column of $D$ is essentially the radiation wavefront of the $m^{\rm th}$ resonance analyzed into the $N$-dimensional channel basis.

The {\it direct scattering matrix} $S_0 = S_0^T$ is symmetric and describes the non-resonant part of scattering that varies slowly with frequency.
The distinction between ``resonant'' and ``non-resonant'' is not always clear; the typical practice in TCMT is to use $H_0$ and $D$ to model a few high-\Q\ resonances in the frequency range of interest, and empirically bundle the contributions from the further-away and low-\Q\ resonances into $S_0$.
As more resonances are included in $H_0$ and $D$, the direct scattering matrix approaches the identity~\cite{2017_Alpeggiani_PRX}.

In a flux-conserving system, time-reversal symmetry requires that $S_0 D^* = -D$ (\cite{2004_Suh_JQE, 2019_Zhao_PRA}); here we assume that absorption or gain can be modeled through an anti-hermitian term in $H_0$ without breaking this condition.
%The amplitudes are normalized such that the magnitude square of resonance amplitude is energy, and that the magnitude square of incoming/outgoing wave amplitude is power.
It follows that the scattering matrix is
\begin{equation}
    \label{eq:chpt1: S_form1}
    S(\w) = \left( I_N - i D \frac{1}{\w - \Hcmt} D^{\dagger} \right) S_0.
\end{equation}
The formalism has been found to be accurate even in the case of several overlapping resonances~\cite{2012_Verslegers_PRL, Peng:2014kl, 2014_Hsu_NL, Zhen:bl}.
%A formalism mathematically equivalent to TCMT is also used in quantum noise theory~\cite{2004_Gardiner_book}.

The similarities between \cref{eq:chpt1: Heidelberg} and \cref{eq:chpt1: S_form1} are evident.
The Hamiltonian matrix $H_0$ describes a closed system and is hermitian in the absence of absorption or gain; $(\w- H_0)$ is analogous to $A_0$ in \cref{eq:chpt1: A0} or $A_0\pr - \Delta$ in \cref{eq:chpt1: Aeff_and_Sigma}.
The positive semidefinite matrix $D^\dagger D / 2$ is analogous to $\Gamma$ in \cref{eq:chpt1: Aeff_and_Sigma}; its diagonal elements are the decay rates of the modes due to radiation into channels in the open environment, and its off-diagonal elements are the dissipative coupling rates between modes, mediated by their interaction with the continuum of propagating channel modes.
Because \cref{eq:chpt1: Heidelberg,eq:chpt1: S_form1} have the same formal structure, TCMT also implies a determinant relation:
\begin{equation}
    \det S(\w) = \det S_0\ \frac{\det (\w - H_0 - iD^\dagger D/2)}{\det (\w - H_0 + iD^\dagger D/2) },
\end{equation} 
so that the poles and zeros of $S$ are the eigenvalues of $\Hcmt$ and $\Hcmt + iD^\dagger D$, respectively.
The determinant of the direct scattering matrix $\det S_0$ is assumed to be smooth and non-vanishing in the frequency range of interest.

Despite the similarities, the two approaches are not equivalent.
TCMT neglects the frequency-dependence of the couplings in $D$ due to openness.
In the effective Hamiltonian formalism, this frequency-dependence changes both the hermitian and anti-hermitian parts of $\hat A_{\rm eff}(\w)$, through the frequency shifts $\Delta(\w)$ and decay rates $\G(\w)$, respectively.
In contrast, for TCMT, even if one were to include a phenomenological dispersion in $D$, it would not account for shifts in the hermitian part of $\Hcmt$.
To correct this, one would need to add an an unknown frequency-dependence in both $D$ and $H_0$.
For many situations, particularly when high-\Q\ resonances are involved, this shortcoming of TCMT is negligible, as these dispersive effects are small over the relevant frequency band.
However, because of this difference, the approximate TCMT \cref{eq:chpt1: S_form1} can possess symmetries that are not present in either the effective Hamiltonian representation of $S(\w)$ [\cref{eq:chpt1: Heidelberg}], or in the scattering Helmholtz PDE [\cref{eq:chpt1: scattering pde}].
We will see an example of this in \cref{sec:chpt4: RSM EP} of the next chapter.

In a typical exposition of TCMT, what we have called $\Hcmt$ usually goes by the name ``effective Hamiltonian''.
For the reasons detailed above, we do not consider TCMT to be exactly equivalent to the effective Hamiltonian formalism of \cref{sec:chpt1: effective hamiltonian}, and therefore we use a different name to refer to the TCMT Hamiltonian in order to avoid ambiguity.

    \section{Implementation of Boundary Conditions \label{sec:chpt2: boundary conditions}}

At this point we have established all of the general properties of the $S$-matrix that we will need in the remaining chapters, and we return to the problem of solving a given scattering geometry.
Earlier, in \cref{sec:chpt1: scattering as BVP}, we transformed the source-free Helmholtz PDE with scattering boundary conditions [\cref{eq:chpt1: scattering pde}] into an inhomogeneous Helmholtz PDE with purely outgoing boundary conditions [\cref{eq:chpt1: scattering pde inhomogeneous}].
Standard numerical methods for solving linear equations can be used in the latter case, assuming that one can implement both the purely outgoing boundary and the source term in a finite computational domain.
In the remainder of this chapter, we address how to do this.

We describe two methods for implementing outgoing boundary conditions in a finite volume: boundary-matching and perfectly matched layers (PMLs).
The former is exact but complicated, and uses the properties of the channel functions.
It results in a frequency-dependent boundary condition that significantly complicates the calculation of resonances and zeros, which in this implementation become nonlinear eigenvalue problems.
The PML method is an exponentially good approximation that is far simpler than boundary-matching, and has the advantage that the resonance and $S$-matrix-zero eigenproblems remain linear in their eigenvalue.
We will use both methods in the numerical calculations of the remaining chapters.

    \subsection{Boundary-Matching \label{sec:chpt2: boundary matching}}

The boundary-matching method~\cite{2005_prog_in_optics} uses the continuity of the field and of its derivative at the LSS.
For the full vectorial wave equation, it would use the appropriate Maxwell interface boundary conditions.
It gives the boundary condition exactly in the form of \cref{eq:chpt1: general boundary conditions}, relating the field at one point on the LSS to its normal derivative at all points on the LSS.
This expression, while analytic, generally depends on the frequency.
For scattering configurations, which have both incoming and outgoing waves at a specified $\w$, this method provides an analytic expression that can be discretized effectively in a finite difference computation.
However, for electromagnetic eigenvalue problems, the dependence of the boundary conditions on $\w$, which is also the eigenvalue, poses a fundamental challenge.
In this case, the discretized wave operator, including the boundary condition, defines a {\it nonlinear eigenvalue problem} (NEVP) of the form $T(\lambda_n)\bx_n={\bf 0}$, and not an ordinary one of the form $M\bx_n = \lambda_n \bx_n$.
NEVPs are nonlinear in their eigenvalue (not in the vector $\bx$) since $T(\lambda)$ is a matrix function of the complex variable $\lambda$.
An ordinary eigenvalue problem corresponds to $T(\lambda) = M - \lambda B$ for constant matrices $M$, $B$.
NEVPs are difficult to solve analytically, but can be handled by numerical packages such as NEP-PACK~\cite{2017_Bezanson_SIAM, 2018_Jarlebring_arxiv}.

The field and its derivatives are all continuous at the LSS, since by definition any scattering inhomogeneity is contained entirely within the scattering region.
Immediately outside the LSS, the outgoing component of the field can be decomposed into the outgoing channel functions, with coefficients $c_m$.
The derivative of the field normal to the LSS, i.e.,~along the outgoing propagation direction, must be equal to the derivative of the channel-function decomposition, by continuity.
Because of this, the $c_m$'s can be expressed in terms of the normal derivative $\partial_{\bf n}\psi$ at the boundary, so that $\psi$ has a definite relation to $\partial_{\bf n}\psi$ at the boundary, in the general form of \cref{eq:chpt1: general boundary conditions} mentioned earlier.
We give examples for three cases:
\begin{enumerate}

\item For a simple one-dimensional system with LSS boundaries at $|x|=a$ (scatterer is entirely in $|x|<a$ space), the outgoing wave outside the LSS is proportional to $\cis[\pm]{k x}$, where $k=\w$ ($c=1$), so
\begin{equation}
    \label{eq:chpt1: one-dim bc}
    \bigg[\psi(x) \mp \frac{1}{ik}\psi\pr(x)\bigg]_{x=\pm a} = 0.
\end{equation}

\item For a quasi-one dimensional perfect (lossless) metallic waveguide with walls at $y=0$ and $y=t$ and with the LSS at $|x|=a$, the channel functions are proportional to $\sin(m\pi y/t)\cis[\pm]{\beta_m(\w) x}$, so that
\begin{equation}
    \left[\psi(x,y) \mp \int \frac{dy\pr}{t} \left( \sum_{m=1}^\infty \frac{2}{i\beta_m(\w)}\sin\Big(\frac{m \pi y}{t}\Big) \sin\Big(\frac{m \pi y\pr}{t}\Big)\right) \ \partial_x\psi(x,y^\prime)\right]_{x=\pm a} = 0.
\end{equation}
For {\it real-valued} $\w$, the propagation constant is
\begin{align}
    \label{eq:chpt1: beta_m}
    \beta_m (\w)&=\sqrt{\left(\frac{\w}{c}\right)^2-\left(\frac{m\pi}{t}\right)^2 + i0^+}.
\end{align}
The square-root branch cut is the conventional one along the negative real axis, such that $\sqrt{-1 + i0^\pm}=\pm i$.
For {\it complex} $\w$, the propagation constant $\beta_m(\w)$ can simply be taken as the analytic continuation of \cref{eq:chpt1: beta_m}.
However, some NEVP solution methods require an integral over a loop in the complex frequency plane~\cite{2009_asakura_jsiam, 2012_Beyn_LA, esterhazy_2014}, which is ambiguous if this loop contains the branch point of $\beta_m(\w)$, which is likely when near a mode cutoff.
This ambiguity can be avoided by traversing the loop twice; further analysis is beyond the scope this thesis.

\item For a two-dimensional finite scatterer in free space, with the LSS at $r=R$, the asymptotic channels are proportional to $\cis{m\theta}H^+_m(k r)$ so that
%\begin{equation}
%    \label{eq:chpt1: 2-dim boundary matching}
%    \psi(R,\theta) = \sum_m c_m \cis{m\theta} H^+_m(\w R),
%\end{equation}
%while the normal derivatives are
%\begin{equation}
%    \label{eq:chpt1: free space boundary matching}
%    \partial_r \psi(r,\theta)|_{r=R} = \w \sum_m c_m \cis{m\theta} H^{+\prime}_m(\w R).
%\end{equation}
%Following the procedure defined above, we solve for the $c_m$'s in terms of $\partial_r \psi(r,\theta)|_R$, and plug it back into \cref{eq:chpt1: 2-dim boundary matching}:
\begin{equation}
    \label{eq:chpt1: free space bc}
    \left[ \psi(r,\theta) - \int \left(\frac{1}{2\pi k}\sum_m\cis{m(\theta-\theta\pr)} \frac{H_m^+(k R)}{H_m^{+\prime}(k R)} \right) \partial_r\psi(r,\theta)\ \, d\theta\pr \right]_{r=R} = 0,
%    \left[ \psi(r,\theta) - \frac{1}{\w}\int \frac{d\theta\pr}{2\pi} \left(\sum_m\cis{m(\theta-\theta\pr)} \frac{H_m^+(\w R)}{H_m^{+\prime}(\w R)} \right) \partial_r\psi(r,\theta) \right]_{r=R} = 0.
\end{equation}

\end{enumerate}

All of these are specific examples of a more general relation, which can be derived from the effective Hamiltonian formalism (see \Cref{chp:app: boundary matching}):
\begin{equation}
    \label{eq:chpt1: general boundary matching}
    \left[\ \psi(\bx) - \oint_{\partial \Omega} \ G^{R,N}_c(\bx,\bx\pr) \nabla \psi(\bx\pr) \cdot d{\bf S}\pr\ \right]_{\bx \in \partial\Omega} = 0.
\end{equation}
$G^{R,N}_c(\bx,\bx\pr)$ is the retarded Green function in the channel space, i.e.,~it is the inverse of $\hat A_c(\w-i0^-)$, subject to Neumann boundary conditions at the LSS.
The one-dimensional and quasi-one-dimensional cases are plausibly consistent with this formula {\it prima facie} (in fact they are).
However, the boundary-matching condition for the two-dimensional free-space case is less obviously related to $G_c^{R,N}$.
\Cref{chp:app: free space} verifies that \cref{eq:chpt1: free space bc} does indeed follow from \cref{eq:chpt1: general boundary matching}.

    \subsection{Perfectly Matched Layers \label{sec:chpt2: PMLs}}

Another method for implementing outgoing boundary conditions, commonly used in resonance calculations~\cite{2018_Lalanne_LPR}, is perfectly matched layers (PMLs)~\cite{1994_Berenger_JCP, 2007_Johnson_PML_note}, which create an effective outgoing boundary through a frequency-independent modification of the wave operator.
The chief advantage of PMLs is that, for non-dispersive scatterers, they turn the self-consistent, boundary-matched NEVP into an ordinary eigenvalue problem, because their construction does not depend on the eigenfrequency.

A naive approach to building a frequency-independent outgoing boundary layer is to surround the LSS with a uniform absorbing layer, characterized by a its depth $d$ and a constant refractive index $n = 1 + in\pr$.
The layer is terminated at its outermost surface with a Dirichlet boundary.
However, a finite depth $d$ is not consistent with purely outgoing waves, since there is necessarily an impedance mismatch, and therefore back-reflection, for any non-zero $n\pr$ at the boundary between vacuum and the absorbing layer.
This is overcome in the limit that $d \to \infty$ followed by $n\pr \to 0$, per the limiting absorption principle~\cite{1971_Schulenberger_ARMA, Cerjan_2015_ps}.
The infinite volume requirement is obviously an impediment for numerical implementation.

PMLs are non-physical absorbing layers which are constructed specifically to be reflectionless at the vacuum interface, regardless of their degree of absorption, unlike the constant index absorbing layer we just described.
The exact limit of purely outgoing radiation requires $d \to \infty$, but it is not accompanied by the requirement of a vanishing absorption per unit length.
Because of this, the reflection from a Dirichlet-terminated PML is exponentially insensitive to the depth of the layer.
Even a sub-wavelength PML depth is often enough to achieve purely outgoing waves with an acceptable accuracy~\cite{2007_Johnson_PML_note, 2011_Oskooi_jcp}.

We briefly summarize the PML construction by using complex coordinate stretching, which analytically continues the wave equation and its solution into the complex plane along a specified contour~\cite{2007_Johnson_PML_note}.
Similar methods are used with the Schr{\"o}dinger equation in the context of quantum mechanics~\cite{2011_Moiseyev_book, 2020_Schneider_natrevphys}.
We will focus on the one-dimensional case; the extension to higher dimensions is trivial so long as the coordinates are separable.

The stretched coordinate $\tilde x$ is
\begin{equation}
    \tilde x(x) = \int_0^x dx\pr\ \alpha(x\pr) \implies d\tilde x(x) =  \alpha(x)dx,
\end{equation}
for some complex function $\alpha$ which is unity everywhere inside the LSS:
\begin{equation}
    \alpha(x) \equiv 1, \quad \text{for } x \in \Omega.
\end{equation}
For simplicity we will take the LSS to be at $x=0$, with the PML to its right, and the scattering region to its left.
The transformed plane-wave solution on the new contour is $\cis{\w \tilde x(x)}$ (remember $c=1$), while the wave operator, expressed in the original coordinates, is
\begin{equation}
    \label{eq:chpt1: wave operator PML}
    \hat A_{\rm PML}(x) \defn \frac{1}{\alpha(x)} \nabla \frac{1}{\alpha(x)} \nabla + \w^2 \e(x).
\end{equation}
Since we are only interested in cases where $\e(x)\equiv 1$ outside the LSS, $\e{\bs(}\tilde x(x){\bs)}\equiv \e(x)$.
A decaying wave corresponds to
\begin{equation}
    \label{eq:chpt1: alpha pml}
    \alpha(x) = 1 + i\frac{\sigma(x)}{\w_0},
\end{equation}
where $\sigma(x)$ is some function with a positive real part which acts like a conductivity, and $\w_0$ is a characteristic frequency  near which the scattering or eigenvalue computation will be done.
Analytically, $\sigma(x)$ does not need to be smooth or small, though a PML in a discretized space is generally more effective when it does not change too rapidly~\cite{2007_Johnson_PML_note}.
By choosing $\sigma(x)$ appropriately, the wave can be made arbitrarily small, though never zero, after propagating any finite distance.
The error induced by truncating the PML with a Dirichlet boundary at a finite $d$ is therefore well-controlled and exponentially small, and we take it to be negligible from here on.

The symmetric form of the transformed Helmholtz equation [using \cref{eq:chpt1: wave operator PML}] is
\begin{equation}
    \label{eq:chpt1: wave equation PML}
    \left\{\nabla \frac{1}{\alpha(x)} \nabla + \w^2 \alpha(x) \e(x)\right\}\psi(x) = 0.
\end{equation}
Define the unconjugated ``inner product''
\begin{equation}
    \left(f,g\right)_{\alpha \e} \defn \int dx\ f(x) \alpha(x) \e(x) g(x),
\end{equation}
where the integral is over the entire computational domain, including the PML, and $f$ and $g$ are any two functions satisfying the Dirichlet boundary condition.
\Cref{eq:chpt1: wave equation PML} is symmetric with respect to this inner product since
\begin{equation}
    \left(f,\hat A_{\rm PML} g \right)_{\alpha\e} = \left(\hat A_{\rm PML} f,g \right)_{\alpha\e},
\end{equation}
which follows from integrating by parts using the Dirichlet boundary condition.
Because of this, the eigenfunctions of $\hat A_{\rm PML}$ satisfy a biorthogonality relation:
\begin{equation}
    \left(\psi_m,\psi_n\right)_{\alpha\e} = \delta_{mn}.
\end{equation}
This is a major advantage of the PML approach over boundary-matching, since in the latter case the eigensolutions of the NEVP do not satisfy a similar biorthogonality relation.
We will return to this point in \Cref{chp:chpt3: cpa ep}.

Sometimes the stretched coordinate will appear explicitly in addition to the scaling function $\alpha$.
For example, in cylindrical coordinates, with a radial PML, the symmetric wave operator is
\begin{equation}
    \left\{\frac{\partial}{\partial r} \frac{\tilde r(r)}{\alpha(r)} \frac{\partial}{\partial r} + \frac{\alpha(r)}{\tilde r (r)}\frac{\partial^2}{\partial \theta^2} + \w^2 \tilde r(r) \alpha(r) \e(r,\theta)\right\}\psi(r,\theta) = 0,
\end{equation}
where $\tilde r(r) = \int_0^r dr\pr \alpha(r\pr)$.

Purely incoming boundary conditions, relevant for computing $S$-matrix zeros and CPA states, can also be implemented with PMLs.
Time-reversing \cref{eq:chpt1: wave equation PML} shows that a ``conjugate PML'', which is constructed from the ordinary PML by taking $ i \sigma \to -i \sigma^*$, will be equivalent to purely-incoming matched boundaries~\cite{2018_Bonnet-BenDhia_PRSA}.

Each eigenfrequency of the matched-boundary NEVP is also an eigenfrequency of an equivalent system with PMLs.
Within the LSS their eigenmodes will be identical.
However, it should be cautioned that the converse is not true; PMLs, both conventional and conjugated, introduce additional ``PML modes'' in the eigenvalue spectrum~\cite{2018_Lalanne_LPR}.
These modes are essential for the completeness of the eigenbasis of $\hat A_{\rm PML}$, however they depend on the specific choice of PML and do not correspond to resonances of the open system, and are therefore unphysical.

    \section{Equivalent Sources \label{sec:chpt2: equivalent sources}}

We describe two methods for implementing a source term for the RHS of \cref{eq:chpt1: scattering pde inhomogeneous}.
The first uses the volume equivalence principle to define a source that is distributed within the bulk of the scattering region, and whose radiated field is $\psi^{\rm out}$.
The second defines an equivalent source, distributed around a surface $\partial {\cal D}$, that in the absence of the scatterer exactly reproduces $\psi^{\rm in}$ in the interior of $\partial {\cal D}$.
The latter has the distinct advantage in that it applies to scattering from both dielectrics and from metallic boundary deformations.

\subsection{Volume Equivalence Principle}

The volume equivalence principle~\cite{2007_Johnson_equivalentsources, 2015_jin_book} relates a source-free wave equation with scattering boundary conditions to a different equation satisfying radiation boundary conditions with a source $j$.
We have actually already used it without naming it in going from \cref{eq:chpt1: scattering pde} to \cref{eq:chpt1: scattering pde inhomogeneous}, which we repeat here for convenience:
\begin{gather}
    \label{eq:chpt1: scattering pde 3}
    \{\nabla^2 + \w^2 \e(\bx,\w)\}\psi^{\rm out}(\bx) = j(\bx), \\
    \label{eq:chpt1: scattering source init again}
    j(\bx) \defn -\w^2\chi(\bx,\w)\psi^{\rm in}(\bx).
\end{gather}
The purely outgoing condition on $\psi^{\rm out}$ can be implemented in either of the two ways given in the previous section.
The total field $\psi$ is recovered from the solution to \cref{eq:chpt1: scattering pde 3} through
\begin{equation}
    \psi(\bx) = \psi^{\rm out}(\bx) + \psi^{\rm in}(\bx).
\end{equation}

This method is straightforward and effective for scatterers that are defined by dielectric inhomogeneities, i.e.,~regions where $\chi\neq0$.
However, it does not apply to scattering from deformations of a metallic boundary, since in this case $\chi(\bx)$ is not finite.
We must use a different method to handle this scenario, such as Huygens' principle, outlined below.

\subsection{Huygens' Principle \label{sec:chpt1: huygens} }

Huygens' principle can be used to define a source $j_{\rm hp}$ that can emulate any target wavefront within a bounded region ${\cal D}$, while generating no extra waves outside it~\cite{2015_jin_book}.
The {\it equivalent source} $j_{\rm hp}$ is distributed along the boundary $\partial{\cal D}$ of the target domain.
We can use an equivalent source for the scattering problem by taking $\psi^{\rm in}(\bx)$ as the target wavefront, and situating ${\cal D}$ entirely within the LSS, while also requiring that it contain every scattering defect, be it a boundary deformation or a dielectric inhomogeneity.

The Helmholtz equation in this case is
\begin{equation}
    \{\nabla^2 + \w^2 \e(\bx,\w)\}\psi_{\rm TF/SF}(\bx) = j_{\rm hp}(\bx),
\end{equation}
subject to purely outgoing boundary conditions.
The ``Total Field/Scattered Field'' solution $\psi_{\rm TF/SF}(\bx)$ is
\begin{equation}
\psi_{\rm TS/SF} = 
    \begin{cases}  
        \psi(\bx) & \bx \in {\cal D}\\
        \psi^{\rm out}(\bx) & \text{otherwise}.
    \end{cases}
\end{equation}

The construction of $j_{\rm hp}$ is straightforward.
Let $\phi(\bx)$ be the target solution of the free-space Helmholtz equation, such as a plane wave:
\begin{equation}
    \{ \nabla^2 + \w^2 \} \phi(\bx) = 0.
\end{equation}
Define a masking function $M(\bx)$ that vanishes everywhere outside the target domain:
\begin{equation}
M_{\cal D}(\bx) = 
    \begin{cases}
        1 & \bx \in {\cal D} \\
        0 & \text{otherwise}.
    \end{cases}
\end{equation}
Then the equivalent source is
\begin{equation}
    \label{eq:chpt1: hp current source}
    j_{\rm hp}(\bx) \defn \{ \nabla^2 + \w^2 \} M_{\cal D}(\bx) \phi(\bx).
\end{equation}
The total field $\psi$ can be reconstructed through
\begin{equation}
    \psi(\bx) = \psi_{\rm TS/SF}(\bx) + [1-M_{\cal D}(\bx)]\phi(\bx).
\end{equation}

\section{Summary of Methods}

We have presented two theoretical frameworks for describing arbitrary linear scattering in terms of the modes of a closed system: the effective Hamiltonian approach~[\cref{eq:chpt1: Heidelberg,eq:chpt1: Aeff_and_Sigma}], and temporal coupled mode theory (TCMT)~[\cref{eq:chpt1: S_form1}], which are nearly equivalent in form.
The effective Hamiltonian approach is rigorous, but difficult to use for calculations; in this thesis we use it primarily to develop the theory of reflectionless scattering modes (RSMs) and their symmetry properties in \Cref{chp:chpt2: rsms}.
TCMT is phenomenological, and is easily used to calculate scattering properties in terms of just a few unknown parameters; we use it in \Cref{chp:chpt2: rsms} to provide a useful intuition about how RSMs in high-\Q\ systems relate to each other for different sets of input channels, and in \Cref{chp:chpt3: cpa ep} to design a \nh, whispering gallery mode resonator, which is absorbing for only one sense of chirality, and reflective for the other.

We have described three ways to represent the incoming wave of a scattering problem, $\psi^{\rm in}(\bx)$:
\begin{enumerate}

    \item as a scattering boundary condition, which is specified asymptotically far from the scattering region;
    
    \item as an equivalent volume current contained entirely within the scattering region; \label{itm:chpt1: equivalent volume}
    
    \item as an equivalent surface current defined on a boundary $\partial{\cal D}$ within the scattering region, such that all scattering events occur in the interior of $\partial {\cal D}$. \label{itm:chpt1: equivalent surface}
           
\end{enumerate}

We have also shown two ways to represent the outgoing boundary condition that $\psi^{\rm out}(\bx)$ must satisfy:
\begin{enumerate}

    \item Boundary-matching, which relates the field to its normal derivative along the last scattering surface (LSS) and depends explicitly on the frequency.
    This can be used in conjunction with any of the representations of $\psi^{\rm in}$ given above.
    
    \item Perfectly matched layers (PMLs), which are reflectionless absorbing layers whose construction does not explicitly depend on frequency.
    They can only be used with $\psi^{\rm in}$-representations \ref{itm:chpt1: equivalent volume} and \ref{itm:chpt1: equivalent surface} above.
           
\end{enumerate}
PMLs are preferred for electromagnetic eigenvalue problems; the frequency-dependence of boundary-matching makes the associated eigenvalue problem a NEVP, which is substantially harder to solve than its linear counterpart.
However, PMLs cannot be used in every case, for example the multipole RSM problem shown in \vref{fig:chpt2: free_space}.

We use all of these these representations, save the scattering boundary condition, in the numerical calculations contained in this thesis.
They are summarized in \cref{tab:chpt1: Summary Table}.

\begin{table}
    \centering
    \begin{tabular}{ c | c | c }
    \toprule
    \diagbox{outgoing\\boundary}{$\psi^{\rm in}(\bx)$} &\makecell{scattering\\bound. cond.\\ \cref{eq:chpt1: scattering bc}}& \makecell{current source\\ \cref{eq:chpt1: scattering source init again} or \\ \cref{eq:chpt1: hp current source}}\\ 
     \midrule
      \makecell{boundary\\matching\\\cref{eq:chpt1: general boundary matching}}  &  \cmark  &  \cmark \\
     \hline
      \makecell{PML\\\cref{eq:chpt1: wave equation PML} \& \\\cref{eq:chpt1: alpha pml}} & \xmark  & \cmark \\
    \bottomrule
    \end{tabular}
    \caption[Summary of methods for representing incident field and outgoing boundary conditions]{Summary of methods for representing the incoming wave $\psi^{\rm in}(\bx)$ [for scattering problems] and the outgoing boundary condition [for both scattering and eigenvalue problems].
    The table indicates cross-compatibility between the methods.
    }
    \label{tab:chpt1: Summary Table}
\end{table}

%We have reduced the full vectorial Maxwell equations to the scalar Helmholtz equation for scatterers with a translational symmetry, which describes the TM polarization for propagation within the plane transverse to the symmetry axis.
%We have described the 

%These are dielectric scatterers that are finite, time-invariant, and local in the spatial degrees of freedom.
%We will also require that the scatterers are linear in the electric field amplitude in this chapter and in \Cref{chp:chpt3: cpa ep}; we will address nonlinear susceptibilities in \Cref{chp:chpt4: saturable cpa,chp:chpt5: saturable scattering}.

\chapter{Reflectionless Scattering Modes \label{chp:chpt2: rsms}}
In this chapter we use the framework discussed in the previous chapter to address the case of reflectionless scattering from finite structures~\cite{sweeney_rsm_2019}.
Of course, reflectionless scattering in hermitian systems with parity symmetry is well known, as exemplified in the Fabry-P{\'e}rot resonator in optics, or the double barrier potential in quantum mechanics.
It is also widely known that when such systems lack parity symmetry, they generically do not have reflectionless resonances.  
These are textbook examples of resonant impedance-matching of waves incident on a structure; such examples are almost exclusively treated in one dimension. 
For the case of impedance matching to finite, lossless, one-dimensional structures, we will show that such states are bidirectional, meaning that they can be accessed from either side of the structure at the same frequency.
We will also present many-channel and higher-dimensional generalizations of this statement.
For a one-dimensional system with gain or loss this bidirectional property is lost, and the impedance-matched reflectionless state is unidirectional.

Much recent attention has been drawn to such unidirectional reflectionless states in ${\cal PT}$-symmetric optics~\cite{2011_Lin_PRL, 2011_Hernandez-Coronado_PLA, 2011_Longhi_JPA, 2012_Ge_PRA, 2012_Jones_JPA, 2013_Feng_nmat, 2014_Ramezani_PRL_2, 2014_Midya_PRA, 2015_Fleury_ncomms, 2016_Rivolta_PRA, 2016_Jin_SR, 2016_Chen_OE, 2016_Yang_OE, 2017_Sarisaman_PRA, 2018_Sarisaman_PRB}, as has been summarized in a recent review~\cite{2017_Huang_nanoph}.
For a one-dimensional, ${\cal PT}$-symmetric structure, one can show~\cite{2012_Ge_PRA} that reflectionless states must also have unit transmission, even though the system has both loss and gain.
While unidirectional resonances appear to be present in all ${\cal PT}$-symmetric structures that support resonances, they apparently disappear at sufficiently high frequency for a fixed value of gain-loss~\cite{2011_Hernandez-Coronado_PLA, 2011_Longhi_JPA, 2012_Ge_PRA, 2012_Jones_JPA, 2013_Mostafazadeh_PRA}.
One-way reflectionless states have also been studied in non-flux-conserving systems that do not possess ${\cal PT}$ symmetry~\cite{2013_Mostafazadeh_PRA, 2013_Castaldi_PRL, 2014_Chen_OE, 2014_Wu_PRL, 2015_Horsley_NP, 2015_Wu_PRA, 2017_Gear_NJP}.
A somewhat different kind of reflectionless state, known as a constant-intensity wave, has been identified as well~\cite{2015_makris_natcomm, 2017_Makris_LSA, 2018_rivet_natphys, 2019_Brandstotter_PRB}, but will not be discussed further here.

%Examples of reflectionless scattering from specific resonant structures can be found in the literature for ${\cal PT}$-symmetric optics~\cite{2011_Lin_PRL,2011_Hernandez-Coronado_PLA,2011_Longhi_JPA,2012_Ge_PRA,2012_Jones_JPA,2013_Feng_nmat,2014_Ramezani_PRL_2,2014_Midya_PRA,2015_Fleury_ncomms,2016_Rivolta_PRA,2016_Jin_SR,2016_Chen_OE,2016_Yang_OE,2017_Sarisaman_PRA,2018_Sarisaman_PRB}, summarized in the review~\cite{2017_Huang_nanoph}, where reflection is zero from one incident direction but not from the other, and the transmission is reciprocal and equal to unity~\cite{2012_Ge_PRA}. 
%While unidirectional reflectionless resonances are present in all ${\cal PT}$-symmetric structures, they apparently disappear at sufficiently high frequency for a fixed value of gain-loss~\cite{2011_Hernandez-Coronado_PLA,2011_Longhi_JPA,2012_Ge_PRA,2012_Jones_JPA,2013_Mostafazadeh_PRA}.
%One-way reflectionless states have also been studied in non-flux-conserving systems that do not possess ${\cal PT}$ symmetry~\cite{2013_Mostafazadeh_PRA,2013_Castaldi_PRL,2014_Chen_OE,2014_Wu_PRL,2015_Horsley_NP,2015_Wu_PRA,2017_Gear_NJP}.
%A somewhat different kind of reflectionless state, known as a constant-intensity wave, has been identified as well~\cite{2015_makris_natcomm, 2017_Makris_LSA, 2018_rivet_natphys, 2019_Brandstotter_PRB}, but will not be discussed further here.

The reflectionless states in the works cited above are special cases of a much more general phenomenon that is related to, but distinct from, resonances and CPA.
One can pose the most general kind of resonant impedance-matching problem in optics and acoustics as follows.
As discussed in \Cref{sec:chpt1: LSS and asymptotics} of the previous chapter, every finite cavity or scattering region will interact with a finite number $N$ of incoming and outgoing asymptotic channels.
$N$ will typically depend on the frequency of the monochromatic input wave. 
%We will assume that the output channels are simply the time-reverse of the input channels in the following discussion.
For such an open system, we can choose a subset of the asymptotic channels as input channels, and search for an input state with no back-reflection into the chosen channels.
Note that in general we do not require that the input and output channels are spatially partitioned into ``left'' and ``right'' groups, and, crucially, our definition of reflectionless states will apply in any dimension, not just one-dimensional or quasi-one-dimensional geometries.
Using this general definition of reflectionless or impedance-matched states, we will show that reflectionless scattering states exist at discrete complex frequencies that we refer to as {\it reflection zeros} ($R$-zeros).
Through parameter-tuning or by imposing symmetry, an $R$-zero can be moved to a real frequency to become a steady-state solution that we refer to as a {\it reflectionless scattering mode} (RSM).
This nomenclature is analogous to how the $S$-matrix zeros, generally complex, are called CPA frequencies when real.

We will put RSMs on a similar footing as the widely-used resonances: for a finite scatterer there is generally a countable infinity of such solutions at discrete complex-valued frequencies for each choice of the input/output sets.
Earlier works recognized these \nh\ reflectionless states for their unidirectional quality, but did not identify them as belonging to a broader class of electromagnetic eigenproblems, since these were not known prior to our work in Ref.~\cite{sweeney_rsm_2019}.
Independently of our work a more limited definition of the complex $R$-zeros for partitioned systems was given in Ref.~\cite{2018_Bonnet-BenDhia_PRSA}.

The unidirectional reflectionless states of one-dimensional ${\cal PT}$-symmetric structures have been characterized in some works as an EP of an unconventional non-symmetric scattering matrix with the transmission coefficients on the diagonal, and reflection coefficients on the off-diagonal~\cite{2007_Cannata_AoP,2013_Feng_nmat,2013_Mostafazadeh_PRA,2014_Wu_PRL,2014_Zhu_OE,2014_Kang_PRA,2017_Huang_nanoph,2018_Jiang_CPB}.
However, we will see using RSM theory that this point of view does not generalize to higher dimension, and that generic (non-degenerate RSMs) are not related to EPs.\footnote{There {\it is} an EP when two RSMs become degenerate, as we will discuss in \Cref{sec:chpt4: RSM EP}}.

We will provide two methods to find the $R$-zeros, one using a filtered scattering matrix and the other its underlying wave operator, and derive an explicit connection between the two.
We will analyze the role of symmetries, showing that flux-conserving cavities are bidirectional in the sense that input and output channels can be interchanged and the resulting state will also be an RSM at the same frequency, as noted above. 
RSMs of cavities with gain and/or absorption are generally unidirectional and do not satisfy this interchange symmetry.  
Non-flux-conserving systems with ${\cal PT}$ symmetry have the property that unidirectional $R$-zeros must occur at complex conjugate frequencies, i.e,~in pairs, unless they are at real frequencies.
We will see that this implies for small ${\cal T}$-breaking that $R$-zeros occur as RSMs on the real axis in ${\cal PT}$-symmetric systems.
This explains the widely observed existence of unidirectional unit-transmission resonances in one-dimensional ${\cal PT}$-symmetric systems, and their disappearance at high frequency as a result of spontaneous ${\cal PT}$ symmetry breaking.
A new type of exceptional point (EP) occurs at this transition, leading to an observable change in the reflection lineshape.
We will also briefly discuss EPs associated with a ${\cal PT}$ symmetry-breaking transition in cavities with both ${\cal P}$ and ${\cal T}$ symmetry, which occur without material gain or loss, unlike the typically studied ${\cal PT}$-symmetric case which has neither ${\cal P}$ nor ${\cal T}$ symmetries.

    \section{General Description of RSMs and $R$-zeros}

The RSM/$R$-zero concept applies to any open system in all dimensions, including systems which do not naturally divide into left and right asymptotic regions.
Any subset of asymptotic channels may be chosen as input with the complementary set being the output, even when the input-only and output-only channels spatially overlap. 
We will show that reflectionless scattering requires a specific monochromatic input wavefront, given by the eigenvector of a filtered scattering matrix with eigenvalue zero.
%We will show that a countably infinite number of $R$-zeros always exist, and in principle any desired subset can be found computationally, for each cavity or structure of interest.
RSMs arise when the parameters of the wave operator are tuned so that a specific $R$-zero occurs at a real frequency.
RSMs may be realized either by tuning parameters of the scatterer which do not break flux-conservation (index tuning), or by adding gain or absorption (gain-loss tuning).
In either case, for a given choice of input/output channels it takes the variation of only one parameter to make a chosen complex $R$-zero into a real RSM, accessible through steady-state excitation.
RSMs generalize to other resonant wave scattering systems, for example sound waves, waves of cold atoms or condensates, and other quantum systems, though we will focus here exclusively on electromagnetics.

$R$-zeros are fundamentally a resonance-like phenomenon, and can be engineered in a cavity or scatterer larger than the relevant wavelength.
They are similar to resonances in that they both occur at a discrete set of complex frequencies, but they differ in their boundary conditions.
For example, in a one-dimensional system, a left-incident $R$-zero solution is purely incoming in the left asymptotic region and purely outgoing in the right, while for resonances the solution is purely outgoing on both sides.
Though the boundary conditions are different between $R$-zeros and resonances, they correspond to the {\it same number of conditions} imposed infinitely far from the scatterer.
In this sense, the existence of $R$-zeros is no more extraordinary than the existence of resonances.

A recent work~\cite{2018_Bonnet-BenDhia_PRSA}, closely related to RSM theory, studied the $R$-zeros of obstacles in a few-mode acoustic waveguide. 
There, the outgoing boundary condition on the right is achieved using a perfectly matched layer (PML), and the incoming boundary condition on the left is achieved using a complex-conjugated PML.
Crucially, the tuning of $R$-zeros to the real axis to realize steady-state solutions (RSMs) was not discussed, so no explicit connection was made to using them as a photonic design tool. 
Moreover, the PML-based approach requires the input channels to be spatially separated from the output channels, limiting its applicability in arbitrary scattering geometries, such as: finite-sized scatterers in free space, where the asymptotic channels are angular-momentum states; planar scatterers illuminated with a finite numerical aperture, where the input channels consist of a subset of incident angles; or multimode waveguides with input and output channels occurring in the same waveguide.
The theory presented here is more general, applicable to any open system and for any choice of input/output channels.

While an RSM with a single input channel can be understood in terms of critical coupling to a cavity~\cite{Haus_book, Yariv:2000dz, 2000_Cai}, the general RSM concept is intrinsically multichannel.
RSMs can be designed to achieve perfect impedance-matching of coherent multichannel inputs or to achieve perfect mode-conversion, and hence has the potential to open up a very promising avenue for design of photonic structures.
In fact, unlike lasing or CPA, which by definition are non-unitary, i.e.,~not conserving energy flux, steady-state RSMs can be flux-conserving and do not require the availability of absorption or gain as a design resource.
We will discuss engineering of passive RSMs in \Cref{sec:chpt2: sym_T}.

    \section{Rigorous Definition \label{sec:chpt2: R-zeros}}

\begin{figure}[t!]
    \centering
    \includegraphics[width=\textwidth]{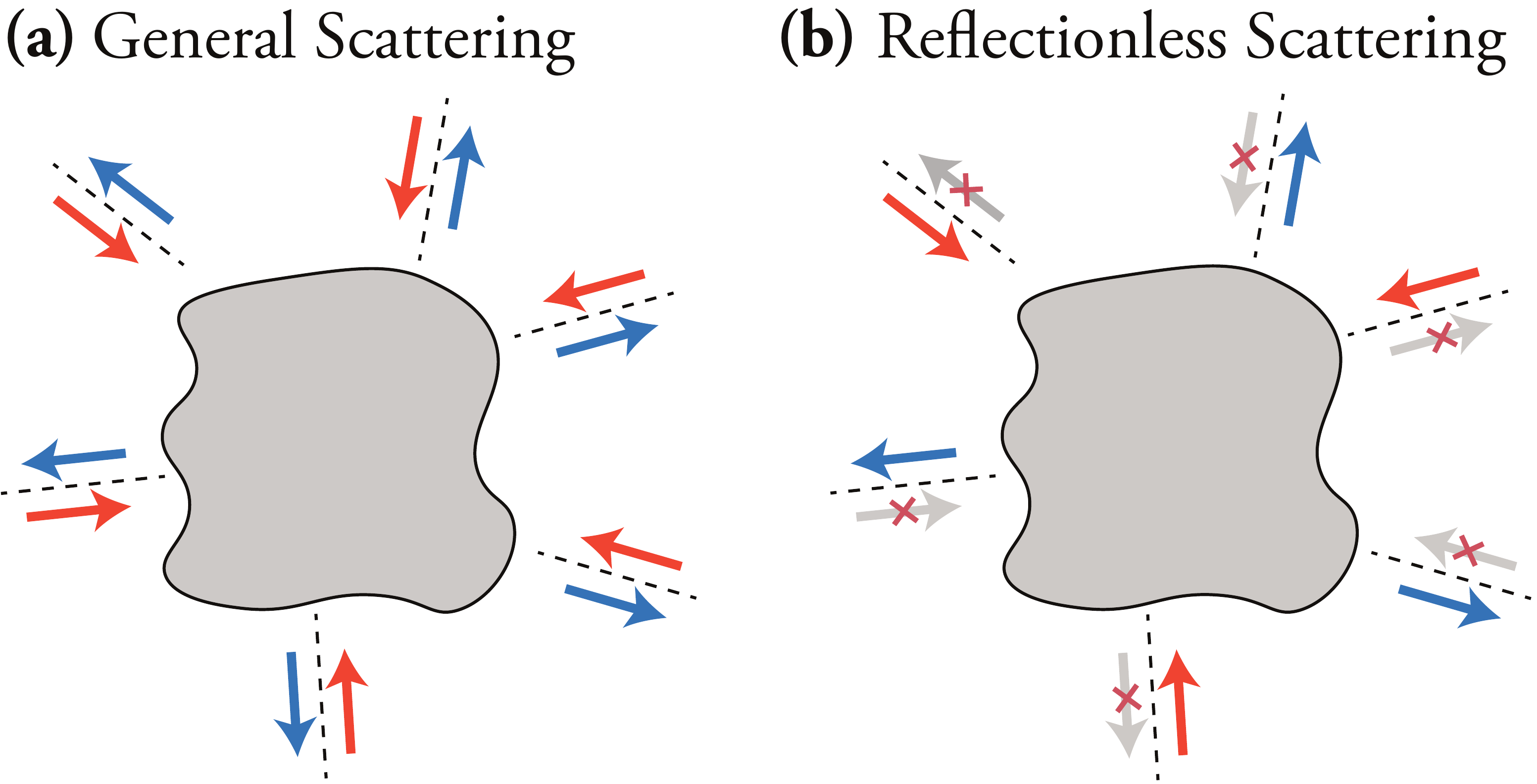}
    \caption[Comparison of general scattering process with reflectionless process]{Schematic depicting a general scattering process (a) and reflectionless process (b).
    A finite scatterer interacts with a finite set of asymptotic incoming and outgoing channels, indicated by the red and blue arrows, respectively, related by time-reversal.
    These channels may be localized in position space (e.g., waveguide channels) or in momentum space (e.g., angular-momentum channels).
    {\bf (a)}:~In the general case without symmetry, each incoming channel will scatter into all outgoing channels. %, including reflection back into the same channel(s).
    {\bf (b)}:~There exist reflectionless states, for which there is no reflection back into a chosen set of incoming channels (the inputs), which in general occur at discrete complex frequencies and do not correspond to a steady-state harmonic solution of the wave equation.
    However, with variation of the cavity parameters, a solution can be tuned to have a real frequency, giving rise to a steady-state reflectionless scattering process for a specific coherent input state, referred to as a reflectionless scattering mode (RSM).
    }
    \label{fig:chpt2: schematic}
\end{figure}

Consider a general scattering problem with $N$ asymptotic channels.
To define an $R$-zero problem, we specify $N_{\rm in}$ of the incoming channels to be ``input channels'' which carry incident flux but no outgoing flux, and are thus reflectionless, with $0<N_{\rm in}<N$.
The complementary set of $N_{\rm out}=N-N_{\rm in}$ outgoing channels (the ``output channels'') carry any outbound flux. 
This is illustrated in \Cref{fig:chpt2: schematic}b.

Let ${\bs \alpha}_{\rm in}$ denote the $N_{\rm in}$-component vector containing the input amplitudes of such a reflectionless incident field, and $R_{\rm in}(\w)$ denote the square $N_{\rm in}$-by-$N_{\rm in}$ {\it generalized reflection matrix}, which is the submatrix of the analytically continued $S(\w)$ defined by the specified input channels.
In other words,
\begin{equation}
    \label{eq:chpt2: Rinij}
    (R_{\rm in})_{i,j} (\w) \defn S_{n_i, n_j}(\w) \qquad \qquad  i,j=1,2,...,N_{\rm in}
\end{equation}
where $\{n_i\}$ is the set of $N_{\rm in}$ channels.
We use the term generalized reflection matrix because typically reflection and transmission are concepts applied to partitioned systems, in which scattering occurs between two sets of asymptotic channels which are spatially separated at infinity.
Here, reflection is defined as any scattering back into the chosen set of channels, which may or may not be spatially overlapping at infinity.
All previous studies have only considered partitioned systems. 
%We say $R_{\rm in}$ is generalized because the input channels are not required to be spatially distinct, so that it is not limited to, say, leftward or rightward scattering.
At the complex frequency $\w_{\rm RZ}$ of an $R$-zero, the absence of reflection can be expressed as
\begin{equation}
    \label{eq:chpt2: RSM definition}
    R_{\rm in}(\w_{\rm RZ}){\bs \alpha}_{\rm in} = {\bf 0}, 
\end{equation}
which is the formal definition of an $R$-zero.
In other words, the $R$-zero frequencies are those at which $R_{\rm in}(\w)$ has a zero eigenvalue, with the corresponding eigenvector ${\boldsymbol \alpha}_{\rm in}$ defining the reflectionless incident wavefront.

The simultaneous absence of reflection in all input channels for the $R$-zero incident wavefront is due to interference: the reflection amplitude of each input channel $i$ destructively interferes with the inter-channel scattering from all the other input channels, $(R_{\rm in})_{ii}{\bs \alpha}_i+\sum_{j\neq i} (R_{\rm in})_{ij}{\bs \alpha}_j=0$, which is precisely \cref{eq:chpt2: RSM definition}.
The transmission into the output channels is not obtained from solving this equation alone, and must be determined by solving the full scattering problem at $\w_{\rm RZ}$.

One way to solve \cref{eq:chpt2: RSM definition} is to find the roots of the complex scalar equation
\begin{equation}
    \label{eq:chpt2: det_Rin_zero}
    \det R_{\rm in}(\omega_{\rm RZ}) = 0,
\end{equation}
and then diagonalize $R_{\rm in}$ at the frequencies $\w_{\rm RZ}$.
Since the generalized reflection matrix can be computed for any open scattering system using the methods outlined in \Cref{chp:chpt1: theory of scattering}, \cref{eq:chpt2: RSM definition,eq:chpt2: det_Rin_zero} provide universal recipes for solving the $R$-zero problem, for each of the $2^{N}-1$ choices of reflectionless input channels.
The case $N_{\rm in}=0$, corresponding to resonance, is excluded from the RSM hierarchy, as here $R_{\rm in}$ and the associated $R$-zeros are not defined.

In the special case with $N_{\rm in}=N$ and $R_{\rm in}=S$, the $R$-zeros are just the $S$-matrix zeros, and RSM corresponds to CPA.
The CPA-RSM case is special among the RSM problems in that only for this case is it necessary to break flux conservation by adding absorption in the medium.
Moreover, for CPA, one must control and (typically) excite all input modes in order to achieve perfect absorption, whereas every other RSM only requires controlling a subset of the scattering channels, making this much more practical to achieve in complex open geometries.

%Even though the case of $N_{\rm in}=N$ leads to a well-defined problem, $R_{\rm in} = S$, we exclude it from the $R$-zero framework since the corresponding complex $\w_{\rm RZ}$ is already understood as a zero of the $S$-matrix~\cite{2010_Chong_PRL}, and for flux-conserving systems is constrained to the upper half-plane.

We will show in Section~\ref{sec:chpt2: rigorous RSM} that $\det R_{\rm in}(\omega)$ is a rational function of $\w$.
Therefore, near each zero, to leading order, $\det R_{\rm in}(\w) \propto (\w - \w_{\rm RZ})$, so that the phase angle $\arg(\det R_{\rm in})$ winds by $2\pi$ along a counterclockwise loop around each $\w_{\rm RZ}$ in the complex-frequency plane.
This shows that each $R$-zero is a topological defect in the complex-frequency plane with a topological charge of $+1$~\cite{1979_Mermin_RMP}.
As such, $R$-zeros are robust: when the optical structure $\e(\bx)$ is perturbed, $\w_{\rm RZ}$ moves in the complex plane but cannot suddenly disappear, similar to topological defects in other systems~\cite{2014_Zhen_PRL,2017_Guo_PRL}.
For an $R$-zero to disappear, it must annihilate with another topological defect with charge $-1$, which we will see is a pole of the $S$-matrix.
For ${\cal T}$-symmetric systems, the zeros and poles can only annihilate on the real-axis, which corresponds to a bound state in the continuum (BIC)~\cite{2016_Hsu_NRM}.
When parameters of the system are tuned such that $n>1$ $R$-zeros meet at the same frequency, they superpose and form a topological defect with charge $+n$ where $\det R_{\rm in} \propto (\w - \w_{\rm RZ})^n$; these are exceptional points (EPs) of a wave operator.
We will return to this in \Cref{sec:chpt4: RSM EP}, and provide an explicit example in \vref{fig:chpt2: PT}.

    \section{Analytic Properties \label{sec:chpt2: rigorous RSM}}

While \cref{eq:chpt2: RSM definition,eq:chpt2: det_Rin_zero} already provide recipes for constructing $R$-zeros, we will further analyze $\det R_{\rm in}(\omega)$ to shed light on their robust existence and to reveal their relation to the wave operator and to resonances.
We adapt the effective Hamiltonian formalism from \Cref{sec:chpt1: effective hamiltonian} of the previous chapter to treat $R$-zeros, and analyze them in a fashion similar to resonances through the relation between the generalized reflection matrix $R_{\rm in}$ and the $S$-matrix.

Let $C_F$ be the set of $N_{\rm in}$ filled input channels, which fixes $C_{\fbar}$, the set of $N_{\rm out}$ filled output channels: % (the remaining incoming and outgoing channels will carry no flux):
\begin{align}
    \label{eq:chpt2: F defs}
    C_F = \{ i:\ \text{channel }i \text{ is input}\}, \qquad \quad &|C_F|=N_{\rm in}\\
   C_{\fbar} = \{ i:\ \text{channel }i \text{ is output}\}, \qquad \quad &|C_{\fbar}|=N_{\rm out}, \\
   N_{\rm in} + N_{\rm out} = N
\end{align}
To extract $ R_{\rm in}$ from $S$, let us define the filtering matrices $F$ and $\fbar$, where $ F:{\mathbb C}^N\to{\mathbb C}^{N_{\rm in}}$ reduces the dimension of the channel space from $N$ to $N_{\rm in}$, i.e., 
\begin{equation}
    \label{eq:chpt2: def of F}
    F \in {\mathbb C}^{N_{\rm in} \times N},\qquad F_{ij} \defn \delta_{c(i),j} {\rm\ \ where\ } c(i) {\rm\ is\ the\ } i^{\rm th} {\rm\ element\ of\ } C_F.
\end{equation}
Likewise $\fbar$ reduces the channel space from $N$ to $N_{\rm out}$.
It follows that 
\begin{gather}
    FF^\dagger =  I_{N_{\rm in}},  \label{eq:chpt2: FFdagger}\\
    \fbar \fbar^\dagger=  I_{N_{\rm out}}, \label{eq:chpt2: barFbarFdagger}\\
    F^\dagger  F+\bar F^\dagger \bar F= I_N. \label{eq:chpt2: sum of filters}
\end{gather}
With these definitions we have
\begin{equation}
    \label{eq:chpt2: Rin_S}
    R_{\rm in}(\w) = FS(\w) F^\dagger.
\end{equation}
Using \cref{eq:chpt2: Rin_S} in \cref{eq:chpt1: Heidelberg},
\begin{gather}
    R_{\rm in} =  I_{N_{\rm in}} - 2 \pi i W_F^\dagger G_{\rm eff} W_F, \label{eq:chpt2: Rin first}\\
    W_F \defn W_p  F^\dagger,
    \qquad W_{\fbar} \defn W_p\bar F^\dagger,
\end{gather}
where $G_{\rm eff}$ is the effective Green function~\cref{eq:chpt1: Aeff_and_Sigma}, and $W_p$ is the coupling matrix of the interior degrees of freedom to the propagating asymptotic channels, as introduced in \Cref{chp:chpt1: theory of scattering}.

Define the retarded self-energy, restricted to the input channels, as
\begin{gather}
    \label{eq:chpt2: RSM self-energy}
    \Sigma^R_F(\w) \defn \Delta_F(\w) - i \G_F(\w), \\
    \Delta_F(\w) \defn {\rm p.v.} \int d \w\pr \, \frac{ W_F(\w\pr) W_F^\dagger(\w\pr)}{\w\pr-\w}, \\
    \Gamma_F(\w) \defn \pi W_F(\w) W_F^\dagger(\w),
\end{gather}
and similarly for $\fbar$ and the output channels.
Using these definitions with \cref{eq:chpt2: sum of filters} leads to the identities
\begin{equation}
    \label{eq:chpt2: F + Fbar matrix id}
    {\mathscr M}_F + {\mathscr M}_{\fbar} = {\mathscr M},
\end{equation}
where ${\mathscr M}$ is one of $\Sigma^R$, $\Delta$, or $\G$.

There is a representation of $R_{\rm in}$ that is analogous to the $K$-matrix representation of $S$ (see \Cref{chp:app: linalg}),
\begin{equation}
    R_{\rm in} = ( I_{N_{\rm in}} + i K_F)/( I_{N_{\rm in}} - i K_F),
\end{equation}    
where
\begin{gather}
    K_F \defn -\pi W_F^\dagger \bar{G}_0 W_F, \\
    \bar G_0 \defn  [A_0\pr - \Delta + i \G_{\fbar}]^{-1}.
\end{gather}
From this follows the determinant relation
\begin{equation}
    \label{eq:chpt2: det(R)}
    \det R_{\rm in}(\w) = \frac{\det{\bs(} A_0\pr(\w) - \Delta(\w) - i[ \G_F(\w) - \G_{\fbar}(\w)]{\bs)}}{\det{\bs(} A_0\pr(\w) - \Delta(\w) + i \G{\bs)}}.
\end{equation}
This equation for $R_{\rm in}$ relates the determinant of an $N_{\rm in}$-by-$N_{\rm in}$ matrix to a ratio of functional determinants, similar to \cref{eq:chpt1: det(S)} for $\det S(\w)$.
In particular, \cref{eq:chpt2: det(R)} shows that $\det R_{\rm in}(\w)$ is a rational function of $\w$, which provides a rigorous basis for the topological-charge interpretation given earlier at the end of \Cref{sec:chpt2: R-zeros}.

\Cref{eq:chpt2: det(R)} is a central result of this chapter.
It relates the $R$-zeros to eigenvalues of a new effective wave operator
\begin{equation}
    \label{eq:chpt2: A_RZ}
    \hat A_{\rm RZ}(\w) \defn \hat A_0(\w) - \hat \Delta(\w) - i[\hat \G_F(\w)-\hat \G_{\fbar}(\w)].
\end{equation}
Whenever $\det  R_{\rm in}(\w_{\rm RZ})=0$ at an $R$-zero, then also $\det \hat A_{\rm RZ}(\w_{\rm RZ}) = 0$.
This defines a nonlinear eigenvalue problem (NEVP) with $\w_{\rm RZ}$ as the eigenvalue
\begin{gather}
    \label{eq:chpt2: det(A)=0}
    \hat A_{\rm RZ}(\omega_{\rm RZ}) \ket{\omega_{\rm RZ}} = 0,
\end{gather}
which can be solved by several standard methods~\cite{2000_Golub_JCAM, 2004_Voss_BIT, 2009_asakura_jsiam, 2011_Su_SIAM, 2012_Beyn_LA}.
In going from \eqref{eq:chpt2: det(R)} to \eqref{eq:chpt2: det(A)=0}, we have re-interpreted the infinite-dimensional matrix as the differential operator that it represents.
For each choice of input channels $C_F$, we expect that there exists a countably infinite set of $R$-zeros at complex frequencies $\{\omega_{\rm RZ}\}$ that satisfy \cref{eq:chpt2: det(A)=0}.
This is similar to resonances, which satisfy the NEVP $\hat{A}_{\rm eff}(\w_p) \ket{\w_p} = 0$.
The denominator of \cref{eq:chpt2: det(R)} for $\det R_{\rm in}(\w)$ is the same as that of \cref{eq:chpt1: det(S)} for $\det S(\w)$, while the numerators differ, implying that the poles (resonances) of $ R_{\rm in}(\w)$ are the same as those of $S(\w)$, while the zeros are different.

A caveat of using the functional determinant \cref{eq:chpt2: det(A)=0} as opposed to finite matrix determinant \cref{eq:chpt2: RSM definition} is that while every $R$-zero corresponds to an eigenmode of $\hat A_{\rm RZ}$, some eigenmodes of $\hat A_{\rm RZ}$ may not be $R$-zeros, which is similar to the case for $S(\w)$, discussed in the previous chapter.
This can be seen from \cref{eq:chpt2: det(R)}, which is not generally zero if both the numerator and the denominator on the RHS are zero, so that we have a simultaneous eigenmode of both $\hat A_{\rm RZ}$ and $\hat A_{\rm eff}$.
This will rarely happen without symmetries in the system, but the bound state in the continuum (BIC) of~\cite{2016_Hsu_NRM} is an example, as pointed out in Ref.~\cite{2018_Bonnet-BenDhia_PRSA} in a more specialized context.
In addition, resonances that do not radiate into certain channels~\cite{2014_Ramezani_PRL_2,2016_Zhou_Optica, 2019_Yin_arXiv}---sometimes referred to as ``unidirectional BICs''---are also such simultaneous eigenmodes.
BICs and unidirectional BICs typically require additional parameter tuning beyond the basic RSM problem~\cite{2016_Hsu_NRM} since they have both no incoming {\it and} no outgoing waves in the dark channels.

\Cref{eq:chpt2: det(R),eq:chpt2: A_RZ,eq:chpt2: det(A)=0} show that $\hat A_{\rm RZ}$ is a sum of operators, including the difference of $\hat \G_F$ and $\hat \G_{\fbar}$, which are associated with the inputs and outputs, respectively.
Since the operators do not in general commute, the eigenvalues of the $\hat A_{\rm RZ}$ will not be a simple function of the eigenvalues of each term.
Nonetheless, since $\hat \G_F$ and $\hat \G_{\fbar}$ are positive semidefinite, this formula still provides an important intuition: for the purposes of reasoning about the zeros of $ R_{\rm in}(\w)$, each input channel acts as an effective ``radiative gain'' to the system (because it contributes a negative semidefinite term to $\hat A_{\rm RZ}$, which then increases the imaginary part of its eigenvalue $\w_{\rm RZ}$), and each output channel as an effective ``radiative loss''.
The balance of these two terms will determine the proximity of an $R$-zero to the real-frequency axis, where it becomes an RSM.
Hence when defining an $R$-zero problem in which few input channels scatter to many output channels, we expect the complex-valued $R$-zero frequency to appear in the lower half-plane for a passive system (similar to resonances).
To realize a steady-state solution we can either add gain in the dielectric function of the operator $\hat A_0$ (gain-loss engineering) or modify the scattering structure to increase the coupling of the input channels (index tuning) so as to move the solution to a real frequency.
In the opposite case of many more input channels than output channels, we expect the $R$-zero frequency to appear in the upper half-plane, requiring the addition of absorption or an increase in the coupling to the output channels to reach the real axis.

As noted, this argument is merely qualitative, since the operator in \cref{eq:chpt2: A_RZ} is the sum of non-commuting terms, so that the imaginary part of the eigenvalue is not simply the sum of contributions from each term.
However this qualitative picture will be confirmed by the approximate coupled-mode analysis of \cref{sec:chpt2: RSM_TCMT} and by finite-difference numerical solutions of several examples (see \cref{fig:chpt2: fp,fig:chpt2: octopus,fig:chpt2: free_space}).
In the important special case in which only one cavity resonance is dominant, the intuitive picture becomes precise.
The imaginary part of the $R$-zero is simply the difference of the total input coupling rate (i.e.,~the sum of input couplings) and the total output coupling rate, as one might guess.
RSM is achieved when these terms are tuned to cancel.

%There is an important difference between the zeros of $S$ and of $R_{\rm in}$.
As noted earlier, flux conservation means that the totally absorbed steady-state CPA cannot be realized without introducing absorption.
On the other hand, flux conservation does not prevent the existence of $R$-zeros on the real axis (RSMs) since the incident flux can be redirected to the $C_{\fbar}$ output channels, except in the case of CPA, where there are no output channels.
Mathematically, $S$ is unitary for lossless systems at real frequencies, with unimodular eigenvalues, but $R_{\rm in}$ is not, and may have a zero even for real frequency.
This is why RSMs can be achieved via pure index tuning (except for CPA), even though lasing cannot; we will give explicit examples later (\cref{fig:chpt2: free_space}, \cref{fig:chpt2: octopus}, and \cref{fig:chpt2: fp}e).

    \section{$R$-zeros as Eigenmodes of a Wave Operator \label{sec:chpt2: RSM_wave_operator}}

The distinguishing feature of the wave operator $\hat A_{\rm RZ}(\w)$, as defined in \cref{eq:chpt2: A_RZ}, is the specific frequency-dependent self-energy, formally defined by \cref{eq:chpt2: RSM self-energy}, that acts on the surface of last scattering and imposes the outgoing boundary conditions for the channels in $C_{\fbar}$ and incoming $C_F$.
Below we describe two implementations, one based on perfectly matched layers (PMLs) and the other based on an extension of the boundary-matching method detailed in \Cref{chp:chpt1: theory of scattering}.

    \subsection{PML-based Boundary Conditions \label{sec:chpt2: BC_PML}}

PMLs can be used to create an effective outgoing or incoming boundary condition, as discussed in \Cref{sec:chpt2: PMLs} of the previous chapter.
So long as the susceptibility is non-dispersive, i.e.,~is independent of frequency, the resonance eigenvalue problem with PMLs is an ordinary eigenvalue problem, instead of the nonlinear eigenvalue problem (NEVP) that arises when boundary-matching is used.
PMLs are still valid with dispersive scattering, but in this case, they will not provide the particular advantage just mentioned.
%An incoming boundary condition, while less common, can be similarly implemented with a Dirichlet boundary and the complex-conjugate of a PML~\cite{2018_Bonnet-BenDhia_PRSA}.

The $R$-zero NEVP can likewise be solved using a mixture of PMLs and their conjugates, however this requires that the incoming and outgoing channels are spatially separated.
Each asymptotic channel is truncated with a Dirichlet boundary which is transverse to the direction of propagation, and padded with a PML if the channel is outgoing, and a conjugated PML if incoming. 
For example, in a quasi-one-dimensional waveguide, a left-to-right $R$-zero is an eigenmode of the wave operator with a conjugated PML on the left and a conventional PML on the right (and vice versa for a right-to-left $R$-zero); this case was recently described in Ref.~\cite{2018_Bonnet-BenDhia_PRSA}. 
Another example, a chaotic six-waveguide junction with an RSM, is shown in \vref{fig:chpt2: octopus}.
As mentioned in \Cref{chp:chpt1: theory of scattering}, both conventional and conjugated PMLs must be used with caution, as they introduce additional modes in the eigenvalue spectrum~\cite{2018_Lalanne_LPR}.

    \subsection{Mode-Matching Boundary Conditions \label{sec:chpt2: BC_MM}}

\begin{figure}[t!]
    \centering
        \includegraphics[width=\textwidth]{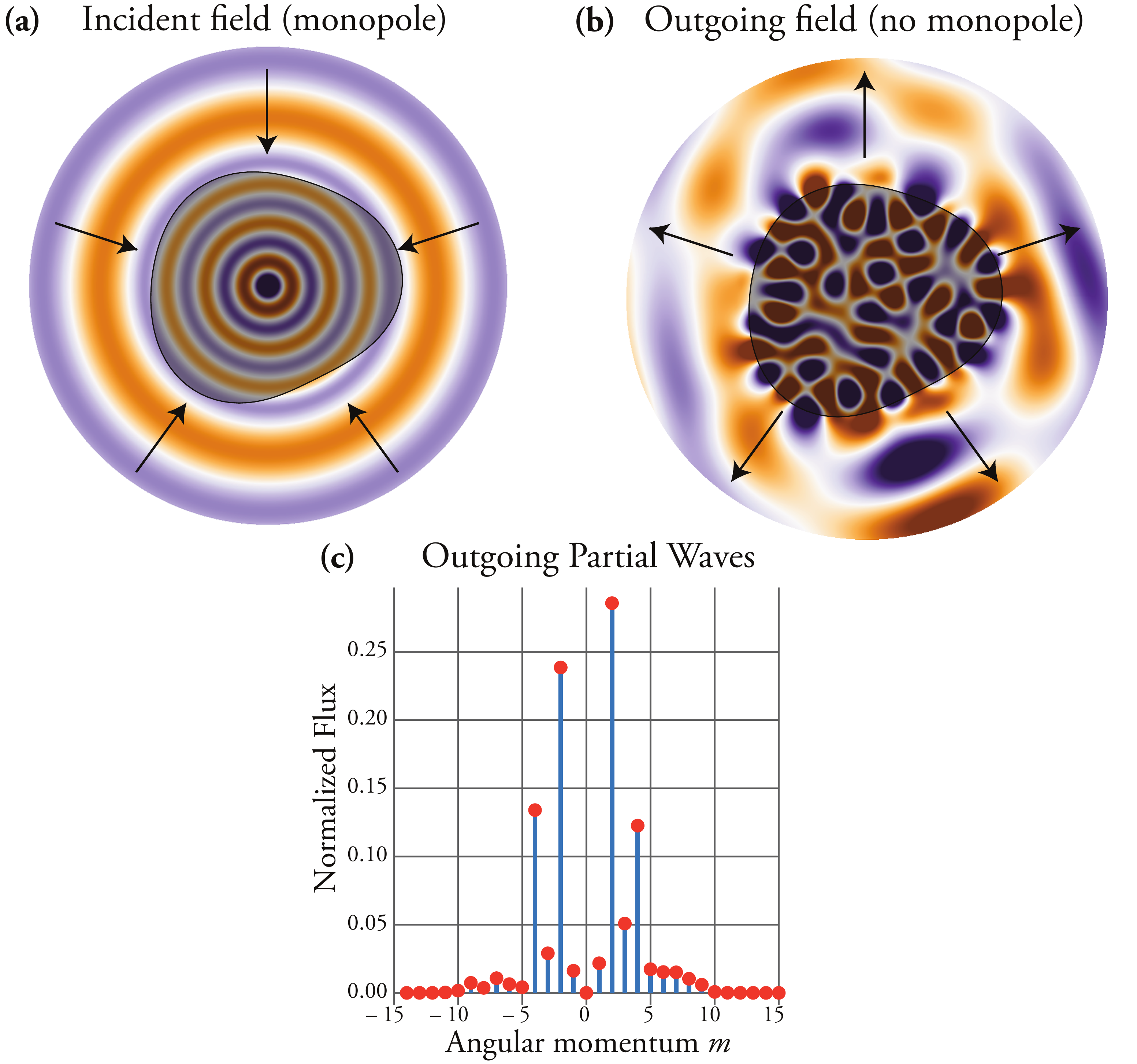}
    \caption[Reflectionless scattering from a deformed disk resonator]{Flux-conserving reflectionless scattering of a monopole scalar wave from a deformed disk dielectric resonator ($n=3$, mean radius $\bar R$), shown as shaded region in (a,b).
    The disk boundary is $r(\theta)/\bar R=1+x \{.8528\cos(2[\theta-.3127])+.8346\cos(3[\theta-.0079])\}$, with $x=.2768$.
    Monopolar RSM was achieved by tuning the dimensionless parameter $x$ until a given complex $R$-zero became real, at frequency $\w_{\rm RSM}\bar R/c=6.4853$.
    {\bf (a)}:~The input field only has an $m=0$ component (pure monopole).
    {\bf (b)}:~The outgoing field contains no $m=0$ component; it is a coherent superposition of many other $m$'s.
    {\bf (c)}:~The outgoing flux, normalized by the net incident flux, for each angular momentum channel $m$. There is none for $m=0$ because the structure was shape-tuned to a monopolar RSM.
    Since the refractive index is real, the RSM we have found is time-reversible, which implies that if we time-reversed the complicated superposition of outgoing multipole radiation, the incident wavefront would interfere perfectly to generate monopole-only outgoing radiation.%The $R$-zero nonlinear eigenvalue problem here was solved using NEP-PACK~\cite{2017_Bezanson_SIAM,2018_Jarlebring_arxiv}.
    }
    \label{fig:chpt2: free_space}
\end{figure}

The PML-based method fails for $R$-zero problems for which the asymptotic input channels are not spatially disjoint from the output channels.
Examples where PMLs cannot be used are a multimode waveguide in which some of the chosen input and output channels are on the same side of the scatterer, and any free-space RSM problem with at least one input, for which the asymptotic channels are incoming and outgoing spherical or cylindrical waves with different angular momenta, as in \vref{fig:chpt2: free_space}.
In these cases, we must do boundary-matching channel by channel, assigning the appropriate incoming or outgoing condition for that geometry.

The mode-matching boundary condition for the $R$-zero problem is a straightforward generalization of \cref{eq:chpt1: general boundary matching}:
\begin{equation}
    \label{eq:chpt2: boundary matching}
    \left[ \psi(\bx) - \oint_{\partial\Omega} [G^{A,N}_F(\bx,\bx\pr)+G^{R,N}_{\fbar}(\bx,\bx\pr)] \nabla\psi(\bx\pr)\cdot d {\bf S}\pr \right]_{\bx \in \partial\Omega} = 0,
\end{equation}
where $G^{A,N}_F$ is the advanced Green function restricted to $C_F$, and $G^{R,N}_{\fbar}$ is the retarded Green function restricted to $C_{\fbar}$, both subject to a Neumann condition at the LSS.
%Presumably the full vectorial wave equation, it would use the appropriate Maxwell interface boundary conditions.
The integral is over the last scattering surface, $\partial\Omega$, which separates the scattering region $\Omega$ from the asymptotic regions $\bar\Omega$. 
\Cref{eq:chpt2: boundary matching} follows from a derivation similar that in \Cref{chp:app: boundary matching}, but using the advanced Green function when inverting $\hat A_c(\w)$ in the input channels, appropriate for incoming radiation.

We summarize the $R$-zero boundary conditions for the same three geometries as in \Cref{chp:chpt1: theory of scattering}.
\begin{enumerate}
    
    \item In one dimension, with the surface of last scattering $|x|=a$:
    \begin{align}
        \label{eq:chpt2: d=1 bc}
        \text{left-incident: } \psi(\pm a)& - \partial_x \psi(\pm a)/ik = 0\\
        \text{right-incident: } \psi(\pm a) & + \partial_x \psi(\pm a)/ik = 0,
    \end{align}
    where $k=\w/c$.

    \item For a two-dimensional perfect (lossless) metallic waveguide with transverse width $t$ and surface of last scattering $|x|=a$:
    \begin{equation}
        \psi(\pm a,y) = \mp \int dy^\prime K_{F_\pm}(y,y^\prime)\ \partial_x\psi(\pm a,y^\prime)
    \end{equation}
    The kernel $K_F$ is
    \begin{equation}
        K_F(y,y\pr) = \sum_{m \in C_F}g^-_m(y,y\pr) + \sum_{m \notin C_F}g^+_m(y,y\pr),
    \end{equation}
    where $C_{F_-}$ is the set of input channels for the left lead, and $C_{F_+}$ for the right, and
    \begin{equation}
        g^\pm_m(y,y\pr) = \pm\frac{1}{i\beta_m^\pm(\w)}\sin\Big(\frac{m \pi y}{t}\Big) \sin\Big(\frac{m \pi y\pr}{t}\Big).
    \end{equation}
    The propagation constant $\beta^\pm_m(\w)$ is the same as before, in \cref{eq:chpt1: beta_m}:
    \begin{align}
        \label{eq:chpt2: beta_m}
        \beta^\pm_m (\w)&=\sqrt{\left(\frac{\w}{c}\right)^2-\left(\frac{m\pi}{t}\right)^2\pm i0^+}.
    \end{align}
    The square-root branch cut is the conventional one along the negative real axis, such that $\sqrt{-1 + i0^\pm}=\pm i$.
    The contribution of each propagating mode to $K_F$ has a sign which depends on whether the mode is designated as input or output, while the non-propagating modes all have the same sign, regardless of $C_F$.
%    For {\it complex} $\w$, the propagation constant $\beta_m$ can simply be taken as the analytic continuation of \cref{eq:chpt2: beta_m}.
%    However, some NEVP solution methods require an integral over a loop in the complex frequency plane~[CITE], which is ambiguous if this loop contains the branch point of $\beta_m^\pm(\w)$, which is likely when near a mode cutoff.
%     The ambiguity can be avoided by traversing the loop twice; further analysis is beyond the scope this thesis.
    
    \item A finite scatterer in two-dimensional free-space, with surface of last scattering $|\bx|=R$, has asymptotic channels specified by angular momentum $m$. The $R$-zero boundary condition is
    \begin{equation}
        \psi(R,\phi) = \sum_{m} \frac{e^{im(\phi-\phi\pr)}}{2\pi k^F_m(R)} \partial_r \psi(R,\phi\pr),
    \end{equation}
    where 
    \begin{equation}
        k^F_m(r) = 
        \begin{cases}
        \partial_r \ln H^+_m(kr),&m\in F\\
        \partial_r \ln H^-_m(kr),&m\notin F.
        \end{cases}
        \label{eq:chpt2: d=2 bc}
    \end{equation}
    $H_m^{\pm}(x)$ are the Hankel functions of the first ($-$) and second ($+$) kind (outgoing and incoming, respectively) of order $m$.

\end{enumerate}

We provide an example in \cref{fig:chpt2: free_space} for which the mode-matching approach must be used: a two-dimensional deformed dielectric resonator in free space, which has been shape-tuned to have an RSM for monopolar input.
The theory implies that one can perfectly impedance-match a specific superposition of any number of coherent input multipoles to the remaining output multipoles.
Thus the RSM-tuned scatterer acts as a perfectly multipole-transforming antenna, with the limitation that one can choose only the entire set of inputs to be transformed, and cannot specify the transformation to be between a specific input and output channel, e.g.~a given multipole input to a given multipole output.
In this example, we were able to find a real frequency at which the monopole input was reflectionless by tuning a single deformation parameter, so that all of the scattered waves were in higher multipoles.
Through further optimization of the shape one could presumably enhance the scattered output into certain desired outgoing channels. 
Note that the scattering here is not perturbative and the output is not simply determined by a single scattering event from a particular multipole of the deformation.

That the single parameter $x$, which controls the overall strength of the boundary deformation, is a candidate for RSM tuning can be understood as follows.
Define a deformation parameter such that $x=0$ corresponds to the circular disk.
In the limit that $x\to 0$, continuous rotational symmetry is restored, the angular momentum scattering channels labeled by $m$ decouple from each other, and the scattering matrix is diagonal.
Therefore for $x=0$, the $R$-zero frequencies are also $S$-matrix zeros, which are constrained by flux conservation to be in the upper half plane.
On the other hand, as the deformation $x$ is made larger, the $m$ channels become increasingly mixed, so that the single monopole input becomes coupled to more and more multipole outputs.
The radiative loss $\G_{\fbar}$ will eventually overtake the radiative gain $\G_F$ and push the $R$-zeros into the lower half-plane.
Therefore by continuity there will be a deformation strength $x_0$ at which the $R$-zero crosses the real axis, becoming an RSM.

    \section{Temporal Coupled Mode Theory of $R$-zeros \label{sec:chpt2: RSM_TCMT}}

The results from TCMT seen in \Cref{chp:chpt1: theory of scattering} can be adapted to $R$-zeros to derive important insights about their behavior in the case of isolated resonances.

One way to derive a TCMT of $R$-zeros is to apply the same manipulations that were used to derive \cref{eq:chpt2: det(R)} for $\det R_{\rm in}(\w)$, but starting with \cref{eq:chpt1: S_form1} for the TCMT $S$-matrix.
However, we will use this as an opportunity to provide a complementary approach: define 
\begin{equation}
    D_{\rm in} \defn F D
\end{equation}
as the coupling coefficients of the $M$ resonances into the $N_{\rm in}$ input channels of $C_F$, and similarly $D_{\rm out}$ as the coefficients to the output channels.
$F$ is the filtering matrix introduced earlier in \cref{eq:chpt2: def of F}.
Using \cref{eq:chpt2: Rin_S}, \cref{eq:chpt1: S_form1}, with $S_0= I_{N}$, and the Woodbury matrix identity~\cite{1950_Woodbury_MR}, we can write the inverse of the generalized reflection matrix as
\begin{equation}
    \label{eq:chpt2: S_in_inverse}
    R_{\rm in}^{-1}(\w) = I_{N_{\rm in}} + i D_{\rm in} \frac{1}{\w - H_{\rm RZ}} D_{\rm in}^{\dagger}.
\end{equation}
The new matrix
\begin{equation}
    \label{eq:chpt2: H_RSM}
    H_{\rm RZ} \defn H_0 + i \frac{D_{\rm in}^\dagger D^{\phantom{\dagger}}_{\rm in}}{2} - i \frac{ D_{\rm out}^\dagger D^{\phantom{\dagger}}_{\rm out}}{2}
\end{equation}
is the $R$-zero Hamiltonian within TCMT, with $(\w - H_{\rm RZ})$ being the analog of $\hat A_{\rm RZ}(\w)$ in \cref{eq:chpt2: A_RZ}.
$R_{\rm in}(\w_{\rm RZ})$ by definition has no inverse, and according to \cref{eq:chpt2: S_in_inverse} this can only happen when $\w$ is an eigenvalue of $H_{\rm RZ}$.
Thus every $R$-zero frequency is necessarily an eigenvalue of $H_{\rm RZ}$.
However the reverse is not true: it is possible for $|| D^{\phantom{\dagger}}_{\rm in}(\w_{\rm RZ} - H_{\rm RZ})^{-1} D_{\rm in}^\dagger||<\infty$, which happens when the eigenmode ${\bf a}$ of $H_{\rm RZ}$ associated with eigenvalue $\w_{\rm RZ}$ satisfies $D_{\rm in} {\bf a} = {\bf 0}$.
Within TCMT, this is the condition for a BIC or a one-sided resonance, consistent with what was mentioned near the end of \Cref{sec:chpt2: rigorous RSM}.

A similar intuition holds for the matrix $H_{\rm RZ}$ as did for $\hat A_{\rm RZ}$.
Outgoing radiation into the $N_{\rm out}$ output channels introduces a lossy matrix $- i D_{\rm out}^\dagger D^{\phantom{\dagger}}_{\rm out}/2$ in the $R$-zero Hamiltonian, while incident light coming from the $N_{\rm in}$ input channels introduces a gain term $+i D_{\rm in}^\dagger D^{\phantom{\dagger}}_{\rm in}/2$.

    \subsection{$R$-zeros of Isolated High-\Q\ Resonances \label{sec:chpt2: high-Q}}

The TCMT analysis of $R$-zeros becomes especially simple for resonances that are well separated compared to their linewidths, which is typical for high-\Q\ cavities.
In this case there is only one dominant resonance in the frequency range of interest, so that $H_0$ is approximated by a complex scalar
\begin{equation}
    H_0 = \w_0 - i\g_0,
\end{equation} 
where $\g_0$ is non-radiative, due solely to absorption or gain within the closed cavity.
The $R$-zero frequency from \cref{eq:chpt2: H_RSM} is
\begin{gather}
    \label{eq:chpt2: omega_RSM_single_mode}
    \w_{\rm RZ} = \left( \w_0-i \g_0 \right) + i \left( \g_{\rm in} - \g_{\rm out} \right), \\
    \g_{\rm in} \defn \sum_{n \in C_F} |d_n|^2/2,
    \quad \g_{\rm out} \defn \sum_{n \notin C_F} |d_n|^2/2,
\end{gather}
where $d_n$ is the coupling coefficient of the isolated mode to the $n$-th asymptotic channel.

As noted above, in this single-mode approximation, which is widely used in the context of high-\Q\ resonant structures, the intuitive understanding of RSMs is explicitly realized. 
The total in-coupling rate for the input channels acts as an effective source of gain, while the total decay rate into the output channels plus the intrinsic absorption in the cavity act as an effective loss.
When these two quantities are equal, the reflectionless state has a real frequency and becomes an RSM.
Thus we see that in this limit, and {\it only} in this limit, is the simple concept of RSM as critical coupling accurate.
For relatively open systems, like multi-waveguide junctions, RSMs will exist according to our general theory and can be found numerically using the general formulation, but do not correspond to any simple condition like critical coupling.

In the isolated resonance approximation there are no multi-resonance effects to shift the RSM frequency and internal cavity fields away from that of the single resonance.
Therefore all $2^N-1$ different $R$-zero boundary conditions move the complex frequency of the corresponding $R$-zero vertically between the purely outgoing solution (resonance) and the purely incoming solution ($S$-matrix zero). 
%It is straightforward to also show that the RSM incident wavefront ${\bs \alpha}_{\rm RSM}$ is simply the phase conjugation of the resonance's radiation wavefront into the designated channels.
%When $\w = \w_{\rm RZ}$, the non-resonant reflection ${\bs \alpha}_{\rm RSM}$ is exactly cancelled by the resonant scattering $D {\bf a}$ back into those channels.
When the frequency is detuned from $\w_{\rm RZ}$, the reflection signal rises as a Lorentzian function with the linewidth being that of the underlying resonance, $\left( \g_{\rm in} + \g_{\rm out} + \g_0 \right)$.

Despite its simplicity, the single-resonance $R$-zero explains the impedance-matching conditions previously found using TCMT in waveguide branches~\cite{2001_Fan_JOSAB}, antireflection surfaces~\cite{2014_Wang_Optica}, and polarization-converting surfaces~\cite{2017_Guo_PRL}: vanishing reflection in a passive system ($\g_0=0$) is achieved at $\w_0$ when the decay rate into the incident channel $\g_{\rm in}$ equals the sum of decay rates into all outgoing channels $\g_{\rm out}$.

\begin{figure}[ht!]
    \centering
    \includegraphics[width=0.9\columnwidth]{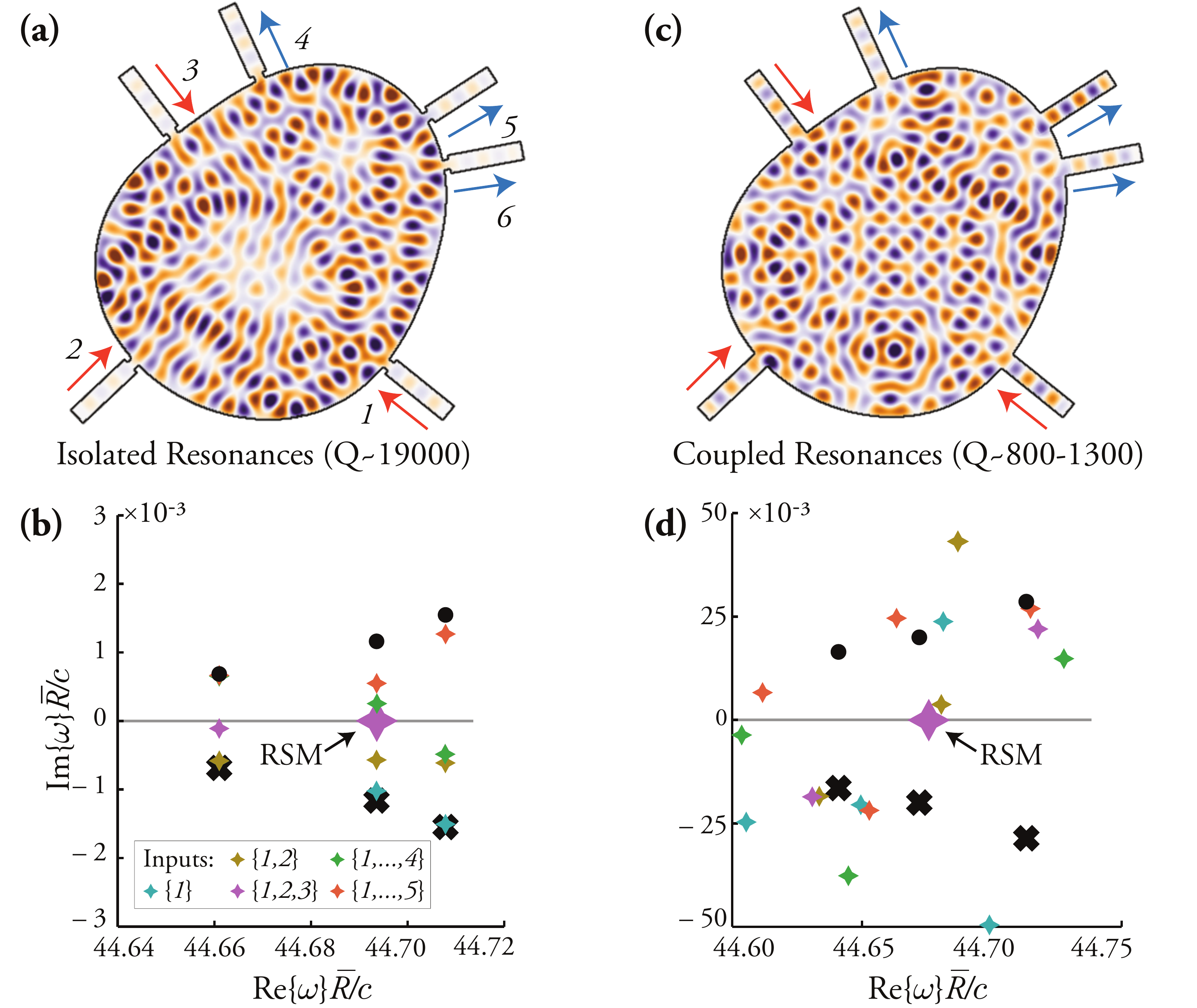}
    \caption[$R$-zeros and reflectionless scattering modes in a chaotic waveguide junction]{A lossless, chaotic resonator (mean radius $\bar R$) and junction for six single-mode waveguides, with constrictions at the entrances to the cavity.
    {\bf (a)}:~Mode profile of a 3-in/3-out RSM in a weakly-coupled, high-\Q\ cavity with well isolated resonances, enforced by small waveguide constrictions.
    Waveguides labeled $1,\ldots,6$; red arrows indicate input channels, and blue, output.
    {\bf (b)}:~Numerically calculated $R$-zero spectrum for cavity in (a).
    Black x's and dots are resonance and $S$-matrix zero frequencies, which are complex conjugates.
    Colored stars are $R$-zeros for various choices of inputs; legend indicates sets of input channels, according to labels in (a).
    $R$-zeros cluster vertically above the resonance and below the $S$-matrix zero frequency, as predicted by single-resonance TCMT approximation \cref{eq:chpt2: omega_RSM_single_mode}.
    The common width of the constrictions for waveguides $\{4,5,6\}$ is slightly tuned to make 3-in/3-out $R$-zero real, creating RSM shown in (a).
    {\bf (c)}:~Mode profile of a 3-in/3-out RSM for same junction as in (a) but with open constrictions, which lowers \Q\ (note change in vertical scale).
    {\bf (d)}:~$R$-zero spectrum for low-\Q\ cavity shown in (c).
    The linewidths of the resonances are now comparable to their spacing. 
    Due to multi-resonance effects, the $R$-zeros are distributed throughout complex plane and are no longer associated with a single resonance.
    Nonetheless, by slightly tuning the constriction width as before, we create a 3-in/3-out RSM, as in the high-\Q\ case.
    }
    \label{fig:chpt2: octopus}
\end{figure}

The single-resonance approximation is typically valid when a cavity has a high quality factor and is weakly-coupled to the input/output channels, so that its resonances are much nearer to the real axis than to each other, and multi-resonance effects can be neglected.
However the general RSM theory applies equally well to low-\Q\ cavities where we expect the above simple picture to break down substantially. 
An example of both limits is given in \vref{fig:chpt2: octopus}, where we study a chaotic resonant cavity connected to six single-mode waveguides, using a finite-difference numerical calculation.
The cavity shown in \cref{fig:chpt2: octopus}a has constrictions at its waveguide junctions to increase the quality factors of the resonances.
In \cref{fig:chpt2: octopus}b we show the $R$-zero spectrum for this cavity, which has the vertical clustering bracketed by the resonance and $S$-matrix zero, as predicted by the single resonance approximation just discussed.
In contrast, the cavity shown in \cref{fig:chpt2: octopus}c has the constrictions opened, which reduces the quality factors by a factor of $\sim 20$.
For this case the single-resonance approximation fails, and the $R$-zero spectrum (\cref{fig:chpt2: octopus}d) is spread out in both the real and imaginary frequency axes.
Strikingly, some $R$-zeros lie below the resonances while others lie above the $S$-matrix zeros in the complex-frequency plane, something which is forbidden within the single resonance approximation.
Nonetheless, in both cases, we were able to tune to a real-frequency RSM for the three-in/three-out boundary condition simply by slightly varying the constrictions of the outgoing waveguides.
This shows the power of the general RSM theory developed here, which works even for open cavities, where multi-resonance effects dominate.

    \section{Symmetry Properties \label{sec:chpt2: Symmetry Properties}}

We now analyze some exact properties of $R$-zeros and RSMs with time-reversal and discrete spatial symmetries.
Earlier works~\cite{2012_Ge_PRA, 2013_Mostafazadeh_PRA} studied the consequences of symmetry in one dimension, and have some overlap with what follows, though the present treatment generalizes to any dimension, and applies to complex-valued $R$-zeros, not only real RSMs.
The different symmetry classes and their properties are summarized in \vref{tab:chpt2: Symmetry Table} and \vref{fig:chpt2: symmetry}.

    \subsection{Time-reversal Transformation (${\cal T}$) and Symmetry\label{sec:chpt2: sym_T}}

\begin{figure}[t!]
    \centering
    \includegraphics[width=\textwidth]{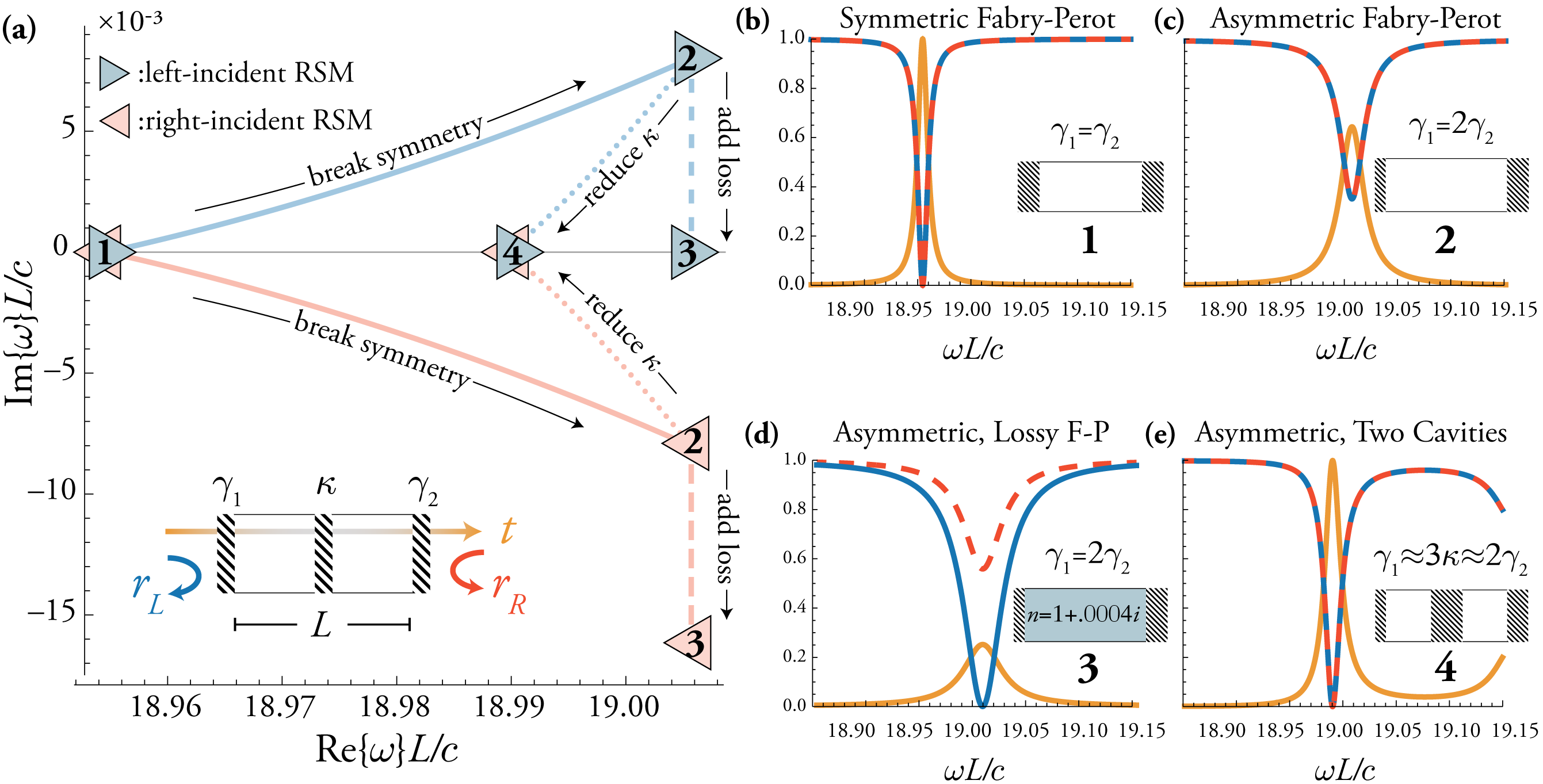}
    \caption[$R$-zeros and reflectionless scattering modes in three-mirror cavity]{Illustration of RSMs and $R$-zero spectrum for simple two- and three-mirror resonators of length $L$ in 1D, consisting of $\delta$-function dielectric mirrors of strengths $\g_1^{-1}, \g_2^{-1}$ and $\kappa^{-1}$, as indicated in the schematic in (a).
    Throughout, we fix $\g_2 \defn c/L$.
    {\bf(a)}:~Blue and red lines $1 \to 2$ indicate the effect of breaking symmetry by varying $\g_1$ from $\g_2 \to 2 \g_2$.
    A bidirectional RSM {\bf(b)} splits into two complex-conjugate $R$-zeros off the real axis and a reflectionless steady-state (RSM) no longer exists, {\bf(c)}.
    Adding absorption to the cavity, indicated by blue and red lines $2 \to 3$ in (a), brings the upper $R$-zero to the real axis (but not the lower one), creating a unipolar left-incident lossy RSM, {\bf(d)}.
    Alternatively, adding a middle mirror and reducing its $\kappa$ from $\infty$ to $\sim 2 \g_2/3$ is sufficient to bring both $R$-zeros back to the real axis ($2 \to 4$ in (a)), creating a bipolar RSM at a different frequency from the symmetric Fabry-P\'erot resonator, without restoring parity symmetry, {\bf(e)}.
    }
    \label{fig:chpt2: fp}
\end{figure}

The action of time reversal (${\cal T}$) is to complex conjugate everything, including the wave operator $\hat A_{\rm RZ}$, and the field $\psi(\bx)$.
This interchanges the input and output channels ($C_F \leftrightarrow C_{\fbar}$):
\begin{equation}
    \label{eq:chpt2: T-symmetry}
    {\cal T}: {\bs(}\w,\e(\bx),C_F{\bs)} \to {\bs(}\w^*,\e^*(\bx),C_{\fbar}{\bs)}.
\end{equation}
If a cavity with dielectric function $\e(\bx)$ has an $R$-zero at frequency $\w_0$ with input channels $C_F$, then the cavity with $\e^*(\bx)$ has an $R$-zero at frequency $\w_0^*$ with complementary input channels $C_{\fbar}$.
The scatterer has time-reversal symmetry if $\e(\bx)=\e^*(\bx)$, i.e.,~when there is no absorption or gain, in which case the ``two cavities'' described above are actually the same.
We conclude that
\begin{itemize}
    \item Complementary $R$-zeros come in complex-conjugate pairs when $\e(\bx)$ is flux-conserving, i.e.,~for passive cavities.
\end{itemize}
This is true regardless of whether spatial symmetries such as parity are present.
The system can be tuned to have an RSM either by index tuning or by gain-loss tuning. 
In the former case, the new cavity still exhibits time-reversal symmetry, so the resulting RSM is {\it bipolar}, by which we mean that a different RSM exists at the same frequency; examples are given in \cref{fig:chpt2: free_space}, \cref{fig:chpt2: octopus}, and \cref{fig:chpt2: fp}(b,e).
We use the term bipolar instead of bidirectional so as to emphasize that the input and output channels can spatially overlap in the general cases we consider.
In the case of gain-loss tuning, the tuned cavity no longer has time-reversal symmetry, and we do not expect another RSM at that frequency; we refer to this as {\it unipolar}.

In \cref{fig:chpt2: fp} we illustrate the concept of tuning to create an RSM out of a complex-valued $R$-zero for the simple case of an asymmetric Fabry-P{\'e}rot cavity.
We start with a symmetric Fabry-P{\'e}rot cavity, which has both ${\cal P}$ and ${\cal T}$ symmetries (this symmetry class will be discussed in more detail later in \Cref{sec:chpt2: sym_P_and_T}); it is well-known that such a cavity has equally-spaced unit-transmission resonances~\cite{1986_Siegman_book}, which we refer to as bidirectional RSMs, since they can be accessed from either side.
We break ${\cal P}$ symmetry but maintain ${\cal T}$ symmetry by making the mirror reflectivities unequal; as a consequence there is no longer an RSM for either direction of incidence at any real frequency.
However, there is now a pair of complex-conjugate $R$-zeros off the real axis, as demanded by ${\cal T}$ symmetry.
We have made the left barrier less reflective than the right one, and observe that the left-incident $R$-zero has moved to the upper half-plane and the right-incident $R$-zero to the lower half-plane, which is consistent with the intuitions regarding effective gain and loss of the input and output channels given near the end of \Cref{sec:chpt2: rigorous RSM}.

One way to recover the RSMs in the asymmetric cavity without restoring parity symmetry is to use gain-loss engineering (non-hermitian tuning).
Adding absorption will reduce the imaginary parts of the $R$-zero frequencies, which can be used to create a left-incident RSM.
A physical argument supporting this conclusion is that when the mirror reflectivities are equal, the prompt reflection from the left mirror is cancelled by the net leftward reflection of waves reaching the interior and internally reflecting an odd number of times before escaping back in the incident direction; a similar statement holds for right-incident waves.
When the left mirror is less reflective than the right one, its prompt reflection is decreased and the internal leftward reflection is increased, so total destructive interference is not possible.
However if one adds the correct amount of absorption to the interior, these two amplitudes can again be balanced and destructive interference in the backwards direction can be restored, though this generally will also require a frequency shift to account for the change in the phase accumulation within the cavity.
Adding loss has the opposite effect on a right input wave, making destructive interference impossible, and leading to a unipolar left-incident RSM.
For a right-incident RSM one would instead have to add the same amount of gain to the interior.
This physical picture is confirmed by \cref{fig:chpt2: fp}(a,d).

Alternatively, we can employ index tuning to create an RSM, in this case by adding a third lossless mirror in the interior.
We can think of the left region and the interior mirror as forming a composite mirror such that for some mirror reflectivity and input frequency, the left and right escape rates are again balanced.
Indeed such a three-mirror system does have an RSM with one-parameter tuning as shown by \cref{fig:chpt2: fp}(a,e).
Moreover, as the system is flux-conserving and reciprocal, it must be a bipolar RSM.

Comparing \cref{fig:chpt2: fp}(a,d,e) illustrates this important difference between the two types of tuning: for the lossless case, both left- and right-incident $R$-zeros maintain their complex conjugate relation as the interior mirror is tuned and hence meet on the real axis.
Note that from the standard point of view, this bidirectional RSM is ``accidental'', i.e.,~not a result of parity symmetry, which has been broken.
Nevertheless the tuning deterministically generates this ``accidental'' behavior that mimics the effect of parity symmetry for this one frequency.

    \subsection{Parity Transformation (${\cal P}$) and Symmetry\label{sec:chpt2: sym_P}}

A parity transformation ${\cal P}$ satisfies ${\cal P}^2 = 1$ and $\det {\cal P} = -1$.
Common examples of parity are reflections in two dimensions, $(x,y) \to (-x,y)$, and inversion in three dimensions, $(x,y,z) \to (-x,-y,-z)$.

Generally, the action of ${\cal P}$ is to leave the frequency unchanged and to map $\e(\bx) \to \e({\cal P} \bx)$, $\psi(\bx) \to \psi({\cal P} \bx)$, relating an $R$-zero/RSM of one structure to that of a structure transformed by ${\cal P}$.
However, these two $R$-zeros are generally not complementary to each other.
We further require the input-output channels to have ${\cal P}$ symmetry, by which we mean that $\e(\bx) = \e({\cal P} \bx)$ in the asymptotic region {\it and} that the input channels are chosen such that $C_F$ maps to $C_{\fbar}$ under ${\cal P}$; we call this a {\it bisected partition} of the channels.
The most common bisected systems would be those that naturally divide into ``left'' and ``right'' and for which the input channels are chosen to be all left or all right channels,
which generalizes the well-studied one-dimensional case.
But there are also other possibilities, e.g.~a partition into clockwise and counterclockwise channels for a finite-sized scatterer in free space.
With a bisected partition $C_F$, the action of ${\cal P}$ is
\begin{equation}
    \mathcal{P}: {\bs(}\w,\e(\bx),C_F{\bs)} \to {\bs(}\w,\e({\cal P} \bx),C_{\fbar}{\bs)}.
\end{equation}
When an $R$-zero with a bisected partition is bipolar, we refer to it as {\it bidirectional}; when it is unipolar, we refer to it as {\it unidirectional}.

The scatterer has ${\cal P}$ symmetry when $\e(\bx) = \e({\cal P} \bx)$, for which we can say that
\begin{itemize}
    \item When the cavity and the channel partition both have ${\cal P}$ symmetry, $R$-zeros are bidirectional, whether or not $\w_{\rm RZ}$ is real.
\end{itemize}
This is in contrast to the case with ${\cal T}$ symmetry, where bipolarity only holds for real-frequency RSMs.

\begin{table}[t!]
    \centering
    \begin{tabular}{c | c  c  c  c  c  c}
    \toprule
     \thead{symmetry\\type} & \thead{$\e(\bx)$\\same as} & \thead{$\w_F$\\pairs with} & \thead{RSM type\\($\w_F\in{\mathbb R}$)} & $N_{\rm RSM}$ & $N_{\rm RSM}^{\rm EP}$ &  \thead{$\mathbb{R}\to \mathbb{C}$\\transition?} \\ 
     \midrule
     none & ---  & --- & unipolar & 1 & 3 & \xmark \\ %\hline
     ${\cal T}$ & $\e^*(\bx)$ & $\w_{\fbar}^*$ & bipolar & 1 & 3 & \xmark \\ %\hline
     ${\cal P}$ & $\e({\cal P} \bx)$ & $\w_{\fbar}$ &  bidirectional & 1 & 3 &  \xmark \\ %\hline
     ${\cal PT}$ & $\e^*({\cal P}\bx)$ & $\omega_F^*$ & unidirectional & 0/1 & 1 &  \cmark \\ \hline
     ${\cal P,T}$ & \makecell{$\e^*(\bx)$,\\$\e({{\cal P}\bx})$} & \makecell{$\w_{\fbar}$,\\$\w^*_{\fbar}$, $\w^*_F$} & bidirectional & 0/1 & 1 & \cmark \\
     \bottomrule
    \end{tabular}
    \caption[Consequences of discrete symmetries for $R$-zeros and reflectionless scattering modes]{Consequences of discrete symmetries for the RSM problem for input states defined by $F$.
    Unipolar means that only one set of input channels has zero reflection at a given frequency and bipolar means that its complement does too.
    When ${\cal P}$ applies, we assume the channels are bisected, i.e., the set of input channels maps to its complement; in this case we use unidirectional instead of unipolar, bidirectional instead of bipolar.
    $N_{\rm RSM}$ is the minimum number of system parameters that must be tuned, consistent with symmetry, to achieve an RSM (make $\w_F$ real), while $N_{\rm RSM}^{\rm EP}$ parameters are necessary for an RSM EP (degenerate, real $\w_{\rm RZ}$).
    }
    \label{tab:chpt2: Symmetry Table}
\end{table}

    \subsection{Parity-Time Transformation (${\cal PT}$) and Symmetry \label{sec:chpt2: sym_PT}}

The action of the joint parity-time-reversal ($\mathcal{PT}$) operator, i.e.,~performing both $\mathcal{P}$ and $\mathcal{T}$ transformations simultaneously, is particularly interesting as it is the case of $\mathcal{PT}$ symmetry that brought recent attention to unidirectional RSMs.
We assume that the asymptotic region has $\mathcal{P}$ symmetry and that the partition $C_F$ is bisected; hence the action of $\mathcal{PT}$ is
\begin{equation}
    \mathcal{PT}: {\bs(}\w,\e(\bx),C_F{\bs)} \to {\bs(}\w^*,\e^*({\cal P} \bx),C_F{\bs)}.
\end{equation}
The scattering region has ${\cal PT}$ symmetry when $\e^*({\cal P} \bx) = \e(\bx)$.
Hence:
\begin{itemize}
    \item ${\cal PT}$-symmetric systems either have unidirectional RSMs with $\w_0 \in {\mathbb R}$, or complex-conjugate pairs of unidirectional $R$-zeros of the {\it same directionality}, with $\w_0 \in {\mathbb C}$.
\end{itemize}
This is in contrast to the case of ${\cal T}$ symmetry alone, which allows real bipolar RSMs or complex-conjugate pairs of $R$-zeros of {\it opposite} polarity/directionality.

It should be noted that physical dielectric materials, which have a causal susceptibility and satisfy the Kramers-Kronig relation~\cite{2012_fox_book}, cannot satisfy an exact ${\cal PT}$-symmetry except at isolated frequencies~\cite{2014_Zyablovsky_pra}.
Nevertheless, in some range around such a frequency, ${\cal PT}$ symmetry will be only weakly broken, and its effects will still be observable.

    \subsubsection{Spontaneous Symmetry-Breaking Transition \label{sec:chpt2: sym breaking}}

This property of the ${\cal PT}$ case implies something quite important.
Often ${\cal PT}$-symmetric scattering systems are studied by beginning with a flux-conserving system (satisfying both ${\cal P,T}$ symmetries separately) and adding gain and absorption anti-symmetrically so as to break ${\cal P}$ and ${\cal T}$ symmetries while preserving ${\cal PT}$.
The initial system has bidirectional RSMs, i.e.,~pairs of left and right RSMs at the same real frequency, but has no other degeneracy.
Thus the left and right RSMs are constrained to remain on the real axis as {\it unidirectional} RSMs as the value of the gain-loss strength increases.

\begin{figure}[t!]
   \centering
   \includegraphics[width=\textwidth]{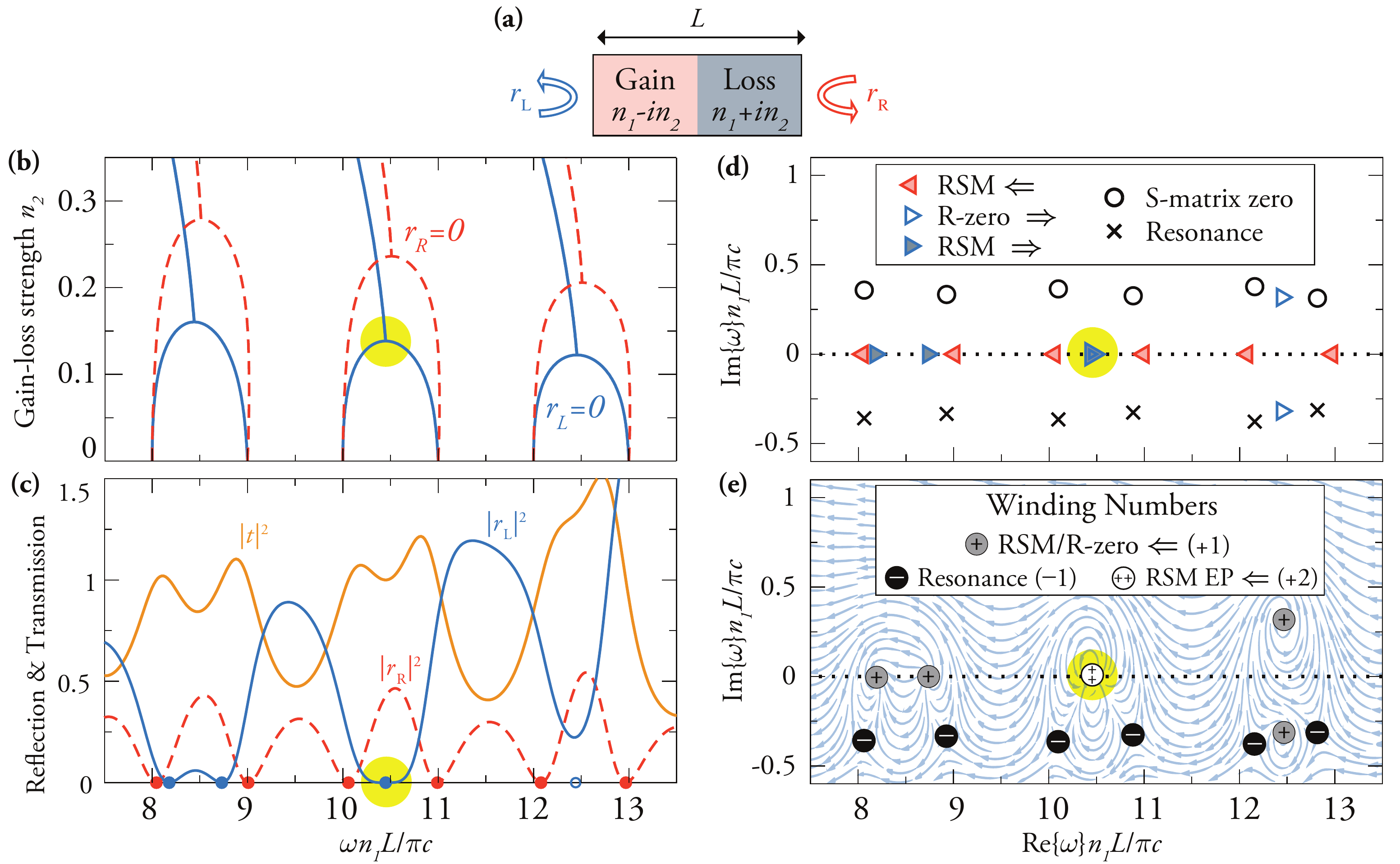}
   \caption[Reflectionless scattering modes and ${\cal PT}$ symmetry]{RSMs in a ${\cal PT}$-symmetric structure.
   {\bf(a)}:~Schematic of structure: an etalon in air, with refractive index $n = n_1 - i n_2$ on the left and $n = n_1 + i n_2$ on the right; here $n_1=2$.
   {\bf(b)}:~Real part of $R$-zero frequencies as the gain-loss strength is increased (solid blue lines: left-incident; dashed red lines: right-incident).
   They are real-valued at small $n_2$.
   After two meet at an RSM EP (yellow highlight), they split into complex-conjugate $R$-zeros.
    {\bf(c)}:~Reflection and transmission spectra when $n_2 = 0.13844$; filled dots mark the RSM frequencies, open dot is real part of complex $R$-zero, which has already crossed the threshold.
   {\bf(d)}:~$R$-zeros/RSMs, $S$-matrix zeros, and resonances in the complex-frequency plane, with same $n_2$ as in (c).
   {\bf(e)}:~Streamlines of the vector field ${\bs(}\re{r_{\rm L}(\w)}, \im{r_{\rm L}(\w)}{\bs)}$, with $+1$ topological charges at $R$-zeros/RSMs, $-1$ at resonances.
   At an RSM EP the topological charges add to $+2$.
   }
    \label{fig:chpt2: PT}
\end{figure}

\begin{figure}[t!]
    \centering
    \includegraphics[width=\textwidth]{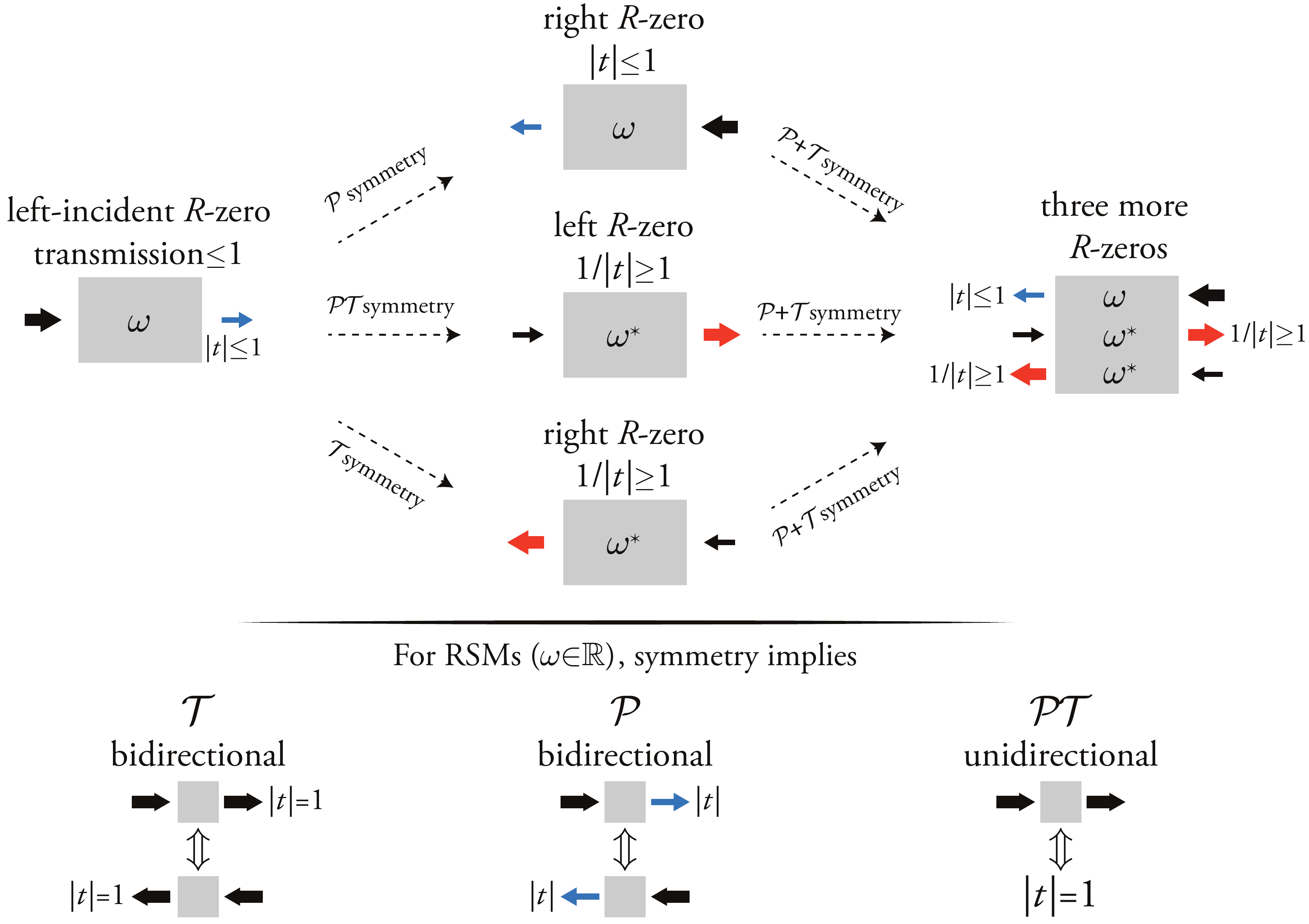}
    \caption[Flow-chart depicting implications of ${\cal P}$, ${\cal T}$, ${\cal PT}$, and ${\cal P,T}$ symmetries on $R$-zeros and reflectionless scattering modes]{Schematic illustrating the implications of symmetry for $R$-zeros and RSMs.
    {\bf(Top)}:~Beginning with the assumption of the existence of a left-incident $R$-zero with with transmission $|t|<1$ at some complex-valued frequency $\w$, there will exist other $R$-zeros with specific properties based on the presence of ${\cal P}$, ${\cal T}$, and/or ${\cal PT}$ symmetry (as shown by dashed arrows).
    The presence of all three implies the existence of three other $R$-zeros.
    {\bf(Bottom)}:~The initial state has {\it real} frequency (is an RSM); the implications of the various symmetries are shown for this case.
    The implications for the ${\cal T}$ and (${\cal P,T}$) cases are the same: bidirectional unit transmission reflectionless states.
    }
    \label{fig:chpt2: symmetry}
\end{figure}

For example, if a left-incident RSM moved off the real axis for infinitesimal gain/loss, it would lack a second left-incident RSM as a complex-conjugate partner, violating this condition.
These unidirectional RSMs are invariant under the ${\cal PT}$ transformation, so they are in the ${\cal PT}$-unbroken phase.
As the gain-loss strength is further increased, eventually each RSM may meet another RSM of the same directionality at a real frequency (see \cref{fig:chpt2: PT}b as an example), above which point the pair of RSMs will generally leave the real axis as complex-conjugate pairs of $R$-zeros, becoming inaccessible with steady-state excitation.
These complex-conjugate pairs of $R$-zeros do not exhibit ${\cal PT}$ symmetry, since one maps not to itself but to its conjugate partner under ${\cal PT}$, so they are in the ${\cal PT}$-broken phase.
The transition is the spontaneous breaking of ${\cal PT}$ symmetry.

This explains in the $R$-zero/RSM framework the widely observed existence of RSMs in one-dimensional systems with ${\cal PT}$ symmetry~\cite{2011_Lin_PRL, 2011_Hernandez-Coronado_PLA, 2011_Longhi_JPA, 2012_Ge_PRA, 2012_Regensburger_Nature, 2013_Feng_nmat, 2014_Ramezani_PRL_2} or anti-$\mathcal{PT}$ symmetry~\cite{2013_Ge_PRA} without fine tuning, and their disappearance at large gain-loss parameters~\cite{2011_Hernandez-Coronado_PLA,2011_Longhi_JPA,2012_Ge_PRA,2012_Jones_JPA,2013_Mostafazadeh_PRA}.
It should be noted that the critical parameters that define the ${\cal PT}$ transitions are different for each RSM pair and directionality, so, as is typical in ${\cal PT}$ scattering systems, the transition does not occur at once for all eigenvalues.
The RSM transitions are also close to, but different from the transitions associated with the full $S$-matrix, at which  a pair of unimodular eigenvalues turn into an amplifying one and an attenuating one, which is another signature of spontaneous ${\cal PT}$ symmetry-breaking in scattering~\cite{Chong:2011ev, 2012_Ge_PRA}.

This behavior is illustrated for the one-dimensional ${\cal PT}$-symmetric structure shown in \cref{fig:chpt2: PT}a: an etalon of thickness $L$ in air where the left half has refractive index $n = n_1 - i n_2$ and the right half has index $n = n_1 + i n_2$.
For a passive etalon ($n_2=0$), bidirectional RSMs exist at real frequencies $\w_{\rm RSM} = m \pi c/n_1L$ with $m\in{\mathbb Z}$.
When the gain-loss strength $n_2$ is increased, pairs of RSMs with the same directionality move toward each other in frequency.
As parity and time-reversal symmetries are individually broken, the right-incident RSMs (for which $r_{\rm R}=0$) and the left-incident RSMs (for which $r_{\rm L}=0$) now occur at different frequencies.
But since the system still exhibits ${\cal PT}$ symmetry, all of these RSM frequencies remain real-valued.
At critical values of $n_2$, a pair of RSMs coalesce. 
As $n_2$ is further increased, the pair of RSMs split into two, leaving the real-frequency axis as complex-conjugate pairs of $R$-zeros. 
The RSM spectrum in the complex-frequency plane, together with the $S$-matrix zeros and poles, is shown in \cref{fig:chpt2: PT}(d,e) for a critical value of $n_2$ where two right-going RSMs meet.
The corresponding reflection and transmission spectra are given in \cref{fig:chpt2: PT}c.

We use this example to illustrate a topological property of $R$-zeros/RSMs and resonances.
\Cref{fig:chpt2: PT}c shows the streamlines of the vector field ${\bs(}\re{r_{\rm L}(\w)}, \im{r_{\rm L}(\w)}{\bs)}$ over the complex plane.
In accordance with the discussion at the end of \Cref{sec:chpt2: R-zeros}, we indeed observe that $\arg(r_{\rm L}(\w))$ winds by $2\pi$ along counterclockwise loops around each $R$-zero or RSM, corresponding to a $+1$ topological charge, while it winds by $-2\pi$ around each resonance, corresponding to a $-1$ topological charge.
When two RSMs meet, they superpose as one topological defect with charge $+2$, as highlighted in yellow.
This is an example of an exceptional point, as introduced in \Cref{sec:chpt0: eps} of \Cref{sec:chpt0: intro}, although not of purely outgoing resonances, nor of purely incoming CPA states, which we will discuss in \Cref{chp:chpt3: cpa ep}.

    \subsection{RSM Exceptional Points \label{sec:chpt4: RSM EP}}

Non-hermitian operators have the property that when two eigenvalues become degenerate in both their real and imaginary parts, then the two associated eigenvectors also coalesce into one, which is an exceptional point (EP) in parameter space.

The $R$-zero/RSM wave operator $\hat A_{\rm RZ}$ of \cref{eq:chpt2: A_RZ} is non-hermitian and is formally similar to the resonance wave operator $\hat A_{\rm eff}$.
It is therefore possible to create {\it RSM exceptional points} where multiple reflectionless states coalesce into one, which is, from a physical point of view, a new kind of EP not previously studied.  
The aforementioned RSM transitions in ${\cal PT}$-symmetric systems are examples of this. 
There is no self-oscillating instability (lasing) when the EP reaches the real axis, since a steady-state RSM EP is compatible with linear response.
At an RSM EP, the lineshape of the reflection intensity will change from its generic quadratic form to a quartic, flat-bottomed lineshape, characteristic of a $+2$ topological charge.
There is generally no qualitative change in the transmission lineshape at a ${\cal PT}$-symmetric RSM EP, since a non-flux-conserving structure lacks a rigid relation between the transmission and reflection coefficients; a different type of flux-conserving RSM EP, to be discussed next, does have a quartic flat-topped transmission lineshape.
Near an EP of any kind, the eigenfrequency typically exhibits a square-root dependence on system parameters.
These effects are shown in \cref{fig:chpt2: PT}, where the RSM EP is highlighted in yellow.
The change in lineshape as a consequence of EP-proximity will be discussed in more detail in \Cref{chp:chpt3: cpa ep} in the context of degenerate CPA.

As mentioned at the beginning of the chapter, RSMs in one-dimensional systems have been characterized in some works as EPs of an unconventional non-symmetric scattering matrix~\cite{2007_Cannata_AoP,2013_Feng_nmat,2013_Mostafazadeh_PRA,2014_Wu_PRL,2014_Zhu_OE,2014_Kang_PRA,2017_Huang_nanoph}, which is {\it not} related to the usual $S$-matrix through a change of basis.
However, those are non-degenerate RSMs, not EPs of the underlying wave operator.
While there is nothing incorrect about this alternative point of view mathematically, in our view it creates confusion and is physically misleading.
First, this type of EP is a an EP of a scattering matrix, not of a wave operator.
It does not imply a change in the lineshape of the resonance, as does an EP of the wave operator.
Encircling such an EP in parameter space does not generate asymmetric state transfer of physical resonances or RSMs, but rather implies some behavior of the eigenstates of the $S$-matrix, which is of lesser interest.
Finally, the identification of a non-degenerate RSM with an unconventional EP only works when the number of channels $N$=2, since the unconventional scattering matrix, with off-diagonal reflection coefficients, does not generalize to arbitrary $N$-channel scattering, which has $N$ reflection coefficients but $N^2-N$ off-diagonal elements.
%As soon as $N$> 2 (i.e.,~even in multi-channel quasi-one-dimensional case) an RSM does not coincide with an EP of $S$.
Hence we believe there is ample reason to avoid this alternative point of view.
We thus adopt a uniform definition of an RSM, which is generically {\it not} an EP of the wave operator, and apply the term RSM EP only to degenerate RSMs which {\it are} an EP of the wave operator.
%Therefore, we do not adopt this convention and reserve the term ``RSM EP'' for the states discussed in this section.

Although the ${\cal PT}$ transition in open systems has often been applied to resonances, their purely outgoing boundary condition does not map into itself under a ${\cal PT}$ transformation, and therefore they lack an exact ${\cal PT}$ symmetry.
In fact, the resonance and anti-resonance boundary conditions map into each other under ${\cal PT}$, which is exploited in the laser-absorber of Refs.~\cite{2010_longhi_pra, Feng:2014gp, Chong:2011ev, 2016_wong_natphot}.
Therefore simply increasing the strength of the gain/loss is not sufficient to create an exact degeneracy of resonances, only an avoided crossing, as is suggested in \cref{fig:chpt2: PT}d, where neighboring resonances have imaginary parts that are visibly different from each other, and as is shown explicitly in Fig.~3 of Ref.~\cite{Chong:2011ev}. 
Within TCMT this is not observed, because TCMT does not account for the frequency-dependent nature of the purely outgoing boundary conditions, and therefore predicts that a ${\cal PT}$-symmetric cavity will have an EP of resonances with single-parameter, gain-loss tuning.
In contrast, the bipartite RSM boundary condition does map into itself under a ${\cal PT}$ transformation, which is why RSM EPs in ${\cal PT}$-symmetric systems are truly accessible through one-parameter tuning.

It is important to recognize that any $R$-zero boundary condition is intrinsically \nh, whether or not the cavity has gain or loss.
Thus, even in a passive cavity, a degeneracy of RSMs will be an EP.
It is not necessary to have ${\cal PT}$ symmetry with gain and loss in the cavity to create an RSM EP, though it enables achieving one with only single-parameter tuning, and makes it unidirectional.
An RSM EP in a generic flux-conserving cavity can also happen, but it requires three-parameter tuning (see \vref{tab:chpt2: Symmetry Table}), and would be bidirectional because of ${\cal T}$ symmetry.
In the next section we discuss a case in which a bidirectional RSM EP can be created in a {\it passive} cavity with only {\it single}-parameter tuning.

    \subsubsection{${\cal P}$ and ${\cal T}$ Symmetry and Spontaneous Symmetry-breaking Transition \label{sec:chpt2: sym_P_and_T}}

The final symmetry class we will discuss is the case of systems with both ${\cal P}$ and ${\cal T}$ symmetries.
The symmetric Fabry-P\'erot cavity, discussed briefly above, is a familiar example. 
Such a ${\cal P,T}$ system simultaneously exhibits all of the symmetry properties discussed so far.
Therefore we can expect bidirectional RSMs on the real-frequency axis without any tuning, with flux conservation implying unit transmission for these RSMs.
However there is another possibility allowed  by symmetry, which to our knowledge has not been pointed out elsewhere: the possibility of bidirectional pairs of $R$-zeros appearing off the real axis as complex  conjugates.
In general, if a behavior is allowed by symmetry, it should actually occur in some systems.

%We have confirmed that t
In fact, this possibility does occur in a physical model similar to that shown in \vref{fig:chpt2: fp}.
The dielectric function consists of three equally-spaced, lossless mirrors, with identical outer mirrors, given by
\begin{equation}
    \e(x) = \delta(x+L)/\g + \delta(x)/\kappa + \delta(x-L)/\g.
\end{equation}
Requiring that the reflection coefficient vanishes leads to a transcendental equation that $k_{\rm RZ} \defn \w_{\rm RZ}/c$ must satisfy:
\begin{equation}
    \label{eq:chpt2: modulus equation}
    \cis{k_{\rm RZ}L} = \frac{k_{\rm RZ}^2 + 2\g^2 \pm 2i\sqrt{k_{\rm RZ}^2 \kappa^2 + 4\g^2\kappa^2-\g^4}}{k_{\rm RZ}^2 - 2i(\g + \kappa) k_{\rm RZ} - 4 \g \kappa}.
\end{equation} 
If $k_{\rm RZ}$ is real (RSM), then the modulus of both sides is unity, which only happens if the expression under the radical is positive; else, $k_{\rm RZ}$ must be complex, and is an $R$-zero.
If $\kappa<\g/2$ then there is a threshold $k_{\rm th}$ below which there are no real $R$-zeros, and above which all of the $R$-zeros are real RSMs:
\begin{equation}
    k_{\rm th} \defn \frac{\g}{\kappa}\sqrt{\g^2-4\kappa^2}.
\end{equation}
%This is evidence of a (${\cal P}$,${\cal T}$) transition.
The other case, with $\kappa>\g/2$, has no such transition, as the modulus of both sides of \cref{eq:chpt2: modulus equation} is always unity for real $k_{\rm RZ}$.
The threshold $k_{\rm th}$ can be increased from $0$ by decreasing $\kappa$, starting from $\kappa=\g/2$.
It is possible to find a value for the internal mirror coupling $\kappa$ such that $k_{\rm th}$ reaches an RSM.
This corresponds to two RSMs coalescing at the symmetry-breaking transition, which is an RSM EP (actually two EPs, one each of right- and left-going RSMs).
Decreasing $\kappa$ further causes the RSMs to move off the real axis as bidirectional complex-conjugate pairs of $R$-zeros.
At the EP, due to the bidirectionality of the RSMs, both transmission and reflection intensity lineshapes will be flattened.

Such degenerate RSM EPs are examples of steady-state EPs in a flux-conserving system, a possibility not discussed in the literature, to our knowledge.
This transition is actually a manifestation of the usual ${\cal PT}$ symmetry-breaking, but without any gain or absorption.
The potential of having exceptional point behavior on the real axis in flux-conserving systems, with single parameter tuning, opens up a number of intriguing avenues for further research and novel experimental investigations.  

\chapter{Exceptional Coherent Perfect Absorption (CPA EP) \label{chp:chpt3: cpa ep}}
cIn this chapter we study exceptional points of coherent perfect absorption (CPA EP)~\cite{2019_Sweeney_PRL}.
These are \nh\ degeneracies of the purely incoming wave operator, in which two perfectly absorbed waves coalesce into a single state.
They are instances of the novel class of RSM EPs discussed in \Cref{sec:chpt4: RSM EP} of the previous chapter, as CPA is the particular choice of RSM in which every channel is incoming.
Such non-hermitian degeneracies can occur at a real-valued frequency without additional noise or non-linearity, in contrast to EPs in lasers, which have amplified spontaneous emission and saturation.
The absorption lineshape for the eigenchannel near the EP is quartic in frequency around its maximum in any spatial dimension.
For absorbing disk resonators, these EPs give rise to chiral absorption: perfect absorption for only one sense of rotation of the input wave.
In general, for the parameters at which a CPA EP occurs, the associated scattering matrix does not also have an EP.
A \nh\ cavity has a non-unitary, \nh\ $S$-matrix, so that a degeneracy of its eigenvalues is generically an EP, which indicates that $S$ is defective and cannot be diagonalized.
When an EP of $S$ is also an eigenfrequency of the incoming wave operator, then the degenerate $S$-matrix eigenvalue is zero.
However, in one dimension, if the $S$-matrix does have a perfectly absorbing EP, then it takes on a universal one-parameter form with degenerate values for all scattering coefficients.

%Most of this chapter will deal two incoming wave solutions merging at a specific frequency, tuned to occur on the real axis.

%This will be discussed at the end of this chapter.

\section{Properties of CPA EP \label{sec:chpt3: properties of CPA EP}}

CPA occurs when an isolated solution to the wave equation with purely incoming boundary conditions has a real eigenfrequency~\cite{Chong:2010ft, Wan:2011bz, Noh:2012wx, Piper:2014js, 2016_Zhou_Optica, Zhao:2016cd, Baranov:2017jv}, as discussed in \Cref{sec:chpt0: lasers and cpa} of \Cref{sec:chpt0: intro}.
Physically, the input state is trapped within the cavity by destructive interference and can only decay by absorption into an active medium within the cavity.
A CPA cavity will perfectly absorb one specific wavefront, given by the eigenmode of the incoming wave operator, at the real frequency which is its eigenvalue.
By flux conservation, the purely incoming and outgoing states cannot have real eigenfrequencies unless there is intrinsic gain or loss; the addition of material loss is necessary to move the frequency of a purely incoming state onto the real axis to achieve CPA.

A CPA EP is the generic degeneracy of such incoming solutions, corresponding to the coalescence of two CPA modes.
Exceptions occur for degenerate but decoupled states~\cite{Piper:2014js}, e.g.~those with different symmetry as mentioned in \Cref{sec:chpt0: eps}, but these cases will be neglected here.
Such absorbing EPs had not been studied prior to our work in~\cite{2019_Sweeney_PRL}, and should be readily observable with setups previously used to investigate resonant EPs~\cite{Peng:2014kl, Chang:2014by, Miao:2016eo}.
Subsequent to that work, we have collaborated with the group of Lan Yang and Washington University in St. Louis on such experiments and very recent results~\cite{2020_yang_cpaep} appear to confirm predictions of our theory.

The novel CPA EP can be directly probed with a steady-state incident wavefront, which is possible precisely because the incoming eigenfrequency is real.
This is in contrast to the often studied resonant EP, which is typically probed indirectly in the sense that the existence of the complex-valued singularities that become degenerate is inferred from their effect on scattering at real frequencies~\cite{Peng:2014kl, Zhen:bl, Zhou:2018dy}.
A complex frequency does not correspond to a physical steady-state solution, but one with an envelope which decays or grows in time, depending on whether it is in the lower or upper half-plane, respectively.
Directly probing a resonant EP in an electromagnetic cavity requires the addition of sufficient gain to bring the degenerate eigenfrequencies to the real axis, and corresponds to lasing at threshold.
But such an amplifying system is not ideal for the study of EPs, due to the large amplified spontaneous emission noise at threshold, and the necessity of including the non-linearity of the medium to stabilize lasing above threshold.

As mentioned earlier in \Cref{sec:chpt0: nh physics}, the solution of an electromagnetic eigenvalue problem, such as a resonance or CPA mode, is also a solution to the more general {\it scattering} problem, albeit a special one.
For example, a CPA mode is the scattering solution in which no flux is scattered from the cavity, and is therefore visible in the scattering spectrum as the vanishing of one eigenvalue of the scattering matrix $S(\w)$.
The signature of CPA EP in scattering is a quartic flattening of the absorption lineshape when probed in the perfectly absorbed eigenchannel of $S$; for ordinary CPA it is quadratic.
Because the eigenchannel input depends on the frequency, the flattening of the absorption line is only visible if the appropriate frequency-dependent input is used; it will not be visible for a fixed frequency-independent input.
%This is only visible if one monitors the frequency-dependent eigenvalues of $S(\w)$, which requires a phase-sensitive measurement of $S$ at each frequency; the flattening is spoiled if one observes the scattered flux for a fixed, frequency-independent input.

\begin{figure}
    \centering
    \centerline{ \includegraphics[width=.9\textwidth]{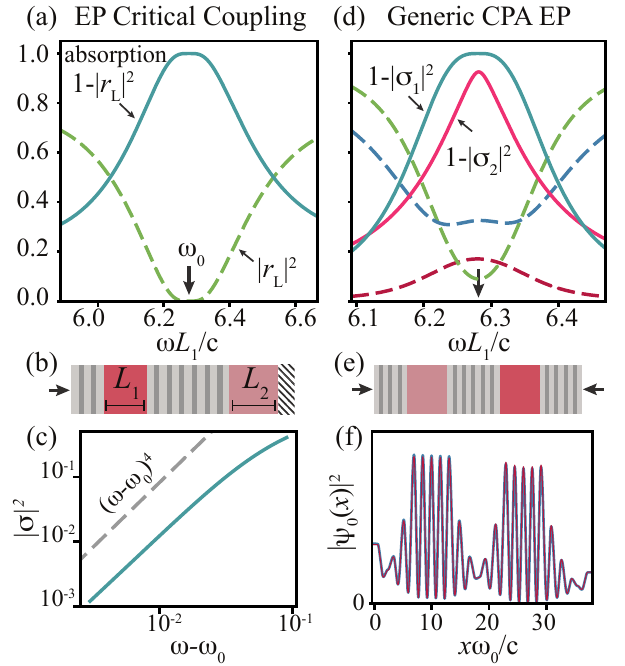} }
    \caption[CPA EP in one-dimension in two cases: one-port critical coupling, generic two-port]{Scattering from coupled cavity structures for two cases, both tuned to CPA EP.
    The end mirrors are asymmetric.
    {\bf (a},{\bf d)}:~Absorption lineshapes of eigenchannels (solid): CPA channel (blue) reaches 100\% absorption at the EP frequency $\wn$.
    In (a,d), CPA lineshape is quartic [verified in {\bf(c)}], while in (d) the non-CPA channel (red) is quadratic.
    Scattering coefflcients $|r_L|^2$, $|r_R|^2$, $|t|^2$ are shown as green, blue, red dashed lines.
    {\bf(b},{\bf e)}:~Schematics of scattering structures: cavities (red) with lengths $L_1, L_2$, and unequal absorption, emitting to free-space through the end mirrors.
    Right mirror is perfect in (b), and permeable but unequal to the left one in (e); parameter values are given in \Cref{tab:chpt4: one dim values}.
    {\bf(f)}:~generic case of unequal coupling yields asymmetric asymptotic values for CPA EP mode $|\psi|^2$, implying that $S$ is not also at an EP, consistent with the discussion in \Cref{sec:chpt4: coincidence of EP of S and CPA EP}.
    }
    \label{fig:chpt4: one dim CPA EP}
\end{figure}

The EP modification of the eigenchannel lineshape, shown for a one-dimensional cavity in \Cref{fig:chpt4: one dim CPA EP}, generalizes to higher dimensional and/or multichannel, quasi-one-dimensional CPA EPs.
Its origin can be understood as follows: near an ordinary CPA frequency $\wn$, an eigenvalue $\sigma(\omega)$ of $S(\w)$ will pass through zero linearly in the detuning $\delta\defn\omega-\wn$, so that $|\sigma(\omega)|^2\propto\delta^2$.
In the vicinity of the parameter values leading to CPA EP, there are two CPA frequencies near each other ($\wn + \delta_1$ and $\wn + \delta_2$), both belonging to the same eigenvalue $\sigma(\omega)$, whose smooth variation implies $\sigma(\omega) \propto \delta_1  \delta_2$.
At CPA EP, both $\delta_{1,2} \to \delta$, and $|\sigma|^2\propto \delta^4$, which is the quartic absorption lineshape.
The other conceivable behavior, where distinct eigenvalues $\sigma_1(\w)$ and $\sigma_2(\w)$ of the $S$-matrix meet at zero, does not correspond to CPA EP, but rather to an EP of $S$ with zero eigenvalue, as noted above.
In this case, the smoothness assumption is violated and the lineshape is not quartic.
At present we neglect this possibility; we will return to it in \Cref{sec:chpt4: EP of S2}.

\section{One-Dimensional CPA EP \label{sec:chpt4: one-dim cpa ep}}

The general properties of a CPA EP described above are exemplified by a one-dimensional electromagnetic structure, consisting of two cavities created by a series of three mirrors (see \cref{fig:chpt4: one dim CPA EP}).
An EP is realized by coupling the two cavities via a central partially reflecting Bragg mirror and by introducing unequal absorption within each cavity.
In this section we will discuss two interesting cases using this setup, and will return to it again in \Cref{sec:chpt4: EP of S2} to discuss a third unusual case.

In the first, shown in \cref{fig:chpt4: one dim CPA EP}a-c, the structure is terminated on the right by a perfect mirror and is accessible only from the left through a partial Bragg mirror.
With only one channel, the $S$-matrix is scalar, comprising one coefficient: the left reflection amplitude $r_L$.
The absorption is $1- |r_L|^2$.
This setup corresponds to the usual critical coupling to a cavity (one-channel CPA), except that the cavity is tuned to an EP of the incoming wave operator and hence the absorption lineshape is quartic.
For this one-channel case, the modified lineshape is evident even with a phase-insensitive measurement of the scattering matrix, since every input is an ``eigenvector'' of the scalar $S$.

The second case, shown in \cref{fig:chpt4: one dim CPA EP}d-f, has unequal Bragg mirrors on the two ends that are both permeable.
This defines a two-channel $S$-matrix, characterized by three scattering amplitudes $r_L$, $r_R$, $t$.
Exciting the absorbing eigenchannel of $S$ requires coherent illumination from both sides with a definite relative intensity and phase \cite{Chong:2010ft}.
The quartic absorption lineshape is evident in \cref{fig:chpt4: one dim CPA EP}d for the $S$-matrix eigenvector whose eigenvalue vanishes at the CPA frequency.
However neither the non-zero eigenchannel, nor the one-sided scattering coefficients ($|r_L|^2$, $|r_R|^2$, $|t|^2$), exhibit such a flat-top profile.

\begin{table}
    \centering
    \begin{tabular}{ l || l l }
         & \textrm{Fig.~a-c}  &  \textrm{Fig.~d-f}  \\
        \toprule
        Grating high index & 2.0                     & 2.0\\
        Grating low index  & 1.5                  & 1.5   \\
        $L_1$        & {\bf1.2500}      & {\bf1.2560}  \\
        $L_2$        & 1.4                  & 1.2566         \\
        $n_1^\prime$        & 2.0 & 2.0 \\
        $in_1^{\prime\prime}$        & {\bf0.0382i} & {\bf0.0043i} \\
        $n_2^\prime$        & 2.0 & 2.0\\
        $in_2^{\prime\prime}$        & {\bf0.0192i} & {\bf0.0472i} \\
        EP frequency $\wn$        & 5.0199             & 5.0012      \\
        \bottomrule
    \end{tabular}
    \caption[Parameters used for in one-dimensional one- and two-port CPA EP in \cref{fig:chpt4: one dim CPA EP}]{Parameters used in each case in \cref{fig:chpt4: one dim CPA EP} (EP-tuned parameters in bold).}
    \label{tab:chpt4: one dim values}
\end{table}

\section{Whispering Gallery Mode CPA EP \label{sec:chpt4: WGM CPA EP}}

We now explore higher dimensional structures, both in free-space and guided wave geometries.
For the case of resonant EPs, there has been extensive study of perturbed and deformed disk resonators in two dimensions, for which the EP of whispering gallery modes (WGMs) directly implies a spatially chiral solution, corresponding to either clockwise (CW) or counterclockwise (CCW) circulations of waves in the disk~\cite{Wiersig:2011hs, Wiersig:2014bq, Cao:2015fv}.
WGMs are resonances of a disk resonator in which the mode is radially concentrated at the circular boundary of the disk, and predominantly propagates azimuthally around the disk, as if repeatedly grazing the disk's edge\footnote{
	The evocative name of these modes comes from the gallery at the base of the dome of St. Paul's Cathedral in London, which has the property that whispers uttered near its circular wall can be heard anywhere else along the wall, even at great distance.
	The conventional explanation of this phenomenon is originally due to Rayleigh.
}.
These strongly chiral resonances have been probed experimentally through asymmetric backscattering and chiral laser emission~\cite{2016_Peng_PNAS}.
We now show that CPA EP in such a system will lead to chiral absorption: perfect absorption for, say, CW input, and substantial backscattering for CCW.
We note that standard CPA in disk and sphere resonators has been studied previously~\cite{Baranov:2017jv, Noh:2012wx}.

\subsection{Free-Space Whispering Gallery Modes \label{sec:chpt4: free space wgm}}

Our first example of a two-dimensional CPA EP is for a perturbed disk in free space, following closely the model introduced by Wiersig~\cite{Wiersig:2011hs} of a dielectric disk perturbed by point scatterers.
Wiersig showed how tuning the parameters of the scatterers could generate resonant EPs (off the real axis) and that the corresponding single eigenfunction would be highly chiral.
Here we adapt this model by adding absorption and showing that one can tune to CPA EP on the real axis.
The parameters that we chose in order to realize a CPA EP are given in \cref{tab:chpt4: free space params}.
%As a first pass at the problem of chirally-dependent absorption, we will adapt the model of Wiersig of a dielectric disk perturbed by two point scatterers~\cite{Wiersig:2011hs}, with parameters chosen to realize an absorbing EP at a real frequency (see \Cref{tab:chpt4: free space params}).
The perturbation from the first point scatterer splits the degenerate WGMs with angular momenta $m =\pm q$ into two standing-wave resonances.
Fine-tuning the second scatterer brings these two resonances back to degeneracy, forming an EP with CCW chirality at a complex frequency.
Finally, introducing a critical degree of absorption brings the absorbing EP to a real frequency.
This free-space system can support an absorbing EP, however it is difficult to implement in practice, and the dissipation is only marginally different for the two chiral inputs.

\begin{table}
    \centering
    \begin{tabular}{ l l}
    	\toprule
        Index of disk                      & 1.5+{\bf0.0021i}  \\
        Radius of disk                    & 1.0                     \\
        Index of scatterers             & 1.5                      \\
        Radius of scatterers          & 0.05                  \\
        Distance of scatterer 1         & 0.04      \\
        Distance of scatterer 2         & {\bf0.0454}                  \\
        Angle between scatterers & {\bf156.30$^\circ$} \\
        EP frequency $\wn$                                & 15.126             \\
        \bottomrule
    \end{tabular}
    \caption[Parameters used in perturbed disk in free-space (\cref{fig:chpt4: free space})]{Parameters used in \cref{fig:chpt4: free space} (EP-tuned parameters in bold).}
    \label{tab:chpt4: free space params}
\end{table}

\begin{figure}
    \centering
    \centerline{ \includegraphics[width=.9\textwidth]{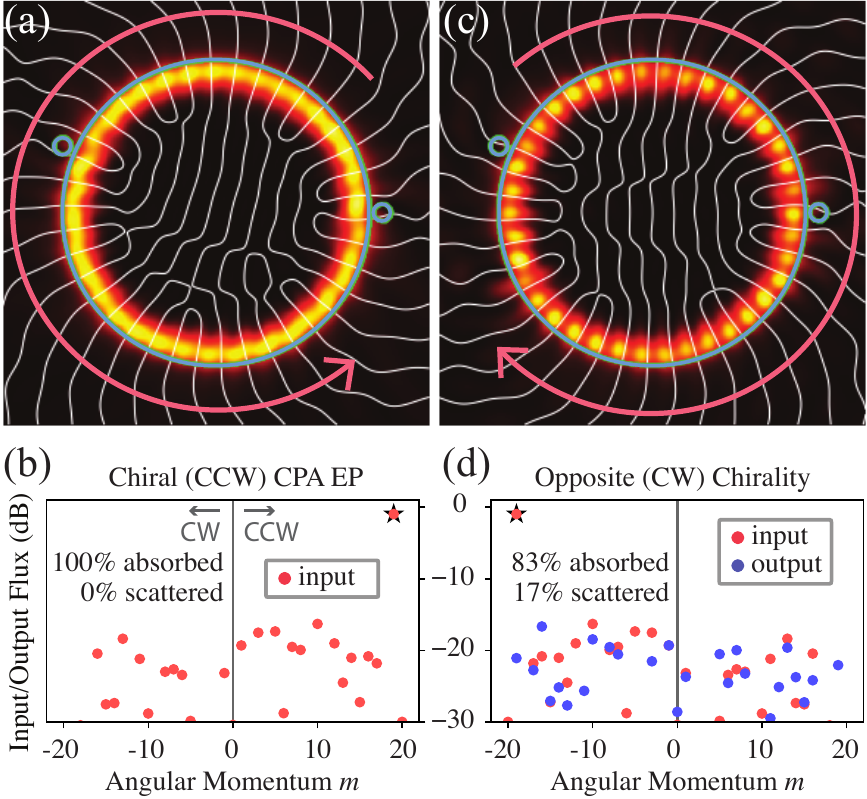} }
    \caption[Chiral CPA EP in a whispering gallery mode resonator perturbed by point scatterers]{Chiral CPA EP of WGMs of an absorbing microdisk perturbed by point scatterers.
    {\bf(a)}:~CCW incident CPA EP mode.
    Intensity plotted as color scale, curves of constant phase (white), disk boundary and scatterers (blue).
    Curvature of phase fronts shows sense of rotation, denoted by arrow.
    Uniform intensity along rim indicates running wave in disk.
    {\bf(b)}:~ Input fluxes carried in each angular momentum channel for CPA EP (CCW) input.
    Dominant channel (denoted by star) carries $80$\% of flux; CPA input is $>99.9$\% absorbed.
    {\bf(c)}:~Total field (incident \& scattered) for reverse chirality input.
    Internal intensity shows standing wave oscillations due to presence of backscattering.
    {\bf(d)}:~As in (b), input fluxes (red), output (blue), for CW input.
    Here we find $\sim 1$\% scattering across many channels, giving a total of $17$\% scattered flux ($83$\% absorption).
    The parameters used here are given in \Cref{tab:chpt4: free space params}.
    }
    \label{fig:chpt4: free space}
\end{figure}

Since the scatterers break the rotational symmetry of the structure, the CPA EP input involves a coherent superposition 
\begin{equation}
    \psi_{\rm in}(r,\phi) = \sum_m c_m \cis{m\phi} H_m^{-}(kr)
\end{equation}
of many angular momenta other than $\pm q$, though at significantly weaker amplitude.
$H_m^{-}$ is the Hankel function of the second kind, and $k \defn \w/c$ is the wavenumber of the incident light.
For the example shown in \Cref{fig:chpt4: free space}, the perfectly absorbed state has $80$\% of its incident flux at $q =19$, with the remaining $20$\% distributed across both CW and CCW at other $m$'s.
Such a wavefront could be feasibly generated for acoustic waves, but an optical implementation will be challenging.
We test the chirality of absorption by exciting the disk with the corresponding CW input through exchanging $c_m \leftrightarrow (-)^m c_{-m}$ in the superposition.
Whereas the original state is $100\%$ absorbed, the state with opposite chirality is only $83\%$ absorbed, which is not dramatically different.
Moreover, if we approximate the CPA input state by solely its dominant component ($m=19$), then both chiralities are equally absorbed (81\%).
This shows that chiral absorption in this system strongly relies on the mutual interference of all the angular momentum channels, and is sensitive to errors in synthesizing the CPA wavefront.
Because of this, we seek a different system to exemplify chirally-dependent absorption.

\subsection{Waveguide-Disk Resonator CPA EP \label{sec:chpt4: waveguide disk CPA EP}}

\begin{figure}
    \centering
    \centerline{ \includegraphics[width=.9\textwidth]{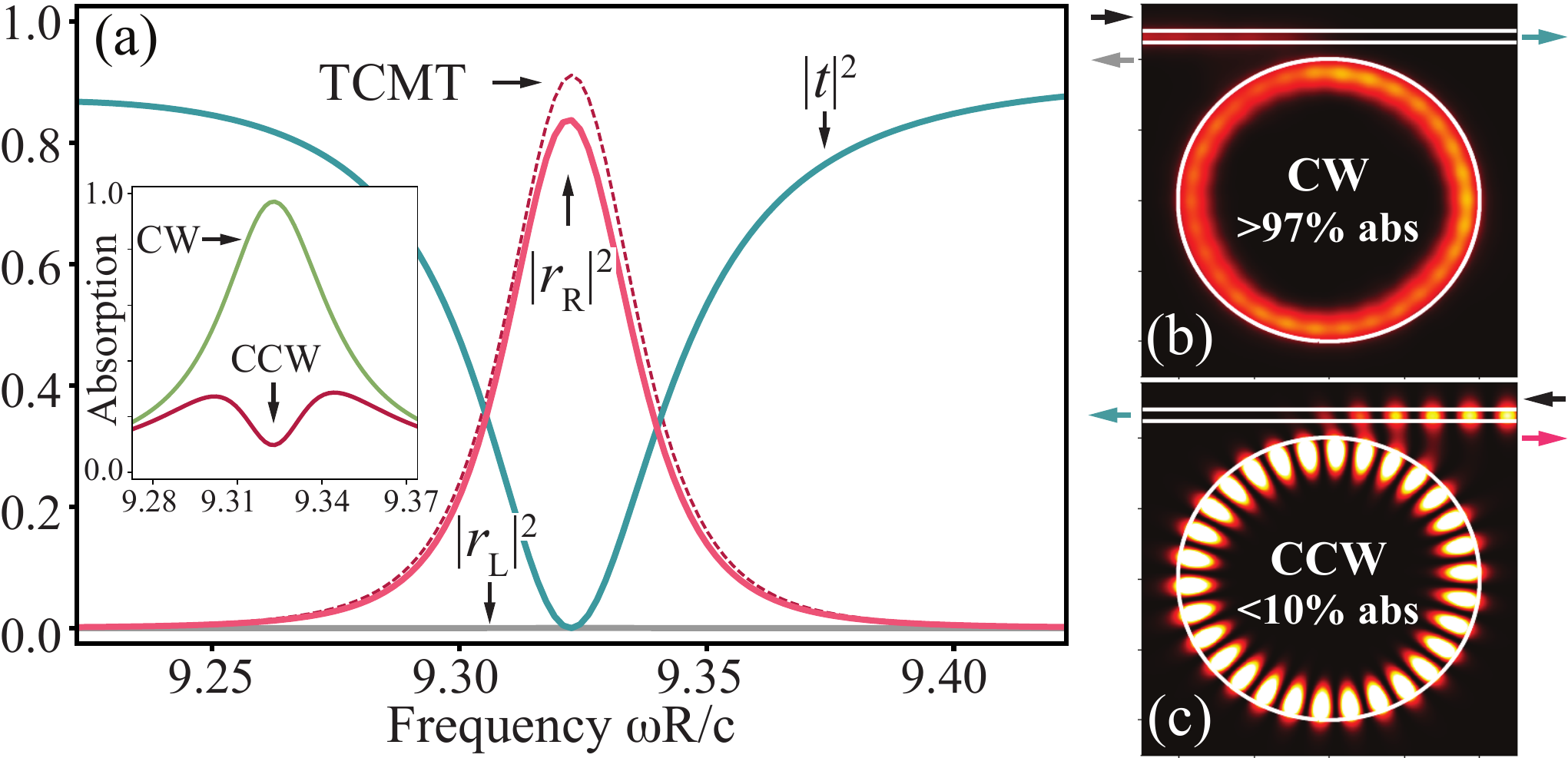} }
    \caption[Chiral absorption from CPA EP waveguide-grated disk system]{Chiral absorption from CPA EP waveguide-grated microdisk system.
    {\bf(a)}:~Reflection $|r_L|^2$ (gray), $|r_R|^2$ (red solid), and transmission $|t|^2$ (blue) for disk with complex azimuthal refractive index grating, tuned to CPA EP.
    TCMT prediction from \Cref{eq:chpt4: back reflection final result} for $r_R$ in dashed red.
    Inset: one-sided absorption spectrum for left illumination (green) and right (red).
    {\bf(b)}:~Intensity of total field (incident \& scattered) for left illumination, corresponding to CPA EP.
    {\bf(c)}:~Same as (b) but for right illumination. Note standing wave of (c) vs running wave of (b) indicates strong coupling between CW and CCW modes only for right incidence, causing chiral absorption.
    The parameters used here are given in \cref{tab:chpt4: wgm disk table}.
    }
    \label{fig:chpt4: waveguide cpa ep}
\end{figure}

We next consider chiral CPA EPs in which a dielectric waveguide or fiber is evanescently coupled to a WGM resonator (see \cref{fig:chpt4: waveguide cpa ep}b,c for geometry).
The goal is to construct an absorbing EP which is mainly chiral within the disk, and is excited almost entirely by a guided mode coming from, say, the left, while at the same time minimizing the absorption for the same input from the right.
The asymptotic regions to the left and right of the disk support both free-space and guided modes, so to achieve waveguide-only CPA we must have the free-space scattering loss rate be much smaller than the waveguide coupling rate.
This makes the light in the disk preferentially escape along the waveguide rather than radiating to free space.
Because of this, using point scatterers as tuning perturbations as we did before is undesirable, since they introduce additional coupling to free space.
Instead of point scatterers, we use an azimuthally varying grating on the real and imaginary parts of the refractive index to promote non-hermitian asymmetric coupling of WGMs.
This configuration is similar to those used to study $\mathcal{PT}$-symmetry breaking and unidirectional invisibility in Refs.~\cite{2011_Lin_PRL, 2012_Regensburger_Nature, 2013_Feng_nmat, Feng:2014gp, Feng:2014gg, Miao:2016eo}.
However, here we introduce only loss and not gain in the grating, and there is no ${\cal PT}$-like discrete symmetry.
Such a symmetry would actually degrade the device's performance, as we will see.

The results of a finite-difference frequency-domain numerical simulation for the setup is shown in \cref{fig:chpt4: waveguide cpa ep}, with parameters summarized in \Cref{tab:chpt4: wgm disk table}.
We see that indeed the absorption in this geometry is strongly chiral, being  $>97\%$ when the disk resonance is excited from the left (CW excitation), but $<10\%$ when it is excited from the right (CCW excitation), as desired.
Note that the $2.7\%$ of the input which is not absorbed for the CW excitation is removed from the disk through free-space radiation, and the CW reflection is truly negligible.
The system satisfies all the requirements for reciprocity, so that the transmission from left to right and vice versa are equal~\cite{2013_Jalas_nphoton}.
Therefore the difference in absorption between left- and right-excitation should be predominantly accounted for by backscattering into the waveguide for the CCW excitation, which it is.
For CCW excitation, $\sim82\%$ is reflected, while only $\sim10\%$ is radiated to free space.
This behavior is possible because of the running-wave versus standing-wave pattern of CW and CCW excitation, respectively.
The CPA EP mode, accessed by illumination from the left, is highly chiral, and hence has no nodes by design; thus it is affected by all of the loss regions equally.
In contrast, the CCW excitation has many nodes which align with the localized lossy regions, thereby minimizing absorption.

The scattering lineshapes are consistent with what would be expected from an absorbing EP.
The reflection lineshape is a squared Lorentzian~\cite{Pick:2017em}, while the transmission lineshape remains Lorentzian (see \Cref{sec:chpt4: tcmt for grated disk} for this derivation). 
Also, the perfectly absorbed eigenchannel of $S(\w)$, which is two-sided except at the CPA frequency, exhibits a quartic lineshape, though this is not shown in \cref{fig:chpt4: waveguide cpa ep}, which focuses on one-sided CW and CCW excitations.

There is some subtlety in adapting the exact CPA EP mode to the problem at hand.
The disk-plus-waveguide setup does not admit CPA solutions which only propagate in along the waveguide.
The exact CPA solutions will require some small free-space flux to excite the disk, just as the corresponding laser would radiate weakly into those free-space channels.
%However, the desired inputs for the system are purely guided modes.
If we excite the system, tuned to an exact CPA EP, solely using the waveguide, then we will find a small but measurable transmission and imperfect absorption.
Since what we are interested in is a chiral absorber without free-space excitation, we adjust the waveguide parameters relative to the CPA EP configuration in order to minimize this small transmission.

This highlights a subtle feature of CPA in general; to achieve CPA, all input/output channels must be controlled.
If some are not, it will be in general impossible to input the required eigenstate, and exact CPA cannot be achieved.
Here we can come quite close because of the smallness of the free-space emission.
We saw in the previous chapter (end of \Cref{sec:chpt2: rigorous RSM}) that reflectionless scattering modes do not have this limitation, if one treats scattering into uncontrolled modes as another loss channel.
Exact RSM can be achieved in experimental configurations in which exact CPA is impossible.

The waveguide parameter that we adjust in this case is its distance from the disk, which controls the coupling rate between the waveguide and the disk's resonances.
Qualitatively, we expect free-space channels to act as a small additional loss with respect to the guided channels.
Therefore by neglecting free-space excitation we are effectively perturbing the critical coupling condition.
We counteract this by increasing the coupling rate to the waveguide by a few percent, until the waveguide transmission becomes negligible. 
Since we are no longer solving the exact CPA problem, then we are no longer guaranteed that all the flux will be absorbed in the grated disk.
The remainder is lost to free-space radiation, and this is the reason why the absorption in \cref{fig:chpt4: waveguide cpa ep} falls slightly short of unity.

The intuitions given above are not enough to fully understand how adding an index-grating leads to chiral absorption, and what the appropriate design parameters are.
To address this, the following section explores in detail the theoretical underpinnings of this design.
Analytic results for the $S$-matrix elements for the disk-plus-grating are derived using TCMT, and were used to generate the optimized results, shown in \cref{fig:chpt4: waveguide cpa ep}.
A good qualitative understanding of the design principles is also achieved, and ${\cal PT}$ symmetry is shown not to lead to optimal chiral response.
Readers not interested at present in understanding these issues in depth may wish to read only the summary on page \pageref{sec:chpt3: summary of analysis}.

\subsection{Calculating Scattering Amplitudes for CPA EP \label{sec:chpt4: scattering amplitude calculation}}

This system is well modeled by temporal coupled-mode theory (TCMT)~\cite{Haus_book, 2003_Fan_JOSAA, 2004_Suh_JQE, 2018_Wang_OL, 2019_Zhao_PRA, 2017_Alpeggiani_PRX}, as introduced in \Cref{sec:chpt1: tcmt}, which takes into account only the CW and CCW angular momentum states of the disk.
These are coupled to each other via the grating, and both are evanescently coupled to the single-mode waveguide.
First we evaluate the effect of the grating perturbation on the closed cavity Hamiltonian $H$ at CPA EP, and then use this within TCMT to evaluate the scattering coefficients as a function of the grating's angular Fourier components.
This provides the design principle implicitly used in \Cref{sec:chpt4: waveguide disk CPA EP} for maximizing the absorption in the disk from one side and the reflection from the other.

\subsubsection{Azimuthally Perturbed Closed-Cavity Hamiltonian}

We will find the TCMT Hamiltonian of the grated disk by treating the grating as a small perturbation of the bare disk.
The relevant resonances are the degenerate CW and CCW WGMs of the bare disk, which have angular momentum quantum numbers $-q$ and $q$, respectively.
The degenerate complex frequency of these modes is $\Omega_0$.

Under a perturbation $V$ the two degenerate eigenvalues of the Hamiltonian $H$ shift by $\delta \Omega_{1,2}$: $H \rightarrow H +V$, $\Omega_0 \rightarrow \Omega_0+\delta \Omega_{1,2}$.
On the other hand, when the Helmholtz wave operator $\hat D \defn -\e^{-1}\nabla^2$ is perturbed by $\delta \hat D$, its spectrum shifts as $\Omega_0^2 \rightarrow (\Omega_0 + \delta \Omega_{1,2})^2 \simeq \Omega_0^2 +2\Omega_0 \delta \Omega_{1,2}$.
By comparing the shifts, it follows that small perturbations in the closed cavity Hamiltonian and the wave operator are related by $V=\delta D /2\Omega_0$.

For the case of the grated microdisk, we limit ourselves to cylindrically separable perturbations of the form $\e(r)\to[1+\rho(r)\tau(\theta)]\e(r)$, for which $\delta \hat D=-\rho(r)\tau(\theta) \hat D$, where $r$ is the radial distance from the center of the disk, and $\theta$ is the azimuth.
The matrix elements of perturbation $V$ in the basis of the unperturbed Hamiltonian are
\begin{equation}
    \label{eq:chpt4: Vmn}
    V_{mn}=\frac{\Omega_0}{2}\int d^2 x\ \phi_m(\bx) \rho(r)\tau(\theta)\psi_n(\bx).
\end{equation}
In deriving this, we used the original eigenvalue equation $\hat D \psi_m = \Omega_0^2 \psi_m$, with $m,n$ are in $\{\pm q\}$.

The operator $\hat D$ is symmetric, therefore the sets of left and right eigenfunctions ($\{\phi\},\,\{\psi\}$, respectively) are equal and biorthogonal with weight $\e(\bx)$, i.e. $\int d^2 x\ \phi_i(\bx)\,\e(\bx)\, \psi_j(\bx) \propto \delta_{ij}$. %, usually written $(\phi_j,\psi_i)=\delta_{ij}$.
The eigenfunctions of $\hat D$ for the unperturbed microdisk are $\psi_m(r,\theta) = R_m(r)\cis{m\theta}$, where the $R_m$'s are functions of the radius only, and so by biorthogonality $\phi_m = \psi_{-m}$.
The matrix elements given by Eq.~\eqref{eq:chpt4: Vmn} can be evaluated in terms of the Fourier components $\tau_n$ of $\tau(\theta)$:
\begin{equation} 
    \label{eq:chpt4: Vmn2}
    V_{mn}= \Omega_0 C_{mn}\tau_{m-n},
\end{equation}
where 
\begin{gather}
    C_{mn} \defn \pi\int_0^\infty dr\,r\, R_m(r) R_{-n}(r) \rho(r) \label{eq:chpt4: Cmn coeffs} \\
    \tau(\theta) \defn \sum_n\tau_n e^{in\theta} \label{eq:chpt4: fourier components}.
\end{gather}
The Hamiltonian of the perturbed, closed-cavity disk, in the degenerate CW/CCW basis, is
\begin{equation}
    \label{eq:chpt4: effective hamiltonian}
    H_{mn} = \Omega_0 \delta_{mn}+ \Omega_0 C_{mn} \tau_{m-n},
\end{equation}
with $m,n$ each being either $-q$ or $q$.

\Cref{eq:chpt4: effective hamiltonian} is the first main result of this analysis section, and is an expression for the degenerate closed cavity Hamiltonian of the disk-plus-grating in terms of the Fourier components of the grating, assuming that the grating is a small perturbation of the bare disk.

    \subsubsection{TCMT for Grated Disk at CPA EP \label{sec:chpt4: tcmt for grated disk}}

%Equipped with an expression for \heff, we now use the formalism presented in~[REF] to compute the scattering matrix coefficients as a function of frequency around the absorbing EP.

In TCMT, the $S$-matrix is related to an the Hamiltonian \heff, which is not necessarily hermitian, by \cref{eq:chpt1: S_form1}:
\begin{equation}
    \label{eq:chpt4: general S-matrix} 
    S(\w) = \left( I - i D \frac{1}{\w - \Hcmt} D^{\dagger} \right) S_0.
\end{equation}
%\begin{equation}
%    \label{eq:chpt4: general S-matrix} 
%    S = \left[ {\mathbb I} - i 2\pi W^\dagger\frac{1}{\omega-(\heffeq-i\pi WW^\dagger)}W \right] S_0.
%\end{equation}
$S_0$ is the ``background'' scattering matrix, i.e., $S$ in the absence of resonances, and $D_{ij}$ is a matrix of coupling coefficients between mode $j$ and asymptotic channel $i$.
For the waveguide-disk system, the background scattering matrix is purely transmitting:
\begin{equation}
    \label{eq:chpt4: S0}
    S_0 = \begin{pmatrix} 
                  0 & 1 \\ 
                  1 & 0
    \end{pmatrix}.
\end{equation}

In the case where there are as many modes as there are channels, $D$ is square.
We assume that each mode couples with the same rate $\g$ to distinct asymptotic channels of the waveguide: the CW mode couples to the right channel, and CCW to the left, so that $D=\sqrt{\g}\,\mathbb{I}$.
Then \Cref{eq:chpt4: general S-matrix} reduces to
\begin{equation}
    \label{eq:chpt4: specific S-matrix}
    SS_0^{-1} = {\mathbb I} - \frac{i\g}{(\w+i\g/2)\mathbb{I} - H},
\end{equation}

Upon applying a perturbation that tunes $H$ to a non-hermitian degeneracy, we can write $H=\Omega_{\mathrm{EP}}{\mathbb I}+N$, where $\Omega_{\mathrm{EP}}$ is the perturbed complex frequency, still degenerate, and $N$ is an unknown nilpotent matrix ($N\neq0$ but $N^2=0$) whose elements will be determined later by comparison with \cref{eq:chpt4: effective hamiltonian}.
Because $N^2=0$, the geometric series expansion of \cref{eq:chpt4: specific S-matrix} naturally truncates after the first two terms:
\begin{equation}
    \label{eq:chpt4: N truncation}
    \begin{pmatrix}
         t(\w) & r_{\rm L}(\w) \\
         r_{\rm R}(\w) & t(\w)
    \end{pmatrix} = 
        \frac{\delta(\w) - i (\g - \G)/2}{\delta(\w) + i (\g+\G)/2}\,\mathbb{I}-\frac{i\gamma}{{\bs(}\delta(\w) + i (\g+\G)/2{\bs)}^2}N,
\end{equation}
where
\begin{equation}
    \label{eq:chpt4: ep detunings}
    \delta(\w) \defn \w - \re{\Omega_{\rm EP}} \qquad\qquad \G/2 \defn -\im{\Omega_{\rm EP}}.
\end{equation}
In evaluating the LHS, we have used the explicit form of $S_0$ from \cref{eq:chpt4: S0}, which is specific to this setup.
\Cref{eq:chpt4: N truncation} fully characterizes the reflection and transmission coefficients as functions of frequency near a CPA EP.

We can relate this to $H$ in \cref{eq:chpt4: effective hamiltonian} by taking $\Omega_{\mathrm{EP}} = \Omega_0(1 + C_{qq} \tau_0)$, and $N_{mn}=(1-\delta_{mn})\Omega_0 C_{mn} \tau_{m-n}$, which vanishes on its diagonal.
Requiring that $N$ is nilpotent puts a strict constraint on the Fourier components of the grating.
First, note that the overlap integrals $C_{mn}$ cannot be chosen to make $N$ nilpotent.
The radial functions $R_m(r)$ are given by Bessel functions of integer order, so that $R_m(r)$ and $R_{-m}(r)$ are related by a sign, as are the overlap integrals $C_{m,-m}$ and $C_{-m,m}$; if one vanishes then so does the other.
Therefore the $C$'s cannot be tuned to make $N^2=0$ but $N\neq0$.
Instead, to achieve nilpotency it necessary to constrain the Fourier components $\tau_{\pm2q}$ of the grating so that {\it only one} of $\tau_{\pm 2q}$ vanishes.
This implies that the grating is \nh, since if $\tau(\theta)\in \mathbb{R}$, then $\tau_{2q}=\tau_{-2q}^*$.
Without loss of generality, we take $\tau_{-2q}=0$, so that all the elements of $N$ vanish except for $N_{q,-q}$.

Plugging this into \cref{eq:chpt4: N truncation} and requiring CPA ($\g=\G$), we determine the lineshapes of the reflection and transmission coefficients at CPA EP:
\begin{gather}
    t(\delta) = \frac{\delta}{\delta+i\Gamma} \\
    r_\mathrm{L}(\delta) = 0 \\
    r_{\rm R}(\delta) = \frac{i}{\Gamma}\frac{\Omega_0 C_{qq}\tau_{2q}}{{\bs(}1-i\delta/\Gamma{\bs)}^2} = \frac{r_{\rm R}(0)}{{\bs(}1-i\delta/\Gamma{\bs)}^2} \label{eq:chpt3: r0 simple}
\end{gather}
where
\begin{gather}
    \delta = \w - \re{\Omega_0(1+C_{qq}\tau_0)} \\
    \Gamma=-2\mathrm{Im}\{\Omega_0(1+C_{qq}\tau_0)\}.
\end{gather}
$\delta$ is the detuning from the degenerate CPA frequency; $\G$ is the HWHM of the transmitted intensity $|t(\w)|^2$.
The amplitudes for transmission and reflection in the CW direction vanish exactly as they would for CPA or critical coupling in the absence of an EP.
The remaining reflection amplitude for CCW incidence at the CPA EP frequency $\delta=0$ is
\begin{equation} 
    \label{eq:chpt4: rd}
    r_\mathrm{R}(0) = - \frac{i}{2}\frac{C_{qq}\tau_{2q}(2+iQ_0^{-1})}{2\mathrm{Im}\{C_{qq}\tau_0\}+Q_0^{-1}(1+\mathrm{Re}\{C_{qq}\tau_0 \})}.
\end{equation}
$Q_0$ is the quality factor of the bare disk, without grating or waveguide:
\begin{equation}
    Q_0 \defn -\frac{\re{\Omega_0}}{2\im{\Omega_0}}.
\end{equation}
For typical WGMs, the $Q$-factors are high, so that we can neglect $Q_0^{-1}$, and the radial wave functions are real to a good approximation, as is $C_{qq}$.
The reflection amplitude at the EP line-center, \cref{eq:chpt4: rd}, takes the remarkably simple form in the high-$Q$ limit
\begin{equation} 
    \label{eq:chpt4: r0}
    |r_\mathrm{R}(0)|^2 = \frac{1}{4}\frac{|\tau_{2q}|^2}{\mathrm{Im}\{\tau_0\}^2}.
\end{equation}
This is the second main result of the analysis, and motivates the grating design: maximize the asymmetry of the CW/CCW couplings by making $\tau_{-2q}$ vanish while simultaneously making $\tau_{2q}$ large, and make the average loss ($\im{\tau_0}$) small but consistent with critical coupling.

The analysis in this section can be extended to include non-separable perturbations that can be decomposed into separable pieces: $\delta\e(r,\theta) = \e \sum_j \rho^j(r)\tau^j(\theta)$.
In this case, the nilpotent matrix becomes $N_{mn}=(1-\delta_{mn})\Omega_0 \sum_j C^j_{mn}\tau^j_{m-n}$, where $C^j_{mn}=\pi\int_0^\infty dr\,r\, R_m(r) R_{-n}(r) \rho^j(r)$ and $\tau^j(\theta)=\sum_n\tau^j_n e^{in\theta}$.
The condition for $N$ nilpotent is that only one of $N_{\pm q,\mp q}$ vanish, say $N_{-q,q}$: $\sum_j C^j_{-q,q}\tau^j_{-2q}=0$, but $\sum_j C^j_{q,-q}\tau^j_{2q}\neq0$.
In this case we no longer need a non-hermitian perturbation to achieve EP, though we must rely on the $C$'s being complex.
The lossless point scatterers used in \cref{fig:chpt4: free space} and in Ref.~\cite{Wiersig:2011hs} exemplify this.

    \subsubsection{Maximizing Asymmetry of Reflection and Absorption with Loss Only}

Clearly if $\tau_0$ is real in \cref{eq:chpt4: r0}, then $r_R$ can be made very large, even exceeding unity, since the \nh\ grating will contain both amplifying and attenuating regions.
The more interesting case is when there is no gain, only alternating regions of loss and no loss.
The simplest experimentally feasible azimuthal grating of this type is piecewise constant, with real and imaginary parts that have the same angular width $\phi$, angular periodicity $2\pi/P$  for integer $P$, and angular offset $\chi$ between them:
\begin{equation} 
    \label{eq:chpt4: tau}
    \tau(\theta)=f(\theta)+if(\theta-\chi).
\end{equation}
Within one period, the piecewise constant function $f(\theta)$ vanishes unless $0<\theta<\phi$, with $f(\theta+2\pi/P)=f(\theta)$.
The offset $\chi$ is determined from the condition that $\tau_{-2q}=0$, which implies $(1+ie^{2iq\chi})f_{2q}=0$ and
\begin{equation}
    \label{eq:chpt4: chi}
    \chi = \left(M-\frac{1}{4}\right)\frac{\pi}{q}.
\end{equation}
Had we instead required the $+2q$ component of $\tau$ to vanish, then the condition would have been $\chi=(M+1/4)\pi/q$.

We can express the reflection from the CCW side in terms of the $f_m$'s according to \cref{eq:chpt4: r0}, and, using \cref{eq:chpt4: chi}:
\begin{equation} 
    \label{eq:chpt4: newr0}
    |r_\mathrm{R}(0)|^2 = \left| \frac{f_{2q}}{f_0} \right|^2.
\end{equation}
The Fourier components of $f$ vanish for $m$ not equal to a multiple of $P$; the non-vanishing components satisfy
\begin{equation}
    \label{eq:chpt4: fm}
    f_{n\cdot P} \propto \frac{P}{2\pi}\int_0^\phi d\theta e^{-inP\theta} = e^{-inP\phi/2}\frac{1}{n\pi} \sin \frac{nP\phi}{2}.
\end{equation}
Plugging this into \cref{eq:chpt4: newr0} and \cref{eq:chpt3: r0 simple} gives
\begin{equation}
    \label{eq:chpt4: back reflection final result}
    |r_\mathrm{R}(\w)|^2 = \left| \frac{ \sin q\phi}{q\phi} \right|^2 \frac{1}{[1 + {\bs(}\delta(\w)/\G{\bs)}^2]^2},
\end{equation}
so long as $NP=2q$, where the integer $N$ is the order of the grating that we are using to couple the $\pm q$ modes.

The asymmetry of the reflection coefficients and of the absorption achieves the maximal value for thin gratings ($\phi\rightarrow0$).
This affirms the intuition that the lossy regions are ``hidden'' in the nodes of the back-scattered field when excited from the non-CPA side.
The thinner the lossy regions, the less of an effect they have for CCW incidence.
On the other hand, for CW incidence the field is a running wave, so the material loss is equally effective regardless of its spatial distribution.

A more general type of grating has different widths, modulation depths, and periodicities for its real and imaginary parts. 
If the real part of the grating has modulation depth $a_r$, periodicity $P_r$, and width $\phi_r$, and similarly the imaginary part has $a_i$, $P_i$, and $\phi_i$, then the conditions for CPA EP are
\begin{gather}
    \left| \frac{\sin q \phi_r}{\sin q \phi_i} \right|=\frac{a_i}{a_r}\frac{P_i}{P_r}, \\
    \chi = \left( M-\frac{1}{4} \right) \frac{\pi}{q} + \frac{\phi_r-\phi_i}{2},
\end{gather}
where integers $P_r$ and $P_i$ must both divide $2q$.
In this case, the non-vanishing (CCW) reflection is
\begin{equation}
    |r_\mathrm{R}(\w)|^2 = \left| \frac{ \sin q\phi_r}{q\phi_r} \right|^2 \frac{\cos^2{\bs(}q(\phi_r-\phi_i){\bs)}}{[1 + {\bs(}\delta(\w)/\G{\bs)}^2]^2},
\end{equation}
which shows that the more restricted grating analyzed earlier is optimal, with the imaginary and real teeth having the same angular size ($\phi_r=\phi_i$).

This condition is not related to gauged ${\cal PT}$-symmetry, which requires that within each period of the grating the real-part is even and the imaginary-part odd, after subtraction of the average imaginary part.
This corresponds to $\phi=\pi/P$, which yields $|r_\mathrm{R}(0)|^2<41\%$ for any grating order $N$; hence, as noted above, imposing ${\cal PT}$ symmetry would prevent us from realizing the far more reflective grating with $|r_R(0)|^2>80\%$.

\begin{table}
    \centering
    \begin{tabular}{ l l}
    	\toprule
        Disk index                       & 2.0  \\
        WGM mode number $q$ & 15 \\
        Grating real index high   & {\bf2.0149} \\
        Grating imag index high  & {\bf0.0153}\\
        Grating width $\phi$        & 2.0$^\circ$ \\
        Offset angle $\chi$          & {\bf8.9400$^\circ$}    \\
        Grating periodicity $P$    & $30=2q$    \\
        Disk radius                      & 1.0                      \\
        Waveguide width             & 0.08                  \\
        Waveguide distance         & 0.16                  \\
        EP frequency $\wn$                               & 9.3230             \\
        \bottomrule
    \end{tabular}
    \caption[Parameters used in waveguide-coupled chiral absorber (\cref{fig:chpt4: waveguide cpa ep})]{Parameters used in \cref{fig:chpt4: waveguide cpa ep} (EP-tuned parameters in bold).}
    \label{tab:chpt4: wgm disk table}
\end{table}

\subsubsection{Summary of Waveguide-WGM Resonator Calculation \label{sec:chpt3: summary of analysis} }

A circular WGM resonator with a \nh\ azimuthal grating, without any gain, can be designed to act as a nearly perfect absorber for one incident direction, and a reflector for the other.
This is possible because of a CPA EP, which requires that the waveguide is critically-coupled to the lossy resonator and that the grating satisfies certain conditions:
\begin{enumerate}
    \item The relative angle between the real and imaginary parts of the index modulation satisfies \cref{eq:chpt4: chi}.
    \item The number of teeth in the grating must divide $2q$, where $q$ is the angular momentum number of the whispering gallery resonance.
\end{enumerate}
When these conditions are met, the transmission coefficient has a Lorentzian-shaped dip on resonance, similar to critical coupling, while one of the reflection coefficients has a squared-Lorentzian peak.
The remaining coefficient is identically zero.
All of this is numerically demonstrated in \cref{fig:chpt4: waveguide cpa ep}, with \Cref{tab:chpt4: wgm disk table} summarizing the parameters used.

\section{Absorbing EPs and Scattering Matrix EPs \label{sec:chpt4: EP of S}}

\begin{figure}
    \centering
    \centerline{ \includegraphics[width=.9\textwidth]{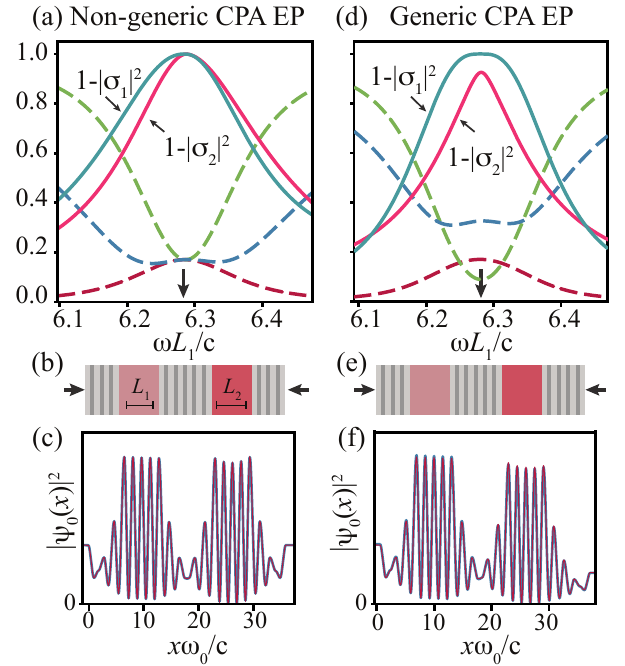} }
    \caption[CPA EP in one-dimension in two cases: generic two-port, symmetric two-port (anomalous case)]{Scattering from coupled cavity structures for two cases, both tuned to CPA EP.
    Panels (d-f) are identical to the earlier \Cref{fig:chpt4: one dim CPA EP}d-f, and are repeated here for comparison to (a-c).
    The end mirrors are symmetric in (a-c), and asymmetric in (d-f).
    {\bf (a)}:~Absorption lineshapes of eigenchannels (solid): CPA channel (blue) reaches 100\% absorption at the EP frequency $\wn$, non-CPA channel (red) nearly does.
    There is an approximate simultaneous EP of $S$, so the lineshape is not quartic.
    Scattering coefflcients $|r_L|^2$, $|r_R|^2$, $|t|^2$ are shown as green, blue, red dashed lines.
    In (a) they become degenerate at $\wn$, as predicted for an EP of $S$-matrix at zero.
    {\bf(b)}:~Schematic of structure: cavities (red) with lengths $L_1, L_2$, and unequal absorption, emitting to free-space through the symmetric end mirrors.
    Parameter values are given in \Cref{tab:chpt4: one dim values2}.
    {\bf(c)}:~CPA EP mode has equal asymptotic magnitude.
    \label{fig:chpt4: one dim CPA EP2}
    }
\end{figure}

We return now to the non-generic case of one-dimensional CPA EP mentioned at the end of \Cref{sec:chpt3: properties of CPA EP}, and illustrated in \cref{fig:chpt4: one dim CPA EP2}a-c.
It is a novel and interesting case, and exhibits qualitatively different behavior from the other two cases shown in \cref{fig:chpt4: one dim CPA EP}.
While the others demonstrated generic scattering behavior near a CPA EP, this case does {\it not} show the generic quartic lineshape, and has a different striking scattering property: at the CPA EP all the scattering coefficients become equal in magnitude.
It is realized in the same type of one-dimensional geometry as before, but with the left and right Bragg mirrors being identical, so that the parity symmetry of the structure is broken only by the small difference between the complex-valued refractive indices of the left and right cavities (see \Cref{tab:chpt4: one dim values2}).
Because of the near-symmetry, this case is anomalous in that a CPA EP {\it and} an EP of the $S$-matrix approximately coincide.
%We will explain this anomaly later in \Cref{sec:chpt4: coincidence of EP of S and CPA EP}.

EPs of the $S$-matrix were identified in~\cite{Chong:2011ev} as a condition for spontaneous symmetry-breaking in ${\cal PT}$-symmetric scattering.
In simplified TCMT models, they can also coincide with resonant EPs or absorbing EPs, so there has been a tendency to ignore the distinction between the two.
Here we show that generically they should not coincide, but under certain symmetry conditions they almost do, leading to qualitatively different scattering behavior than the previous cases, where the $S$-matrix is not defective when the relevant wave operator is.
We emphasize that the novel EPs associated with CPA and RSMs are not in general associated with EPs of the $S$-matrix.

%For now, w
%In this section we address the distinction between an EP of the incoming wave operator and an EP of the scattering matrix, showing that a CPA EP and EP of $S$ do not occur simultaneously without tuning additional parameters.
%We then show that symmetry of the outcouplings of each resonance constrains the two kinds of EP to occur together, removing the need for further parameter-tuning, which is exactly the case presented in \Cref{fig:chpt4: one dim CPA EP2}a-c.

\begin{table}
    \centering
    \begin{tabular}{ l || l l l }
         & \textrm{Fig.~a-c}  &  \textrm{Fig.~d-f} \\
        \toprule
        Grating high index                      & 2.0                     & 2.0\\
        Grating low index                    & 1.5                  & 1.5\\
        $L_1$     & 1.2566         & {\bf1.2560}      \\
        $L_2$            &1.2566               & 1.2566            \\
        $n_1^\prime$    & {\bf1.9981} & 2.0 \\
        $in_1^{\prime\prime}$    & {\bf0.0037i}   & {\bf0.0043i}\\
        $n_2^\prime$    & 2.0      & 2.0\\
        $in_2^{\prime\prime}$   & {\bf0.0609i}   & {\bf0.0472i} \\
        EP frequency $\wn$            & 5.0022          & 5.0012           \\
        \bottomrule
    \end{tabular}
    \caption[Parameters used for in one-dimensional generic and anomalous two-port CPA EP in \cref{fig:chpt4: one dim CPA EP2}]{Parameters used in each case in \cref{fig:chpt4: one dim CPA EP2} (EP-tuned parameters in bold).}
    \label{tab:chpt4: one dim values2}
\end{table}

\subsection{Distinctness of the EP Classes \label{sec:chpt4: EP of S2}}

Every eigenstate of the wave operator with incoming boundary conditions also corresponds to an eigenvector of $S$ with eigenvalue zero.
However, the coalescence of two incoming states does {\it not} simultaneously generate an EP of $S$, as we now prove.

For simplicity, consider an arbitrary one-dimensional cavity described by the Helmholtz equation:
\begin{equation} 
    \label{eq:chpt4: Helmholtz}
    \{\nabla^2 + \e (x) k_j^2 \} \psi_j(x) = 0,
\end{equation}
where $\e (x)$ is the dielectric function of the medium, $k_j = \w_j/c$, and $\w_j$ are the discrete complex eigenfrequencies with purely incoming boundary conditions.
Consider two eigenfrequencies, $\omega_1,\omega_2$, initially with different values and linearly independent solutions, $\psi_1 (x),\psi_2 (x)$.
Further assume that tuning $\e (x)$ causes these two solutions to coalesce at $\omega_0$: $\psi_1,\psi_2 \rightarrow \psi_0$.
In this limit,
\begin{equation}
    \label{eq:chpt4: bulk integral}
    -2i \omega_0 \int_{\mathrm{cav}} dx\, \psi_0 (x) \e (x) \psi_0 (x) = c_0\, \hat{s}_0 \cdot \hat{s}_0,
\end{equation}
where $\hat{s}_0$ is the normalized eigenvector of the $S$-matrix corresponding to $\psi_0$ (i.e. with eigenvalue zero), and $c_0$ is a system-specific constant.
We will prove this identity later, in \Cref{sec:chpt4: integral relation}.
The integral on the left hand side of \Cref{eq:chpt4: bulk integral} does not vanish, since solutions of the wave equation with either purely incoming or outgoing boundary conditions do not satisfy any simple biorthogonality relation over the scattering region~\cite{Leung:1994fq, 2011_Moiseyev_book} (this is a nuanced point that we will return to in \Cref{sec:chpt4: self-orthogonality}).
Hence at an EP of the incoming wave operator, integrals of this type are non-zero.
On the other hand, the RHS of \Cref{eq:chpt4: bulk integral} is proportional to the biorthogonal norm of the eigenvector of the symmetric $S$-matrix with eigenvalue zero; as such it vanishes iff $S$ is {\it also} at an EP~\cite{Trefethen:2005wt}.
A non-vanishing LHS implies that CPA EP does not correspond to an EP of $S$; indeed for the generic case of CPA EP shown in \Cref{fig:chpt4: one dim CPA EP2}d-f, the $S$-matrix has a distinct second eigenvector which is not perfectly absorbed.
This proof generalizes to higher dimensional scattering geometries using Green's theorem.

Conversely, one can construct a scattering structure which has an EP of $S$ with a degenerate eigenvalue zero, which is a specific case of the type mentioned in \Cref{sec:chpt0: eps}~\cite{Chong:2011ev, 2012_Ge_PRA, Ambichl:2013gq}.
But this does not imply CPA EP.
It has the special property that for a reciprocal two-channel system, the scattering coefficients satisfy the relation
\begin{equation}
    r_L(\w_{\rm EP})= - r_R(\w_{\rm EP}) = \pm it(\w_{\rm EP}),
\end{equation}
where $\w_{\rm EP}$ is the EP frequency.
This follows from the three conditions ${\rm det} S=0$, ${\rm Tr} S=0$, and $S=S^T$.
Hence
\begin{equation}
    \label{eq:chpt4: EP_S_feature}
    |r_L(\w_{\rm EP})|^2 = |r_R(\w_{\rm EP})|^2 = |t(\w_{\rm EP})|^2
\end{equation}
is a signature of an EP of $S$ at zero that can be observed with standard phase-insensitive, one-sided reflection and transmission measurements.

The scattering behavior of the structure shown in \cref{fig:chpt4: one dim CPA EP2}a-c, which differs from that of \cref{fig:chpt4: one dim CPA EP2}e only by having identical Bragg mirrors at either end, shows precisely the triple degeneracy of the scattering coefficients characteristic of an EP of $S$ at zero.
This is surprising, since its parameters were chosen to be at CPA EP, and not at an EP of $S$.

\subsection{Coincidence of EPs for Symmetric Outcoupling \label{sec:chpt4: coincidence of EP of S and CPA EP}}

To understand why for this nearly symmetric structure a CPA EP and an EP of $S$ coincide, we turn to TCMT. %, which will provide an analytic but approximate relationship between the eigenfrequencies of the wave operator and the $S$-matrix.
In this section we will show that, within TCMT, if the two cavities have equal out-coupling rates, then CPA EP {\it does} imply a simultaneous EP of the $S$-matrix.
Thus, essentially the same experimental setup can test the properties of these two different types of absorbing EPs.

If TCMT were exact, the two eigenvalues of $S$ would coincide precisely at $\w_{\rm EP}$ and would not be analytic there, leading to a complicated, non-quartic behavior near CPA. 
Due to its approximate nature, we instead find a slight displacement of the EP of $S$ from CPA EP, though this is not visible in the results of \cref{fig:chpt4: one dim CPA EP2}a-c.

We slightly generalize \cref{eq:chpt4: N truncation} from the earlier analysis of the grated disk, which assumed equal outcoupling $\g$, to
\begin{equation}
    SS_0^{-1} = \Lambda - \frac{i\gamma}{(\omega-\Omega_{\mathrm{EP}}+i\gamma/2)^2}N.
\end{equation}
$S_0$ is the non-resonant part of the scattering matrix, $\Lambda$ is some diagonal matrix, and $N$ is nilpotent.
This makes $SS_0^{-1}$ manifestly defective (exceptional), since $(SS_0^{-1}-\Lambda)$ is also nilpotent.

For one-dimensional structures there is no non-resonant coupling of left and right channels, and $S_0\propto \mathbb{I}$.
Therefore in the geometry of \cref{fig:chpt4: one dim CPA EP2}a-c, which has symmetric outcoupling, an EP of the wave operator (in the TCMT approximation this means an EP of \heff) implies a simultaneous EP of $S$, which is what we wanted to show.

\subsection{Integral Relation for Helmholtz EP \label{sec:chpt4: integral relation}}

In this section we derive \Cref{eq:chpt4: bulk integral}, which we used earlier in \Cref{sec:chpt4: EP of S2} to show that EPs of $S$ and the wave operator do not coincide. 
This relation between the unconjugated inner product of the wave operator eigenfunctions and the unconjugated ``norm'' of the $S$-matrix eigenvectors was previously noted in~\cite{YaZeldovich:1961wz, Lai:1990tn, Chang:1996vz}.
We will also discuss its consequences for biorthogonality and self-orthogonal EPs in open geometries.

For simplicity, we focus on the scalar Helmholtz operator in one dimension ($c=1$, $\omega=k$), over a domain of length $2L$, and with purely incoming boundary conditions (appropriate for CPA):
\begin{equation}
    \label{eq:chpt4: incoming helmholtz}
    \left\{ \nabla^2 + \varepsilon(x) \omega_m^2\right\} \psi_m =0\qquad \left[\nabla\psi_m=\mp i\omega_m\psi_m\right]_{\pm L}.
\end{equation}

Take two (nearby) eigensolutions $\psi_{1,2}$ of \cref{eq:chpt4: incoming helmholtz}, with eigenvalues $\omega_{1,2}$.
Consider the integral
\begin{equation}
    \int dx\,\psi_2 \{\nabla^2+\varepsilon\omega_1^2\}\psi_1 = 0.
\end{equation}
Integrating by parts twice, applying the boundary conditions, and dividing by a common factor of $(\omega_2-\omega_1)$ gives
\begin{equation}
    \label{eq:chpt4: bulk-boundary}
    c_0\hat s_1\cdot \hat s_2 = -i(\omega_2+\omega_1)\int dx\,\psi_2\,\varepsilon\,\psi_1,
\end{equation}
where $c_0^{2}=[\psi_1^2(-L)+\psi_1^2(L)][\psi_2^2(-L)+\psi_2^2(L)]$.
The vectors $\hat s_{1,2}\propto{\bs(}\psi_{1,2}(-L),\psi_{1,2}(L){\bs)}$ are the normalized $S$-matrix eigenvectors at $\omega_{1,2}$ with eigenvalue equal to zero.

The dielectric function can be parametrically deformed to bring about an accidental degeneracy (EP), so that $\omega_{1,2} \to \omega_0$, and $\psi_{1,2} \to \psi_0$, in which case
\begin{equation}
    \label{eq:chpt3: unconv ip}
    c_0 \hat s_0\cdot \hat s_0 = -2i\omega_0\int dx\,\psi_0\,\varepsilon\,\psi_0,
\end{equation}
which is what we wanted to show.

\subsubsection{Self-Orthogonality \label{sec:chpt4: self-orthogonality}}

For \nh\ closed systems described by symmetric linear operators, the eigenmodes satisfy a biorthogonality condition~\cite[884]{1981_morse_book}, with the consequence that when two modes coalesce at an EP, each becomes ``orthogonal'' to itself.
This remarkable feature of EPs in closed systems does not easily generalize to open \nh\ systems.

For closed systems, the argument for biorthogonality and EP self-orthogonality is as follows.
The Helmholtz equation, subjected to a closed boundary condition, such as Dirichlet ($\psi(\pm L)=0$), defines an ordinary eigenvalue problem $\hat D \psi_m = k_m^2\psi_m$, where $\hat D \defn -\e^{-1}\nabla^2$.
Define the inner product as
\begin{equation}
    \label{eq:chpt4: inner product}
    \left(\psi,\phi\right) \defn \int_{-L}^Ldx\,\psi (x)\,\e (x)\,\phi(x).
\end{equation}
$\hat D$ is symmetric with respect to this inner product since $(\psi,\hat D\phi)=(\hat D \psi,\phi)$ for any $\psi$, $\phi$ satisfying the  boundary conditions, by integration by parts.
The biorthogonality relation between eigenfunctions, $(\psi_m,\psi_n)\propto\delta_{mn}$, follows.
This also shows that a {\it closed} system EP has a self-orthogonal eigenvector $(\psi_{\rm EP},\psi_{\rm EP})=0$.

For open boundary conditions, such as in \cref{eq:chpt4: incoming helmholtz}, $\hat D$ does not define an eigenvalue problem in the usual way, since the boundary conditions are frequency-dependent, so that the problem is defined self-consistently in its eigenvalue.
Consequently $(\psi,\hat D\phi)\neq(\hat D \psi,\phi)$ since in general $\psi$ and $\phi$ have distinct eigenfrequencies, and therefore satisfy distinct instances of the open boundary condition.
Because of this, biorthogonality does not hold with respect to the usual inner product, and we do not have EP self-orthogonality.

By modifying the usual inner-product for open systems, one may recover biorthogonality, and therefore EP self-orthogonality.
For example, a new inner product defined by the subtraction of the two sides of \cref{eq:chpt4: bulk-boundary}:
\begin{equation}
    ((\psi_2,\psi_1)) \defn \int_{-L}^Ldx\,\psi_2(x)\,\e(x)\,\psi_1(x)-\frac{ic_0\hat s_2\cdot \hat s_1}{\omega_2+\omega_1},
\end{equation}
makes eigenstates biorthogonal by construction, and EP self-orthogonality follows.
This is the procedure followed in \cite{YaZeldovich:1961wz, Lai:1990tn, Chang:1996vz}.
Alternatively, a field or coordinate transformation which suppresses the boundary terms of \cref{eq:chpt4: bulk-boundary}, such as is used in the complex scaling implementation of perfectly matched layers, also leads to biorthogonality provided that the transformation is also appropriately applied to $\e(x)$, as is done in \cite{Pick:2017em, 2011_Moiseyev_book}.
In this way one may still speak of biorthogonal resonant or CPA states, and self-orthogonal EPs; such inner product redefinitions are conventional in the literature.

However, this convention is misleading here, as it obscures the relationship between the relevant (non-vanishing) integral over the cavity region and the self-norm of the $S$-matrix eigenvectors, \cref{eq:chpt4: bulk-boundary}, that proves that an EP of $S$ does not coincide in general with an EP of the wave operator.
Another potential confusion with the inner product defined as in \cref{eq:chpt3: unconv ip}, arises in the context of steady-state {\it ab initio} laser theory (SALT).
There, one introduces a complete set of ``constant-flux'' (CF) states, which describe purely outgoing solutions at a fixed external frequency, $\w$.
Since this boundary condition is eigenvalue-independent, the CF states {\it do} satisfy a simple biorthogonality relation over the cavity, unlike the resonances.

% but this is not a useful convention to adopt here, as it obscures the important point that an EP of the wave operator is generally not an EP of $S$.

\chapter{Saturable Coherent Perfect Absorption \label{chp:chpt4: saturable cpa}}
In the previous two chapters we explored various electromagnetic eigenvalue problems that lead to unusual \nh\ scattering states when the parameters of the scatterer are appropriately tuned, such as CPA.
The dielectric response was assumed to be instantaneous, $\e (\bx,t-t\pr) = \delta(t-t\pr) \e(\bx)$, and therefore spectrally uniform, i.e., non-dispersive: $\e(\bx,\w) = \e(\bx)$.
It was also assumed to be linear, meaning independent of the electric field.
These simplifying assumptions are mathematically convenient, but are not physically realistic much of the time.
The underlying source of the macroscopic dielectric function is the response of the constituent atoms' charge distribution to an external electric field.
The response is not instantaneous and is only linear in the electric field for small amplitudes. %can exhibit its own complicated resonances, distinct from the optical resonances of the macroscopic scatterer.
The dielectrics considered in the rest of this thesis are made of two types of material: the cavity substrate and the gain/loss defect atoms, which typically have partial or complete spatial overlap.
The total susceptibility is simply the sum of each constituent susceptibility.

A more realistic dielectric model for the amplifying or attenuating part of the medium than a constant complex refractive index is the two-level susceptibility discussed in the introductory \Cref{sec:chpt0: saturable two-level media}, at least for some \nh\ materials.
This is parameterized by an atomic resonance frequency $\w_a$, a dephasing rate $\gp$, a pump parameter $D_0$, and a pump profile $F(\bx)$.
The response is negligible outside a frequency window of width $2\gp$, centered on $\w_a$.
The part of the susceptibility that is not due to the two-level medium, denoted by $\e_c(\bx)$, is assumed to be linear and non-dispersive.
This is reasonable so long as its true dispersion is small within the gain/loss band of frequencies of the two-level system.

The pump parameter $D_0$ is the inversion density, defined as the difference between the density of excited and ground-state atoms.
In a physical medium with an ensemble of embedded defects, and without any external forcing, the atoms will be thermally distributed between the two levels. 
%Typically most of the atoms will be in the ground state
%can be taken to be in the ground state to a good approximation.
%This is because the energy gap for applications in photonics is far larger than room-temperature thermal energies.
Therefore a two-level atomic medium with no external forcing, such as a pump beam or injected current, corresponds to a {\it negative} pump parameter $D_0$, since the excited-state population is less than the ground-state.
The case of $D_0=0$, the transparency point, where the atoms neither absorb nor amplify on average, requires external pumping, so that the ground and excited state populations are equal.
Throughout this chapter we will assume a setup in which the absorbing cavity is subject to a pump which can tune the negative value of $D_0$ above its ground state value to some desired point.

The terminology that we will use is that a {\it passive} cavity is one where $D_0=0$, so that the two-level component of the susceptibility is transparent in a mean-field sense, which requires external pumping to achieve.
This is also called a {\it cold} or {\it transparent} cavity~\cite{1986_Siegman_book}.
It is unfortunate that this standard terminology suggests that a two-level medium does not influence the resonances even when a cavity is not externally pumped ($D_0<0$), when the absorption of the atoms increases the resonance width and can have other effects as well.

In this chapter we will extend the previous theory of linear CPA to include the dispersion and saturation of a two-level absorbing medium.
We will generalize the CPA theorem from \Cref{sec:chpt0: lasers and cpa} to apply to such a medium.
The result is a theory of ``saturable CPA'', which exhibits two new features over its linear counterpart:
\begin{enumerate}

    \item The two-level susceptibility is dispersive even without saturation, i.e., in the limit of vanishing input power.
    Because of this, there is a shift of the CPA frequency at pump $D_0^{\rm CPA}<0$ relative to the equivalent lasing frequency at pump $D_0^{\rm las}>0$, as well as a change in the magnitude of the pump: $|D_0^{\rm CPA}|\neq D_0^{\rm las}$.
    Thus when dispersion is taken into account, interchanging gain with loss does not make lasing into CPA.
    This does not violate the CPA theorem from \Cref{sec:chpt0: lasers and cpa}, which is a statement about Maxwell's equations under time-reversal ($n \to n^*$); there is some confusion about this in the literature~\cite{longhi_2011_pra}.
    The difference between CPA and lasing becomes very large in the ``bad-cavity'' limit, in which the radiative loss rate of the cavity approaches the intrinsic dephasing rate, $\gp$, of the medium.
    We will discuss this in detail later in \Cref{sec:chpt4: dispersion}.
    %I show that this a manifestation of an avoided crossing between a cavity mode and a hybrid atom-cavity mode.
            
    \item \label{itm:cpt5: saturation} A cavity with sufficiently large (``over-damped'') two-level absorption can perfectly absorb a particular incident wavefront {\it at a specific power}.
    It is not necessary to fine-tune the pump parameter, only the frequency and amplitude of the incident light.
    This is in contrast to linear CPA, for which {\it any} input power can be totally absorbed, but only for a specific degree of material absorption.
    %This suggests a device that acts as a ``coherent delimiter'': it absorbs a particular coherent wavefront for input power less than some maximum, and acts as an attenuator for greater powers.

\end{enumerate}

\begin{figure}[t]
    \centering
    \centerline{ \includegraphics[width=\textwidth]{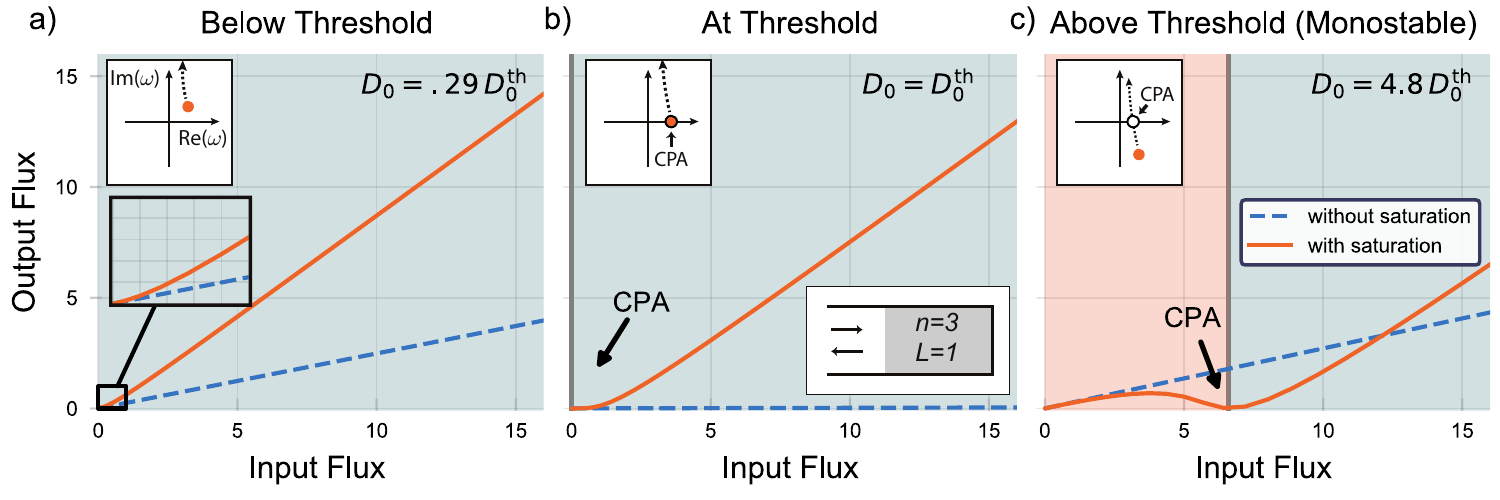} } 
    \caption[Input-output curves for saturable absorber in three regimes: under-damped, critically-damped, over-damped]{Comparison of input-output curves for linear (dashed blue) and saturable CPA (solid orange) for three values of $D_0$.
    The structure is the one-sided, one-dimensional cavity shown in the lower inset in (b), with $\w_a=40.45$, $\gp=2$.
    Incident frequency is $\w=40.292$, which is the saturable CPA threshold frequency; $D_0^{\rm th}=-0.0519$ is the CPA threshold pump.
    Blue-shaded regions indicate the $S$-matrix zero is under-damped (for $S$ linearized around saturable solution); red-shading for over-damping.
    {\bf (a)}:~$|D_0|<|D_0^{\rm th}|$. Saturable CPA does not occur for any input power; saturation quickly causes the output to deviate from linear case (lower inset).
    Upper inset: Argand diagram of trajectory of $S$-matrix zero as input is increased; it never crosses real axis.    
    {\bf (b)}:~$D_0=D_0^{\rm th}$, linear CPA, indicated by zero slope of dashed blue line.
    As with (a), saturation causes orange curve to deviate from blue as input is increased and spoils the CPA condition.
    {\bf (c)}:~$|D_0|>|D_0^{\rm th}|$, over-damped with respect to linear CPA.
    At some finite input, saturation restores CPA condition and linearized $S$-matrix zero crosses real axis.
    In all cases, solutions are in monostable regime.
    }
    \label{fig:chpt4: lin v sat}
\end{figure}

This second feature is to some extent intuitive, and is illustrated in \cref{fig:chpt4: lin v sat}.
If the two-level medium has more atoms in the ground state than in the excited state, corresponding to $D_0<0$, then some of the absorbed photon energy will not be re-emitted as light, but will instead be irreversibly lost to unobserved degrees of freedom, e.g. as phonons.
%This loss rate is phenomenologically captured in the parameter $D_0$.
With a sufficiently high density of ground-state atoms, the scatterer will be over-damped in the vicinity of $\w_a$, and any zeros of the scattering matrix, corresponding to complex eigenfrequencies of the incoming wave operator, will be in the lower half-plane.
The ground state will be partially depleted in regions where the electric field intensity is high, suppressing the local absorption rate of the medium, which is the saturation of absorption.
A cavity which is under-damped, even without saturation, will remain so as the field intensity is increased and saturation becomes important (see \cref{fig:chpt4: lin v sat}a).
On the other hand, an initially over-damped cavity can be made critically damped as the intensity of the incident field is increased from zero, so that, with fine-tuning of the shape and frequency of the incoming wavefront, CPA can be achieved (\cref{fig:chpt4: lin v sat}c).
As the intensity is increased further, the zero of the scattering matrix, linearized around the saturated solution, moves into the upper half-plane, and the saturated cavity becomes under-damped.
The case of a cavity which is initially critically-damped without saturation, i.e., linear CPA, divides these two kinds of behaviors (\cref{fig:chpt4: lin v sat}b).

Uniqueness of scattering solutions is not guaranteed because the two-level susceptibility is nonlinear in the field.
When there is only one solution for a given set of parameters, the solution is monostable.
This is the case depicted in \cref{fig:chpt4: lin v sat}.
It is also possible for the same set of parameters to support more than one solution, as is the case with optical bistability~\cite{1982_Abraham_repprog}.
In this case, a curve such as that in \cref{fig:chpt4: lin v sat}c will become multi-valued in a transition region between high and low input flux.
In this chapter all the solutions will be monostable; more will be said about bistability in \Cref{chp:chpt5: saturable scattering}.

    \section{CPA Theorem for a Saturable Two-Level Medium}

The above heuristic argument for the existence of saturable CPA does not provide a method for finding the CPA frequencies or incident wavefronts.
For that, we seek an analog of the linear CPA theorem which constructs these quantities by relating the outgoing solutions of one structure to the incoming solutions of another, through time-reversal.
Specifically, as noted above, for a nondispersive linear structure with an amplifying refractive index $n\pr(\bx)-in\prpr(\bx)$ and purely outgoing radiation, there is a totally absorbed, purely incoming state of the lossy structure with index $n\pr(\bx)+in\prpr(\bx)$.
The wavefront of the CPA state is the time-reverse of the outgoing state, and their frequencies are identical.

The saturable CPA equation for the dispersive two-level medium considered here is
\begin{equation}
    \label{eq:chpt4: saturable cpa equation}
    \{\nabla^2 + \left[ \e_c(\bx) + \chi_{\rm nl}(\bx,\w_\mu,\psi_\mu) \right]\left(\w_\mu/c\right)^2\} \psi_\mu(\bx) = 0,
\end{equation}
subject to purely incoming boundary conditions, where the nonlinear and frequency-dependent part of the susceptibility is
\begin{gather}
    \label{eq:chpt5: two-level sus}
    \chi_{\rm nl}(\w,\bx,\psi) = \g(\w) D(\w,\bx,\psi) , \\
    \label{eq:chpt5: two-level sus gamma} 
    \g(\w) \defn \frac{\gp}{\w - \w_a + i\g_\perp}, \\
    \label{eq:chpt5: two-level sus D}
    D(\w,\bx,\psi) \defn \frac{D_0 F(\bx)}{1+h(\w,\bx,\psi)}, \quad h(\w,\bx,\psi) \defn |\g(\w)\psi(\bx)|^2.
\end{gather}
$D_0 F(\bx)$ is the real-valued inversion profile that the system relaxes to when not excited by an external source, and $D_0 \in {\mathbb R}$ is the pump parameter, which is negative for absorption and positive for amplification.
As noted above, the naive expectation is that by swapping gain for loss, $D_0 \to -D_0$, a lasing system becomes one that supports CPA and vice versa, but this is incorrect.
The correct application of the CPA theorem is that given a lasing solution in a cavity with two-level susceptibility $\g(\w)D(\w,\bx,\psi)$, there is a totally absorbed incoming solution for a different cavity with $\g^*(\w)D(\w,\bx,\psi)$.
This transformation cannot be brought about solely by tuning $D_0$, which is the parameter that controls gain and loss in a physical setup, since $\g(\w)\not \propto \g^*(\w)$, due to dispersion.
Therefore the CPA theorem, as originally stated, is of limited usefulness in a two-level medium where $\gp$ is small enough so that dispersion is not negligible, and where only the pump parameter $D_0$ is externally controlled, while $\w_a$ and $\gp$ are fixed intrinsic properties of the medium.

We will extend the CPA theorem to include saturating two-level media.
The new theorem will map saturable CPA in one physically realizable cavity to single-mode lasing in another, fictitious cavity; it will provide a construction of saturable CPA solutions from the well-verified SALT solutions for lasing.
The CPA theorem for a saturable two-level medium is as follows:
\begin{description}
    \item {\bf Saturable CPA theorem:} The specific input state which is totally absorbed by a two-level medium is the time-reverse of the threshold lasing mode for the same passive (empty) cavity, but with the active two-level medium characterized by $D_0 \to -D_0$ {\it and} $\gp \to -\gp$.
\end{description}
Its derivation follows from the observation that under this transformation $\g(\w) \to -\g^*(\w)$, while $h(\w,\bx,\psi)$ is unchanged, so that $\chi(\w,\bx,\psi) \to \chi^*(\w,\bx,\psi)$, and the original CPA theorem applies.
The fictitious lasing cavity is unphysical, having a negative ``gain bandwidth'' and intrinsic ``dephasing'' rate of $-\gp$.
The saturated CPA mode is thus the time-reverse of the threshold lasing mode of a different, fictitious cavity, with linear gain equal to the loss of the saturated cavity.

A major feature of the saturable CPA theorem is that no additional computational work is necessary to find the saturated CPA modes beyond what is required to find the lasing modes.
The two are related to each other in a one-to-one fashion.
This has several important consequences:
\begin{enumerate}

    \item Saturable CPA, like laser oscillation, can exist in any dimension, and in any geometry.

    \item As noted above, there exists a CPA threshold pump $D_0^{\rm th\ CPA}<0$, which is distinct from the lasing threshold ($|D_0^{\rm th\ CPA}| \neq |D_0^{\rm th\ las}|$).
When there is a sufficiently higher density of ground-state atoms than excited ones, ($D_0<D_0^{\rm th\ CPA}$), then there exists a totally absorbed incident wavefront with a specific amplitude.
Otherwise, if $D_0>D_0^{\rm th\ CPA}$, then no such solution exists.
This is different from the result of linear CPA, where purely incoming solutions exist only for a single $D_0$, but can have any amplitude in principle (see \cref{fig:chpt4: lin v sat}b).

    \item As the absorption is increased ($D_0$ is decreased), the amplitude of the totally absorbed state tends to increase, though it is possible for the amplitude to be non-monotonic in $D_0$, as with lasing~\cite{2012_Liertzer_PRL, 2014_Brandstetter_NatComm}.
For high-\Q\ cavities, the incident power is approximately proportional to $|D_0 - D_0^{\rm th\ CPA}|$~\cite{ge_2010}.

    \item Multi-mode saturable CPA solutions can also exist, since multimode lasing solutions do.
However, unlike with laser oscillation, all combinations of valid modes can be considered, not just the one that maximizes saturation.
This is because saturable CPA is an engineered state, and not a self-organized state like lasing.
For lasing, typically only the multimode state with the highest number of modes is stable against weak perturbations.
Here we focus exclusively on single-mode saturable CPA.

\end{enumerate}

Despite the connection between laser oscillation and saturable CPA, they also have several important differences.
As noted, laser oscillation is a self-organizing phenomenon, and requires only sufficient pumping of the gain medium, while saturable CPA requires fine-tuning of the external driving field both in amplitude and in spatial structure.
The linear theory of laser oscillation, which is valid when field amplitudes are not too large, predicts a singular cavity response function which diverges at the lasing frequency, leading to infinite field amplitudes within the cavity, and invalidating the linear approximation.
In fact the nonlinearities present in the gain medium regularize the divergence and predict a finite response of the gain cavity with a definite amplitude.
The lasing frequencies and output powers are realized by the system itself without external parameter-tuning, and in this sense lasing is self-organized.
On the other hand, the time-reversed phenomenon of CPA requires synthesizing a specific incident wavefront, with the correct frequency and amplitude, and is in this sense fine-tuned.

    \subsection{The Coherent Delimiter \label{sec:chpt5: delimiter}}

Below, we describe the operation of a hypothetical device that completely suppresses a specific optical wavefront with intensity smaller than some threshold, while only partially absorbing it when it has greater intensity.
Such a coherent delimiter relies on the saturable CPA properties of an absorbing two-level medium.

A real two-level system has a definite number of gain/loss atoms.
If all of them were in the ground state, the device would be characterized by some device-specific $D_0^{\rm min}<0$, which is a lower-bound on the inversion density.
Therefore saturable CPA is only attainable over some finite range of negative pump values $D_0^{\rm min} \le D_0 \le D_0^{\rm th}$.
At each different pump within this range, a different input power is perfectly absorbed.

This suggests the following optical device, the coherent delimiter: an absorbing two-level cavity with light incident at each of its ports, with the correct relative phases and amplitudes required for CPA.
The pump parameter $D_0$ is determined by some external mechanism, such as a pump beam, which is controlled by a feedback mechanism that tracks the light reflected from the cavity.
When the pump has the correct value for the given input power, there is no reflection from the cavity at all.
However, if $D_0$ is mismatched to the input power, which is detected by an increase in the reflected intensity, then the pump control can be changed until the reflected intensity is again negligible.
This can only maintain CPA over some bounded range of $D_0$'s as argued above, and therefore over some finite range of input powers, starting from vanishing input power associated with $D_0^{\rm th}$, and increasing to some maximum input power associated with $D_0^{\rm min}$.
Within this range, no light is reflected, while outside this range, the device will act is a saturable absorber.

\begin{figure}[t]
    \centering
    \centerline{ \includegraphics[width=\textwidth]{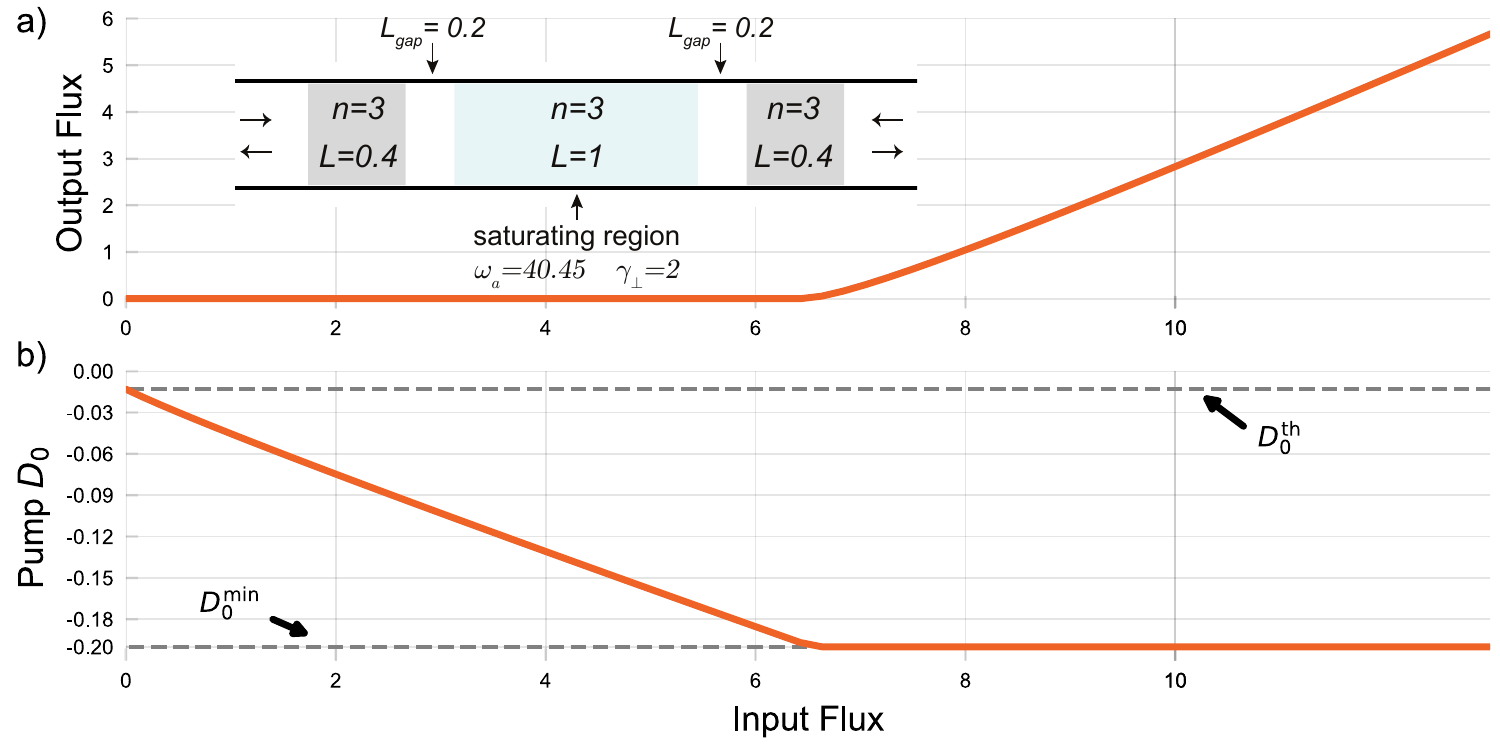} } 
    \caption[Coherent delimiter input-output curve]{Input-output curves of a one-dimensional, two-sided coherent delimiter, with structure given in inset.
    The saturable CPA threshold pump is $D_0^{\rm th}=-.01315$, and the incident frequency is fixed at the threshold $\w_{\rm CPA}^{\rm th}=40.751$ for all inputs.
    The fixed relative phase between the beams is also determined by threshold CPA.
    {\bf (a)}:~Below some input threshold, no light is scattered from device; above there is partial scattering.
    The behavior is complementary to a standard limiter.
    {\bf (b)}:~The pump parameter $D_0<0$ is continuously tuned between the upper limit of CPA threshold (requiring an externally-applied forcing, like a pump laser), and its ground-state value $D_0^{\rm min}<D_0^{\rm th}$ (no external forcing), depending on the intensity of incident light.
    }
    \label{fig:chpt4: delimiter}
\end{figure}

This action is complementary to a limiter, hence ``delimiter'', and holds only for appropriately tuned coherent inputs.
\Cref{fig:chpt4: delimiter} shows the input-output characteristic curve for a one-dimensional, two-channel version of this device.
As is typical with other CPA phenomena, the single-channel case is simpler in that it does not require control over the relative phase of multiple channels.
The one-channel version would be easier to design and fabricate, while the two-channel version has an additional control knob through the relative phase of the input beams.
Such a device could be used to realize an optical implementation of the ReLU nonlinear function, which is used widely as an activation function in machine learning~\cite{2000_hahnloser_nature}.
This would have obvious benefits for all-optical or hybrid-optical implementations of neural networks~\cite{2017_shen_2017}.

Such a device requires that the entire range of operation has a unique stable solution (monostable).
We will show in \Cref{chp:chpt5: saturable scattering} that when $D_0$ is very negative, there can exist an interval of input powers that do not have a unique scattering solution (multistability).
This defines a theoretical upper bound on the operating range of a coherent delimiter.

    \subsection{Summary of Saturable CPA Algorithm \label{sec:chpt5: algorithm summary}}

The algorithm for solving saturable CPA closely mirrors the solution method for SALT in the single-mode regime, which is reviewed elsewhere~\cite{esterhazy_2014,ge_2010}.
Because of this, we will give only a brief summary of the saturable CPA algorithm here.

To compute the saturable CPA modes and frequencies for a cavity with a two-level absorbing medium specified by $\w_a$, $\gp$, $D_0$, and $F(\bx)$, first solve the equivalent SALT equation but with $\gp\to-\gp$ and $D_0\to-D_0$.

SALT solvers implement different iterative self-consistent solutions of the saturable wave equation \vref{eq:chpt0: salt} with outgoing boundary conditions.
For multimode lasing the solver must monitor multiple frequencies for the onset of new lasing modes.
An open source efficient SALT solver can be found in Ref.~\cite{maxwell_salt_git}.

Assuming that one is interested in monochromatic CPA, then SALT can be solved for only a single mode, which substantially simplifies the calculation from the usual case where the number of lasing modes is not known {\it a priori}. 
%The SALT algorithm first computes the threshold pump $D_0^{\rm th}$; if $|D_0| < D_0^{\rm th}$, then saturable CPA is not possible.
This gives the threshold ``lasing'' pump $D_0^{\rm th}$, the purely outgoing mode $\psi_{\rm SALT}(\bx)$ and frequency $\w_{\rm SALT}$.
If $D_0<-D_0^{\rm th}<0$, then the final step of the saturable CPA algorithm is to time-reverse the SALT answer: $\psi_{\rm CPA}(\bx) = \psi^*_{\rm SALT}(\bx)$, and $\w_{\rm CPA} = \w_{\rm SALT}$.
Otherwise, if $D_0>-D_0^{\rm th}$, then saturable CPA is not possible.

%The SALT algorithm~[CITE] for computing the frequencies and amplitudes of a laser oscillator as a function of the pump parameter $D_0$ has two parts: the threshold and above threshold calculations.
%Above threshold, the full nonlinear SALT PDE must be solved self-consistently for the lasing frequencies and modes.
%The solution for a given $D_0$ is assumed to be similar to the solution at a slightly smaller pump $D_0-\delta D_0$, so long as neither is in the immediate vicinity of the threshold of a new lasing mode.
%In this way, one can adiabatically increase $D_0$ from its threshold value, at each step using the solution from the previous step to seed the search for a self-consistent solution.
%The initial seed for the first above-threshold pump is the threshold lasing mode computed from the first part of the algorithm.

    \section{Dispersion and Saturable CPA at Threshold \label{sec:chpt4: dispersion}}

The core of the saturable CPA calculation is the SALT algorithm, which has two parts: the computation of threshold quantities, and the solution of the above-threshold nonlinear SALT PDE [\cref{eq:chpt0: salt}], which is bootstrapped from the threshold calculation by slowly changing $D_0$.
The dispersion of the two level susceptibility, encoded in $\g(\w)$ of \cref{eq:chpt5: two-level sus gamma}, has a qualitatively different effect on the first part of this algorithm as applied to saturable CPA than it does for lasing.
This is most evident for bad cavities, in which the loss rate of a passive cavity resonance is larger than the intrinsic dephasing rate, $\gp$, of the two-level medium.
Saturation is not important for this phenomenon, since in the threshold calculation the overall field amplitude is vanishingly small.
The rest of this chapter is devoted to explaining this nontrivial effect of dispersion in saturable CPA.
We will show that it is due to an avoided crossing of the usual $S$-matrix zeros of the cavity with a second set of zeros that are associated with the atomic polarization, which has not been described before.

\begin{figure}[t]
    \centering
    \centerline{ \includegraphics[width=\textwidth]{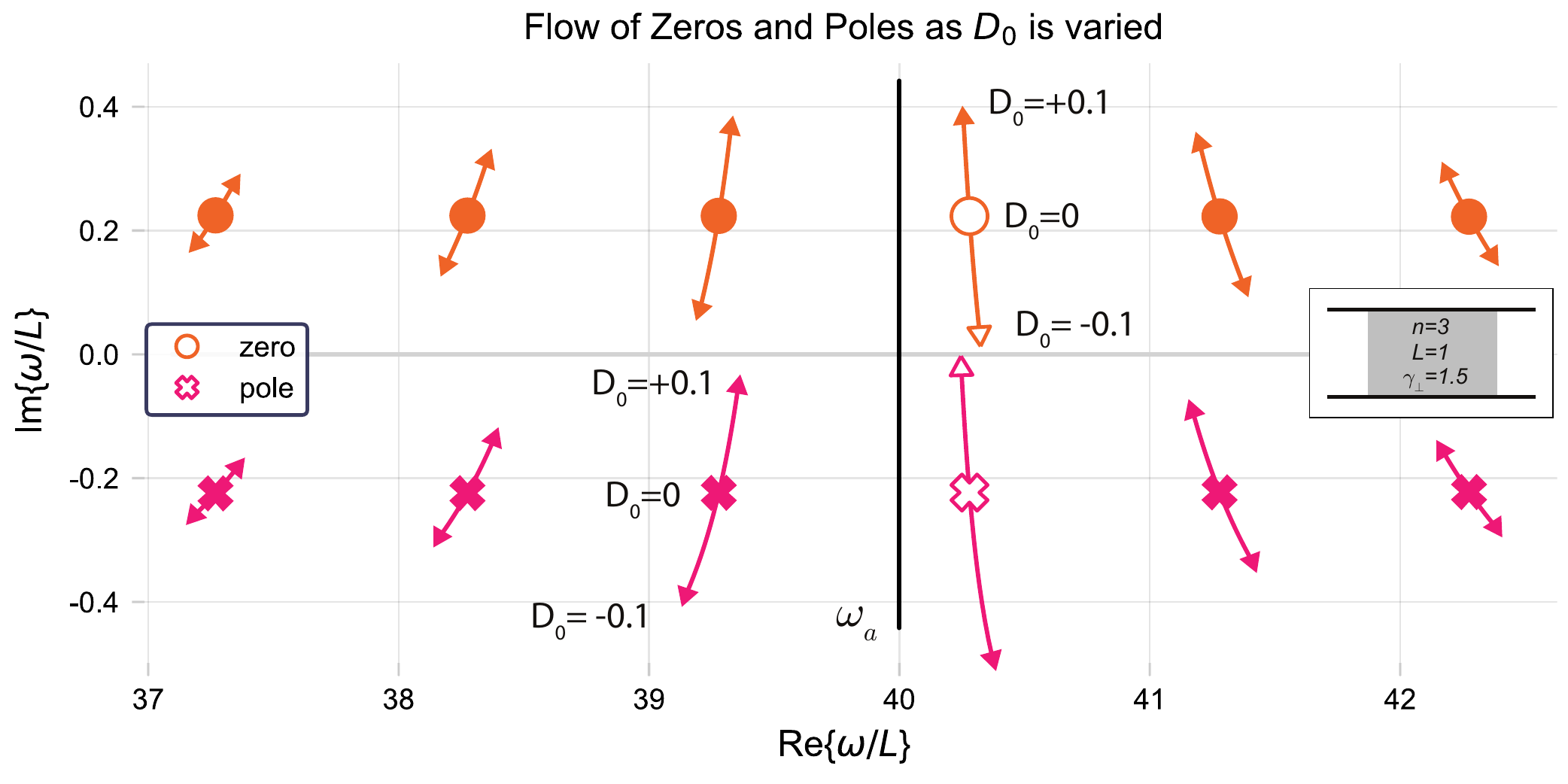}  }
    \caption[Flow of $S$-matrix zeros and poles as pump $D_0$ is varied, showing both line-pulling and line-pushing]{The flow of complex $S$-matrix poles (red) and zeros (orange) in a cavity with dispersive two-level susceptibility.
    Pump parameter $D_0$ is varied over the range $-.1<D_0<.1$ for the cavity shown in the inset ($n=3$, $L=1$, $\gp=1.5$).
    Passive cavity ($D_0=0$) poles shown as x's, zeros as circles.
    As cavity is actively pumped above transparency ($D_0>0$), poles and zeros flow upward and are pulled inward toward the atomic frequency $\w_a=40$.
    Threshold lasing occurs when the first pole reaches the real axis (open triangle), at a frequency that is between the atomic and passive cavity frequencies (line-pulling).
    When the pump is turned down from transparency ($D_0<0$), the zeros and poles flow downward and away from $\w_a$, which is line-pushing.
    Linear CPA becomes possible whenever an $S$-matrix zero crosses the real axis.
    Note that the location of the pole just at the threshold for laser oscillation $D_0=.1$ (open triangle) is different from the location of the associated zero at $D_0=-.1$, which is not quite at CPA.
    }
    \label{fig:chpt4: dispersion simple}
\end{figure}

For lasing, the threshold calculation in SALT aims to compute the self-oscillating mode in the limit that its amplitude vanishes, so that it is sufficient to take $h(\w,\bx,\psi)\equiv 0$.
The resulting equation is a resonance eigenvalue problem with complex-valued eigenfrequencies and with a dispersive susceptibility given by $\g(\w)D_0F(\bx)$.
When the two-level medium is transparent ($D_0=0$), the complex eigenfrequencies are the passive cavity resonances, required by causality to be in the lower half-plane.
As $D_0$ is increased from zero, the resonance poles move in the complex plane, usually upwards and towards the center of the gain band, $\w_a$.
This is known as {\it line-pulling}, and is a consequence of the two-level dispersion for lasing.
For high-\Q\ cavities, it is characterized by
\begin{equation}
    \label{eq:chpt4: line-pulling}
    (\w_{{\rm las},\,n}-\w_a) = (\w^r_n-\w_a) \left( \frac{1}{1+\g^r_n/\gp} \right),
\end{equation}
where the cavity resonance is at $\w^r_n-i\g^r_n$~\cite{1986_Siegman_book}.
In this form, it is evident that the lasing frequency is somewhere between the center of the gain band $\w_a$ and the cavity resonance (see \cref{fig:chpt4: dispersion simple}).
At some threshold value of $D_0$, a resonant pole crosses the real axis for the first time and the cavity becomes capable of self-oscillation.
This defines the threshold pump parameter, the first threshold lasing frequency, and the threshold spatial-profile of the lasing field.

When using SALT in the saturable CPA algorithm, $\gp\to-\gp$, and what was previously line-pulling becomes {\it line-pushing}:
\begin{equation}
    \label{eq:chpt4: line-pushing}
    (\w_{{\rm CPA},\,n}-\w_a) = (\w^r_n-\w_a)\left( \frac{1}{1-\g^r_n/\gp} \right).
\end{equation}
The CPA frequency is {\it further} from $\w_a$ than from the passive cavity resonance, in contrast to lasing.
For good cavities ($\g^r_n<\gp$) the threshold CPA frequency is on the same side of $\w_a$ as the passive cavity resonance, while, remarkably, for bad cavities ($\g^r_n>\gp$) it is on the opposite side, since the denominator of \cref{eq:chpt4: line-pushing} becomes negative.
Even more puzzling is what happens when $\g^r_n = \gp$, which was previously noted, but not solved, by Longhi in Ref.~\cite{longhi_2011_pra}.

Another way to derive line-pushing instead of line-pulling for saturable CPA uses a constant flux (CF) construction.
Threshold constant flux (TCF) states are a complete, biorthogonal basis set for analyzing problems with purely outgoing solutions~\cite{ge_2010}, often used for the solution of the SALT equations.
A related set of normalized states exist for incoming flux, the incident TCF states~\cite{2014_cerjan_pra}.
They satisfy the eigenvalue equation
\begin{equation}
    \label{eq:chpt4: incident CF definition}
    \{\nabla^2+[\e_c(\bx)+\beta_n F(\bx)]\w^2\}v_n(\bx)=0,
\end{equation}
with purely incident boundary conditions, which in one dimension are 
\begin{equation}
    \label{eq:chpt4: incident boundary conditions 1d}
    [\partial_x v_n\mp i\w v_n]_{\pm a}=0,
\end{equation}
for a domain of length $2a$ centered at the origin.
This differs from the usual TCF construction by using incoming instead of outgoing boundary conditions.
The complex eigenvalues are $\beta_n$, which are implicit functions of the real-valued frequency $\w$.
This contrasts with the usual complex resonance equation where the $\w_n$ are complex eigenfrequencies.
The threshold saturable CPA mode satisfies
 \begin{equation}
    \label{eq:chpt4: TCF threshold}
    D_{0,n}^{\rm th}\gamma(\w_n^{\rm th}) = \beta_n(\w_n^{\rm th}),
\end{equation}
which is similar to the threshold lasing condition.
This is a single complex equation in two real quantities, $D_{0,n}^{\rm th}$ and $\w_n^{\rm th}$, which can therefore be implicitly solved for each TCF index $n$.
The threshold CPA solutions generated this way correspond to different cavity zeros flowing to the real axis; there is one threshold pump and frequency for each passive cavity zero.
The real and imaginary parts of \cref{eq:chpt4: TCF threshold} can be solved for the threshold quantities:
\begin{gather}
    \label{eq:chpt4: CPA threshold pump}
    D_{0,\,n}^{\rm th} = -\im{\beta_n(\w_n^{\rm th})} \left[ 1+\left(\frac{\w_n^{\rm th}-\w_a}{\gp}\right)^2 \right] \\
    \label{eq:chpt4: CPA threshold freq}
	\frac{\w_n^{\rm th}-\w_a}{\gp} = -\frac{\re{\beta_n(\w_n^{\rm th})}}{\im{\beta_n(\w_n^{\rm th})}}.
\end{gather}
These are formally equivalent to the TCF conditions for threshold lasing if we replace $\beta_n$ with $\eta_n\equiv \beta_n^*$\footnote{
This requires that $\e_c(\bx)$ is real, which was explicitly violated in previous chapters.
However, in \Cref{chp:chpt4: saturable cpa,chp:chpt5: saturable scattering} we assume that $\e_c(\bx)$ is real, so that the two-level susceptibility is responsible for any material amplification or absorption.}, which have been shown to reproduce line-pulling in the high-\Q\ limit~\cite{ge_2010}.
With this replacement, both equations pick up a minus sign relative to their lasing counterparts, consistent with line-pushing instead of line-pulling for saturable CPA in the same high-\Q\ limit.

\begin{figure}[t!]
    \centering
    \centerline{ \includegraphics[width=.95\textwidth]{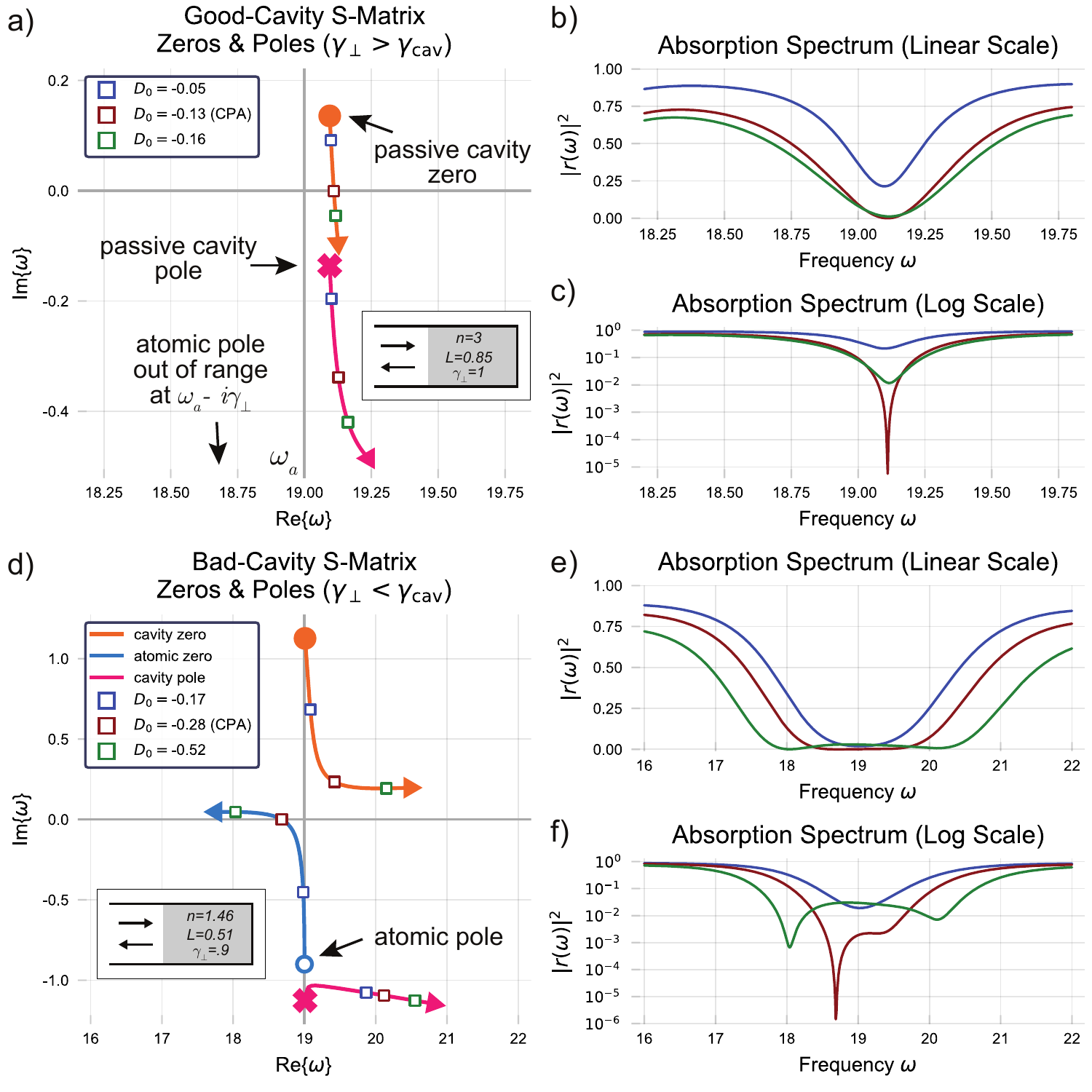} } 
    \caption[Flow of $S$-matrix zeros and poles as pump parameter is varied in both the good- and bad-cavity limits; flow of cavity polariton zero]{Flow of the $S$-matrix zeros and poles as pump parameter $D_0$ is varied for good and bad cavities (shown in insets of (a,c)).
    Arrowheads on trajectories indicate direction of flow as $D_0$ becomes more negative.
    {\bf(a)}:~Good-cavity zeros and poles flow downward, starting from passive cavity positions, which are complex-conjugates of each other.
    CPA corresponds to the open red square, where the zero is real.
    {\bf(b},{\bf c)}:~Spectrum of reflection coefficients from (a) for three values of $D_0$, shown as color-coded open squares (absorption is $1-|r(\w)|^2$).
    Blue is under-damped, green over-damped, and red is critically damped.
    The width of the dip increases as the pole gets further from the real axis, while the minimum is zero only at CPA.
    {\bf(d)}:~Similar to (a), but in bad cavity limit.
    The blue trajectory is a new zero, not seen in (a), attributed to the singularity of the atomic susceptibility (shown as open circle).
    CPA is still achieved (open red square), but is not connected to passive cavity zero (orange trajectory).
    {\bf(e},{\bf f)}:~Similar to (b,c).
    When trajectories of (d) are near closest approach, the absorption spectra are significantly flattened, similar to~\Cref{chp:chpt3: cpa ep}.
    }
    \label{fig:chpt4: dispersion}
\end{figure}

We return to the question of line-pushing in the bad-cavity limit.
\Cref{fig:chpt4: dispersion}a shows the flow of the complex-valued zeros and poles of the $S$-matrix as $D_0$ is decreased from transparency, for a simple one-channel cavity tuned so that the passive cavity resonance is centered slightly above the atomic frequency $\w_a$, and its linewidth is several times smaller than the absorption bandwidth $\gp$ (good cavity limit).
Within the range of the plot there is only one relevant cavity zero and the effect of dispersion is small.
The absorption line near the CPA frequency shows only a single dip.

In contrast, \cref{fig:chpt4: dispersion}d shows the same flow in the bad cavity limit, where the passive cavity linewidth is larger than the absorption bandwidth $\gp$.
There is still only one relevant passive cavity resonance ($S$-matrix pole), yet there are now two $S$-matrix zeros visible.
In the limit that $D_0\to0$, one of the zeros converges to the pole of the atomic susceptibility $\w_a-i\gp$, so that in this limit it can be predominantly associated with the material polarization degrees of freedom rather than the cavity electromagnetic degrees of freedom.
When $D_0=0$, the atomic and electromagnetic degrees of freedom are fully decoupled, and there are no additional zeros in the electromagnetic wave equation.
As the absorption is increased by making $D_0$ more negative, this ``polarization zero'' or ``cavity polariton zero'' flows upwards in the complex-frequency plane, eventually turning to the left in an avoided crossing with the cavity zero.
At large negative $D_0$, changes in the absorption cause a predominantly dispersive shift of the zeros, meaning that their frequencies shift while their linewidths are relatively constant (see \cref{fig:chpt4: dispersion}f and \vref{fig:chpt4: dispersion ep}).
The threshold CPA frequency in this case is associated with the polarization zero instead of the cavity zero, which explains how it occurs on the other side of $\w_a$ from the passive cavity resonance.
At the avoided crossing and above, the ``polarization'' zero and ``cavity'' zero are strongly hybridized, and can no longer be associated mainly with one or the other.
The signature of dispersive CPA in the bad-cavity limit is the appearance of {\it two} absorption peaks; one due to CPA, and the other due to an additional zero close to the real axis (see log-plot in \cref{fig:chpt4: dispersion}f).
Note also that the cavity pole in the bad cavity limit flows slightly {\it upwards} as $D_0$ is decreased before being deflected in its own avoided crossing.
This is in contrast to the usual good-cavity case (\cref{fig:chpt4: dispersion}a), for which there are no visible polarization zeros (they exist, but out of the range of the plot, and their effects are not physically relevant), and the cavity poles and zeros both flow predominantly downward to the lower half-plane as the absorption is increased, with $\w_{\rm CPA}^{\rm th}$ being only slightly pushed away from $\w_a$.

Hermitian symmetry implies that there also exist cavity polariton poles, associated primarily with the polarization degrees of freedom, which are distinct from the cavity poles.
When $D_0=0$ the atomic medium becomes transparent in a mean-field sense, and hermiticity is restored, with the consequence that $S$-matrix poles are the complex-conjugates of the zeros.
Therefore, if there were truly a zero at $\w_a-i\gp$ as $D_0\to0$, then there would also be a pole at $\w_a+i\gp$, which is forbidden by causality.
More obviously, at $D_0=0$, the poles and zeros can have no parametric dependence on the atomic parameters, since $\chi_{\rm nl}\equiv 0$.
Yet for infinitesimal $D_0$ there {\it is} a cavity zero in the lower half-plane, as evident in \cref{fig:chpt4: dispersion}d.
It must therefore be annihilated by a pole when $D_0\to0$.
Therefore when $D_0\neq0$, we expect not only a polarization zero but also a polarization pole, which is presumably the cause of the avoided crossing in \cref{fig:chpt4: dispersion}d.
We will see next that in fact that the polarization zeros and poles exist in a one-to-one fashion with the cavity poles and zeros, which are countably infinite, except when $D_0=0$.

To understand this behavior, we analyze the dispersive susceptibility $\chi_{\rm nl}(\w,\bx,0)$ from \cref{eq:chpt5: two-level sus}, which has a singularity at $\w_a-i\gp$.
Because of the divergence at this point, the atomic susceptibility attains \textit{any} large value in its neighborhood, regardless of the non-zero value of $D_0$.
In particular, there exists a complex frequency $\w_n^{\rm \pi z}$ at which $\chi_{\rm nl}(\w_n^{\rm \pi z},\bx,0)$ takes the value that in an equivalent dispersionless system would pull the $n^{\rm th}$ passive-cavity $S$-matrix zero from $\w_n^{\rm z}$ to $\w_n^{\rm \pi z}$.
The frequencies $\w_n^{\rm \pi z}$ are the polarization zeros.
A similar statement holds for the polarization poles.
In summary, for $D_0\neq0$, there exists a countably infinite set of poles and zeros of the $S$-matrix, in addition to the ordinary cavity poles and zeros, which possess $\w_a-i\gp$ as a limit point, and which all flow to said point as $D_0\to0$, annihilating each other at $D_0=0$.
A way to state this is that the atomic and photonic degrees of freedom are coupled through $\chi_{\rm nl}$, so that all the quasi-normal modes of both systems become hybrid atomic-photonic modes, doubling the number of poles and zeros appearing in the electromagnetic wave equations (except at $D_0 = 0$).

    \subsection{Finding Cavity Polariton Zeros with TCF States}

In this section we give semi-analytical expressions for the complex trajectories of the cavity and polarization zeros as $D_0$ is decreased from transparency, using the incoming TCF states.

It is difficult to find the polarization poles and zeros using the resonance eigenvalue problem
\begin{equation}
    \label{eq:chpt4: resonance}
    \{\nabla^2+ [\e_{\rm c}(\bx) + \chi_{\rm nl}(\w_n,\bx,0) ]\w_n\}\psi_n(\bx)=0,
\end{equation}
subject to purely outgoing or incoming boundary conditions, respectively.
It is a nonlinear-eigenvalue problem (NEVP), where the nonlinear dependence on $\w_n$ cannot be removed by the use of PMLs, unlike what was described in \Cref{chp:chpt1: theory of scattering}, \Cref{sec:chpt2: PMLs} for dispersionless susceptibilities.
Furthermore the divergence at $\w_a-i\gp$ and the infinite density of poles/zeros in its vicinity renders the usual methods for solving NEVPs ineffective, such as the contour integral of Refs.~\cite{2012_Beyn_LA, esterhazy_2014}.
One method for finding the polarization zeros which is free of this problem is nonlinear root-finding applied to the analytically-continued incoming TCF eigenvalue problem \cref{eq:chpt4: incident CF definition}.
Analytic continuation means that we take $\w$ to be complex instead of real, and then seek solutions to
\begin{equation}
    \label{eq:chpt4: cf res}
    \beta_n(\w_n^{\rm \pi z}) = \g(\w_n^{\rm \pi z})D_0,
\end{equation}
where $\w_n^{\rm \pi z}$ is in the vicinity of $\w_a-i\gp$.
There is no diverging quantity in \cref{eq:chpt4: incident CF definition}, and the $\beta_n$'s remain well-separated, even though the $\w_n^{\rm \pi z}$'s are tightly clustered, so that this method is free of the difficulties posed for the NEVP resonance methods.
The same method, but using the outgoing the TCF solutions, applies to the polarization poles.

We can use incoming TCF states to construct a theory of the cavity and polarization zeros and their anticrossing.
On the one hand, the two real-valued threshold CPA quantities $\w_n^{\rm th}$ and $D_{0,n}^{\rm th}$ are determined from satisfying the complex-valued \cref{eq:chpt4: TCF threshold}.
On the other hand, if we fix $D_0$ but promote $\w$ to be a complex variable, then the analogous equation determines the $S$-matrix zeros with two-level dispersion:
\begin{equation}
    \label{eq:chpt4: tcf zero}
    \beta_n(\w_n^{\rm z, \pi z}) =  \frac{D_0 \gp}{\w_n^{\rm z, \pi z} - \w_a + i\gp},
\end{equation} 
where $\w_n^{\rm z}$ are the active cavity zeros, and $\w_n^{\rm \pi z}$ are the active cavity polariton zeros.
We seek an analytic expression for $\w_n^{\rm \pi z}(D_0)$, that is, for the trajectories over the complex plane of the cavity and polarization zeros as the two-level absorption is increased.
%For the passive cavity zeros $(\w_n^{\rm z})^0$, $D_0=0$ and \cref{eq:chpt4: tcf zero} reproduces \cref{eq:chpt4: cf res}.

The TCF eigenvalues can be expanded in a Taylor series about each of the $(\w_n^{\rm z, \pi z})^0 \defn \w_n^{\rm z, \pi z}(D_0\to 0)$:
\begin{align}
    \beta_n(\w) &= \cancelto{0}{\beta_n{\bs(} (\w_n^{\rm z})^0{\bs)}} + \beta_n^\prime{\bs(} (\w_n^{\rm z})^0{\bs)}{\bs(}\w-(\w_n^{\rm z})^0{\bs)} + \ldots \label{eq:chpt4: taylor cavity} \\
    \beta_n(\w) &= \beta_n{\bs(} (\w_n^{\rm \pi z})^0{\bs)} + \beta_n^\prime{\bs(} (\w_n^{\rm \pi z})^0{\bs)}{\bs(}\w-(\w_n^{\rm \pi z})^0{\bs)} + \ldots \label{eq:chpt4: taylor polarization}
\end{align}
The leading order term of \cref{eq:chpt4: taylor cavity} vanishes by the definition of the passive cavity zeros  ($D_0=0$) and \cref{eq:chpt4: tcf zero}.
For the cavity zero, $(\w_n^{\rm z})^0$ is just the passive cavity zero, while for the polarization zero, $(\w_n^{\rm \pi z})^0 = \w_a-i\gp$, as was argued above.

For small displacements of the complex frequency, so that the series truncation is valid, \cref{eq:chpt4: tcf zero,eq:chpt4: taylor cavity} for $\w_n^{\rm z}$ imply the quadratic:
%Assuming that the zeros of the active cavity are not too far displaced from the passive cavity zeros, so that the series truncation is valid, these two equations taken together give a quadratic in $\w_n^{\rm z}$:
\begin{equation}
    \label{eq:chpt4: quadratic for wz}
    {\bs(}\w_n^{\rm z} - \w_a + i\gp{\bs)}{\bs(}\w_n^{\rm z}-(\w_n^{\rm z})^0{\bs)}  =  \frac{D_0 \gp}{\beta_n^\prime{\bs(} (\w_n^{\rm z})^0{\bs)}},
\end{equation}
with the solution
\begin{equation}
\label{eq:chpt4: quadratic solution cavity}
    \w_n^{\rm z}(D_0) = \frac{(\w_n^{\rm z})^0 + \w_a - i\gp}{2} + \sqrt{ \left(\frac{(\w_n^{\rm z})^0 - \w_a + i\gp}{2}\right)^2 + \frac{D_0 \gp}{\beta_n^\prime{\bs(} (\w_n^{\rm z})^0{\bs)}}}.
\end{equation}
For now we neglect the other choice of sign, as it does not necessarily correspond to a small shift in frequency as $D_0 \to 0$.
A similar analysis, but using \cref{eq:chpt4: tcf zero,eq:chpt4: taylor polarization} for $\w_n^{\rm \pi z}$ imples:
%Assuming that the zeros of the active cavity are not too far displaced from the passive cavity zeros, so that the series truncation is valid, these two equations taken together give a quadratic in $\w_n^{\rm z}$:
\begin{equation}
\label{eq:chpt4: quadratic solution polarizatio}
    \w_n^{\rm \pi z}(D_0) = \left(\w_a - i\gp\right) - \frac{\beta_n{\bs(} (\w_n^{\rm \pi z})^0{\bs)}}{2\beta\pr_n{\bs(} (\w_n^{\rm \pi z})^0{\bs)}}+ \sqrt{ \left[\frac{\beta_n{\bs(} (\w_n^{\rm \pi z})^0{\bs)}}{2\beta\pr_n{\bs(} (\w_n^{\rm \pi z})^0{\bs)}}\right]^2 + \frac{D_0 \gp}{\beta\pr_n{\bs(} (\w_n^{\rm \pi z})^0{\bs)}}}.
\end{equation}
As before, we choose the sign that corresponds to a small shift in frequency as $D_0 \to 0$.
\Cref{eq:chpt4: quadratic solution cavity,eq:chpt4: quadratic solution polarizatio} are the main result of this analysis.
They show that there is an avoided crossing behavior in both the cavity and polarization zeros as $D_0$ is varied.

In \cref{fig:chpt4: dispersion}d it was apparent that the polarization zero flowed upwards as absorption was added to the system ($D_0$ became more negative), in contrast to the typical behavior of the cavity zero, which flowed downward.
We can verify that this is consistent with \cref{eq:chpt4: quadratic solution cavity,eq:chpt4: quadratic solution polarizatio} for $\w_n^{\rm z}(D_0)$ and $\w_n^{\rm \pi z}(D_0)$, respectively, by analyzing their behavior for small $D_0$.
The leading order behaviors for $\w_n^{\rm z, \pi z}(D_0)$ are
\begin{gather}
    \w_n^{\rm z}(D_0) = (\w_n^{\rm z})^0 + D_0\frac{\gp}{\beta\pr_n{\bs(} (\w_n^{\rm \pi z})^0{\bs)}[(\w_n^{\rm z})^0 - \w_a + i\gp]} + {\cal O}(D_0^2) \label{eq:chpt4: small flow cavity zero}\\
    \w_n^{\rm \pi z}(D_0) = (\w_a - i\gp) + D_0\frac{\gp}{\beta_n(\w_a - i\gp)} + {\cal O}(D_0^2)
\end{gather}

We are interested in the case where the detuning of the cavity zero with the absorption line center is negligible, so that $\re{(\w_n^{\rm z})^0}-\w_a \simeq 0$.
For the cavity zero, $\w_n^{\rm z}$, this implies
\begin{equation}
    \w_n^{\rm z}(D_0) = (\w_n^{\rm z})^0 - iD_0\frac{\gp}{[\g_n + \gp] \beta\pr_n{\bs(} (\w_n^{\rm \pi z})^0{\bs)}}  +  {\cal O}(D_0^2),
\end{equation}
where $\g_n \defn \im{(\w_n^z)^0}$ is the passive cavity linewidth of the relevant resonance/zero that we are considering.

The constant $\beta_n^\prime{\bs(} (\w_n^{\rm z})^0{\bs)}$ can be semi-analytically evaluated, using conjugate PML's to enforce the incoming boundary conditions.
To see this, focus on a single TCF eigenpair $(v_n(\w,\bx),\beta_n(\w))$, and make a small variation in the frequency $\w$ of the TCF eigenvalue \cref{eq:chpt4: incident CF definition}.
The sum of the terms proportional to the arbitrary small variation must vanish.
Projecting this sum onto $v_n(\w,\bx)$ and integrating gives the exact relation
\begin{equation}
    \label{eq:chpt4: beta prime}
    \beta^\prime_n(\w) = -\frac{2}{\w} \left[ \beta_n(\w)  + \frac{\int d^dx\, \alpha(\bx) \e(\bx) v_n^2(\w,\bx) }{\int d^dx\, F(\bx) v_n^2(\w,\bx)} \right],
\end{equation}
which to our knowledge has not been noted previously, and can be used in gradient-based root-finding methods.
The function $\alpha(\bx) \defn \prod_{j=1}^d \alpha_j(\bx)$ is the product of all the complex-scaling functions $\alpha_j(\bx)$ used to define the conjugate PML.
If boundary-matching is used for the incoming boundary conditions instead of conjugate PMLs, then the change in the boundary condition when $\w$ is varied must also be taken into account, and \cref{eq:chpt4: beta prime} must be accordingly modified.

Using this in \cref{eq:chpt4: quadratic solution cavity}, with $\beta_n{\bs(}(\w_n^{\rm z})^0{\bs)}=0$, gives
\begin{gather}
    \label{eq:chpt4: tcf cavity zeros}
    \w_n^{\rm z}(D_0) = (\w_n^{\rm z})^0 + iD_0\frac{\gp}{2(\g_{\rm cav} + \gp)} \left[ \frac{\w \int d^dx\, F(\bx) v_n^2(\w,\bx)}{\int d^dx\, \alpha(\bx) \e(\bx) v_n^2(\w,\bx) } \right]_{(\w_n^{\rm z})^0} + {\cal O}(D_0^2)
\end{gather}
The bracketed term on the RHS is mostly real, with a positive real part, even for passive cavities with moderate-\Q.\footnote{We are envisioning that the cavity has no amplifying regions, so that $F(\bx)$ is everywhere non-negative.}
Therefore, for $D_0<0$, the flow of the cavity zero is indeed predominantly downward in the complex frequency plane, which is consistent with \cref{fig:chpt4: dispersion}d and with the usual intuition.

Meanwhile, the flow of the polarization zeros is given by
\begin{equation}
    \w_n^{\rm \pi z}(D_0) = (\w_a - i\gp) + D_0\frac{\gp}{\beta_n{\bs(} \w_a - i\gp{\bs)}} + {\cal O}(D_0^2).
\end{equation}
Generally, $\beta_n(\w)$ is the linear susceptibility that would be needed to bring a passive cavity zero in the upper half-plane to the (complex) frequency $\w$.
In this case the frequency is the atomic pole, which is in the lower half-plane, indicating that $\beta_n$ will be lossy, i.e.,~predominantly positive imaginary.
Therefore, for $D_0<0$, the polarization zero flows upwards, which verifies what was seen in \cref{fig:chpt4: dispersion}d.

\begin{figure}[t!]
    \centering
    \centerline{ \includegraphics[width=\textwidth]{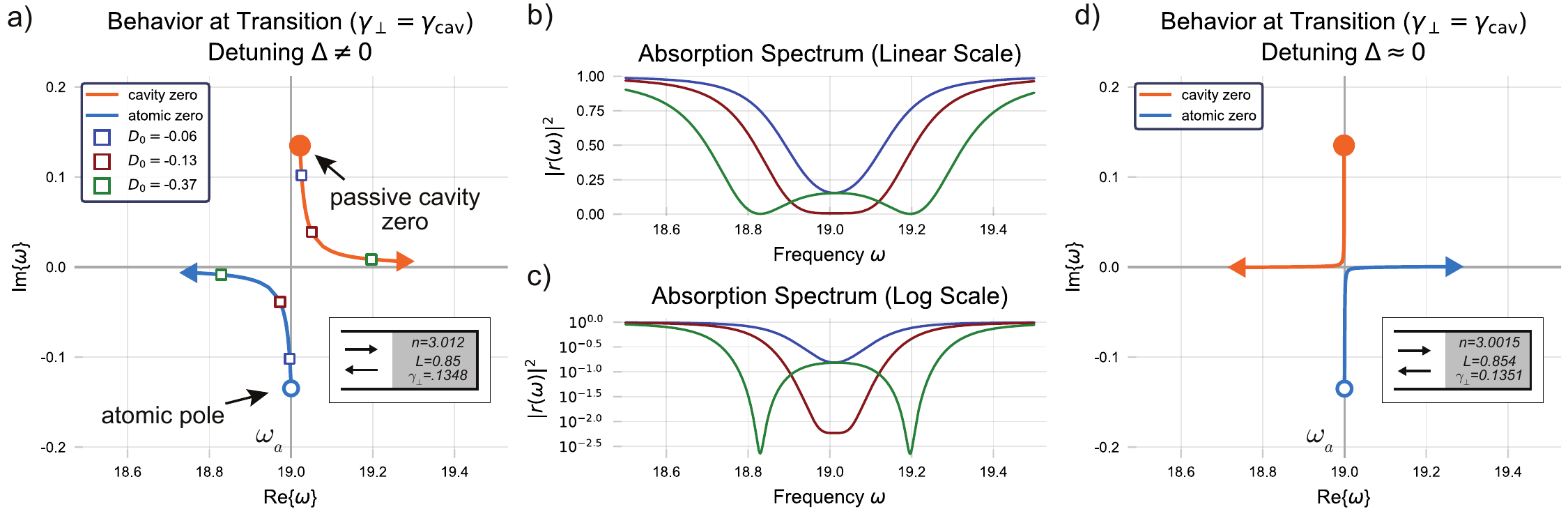} } 
    \caption[Flow of cavity and cavity polariton zeros at the good-cavity/bad-cavity transition]{Flow of $S$-matrix zeros at the good-cavity/bad-cavity transition for two cases.
    {\bf (a}-{\bf c)}:~Similar to \Cref{fig:chpt4: dispersion}, but with $\gp=\g_{\rm cav}$, and there is no CPA.
    The passive cavity resonance is slightly detuned above the atomic frequency, leading to a broad avoided crossing.
    The effect of dispersion as $D_0$ becomes more negative is apparent in the splitting of the absorption in (b,c).
    {\bf (d)}:~Similar to (a) but with very small detuning (not visible).
    The avoided crossing is very narrow, nearly at an EP degeneracy.
    The absorption curves are not shown because of their similarity to (b,c).
    }
    \label{fig:chpt4: dispersion ep}
\end{figure}

\Cref{eq:chpt4: quadratic solution cavity,eq:chpt4: quadratic solution polarizatio} describe the avoided crossing seen in \cref{fig:chpt4: dispersion}, and also suggest an intriguing possibility: a polaritonic CPA EP, with strongly hybridized optical and atomic degrees of freedom. 
If we assume that the atomic pole $\w_a-i\gp$ is not too far from the passive cavity zero, compared to the typical frequency scale over which the TCF eigenvalues $\beta_n$ and their derivatives appreciably change, then the series truncation of the $\beta_n$'s is valid for both solutions of \cref{eq:chpt4: quadratic solution cavity}.
That is, both $\pm$ signs of the radical are allowed, whereas previously we neglected the minus solution.
In this case, by engineering the passive cavity resonant frequency to be $\w_a$, and linewidth to be $\gp$, then the first term under the radical evaluates to $-\gp^2$.
For a high-\Q\ cavity, $(\w_n^{\rm z})^0$ is mostly real, as are the integrals in \cref{eq:chpt4: beta prime}.
Because of this, near some critical (real) value $D_0^{\rm cr}$, there is a narrowly avoided crossing of complex $S$-matrix zeros whose gap size is related to the $Q^{-1}$ of the resonance:
\begin{equation}
   D_0^{\rm cr} = -\frac{2}{\gp }{\rm Re}\left\{\frac{\int d^dx\, \alpha(\bx) \e(\bx) v_n^2(\w,\bx)}{\w \int d^dx\, F(\bx) v_n^2(\w,\bx)}\right\}_{\w=(\w_n^{\rm z})^0}
\end{equation}
This case is shown in \cref{fig:chpt4: dispersion ep}d.
If the linewidth condition is met ($\im{\w_n}=\gp$), but the cavity is slightly detuned from the atomic frequency ($\re{\w_n} \neq \w_a$), then the avoided crossing becomes much broader, as shown in \Cref{fig:chpt4: dispersion ep}a.
Presumably the zeros can be brought to an exact EP degeneracy with finer tuning.

Similar formulas hold for the $S$-matrix poles, but with $i\gp\to-i\gp$, $\beta_n\to\beta_n^*$, $v_n \to v_n^*$, and $(\w_n^{\rm z})^0 \to [(\w_n^{\rm z})^0]^*$, so that polaritonic EP resonances should also be possible in the lower half-plane.

\chapter{Scattering with Saturable Absorption and Amplification \label{chp:chpt5: saturable scattering}}
In both linear and saturable Coherent Perfect Absorption (CPA), a specific wavefront is completely absorbed and no flux is scattered.
The entire wavefield is purely incident in the asymptotic regions.
Because of this, CPA can be constructed as the solution of a homogeneous equation, meaning source-free, of the form $\hat L \psi = 0$.
$\hat L$ is some differential operator, not necessarily linear in the wavefield $\psi$, with incoming boundary conditions.
This was the point of view taken in \Cref{chp:chpt3: cpa ep,chp:chpt4: saturable cpa}.

An alternative formulation of CPA is as a special solution to an inhomogeneous scattering problem in which the scattered component of the wave destructively interferes with the unscattered, outgoing portion of the exciting wave, exactly canceling it.
In the case of {\it linear} CPA, these two points of view are related in a one-to-one fashion: the purely incident eigenproblem, with eigenfunction $\psi$ and complex eigenfrequency $\w_{\rm z}$, is equivalent to the scattering matrix $S(\w_{\rm z})$ having a zero eigenvalue, with an eigenvector given by $\psi$ analyzed into the incoming channel basis.
However, this equivalence breaks down for {\it saturable} CPA, whose construction was given in the previous chapter, since the nonlinear \cref{eq:chpt4: saturable cpa equation} does not generally have a unique solution, and the $S$-matrix is not meaningfully defined.

The basic problem addressed in this chapter is to find all the steady-state solutions of the scattering problem when some part of the scatterer behaves as a saturable two-level medium.
Specifically, we seek all solutions to the steady-state inhomogeneous scalar Maxwell-Bloch equations, without any assumptions about the geometry or spatial dependence of the fields. 
As in the Steady-State {\it ab initio} Laser Theory (SALT)~\cite{tureci_2006, tureci_2007, tureci_2008, ge_2010, esterhazy_2014, cerjan_2015}, the coupled field equations can be reduced to a single nonlinear equation for the electric field, $\vph(\bx)$:
\be
    \label{eq:chpt5: intro equation}
    \left\{\nabla^2\ + \left[ \e_c({\bx}) +\frac{\g(\w) D_0 F(\bx)}{1+\G(\w)|\vph(\bx,\w)|^2} \right] \w^2 \right\} \vph(\bx) = j({\bx}),
\ee
subject to outgoing boundary conditions.
For later convenience, we have written the equation in terms of an equivalent source $j(\bx)$ and used outgoing boundary conditions; we could also have used the scattering boundary conditions described in \vref{eq:chpt1: scattering bc} without a source term.
We will assume that $j(\bx)$ is defined on a surface containing the scatterer, constructed from Huygens' principle (\Cref{sec:chpt1: huygens}); the volume equivalence principle will be of limited use here, since it relies on $\chi(\bx)$, which itself depends on the unknown total field $\vph(\bx)$.
The salient difference between \cref{eq:chpt5: intro equation} and the SALT \cref{eq:chpt0: salt} is the source term $j(\bx)$ on the RHS.
%As previously shown~[CITE], this is sufficiently general to describe scattering, since the source can emulate any incident wavefront through the equivalent source method. 

Previous works, some dating back to the 1970's, have studied optical scattering from systems with saturating nonlinearities~\cite{1969_szoke_apl, 1978_bonifacio_pra, 1978_bonifacio_nc, 1979_Schwendimann_jpa, 1979_agrawal_pra, 1981_gronchi_pra, 1984_LUGIATO_prog,  longhi_2011_pra}, and dispersive Kerr nonlinearities~\cite{1979_Schwendimann_jpa, 1984_LUGIATO_prog, 2000_centeno_prb, 2002_soljacic_pre, 2002_Mingaleev_jo,  2003_Soljacic_ol}, with a special focus on optical bistability, as reviewed in Ref.~\cite{1982_Abraham_repprog}.
This phenomenon, a consequence of nonlinearity, is the simultaneous existence of two stable steady-states, or {\it solution branches}, both consistent with the same input field.
The state of the system exhibits hysteresis: it is not uniquely determined by the input alone, but also by its history.
The onset of bistability is a constraint on the operating range of a device in the vicinity of the saturable CPA point, such as the CPA delimiter described previously [\Cref{sec:chpt5: delimiter}], since in the bistable regime fluctuations in the pump or injected amplitude can drive the device to the non-CPA solution branch.
Typically, the system can be brought from one state to the other by executing a loop in some control parameter, such as the input power.
Most previous studies greatly simplify the problem by considering only running waves (for example, in ring resonators), thereby suppressing the spatial-dependence of the nonlinearity.
Some studies of the Kerr nonlinearity -- not saturating -- do have spatial complexity~\cite{2000_centeno_prb, 2002_soljacic_pre, 2002_Mingaleev_jo, 2003_Soljacic_ol}, but the scale of resonant structures is similar to the wavelength of the input light, again minimizing the effect of spatial complexity within the nonlinear region.
A study from Longhi~\cite{longhi_2011_pra}, especially relevant to this (and the previous) chapter, addressed optical bistability for the SALT-like saturating nonlinearity, but without spatial complexity.
One study that specifically accounted for arbitrary spatial dependence is Cerjan, {\it et al.}~\cite{2014_cerjan_pra}, which addressed \cref{eq:chpt5: intro equation} without any assumptions about the geometry, using a constant flux (CF) basis.
However, that work's main focus was on lasing in the presence of an injected signal, and it did not account for bistability.

In this chapter, we give a treatment of the steady-state saturable amplifier/absorber described by \cref{eq:chpt5: intro equation} in its full spatial complexity, including spatial-hole burning to all orders.
We will also generalize some results known about optical bistability in saturable absorbers without space dependence to the case with arbitrary spatial complexity.
We present an iterative solution method valid in both the monostable and bistable regimes, and for all solution branches, stable and unstable, which makes no simplifying assumptions about the geometry, strength of the nonlinearity, or input power.
While this iterative method works in general, it can be challenging to find all the solution branches within the bistable regime.
Therefore we will develop a {\it self-consistent single-pole approximation} (SPA), valid for isolated resonances with high quality factors (high-\Q), that gives all solution branches in terms of just one, which can be computed from the above-mentioned iterative method. 
%The \scSPA\  is valid for isolated resonances with high quality factors (high-\Q), which can exhibit internal resonances that nontrivially enhance hole-burning.
While the \scSPA\ is very accurate, it is not particularly amenable to analysis, and therefore is of limited use in drawing general conclusions about bistability in the context of arbitrary space-dependence.
Therefore we will also develop the {\it \iSPA}, which we will use to show that the number of saturable solutions is at most three, with the new solutions emerging and disappearing at saddle-node bifurcations over some finite range of input powers, outside of which the solution is unique (monostable), which generalizes the conventional bistability results.
Finally, we discuss the failure of the \iSPA\ as an approximation that is not well-controlled, in contrast to the \scSPA, which is.
We show that, despite its poor numerical performance, the conclusions we drew about bistability from the \iSPA\ remain valid.
Furthermore, it provides a useful starting approximation that can be refined by the other solution methods described above.

In previous literature, ``single pole approximation'' (SPA) has been used as the name for two different, but related, approximations used to solve the SALT equations for lasing.
The first refers to an approximation of the electromagnetic Green function~\cite{tureci_2006}, while the second refers to an approximation of the inverse Green function~\cite{ge_2010}.
We will call the former the \scSPA, and sometimes just the SPA, while we will call the latter the \iSPA.

    \section{Fixed Point Solution of Saturable Scattering}

In this section we describe a general iterative algorithm for solving saturable scattering [\cref{eq:chpt5: intro equation}], which works specifically for a scattering configuration, but does not apply to CPA or lasing.
The numerical examples shown below provide empirical evidence that the algorithm can be rapidly convergent and robust, though at present it is not known whether this is always true.

Consider a two-level saturable scatterer, specified in what by now should be a familiar fashion: a flux-conserving cavity dielectric function $\e_{\rm c}(\bx)$, which is real and linear, and an additional nonlinear susceptibility
\begin{gather}
    \label{eq:chpt5: nonlinear sus}
    \chi_{\rm nl}(\bx,\w,\vph) \defn \frac{\g(\w) D_0 F(\bx)}{1+\G(\w)|\vph(\bx,\w)|^2}, \\
    \g(\w) \defn \gp/(\w-\w_a+i\gp) \qquad \qquad \G(\w) \defn |\g(\w)|^2.     \label{eq:chpt5: gammas}
\end{gather}
$F(\bx)$ is the spatially-dependent pump profile, $D_0$ the pump parameter (related to the relaxation value of the two-level inversion), $\g(\w) \defn \gp/(\w-\w_a+i\gp)$ and $\G(\w) \defn |\g(\w)|^2$ describe the gain curve, and $\vph(\bx,\w)$ is the electric field profile.

A steady-state incident wavefront is specified asymptotically far from the cavity, which, in the absence of the scatterer, defines a field $\vph_{\rm in}(\bx,\w)$.
This can be represented as an equivalent current $j(\bx,\w)$ which is localized at some bounding surface $\partial \Omega$ enclosing the scatterer, as described in \Cref{chp:chpt1: theory of scattering}.
The steady-state scattering solutions must satisfy the nonlinear Helmholtz equation:
\be
    \label{eq:chpt5: fundamental equation}
    \left\{ \nabla^2  + \left[\e_{\rm c}(\bx) + \chi_{\rm nl}(\bx,\w, \vph) \right]\w^2 \right\} \vph(\bx,\w) = j(\bx,\w),
\ee
subject to radiating boundary conditions.
To be definite we assume the boundary conditions are implemented as perfectly matched layers (PML's), as described in \Cref{sec:chpt2: PMLs}.
Within the volume bounded by $\partial \Omega$, $\vph(\bx,\w)$ is the total field, which is a superposition of the incident and scattered fields, while outside it is only the outgoing scattered portion.

\Cref{eq:chpt5: fundamental equation} is formally similar to SALT, but it differs importantly in having a non-zero RHS and a fixed frequency $\w$.
In principle, the various numerical solution methods for SALT will also solve this problem:
projection onto a CF basis~\cite{ge_2010, tureci_2008, 2014_cerjan_pra}, and a nonlinear root-finding applied to the discretized wave equation~\cite{esterhazy_2014}.
Each of these has drawbacks in the context of scattering.
The CF method is plagued by an ambiguity in the representation of the incoming field in the CF bases, which makes the solution difficult to truncate in CF-space~\cite{2014_cerjan_pra}.
The root-finding method works well, but relies heavily on the proper construction of the Jacobian, which can be costly, both computationally and in hours spent programming.

For scattering, $\w$ is a fixed external parameter, unlike in SALT.
This independence of \cref{eq:chpt5: fundamental equation} on $\w$ opens a new class of solutions that were not available before: fixed point methods.
These have the advantage of being derivative-free, and do not require special care in choosing the CF representation of the incident wave.
%The proposed method for finding a solution to the nonlinear scattering problem is a fixed-point algorithm.

We now describe the proposed iterative solution to \cref{eq:chpt5: fundamental equation}.
Define the discrete nonlinear map $M$ by
\be
    \label{eq:chpt5: nonlinear discrete map}
    M{\bs (}f(\bx,\w){\bs )} \defn \int d^dx\pr G_f(\bx,\bx\pr;\w) j(\bx\pr,\w).
\ee
$G_f(\bx,\bx\pr;\w)$ is the Green function of $\hat L_f$, the linear Helmholtz operator saturated by the field $f(\bx,\w)$,
\be
    \label{eq:chpt5: saturated helmholtz operator}
    \hat L_f \defn \nabla^2 + \left[ \e_{\rm c}(\bx) + \chi_{\rm nl}(\bx,\w,f) \right]\w^2,
\ee
so that $G_f$ by definition satisfies
\be
\label{eq:chpt5: green's function}
    \hat L_f G_f(\bx,\bx\pr;\w) = \delta^d(\bx-\bx\pr).
\ee
$\chi_{\rm nl}$ is the saturating susceptibility given earlier in \cref{eq:chpt5: nonlinear sus}.
A solution to the saturable scattering problem is a fixed point $\vph_*(\bx,\w)$ of the map $M$:
\be
\label{eq:chpt5: fixed point}
    M{\bs (} \vph_*(\bx,\w) {\bs )} = \vph_*(\bx,\w),
\ee
since $\hat L_\vph M{\bs (} \vph(\bx,\w) {\bs )} = j(\bx,\w)$ by construction.

The solution method is to iterate $M$, starting with an initial field $\vph_0(\bx,\w)$:
\be
    \label{eq:chpt5: iterative solution method}
    \vph_{n+1}(\bx,\w) = M{\bs (}\vph_n(\bx,\w)\bs{)} = M\circ M\cdots M{\bs(}\vph_0(\bx,\w){\bs)}.
\ee
If this converges, then $\lim_{n\rightarrow\infty} \vph_n =\vph_*$, and some finite number of iterations will suffice to yield a solution to \cref{eq:chpt5: fundamental equation} to within some tolerance.
More sophisticated fixed-point schemes with better convergence properties, such as Anderson acceleration~\cite{2011_homer_siam}, can be straightforwardly applied.

Each step of the algorithm can be summarized as follows: saturate the Green function with a trial guess and use it to compute a new field due to the equivalent source $j$.
In each step, the new field from the current iteration becomes the trial guess for the next iteration.
The reason this can be done for saturable scattering, but not lasing or CPA, is because those problems do not have a source term $j$, and, as was mentioned in \Cref{chp:chpt1: theory of scattering}, are over-constrained and can only be satisfied at isolated $\w$'s.
In contrast, for saturable scattering, $\w$ is arbitrary, fixed externally, and does not change from iteration to iteration.

\begin{figure}[t]
    \centering
    \centerline{ \includegraphics[width=\textwidth]{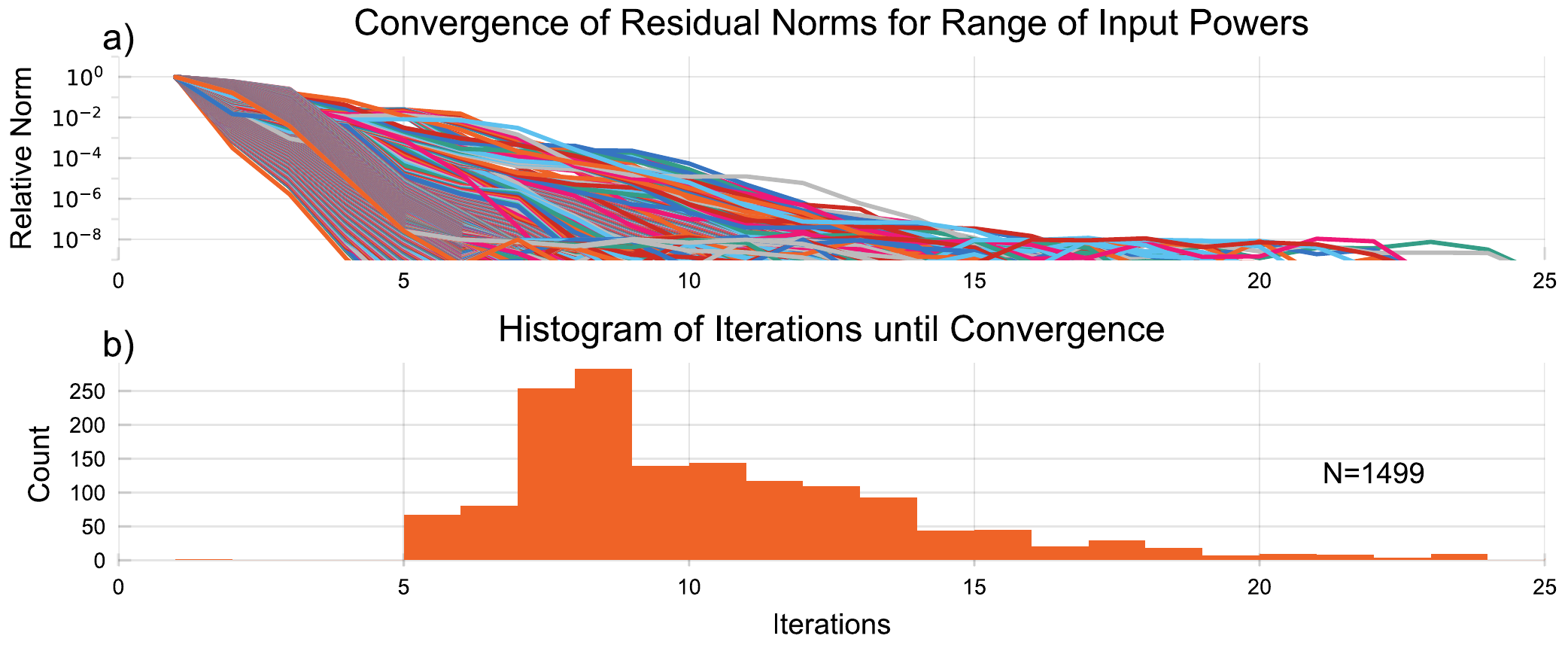} } 
    \caption[Convergence histories of iterative solutions applied to $1499$ different input amplitudes]{Convergence histories of $N=1499$ nonlinear scattering solutions with different input amplitudes.
    The fixed-point method \cref{eq:chpt5: iterative solution method} was used with Anderson acceleration parameter $m=1$.
    {\bf (a):} The decay of the relative maximum norm (log-scale), defined by $\max_{\bx}|f_n(\bx)|/\max_{\bx}|f_0(\bx)|$ as a function of iteration $n$.
    Here $f_n(\bx) \defn \vph_{n+1}(\bx)-\vph_n(\bx)$ measures the size of the change under an application of $M$.
    The linear decay on the log plot indicates exponential convergence.
    {\bf (b):} Histogram of the number of iterations until convergence for each of the histories of panel (a).
    The convergence criterion was $\max_{\bx}|f_n(\bx)|<10^{-11}$, and the median number of iterations until convergence was $7.14$.
    All $1499$ different solutions converged in under $25$ iterations.
    In both panels, the scatterer is the compound structure shown in \cref{fig:chpt5: structure and solutions}a.}
    \label{fig:chpt5: convergence_histories}
\end{figure}

\begin{figure}[t]
    \centering
    \centerline{ \includegraphics[width=\textwidth]{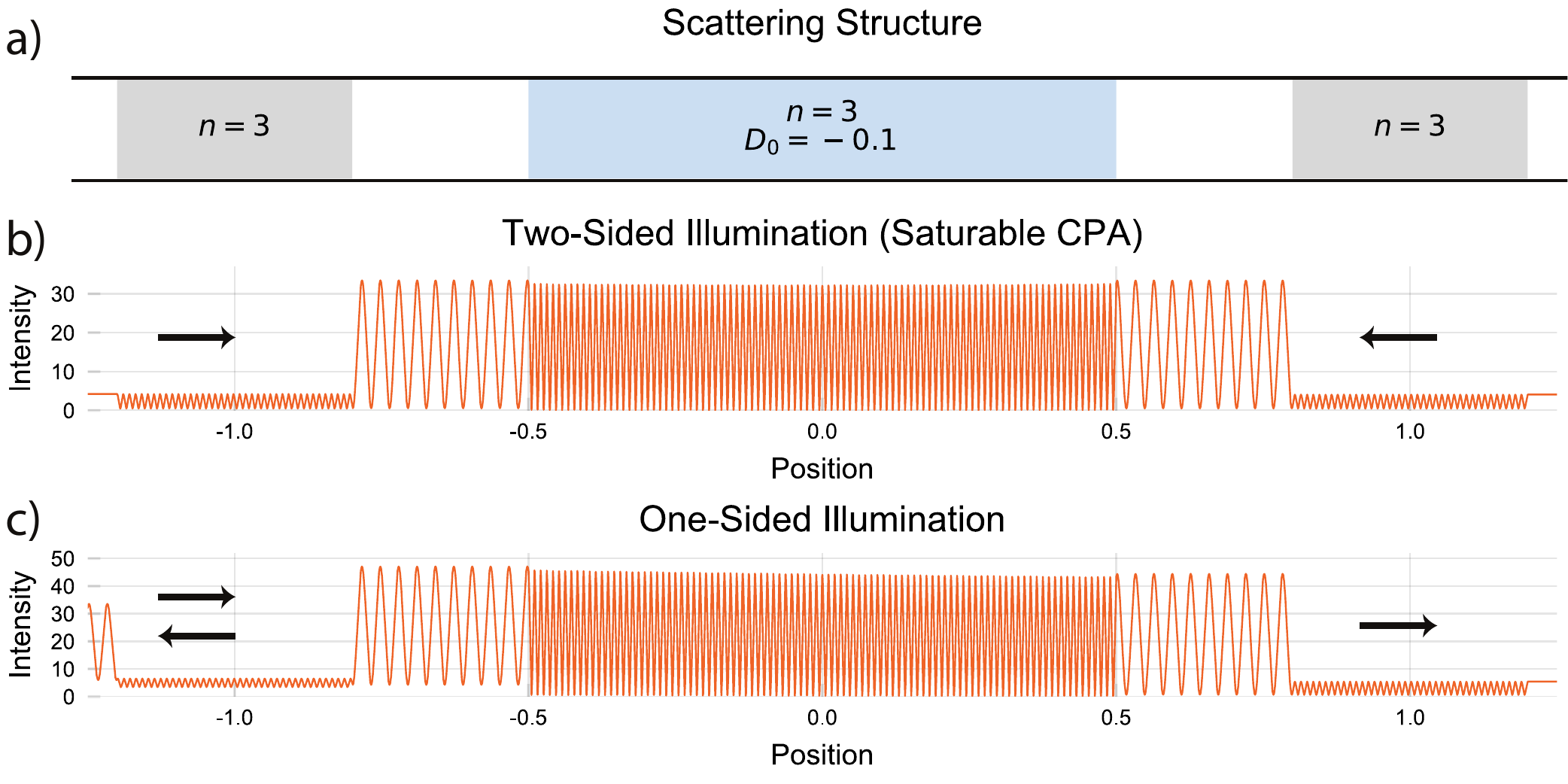} }
    \caption[Two-sided and one-sided saturable scattering solutions from one-dimensional compound structure]{Saturable scattering from one-dimensional compound structure.
    {\bf (a):} Schematic of structure: a symmetric configuration of three coupled-cavities (refractive index $n=3$) in air ($n=1$).
    The central cavity, of length $L$, hosts a dissipative two-level medium with $D_0=-0.1$, $\gp=5$, $\w_a/c=40.18/L$.
    The outer cavities have length $0.4L$, and are separated from the central cavity by a gap of $0.3L$.
    {\bf (b):} Scattering with a two-sided incoming wave at frequency $\w=100.416$.
    The incident field was constructed to be the same as the saturable CPA field, and therefore none of the flux is scattered.
    {\bf (c):} Similar to panel (b), but with only left-incident flux, which is equal to the sum of fluxes from both side in (b).
    }
    \label{fig:chpt5: structure and solutions}
\end{figure}

This method appears to be rapidly convergent and stable.
In cases where it is not, we can improve convergence by starting with a solution to the linear problem, i.e.,~without saturation, and turn on the effect of saturation adiabatically, either by ramping up the injected amplitude from zero, or by slowly changing $D_0$ from transparency.
\Cref{fig:chpt5: convergence_histories} shows the convergence histories of the iterative solution applied to the scatterer depicted in \cref{fig:chpt5: structure and solutions}a with one-sided illumination, for $1499$ different input powers.
The median number of steps to reach convergence to an absolute tolerance of $10^{-11}$ was $\sim7$.

Of course, there is no guarantee that a given choice of $\vph_0(\bx,\w)$ will converge.
Furthermore, it is not obvious that every fixed point of $M$, i.e.,~solution of \cref{eq:chpt5: fundamental equation}, is also the limit of some iterative sequence.
That depends on whether each fixed point is stable under the discrete map $M$; if there exist unstable fixed points then the proposed algorithm would fail even in principle to find those solutions.
Whether this happens is at present an open question.
The numerical evidence presented throughout the rest of this chapter suggests that the algorithm has robust convergence properties, and there is no suggestion of ``missed'' unstable solutions.

There are two important things to note in implementing this algorithm.
First, the full saturated Green function $G_\vph(\bx,\bx\pr;\w)$ need not be found, obviating the computationally expensive task of inverting $\hat L_\vph$.
Instead, only the solution to $\hat L_{\vph_n} \vph_{n+1}(\bx,\w) = j(\bx,\w)$ is needed, where $\hat L_{\vph_n}$ and $j(\bx,\w)$ are known {\it a priori}.
This is a far less costly operation, with known accelerations due to the sparsity of the discretized $\hat L_\vph$'s~\cite{2001_mumps_1,2019_mumps2}.
Second, the formulation of the scattering problem in terms of an equivalent source, PML's, and Green functions is only one prescription, and is not unique.
The iterative method described above can be generalized as follows: given a program $P$ that takes as inputs a specified incoming field $\vph_{\rm in}(\bx,\w)$ and a differential operator $\hat L$ and returns the total field $\vph(\bx,\w)$ within the scattering region [$P\{\hat L, \vph_{\rm in}(\bx,\w)\} \mapsto \vph(\bx,\w)$], then the map $M(\vph(\bx,\w)) = P\{\hat L_\vph, \vph_{\rm in}(\bx,\w)\}$, which first constructs the saturated Helmholtz operator $\hat L_\vph$ and then runs $P$ on it, is equivalent to the map defined in \cref{eq:chpt5: nonlinear discrete map}.
This means that the iterative saturable scattering solution can be implemented with whatever methods are most efficient and best suited for the problem at hand.
For example, instead of the specific prescription described here, one could use purely outgoing boundary-matching conditions~\cite{2005_prog_in_optics} (\Cref{sec:chpt2: boundary matching}), or the Recursive Green Function method~\cite{Thouless_1981_jpc, 1981_lee_prl, 1985_mackinnon_zpb}.

The solutions found by this method are constrained only by the assumptions of single-mode SALT, namely, that the inversion field $D(\bx)$ is static, though not known {\it a priori}.
It is valid for arbitrary geometries in any dimension, for any far-field exciting wavefront, and for both amplifying and absorbing two-level media.
The example in \vref{fig:chpt5: structure and solutions} demonstrates this for a compound one-dimensional scatterer made of three cavities: two lossless cavities on either side and a passive, absorbing central cavity.
The parameters $D_0$ and the total input power are the values consistent with saturable CPA, which require two-sided input.
\Cref{fig:chpt5: structure and solutions}b reconstructs this two-sided saturable CPA solution, but from the scattering perspective, and not as a purely incident eigenproblem; both incident and outgoing waves are allowed, though in this case the outgoing amplitude is zero.
\Cref{fig:chpt5: structure and solutions}c is the scattering solution with the same $D_0$ and total input power as saturable CPA, but with one-sided illumination.

\begin{figure}[h!]
    \centering
    \centerline{ \includegraphics[width=\textwidth]{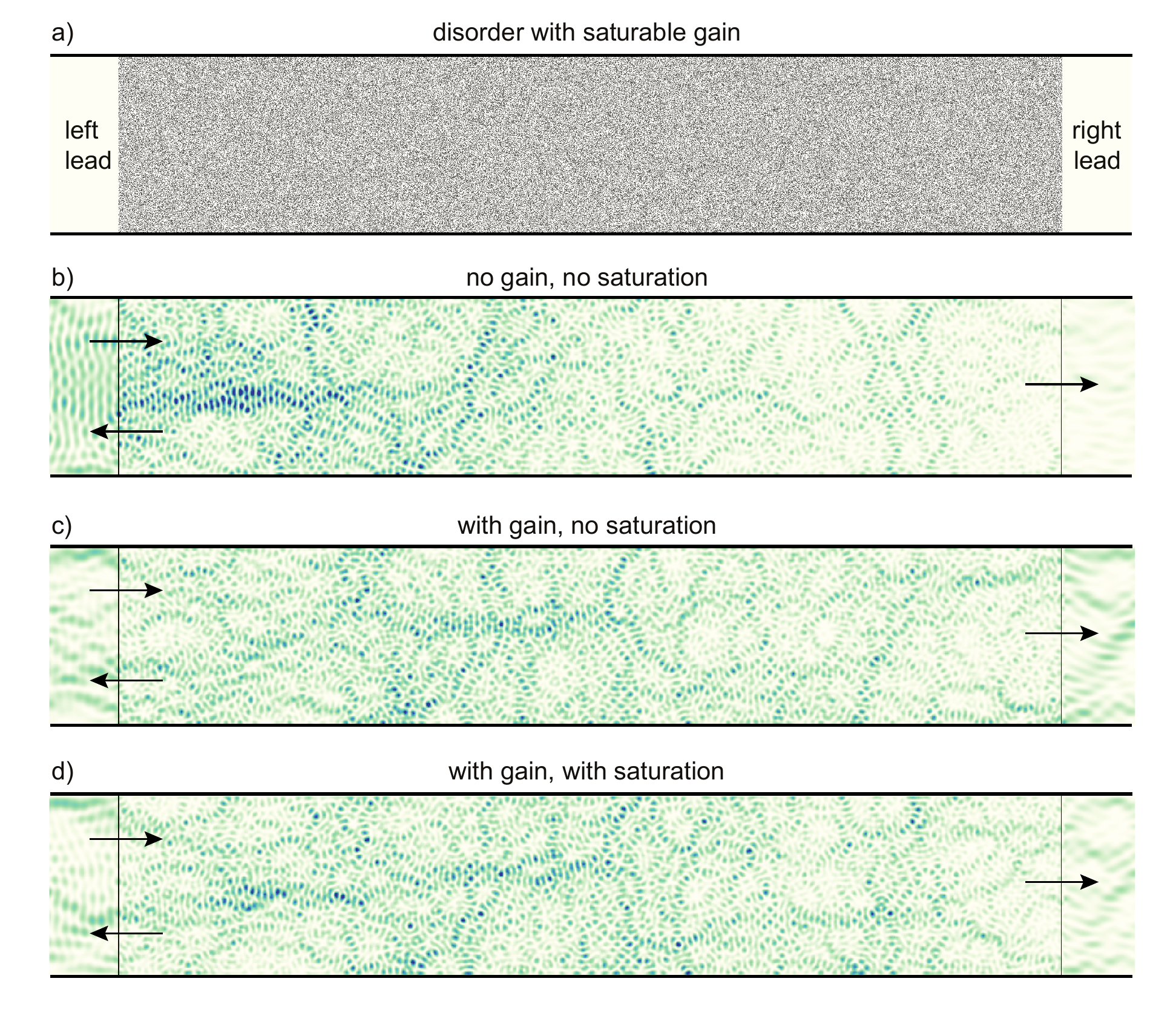} }
    \caption[Propagation through a disordered waveguide in three scenarios: hermitian, linear amplifying, saturable amplifying]{Rightward propagation through a disordered $20$-mode waveguide, about $4$ mean-free paths deep, depicted schematically in {\bf (a)}, in three scenarios:
    {\bf (b):} a waveguide with neither gain nor saturating nonlinearity.
    The transport is diffusive, with the field intensity decaying approximately linearly from left to right within the sample, with $\sim21\%$ transmitted flux.
    {\bf (c):} a waveguide with gain, but no nonlinearity.
    The gain competes with diffusive decay, giving rise to a spatial intensity distribution which is more uniform than in (b).
    The transmitted flux is $\sim190\%$ the injected flux.
    The overall intensity is significantly higher than in (b), but the plot has been scaled by the average intensity for the sake of visual comparison.
    {\bf (d):} a waveguide with both gain and saturation.
    Not only is the transmitted flux reduced from (c) to $\sim67\%$, but the spatial distribution is something of a hybrid between (b) and (c): similar to (c) in the center of the sample, but more like (b) at the end.
    }
    \label{fig:chpt5: disordered waveguide}
\end{figure}

Another example demonstrating the broad applicability of this method is shown in \cref{fig:chpt5: disordered waveguide}: diffusive transport through a two-dimensional disordered waveguide with saturable two-level amplification.
To our knowledge, this kind of calculation has never been done before. 
% SAY THIS IS A CALCULATION THAT WE ARE NOT AWARE THAT ANYONE HAS EVER DONE.
To illustrate the influence that saturation has on both the overall transport and on the intensity distribution within the disordered region, we have computed the scattering solution for three configurations.
The first has neither gain nor saturation (\cref{fig:chpt5: disordered waveguide}b), and shows the expected decay of the intensity across the sample (approximately linear in the diffusive regime~\cite{2007_akkermans_book}).
The second has gain but no saturation (\cref{fig:chpt5: disordered waveguide}c).
The intensity no longer decays as it traverses the sample, since the loss in transmission due to scattering is approximately balanced by amplification.
The final calculation accounts for both gain and saturation (\cref{fig:chpt5: disordered waveguide}d), with the intensity profile being something of a hybrid between the two linear cases, which is somewhat expected:
where the intensity is low, it should behave more like the case with gain but no saturation, and where high, like the case with neither.
The dependence of the intensity distribution on input power is consequential for wavefront shaping in amplifying nonlinear disordered media.

    \section{Bistability}

\begin{figure}[t]
\centering
    \centerline{ \includegraphics[width=\textwidth]{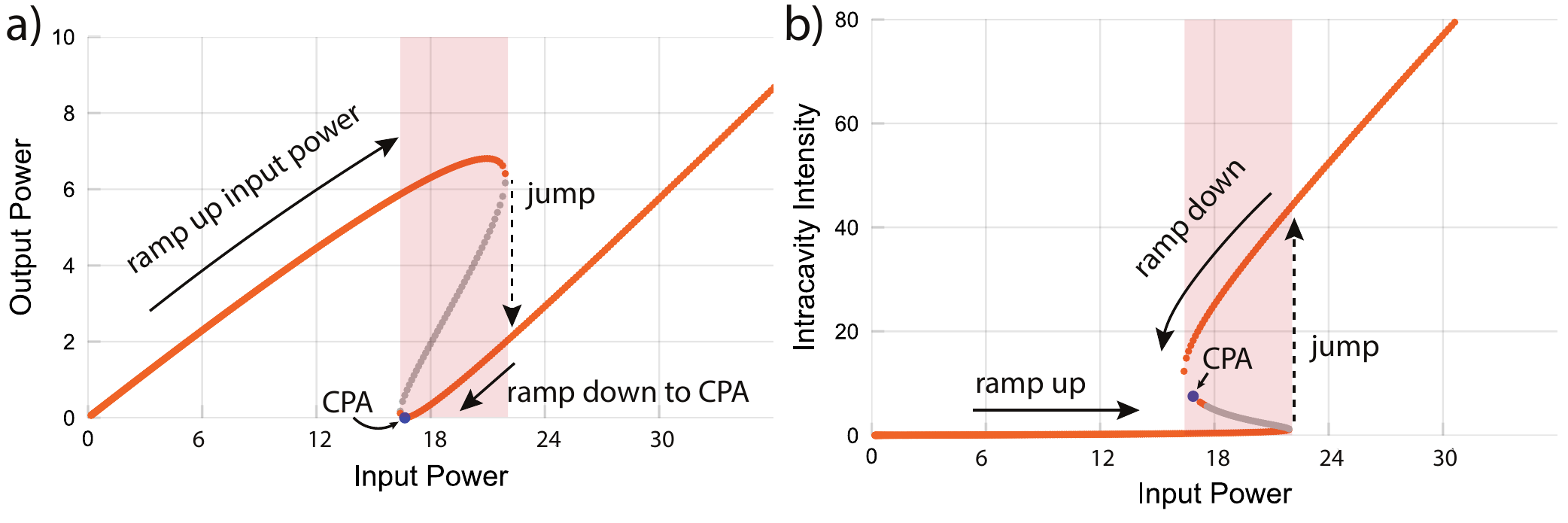} }
     \caption[Bistability curves for two-sided illumination]{Bistable nonlinear scattering.
    The scatterer is the lossy symmetric compound structure described in \cref{fig:chpt5: structure and solutions}a on \pageref{fig:chpt5: structure and solutions}, and is illuminated from both sides with the correct relative phase and amplitude for saturable CPA, but with variable overall input power ($x$-axis).
%    The input power is normalized to the power required for saturable CPA.
%    Red shaded region is the empirically-determined bistable region.
    {\bf (a):} Input-output curve, showing bistability in the red-shaded region. %The fraction of the two-sided input power that is scattered to the right.
    The unstable solution branch is shown in gray.
%%    It vanishes at the saturable CPA input power, which is in the bistable region.
    CPA (blue dot) cannot be reached directly by ramping up input power from zero; power must be increased beyond critical point where the high-output branch ceases, so that the system jumps to the lower branch; then power must be reduced until CPA is achieved (indicated by arrows).
    CPA point is near the edge of the bistability region, but still within it.
    All three solution branches were computed using the fixed point \cref{eq:chpt5: fixed point}, initialized by the results of the \scSPA\ and the \iSPA\ (see summary of algorithm in \Cref{sec:chpt5: summary of algorithm}).
    The total fractional absorption varies from $\sim20\%$ to 100\% between branches in the region of bistability.
    {\bf (b):} Similar to (a), but showing intensity integrated over the saturating region.
    The low intensity branch corresponds to high reflection and low absorption, while the high intensity branch has high absorption, making CPA possible.
    Arrows indicate the same path as in (a).
    }
    \label{fig:chpt5: two-sided fields}
\end{figure}

In the previous section we constructed a solution $\vph(\bx,\w)$ to the nonlinear scattering problem \cref{eq:chpt5: fundamental equation}, but at the moment we have no indication whether this is the unique solution, or whether there could be others consistent with the same input.
For a two-level saturable medium, it is known that there can exist two stable, steady-state solution branches for the same input field, accompanied by a third steady-state solution which is unstable~\cite{1982_Abraham_repprog}; this is optical bistability for saturable absorbers.
The iterative solution method described above can be used to find all of the solutions, if it could be applied to an appropriate set of initial guesses.
In fact, \cref{fig:chpt5: two-sided fields} shows all solution branches computed using the iterative method for a two-sided  input with variable power for the cavity shown in \vref{fig:chpt5: structure and solutions}.
However, the method itself does not tell us when we are in the region of bistability, and hence when we should look for other solutions.
Nor does it furnish us with appropriate guesses that will converge to the remaining solutions.

In the following sections we will develop a series of approximations that can be used to indicate whether other solutions exist, and to provide trial guesses that are likely to converge to the remaining solution branches.
The solutions we find solve the steady-state SALT-like scattering equation \cref{eq:chpt5: fundamental equation}, but are not necessarily stable to small perturbations.
The usual phenomenology for saturable absorbers is that there are two stable solution branches, one with high intracavity intensity, one with low, and an unstable branch with an intermediate intensity.
We will assume this remains true even for spatially-complex saturable absorption and amplification, though none of the results in this chapter depend on that assumption, and the question of dynamical stability of spatially-complex saturable absorption and amplification, while important, is not addressed here.

%For that we must make some simplifying assumptions, while still preserving arbitrary space-dependence.
%The implicit assumption is that only two of the solution branches are stable, with the third being unstable to perturbations, which is a standard result from previous theories of bistability that do not account for spatial complexity~[CITE].

%PUT THIS LATER, AFTER HAVE INTRODUCED EXISTENCE OF THREE SOLUTIONS. SAY "ASSUME USUAL PHENOMENOLOGY."

%The iterative approach will furnish all the solutions when applied to an appropriate set of initial guesses, but does not by itself indicate whether a solution is unique, nor does it systematically provide any additional  guesses that might converge to a different fixed point, should it exist.

%PERHAPS IT GOES HERE: WE ARE WORKING IN THE OPPOSITE LIMIT OF LOTS OF TECHNIQUES FOR OPTICAL BISTABILITY, HIGH-Q VS LOW-Q? LOTS OF ABSORPTION COMPARED TO OUTPUT. DISSIPATION MUCH HIGHER THAN OUTCOUPLING LOSS.
%FOR STRONG INPUT, STRONGLY UNDERDAMPED.
%Throughout the chapter I refer to bistability instead of multistability, even when there are three simultaneous steady-state solution branches.

    \subsection{Self-Consistent Single Pole Approximation (SPA) \label{sec:chpt5: self-consistent SPA}}

In this section we develop the self-consistent single pole approximation (SPA), valid for high-\Q\ resonances, which, given a saturable scattering solution $\vph(\bx)$, can be used to find an additional solution $\vph\pr(\bx)$ within the bistable region.
If an additional solution is not found, then likely $\vph(\bx)$ is the unique, monostable solution.
%In this section we show that by adding an additional outgoing wave to a known solution $\vph(\bx,\w)$, it is possible in the single pole approximation (SPA), valid for high-\Q\ resonances, both to determine whether $\vph(\bx,\w)$ is the unique solution, and, if not, to construct the remaining solutions.

Begin with a known solution to the nonlinear scattering problem $\vph(\bx)$.
For the sake of brevity we suppress the explicit dependence on the frequency $\w$.
Any other solution $\vph\pr(\bx)$ will deviate from this by a purely outgoing field $\z(\bx)$: 
\be
    \z(\bx) \defn \vph\pr(\bx) - \vph(\bx) .
\ee
Since both $\vph(\bx)$ and $\vph\pr(\bx)$ are solutions to the same nonlinear scattering problem, it follows that
\be
    \label{eq:chpt5: basic multistable equation}
    \left\{ \nabla^2 + \left[\e_{\rm c}(\bx) + \chi_{\rm nl} (\bx,\vph+\z) \right] \w^2 \right\}  (\vph(\bx)+\z(\bx) {\bs)} =  \left\{ \nabla^2 + \left[\e_{\rm c}(\bx) + \chi_{\rm nl} (\bx, \vph )  \right]\w^2\right\} \vph(\bx).
\ee
\Cref{eq:chpt5: basic multistable equation} can be ``solved'' in the form of a self-consistent equation for $\z(\bx)$:
\be
    \label{eq:chpt5: gf multistable equation}
    \z(\bx)  = \int d^dx\ G_c(\bx,\bx\pr) [\chi_{\rm nl} (\bx\pr,\vph) \vph(\bx\pr) -\chi_{\rm nl} (\bx\pr, \vph+\z)  {\bs (} \vph(\bx\pr) + \z(\bx\pr) {\bs )}],
\ee
where the cavity Green function $G_c(\bx,\bx\pr)$ is the inverse of the passive cavity Helmholtz operator, i.e., without the two-level medium:
\be
    \{ \nabla^2 + \e_c(\bx)\w^2\}G_c(\bx,\bx\pr) = \delta(\bx-\bx\pr).
 \ee

Using the definition of $\chi_{\rm nl}$ [\vref{eq:chpt5: nonlinear sus}] in \cref{eq:chpt5: gf multistable equation}, putting the resulting terms under a common denominator, and simplifying the numerator through some cancellation, gives
\be
    \label{eq:chpt5: gf multistable reduced}
    \z(\bx) = D_0 \g \int d^dx\ G_c(\bx,\bx\pr) F(\bx\pr) \frac{\G \vph(\bx\pr) |\z(\bx\pr)|^2 + \G \vph^2(\bx\pr)\z^*(\bx\pr)-\z(\bx\pr)}{{\bs (}1+\G|\vph(\bx\pr)|^2{\bs )}{\bs (}1+\G|\vph(\bx\pr)+\z(\bx\pr)|^2{\bs )}},
\ee
which obviously has the solution $\z(\bx)\equiv0$, since $\vph(\bx)$ itself was assumed to be a solution.

The field $\z(\bx)$ must itself be purely outgoing in the asymptotic regions, since $\vph\pr(\bx)$ is supposed to be a distinct solution when illuminated by the same asymptotic input.
Therefore $\z(\bx)$ can be expanded in the constant flux (CF) basis descibed in~\cite{ge_2010}:
\be
    \label{eq:chpt5: CF expansion zeta}
    \z(\bx) = \sum_n a_n u_n(\bx) \qquad a_n = \int d^d x\, u_n(\bx) F_s(\bx) \z(\bx).
\ee
The CF basis functions satisfy
\be
    \label{eq:chpt5: CF equation}
    \left\{\nabla^2 + \left[\e_{\rm c}(\bx) + \eta_n F_s(\bx) \right]\w^2 \right\} u_n(\bx) = 0,
\ee
subject to purely outgoing boundary conditions.
We assume that they are normalized according to
\begin{equation}
    \int d^d\bx\, u_\mu(\bx) F_s(\bx) u_\mu(\bx) = 1.
\end{equation}
Typically, $F_s(\bx)$ is the spatial profile of the unsaturated pump~\cite{ge_2010}, i.e.,~$F(\bx)$.
However, for the purposes of a CF expansion, it does not have to be the same as $F$, so long as $F_s$ does not vanish anywhere inside the actively pumped region; it can instead be chosen for convenience.
In this case, the most convenient choice will be that $F_s$ is the saturated pump, as we will see below. 
%The function $F_s(\bx)$ is an arbitrary function that does not vanish anywhere inside the scatterer. 
%For now, it is unrelated to the pump function $F(\bx)$, though later we will see that there is in fact a natural choice for it.
In the CF basis, the Green function has the spectral representation
\be
\label{eq:chpt5: spectral representation of Gc}
    G_c(\bx,\bx\pr) = -\sum_\m \frac{u_\m(\bx) u_\m(\bx\pr)}{\eta_\mu}.
\ee
This equation obviously has the most significant contributions from the terms with the smallest CF eigenvalues $\eta$.

In the \scSPA, we assume that the largest contribution to the Green function, which we here denote with index zero, dominates the rest:
\be
    \label{eq:chpt5: spa green fn}
    G_c(\bx,\bx\pr)\simeq G_{\rm SPA}(\bx,\bx\pr) \defn -\frac{u_0(\bx)u_0(\bx\pr)}{\eta_0}.
\ee
In other words, we approximate the Green function as having a single CF pole, hence the name of the approximation.
The validity of this approximation is determined by the condition $|\eta_0|\ll\eta_i$, $\forall i\neq0$, and is generally equivalent to the condition that the frequency $\w$ is close to one isolated  resonance.
Using the SPA, the CF expansion of $\z(\bx)$ [\cref{eq:chpt5: CF expansion zeta}] gives 
\begin{equation}
    \z(\bx)=a_0 u_0(\bx).
\end{equation}
Using a polar representation of the complex scalar $a_0=\rho \cis{\theta}$, \cref{eq:chpt5: gf multistable reduced} can be rewritten as
\be
    \label{eq:chpt5: self-consistent rho}
    \rho =  C_0(\rho,\theta)\cis{\theta} - C_1(\rho,\theta)\cis[-]{\theta} ,
\ee
where
\be
    \label{eq:chpt5: self-consistent Cs}
    a_0\defn\rho \cis{\theta}, \qquad C_0(\rho,\theta) \defn \frac{I_0(\rho,\theta) - \eta_0/D_0\g}{I_1(\rho,\theta)}, \qquad C_1(\rho,\theta) \defn \frac{ I_2(\rho,\theta)}{I_1(\rho,\theta)}
\ee
and
\begin{align}
    \label{eq:chpt5: self consistent I0}
    I_0(\rho,\theta) &\defn \int d^dx\ \frac{F(\bx)}{ {\bs (}1+\G|\vph(\bx)|^2 {\bs )}} \frac{u_0^2(\bx)}{{\bs (} 1+\G|\vph(\bx)+\rho\cis{\theta}u(\bx)|^2 {\bs )} } \\
    I_1(\rho,\theta) &\defn \int d^dx\ \frac{F(\bx)}{ {\bs (}1+\G|\vph(\bx)|^2 {\bs )}} \frac{|u_0(\bx)|^2 u(\bx) \vph(\bx)}{{\bs (} 1+\G|\vph(\bx)+\rho\cis{\theta}u(\bx)|^2 {\bs )} } \\
    \label{eq:chpt5: self consistent I2}
    I_2(\rho,\theta) &\defn \int d^dx\ \frac{F(\bx)}{ {\bs (}1+\G|\vph(\bx)|^2 {\bs )}} \frac{|u_0(\bx)|^2 \vph^2(\bx)}{{\bs (} 1+\G|\vph(\bx)+\rho \cis{\theta} u_0(\bx)|^2 {\bs )} }.
\end{align}
In writing this, we have assumed that $a_0 \neq 0$, which is appropriate if we want to find {\it new} scattering solutions, distinct from the original $\vph(\bx)$, so that we were able to divide both sides of  \cref{eq:chpt5: gf multistable reduced} by $|a_0|=\rho$.
In terms of the functions $C_{0,1}(\rho,\theta)$, the real and imaginary parts of \cref{eq:chpt5: self-consistent rho} can be solved implicitly for $\rho$ and $\theta$:
\begin{gather}
    \label{eq:chpt5: self-consistent SPA theta}
    \theta = -\arctan \frac{\im{C_0(\rho,\theta)-C_1(\rho,\theta)}}{\re{C_0(\rho,\theta)+C_1(\rho,\theta)}} \\
    \label{eq:chpt5: self-consistent SPA rho}
    \rho =  \re{C_0(\rho,\theta) - C_1(\rho,\theta)}\cos{\theta}-\im{C_0(\rho,\theta) - C_1(\rho,\theta)}\sin{\theta}.
\end{gather}
The \scSPA\ solution is
\begin{equation}
    \vph\pr_{\rm SPA}(\bx) = \vph(\bx) + \rho\cis{\theta} u_0(\bx),
\end{equation}
with $\theta$ and $\rho$ satisfying \cref{eq:chpt5: self-consistent SPA theta,eq:chpt5: self-consistent SPA rho}.

We can solve for $\theta$ and $\rho$ using a variety of standard nonlinear root-finding methods, such as Newton-Raphson or trust region.
However, it is also amenable to iteration in the form
\begin{gather}
    \theta_{n+1} = -\arctan \frac{\im{C_0-C_1}}{\re{C_0+C_1}} \bigg|_{(\rho_n,\theta_n)} \\
    \rho_{n+1} =  \left(\re{C_0 - C_1}\cos{\theta}-\im{C_0 - C_1}\sin{\theta}\right) \big|_{(\rho_n,\theta_n)}\,.
\end{gather}
Convergence is predicated on the existence of solutions with $\rho\neq0$, therefore a failure to converge can be taken as evidence that the given solution $\vph(\bx)$ is the unique solution to the nonlinear scattering problem, up to a global phase.

\begin{figure}[t!]
\centering
    \centerline{ \includegraphics[width=\textwidth]{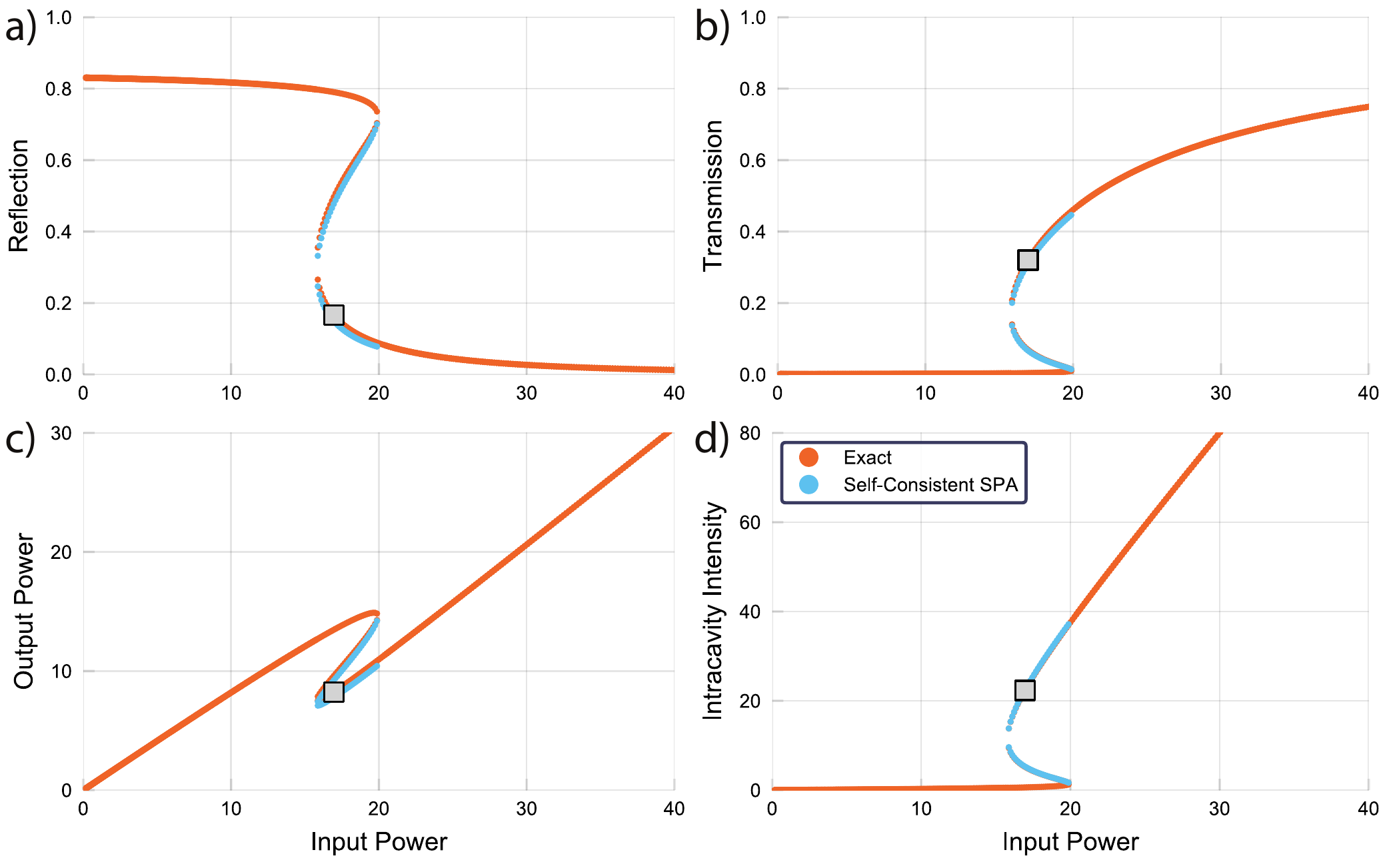} }
    \caption[Bistability curves for one-sided illumination, comparison of exact solution and self-consistent single pole approximation]{Bistable nonlinear scattering, similar to \vref{fig:chpt5: two-sided fields} but for one-sided input.
    The scatterer is the lossy symmetric compound structure described in \cref{fig:chpt5: structure and solutions}a on \pageref{fig:chpt5: structure and solutions}, and is illuminated from the left with variable input power ($x$-axis).
    CPA is not achieved here since the illumination is one-sided.
    In all four panels, blue is result of \scSPA\ [\cref{sec:chpt5: self-consistent SPA}], which agrees very well with the exact solution from the iterative method (orange) computed from \cref{eq:chpt5: fixed point}, and gray square denotes the specific input and output powers associated with \cref{fig:chpt5: structure and solutions}c.
    {\bf (a):} The reflection coefficient goes from high to low in a narrow region around the saturable CPA power, exhibiting bistability.
    An explanation for this is provided in the main text.
%    Orange is the exact reflection as computed from , and blue is computed from the self-consistent single pole approximation (SPA)
    {\bf (b):} Similar to (a), but for transmission.
    In this case the transmission jumps by $18{\rm\ dB}$ in going from the lower to the upper branch as the input power is varied through the bistable-monostable transition.
    {\bf (c)} \& {\bf (d):}  The total output power and the integrated intensity inside the saturating region.
    Note that because of absorption in the cavity, $|t|^2 + |r|^2 \neq 1$.
    }
    \label{fig:chpt5: one-sided fields 1}
\end{figure}

\Cref{fig:chpt5: one-sided fields 1} summarizes all branches of the nonlinear scattering solutions from the same structure as in \cref{fig:chpt5: structure and solutions}a, subject to one-sided illumination with variable input power.
As a reminder, that compound structure was made from three cavities in a symmetric configuration, with the middle cavity containing a lossy, two-level medium.
For comparison, the \scSPA\ prediction is also shown.
The initial solution branch $\vph(\bx)$ was taken to be the solution with the lowest intracavity intensity.
Over the entire range of bistability, the agreement between the \scSPA\ prediction for $a_0$ and the equivalent CF coefficient of the exact solutions on the remaining two branches was better than 1\%, an excellent agreement given that no assumptions were made beyond stationary inversion and isolated resonance.
%\Vref{fig:chpt5: one-sided comparison} compares the exact wavefields with the \scSPA\ predictions, again showing good agreement.

\Cref{fig:chpt5: one-sided fields 1} show that there is a rapid change in the solutions over some range of input powers near the power required for saturable CPA.
The input-output characteristic curve and the intracavity power (Figs.~\ref{fig:chpt5: one-sided fields 1}c--d) have kinks that have been understood for systems without spatial complexity, such as ring-cavities~\cite{1982_Abraham_repprog}.
A novel observation, enabled by the solution methods presented here, is that the reflection goes from a high value to nearly zero as it passes through the multistable regime, while the transmission exhibits the complementary behavior of going from nearly zero to a high value.
The asymptotic behavior of the transmission is generic for this system: at low input powers, where saturation is negligible, the very lossy central cavity absorbs nearly all flux traversing it, so none of it is transmitted.
For high input powers, for which the cavity is so saturated that the effective pump profile becomes negligibly small everywhere, the solution is well-described by scattering from a linear, lossless symmetric structure, with perfectly transmitting resonances.
This latter model also explains the disappearance of the reflection at high input powers.
However, the sharp boundary between the two behaviors cannot be understood from comparison to simplified linear models.

    \subsubsection{The Self-Consistent Single Pole Approximation in the high-\Q\ limit}

The \scSPA\ does not have even a semi-analytic solution in general, which is its shortcoming.
However, in the high-\Q\ limit we can simplify the \scSPA\ equation for $\theta$ [\cref{eq:chpt5: self-consistent SPA theta}], and show that $\zeta(\bx)$ is either in- or out-of phase with $\vph(\bx)$.

In the high-\Q\ limit, the relevant CF states are nearly standing waves, and the solution $\vph(\bx)$ has a much larger standing wave than traveling wave component inside the cavity.
We can formalize this by saying
\be
    \label{eq:chpt5: high-Q expression in SPA}
    \vph(\bx) = \cis{\alpha}\sigma(\bx) + \tau(\bx), \qquad u(\bx) = \re{u(\bx)}{\bs (}1+iv(\bx) {\bs )},
\ee
where $\sigma(\bx)$, $v(\bx)$ are real functions, $|v(\bx)|<<1$, and contributions to the integrals $I_{1,2,3}$ [\crefrange{eq:chpt5: self consistent I0}{eq:chpt5: self consistent I2}] from $\tau(\bx)$  are negligible in comparison to the contributions from $\sigma(\bx)$.
The phase factor $\cis{\alpha}$ is unconstrained except for $\alpha\in{\mathbb R}$ --- a manifestation of the global $U(1)$ gauge invariance of \cref{eq:chpt5: fundamental equation}, and will depend on the specific phase of $\vph(\bx)$.
For a given solution $\vph(\bx)$, the phase $\alpha$ can be estimated by 
\be
    \label{eq:chpt5: extract phase}
    \alpha \approx \frac{1}{2}\arg \int d^dx\ \vph^2(\bx) F(\bx),
\ee
since the traveling wave component inside the cavity self-averages to a negligibly small value upon integration (``self-cancels'').

%The \scSPA\ $\theta$-\cref{eq:chpt5: self-consistent SPA theta} can be solved explicitly in the high-\Q\ limit.

Additionally, if we assume that the resonance in question is near the center of the gain band of the two-level medium, then the phase of $\eta_0$ and $\g$ are nearly identical, so that $(\eta_0/D_0\g) \in {\mathbb R}$.
These assumptions imply that $C_0=\cis[-]{\alpha}\re{C_0}$ and $C_1=\cis{\alpha}\re{C_1}$.
Plugging this into \cref{eq:chpt5: self-consistent SPA theta} implies that 
\begin{equation}
    (\theta-\alpha)\in\{0,\pi\},
\end{equation}
so that the other solution branches are either in-phase or out-of-phase with the given solution, $\vph$.
The magnitude of the SPA CF coefficient $a_0$ [\cref{eq:chpt5: self-consistent SPA rho}] must still be solved for self-consistently even in this high-\Q\ limit.

    \subsection{Inverse Single Pole Approximation and Bistability Criterion \label{sec:chpt5: inverse SPA}}

The shortcoming of the \scSPA\ is that it cannot be solved explicitly: even if $\z(\bx)$ is assumed to be in- or out-of-phase with the solution $\vph(\bx)$, so that $\theta$ is fixed, the amplitude $\rho$ {\it still} appears in the denominators of the integrals $I_{0,1,2}$ in \crefrange{eq:chpt5: self consistent I0}{eq:chpt5: self consistent I2}.
Because of this, it does not provide an appropriate initial estimate for its own solution, and sheds little light onto the limits of bistability, or even how many solutions can be expected.

In this section we present the {\it inverse} SPA, which is the single pole approximation applied to the {\it inverse} Green function, i.e.,~applied directly to \vref{eq:chpt5: basic multistable equation}.
It results in a complex cubic polynomial in the CF coefficient $a_0$ and its complex conjugate, which can be semi-analytically solved in the high-\Q\ limit in terms of overlap integrals between the CF state $u_0(\bx)$ and the original solution $\vph(\bx)$.
We will see that according to the \iSPA, the onset of bistability is a saddle-node bifurcation, that is, two solution branches are created or annihilated pairwise as the input power is increased.
This extends the known results of conventional bistability theory to cases with arbitrary space-dependence.

This approximation has the advantage that it provides a rough estimate for the location of the bistability region and for the other solution branches.
Its estimates are useful for initializing the \scSPA, which is in fact what was done in \cref{fig:chpt5: structure and solutions,fig:chpt5: one-sided fields 1,fig:chpt5: two-sided fields}.
However, it is not a well-controlled approximation, as we will see in \Cref{sec:chpt5: failure of ispa}, and can have poor predictive power, even for isolated resonances.
The \iSPA\ has been used widely for lasing, and the considerations here suggest that we need to reexamine its accuracy in that context.
%However, the \iSPA\ has the fundamental shortcoming that it is not a well-controlled approximation and has poor predictive power, even for isolated resonances.

The \iSPA\ follows from first multiplying \cref{eq:chpt5: basic multistable equation} by the denominator $1+\G|\vph(\bx)+\z(\bx)|^2$, and then projecting onto the CF state $u_q(\bx)$.
The result is an infinite collection of infinite polynomial equations in the unknown CF coefficients and their conjugates:
\be
    \label{eq:chpt5: full nonlinear CF coefficient equation}
    A_q^{nm,p}\ a_n a_m a_p^* + B_q^{n,m}\ a_n a_m^* + C_q^{nm,}\ a_n a_m +  E_q^{,n}\ a_n^*  +  G_q^{n,}\ a_n = 0,
\ee
which holds $\forall q$, with implicit summation over repeated indices.
We use the notation that upstairs indices preceding the comma pair with CF amplitudes, while those following pair with conjugate amplitudes.
The downstairs index $q$ refers to the CF state that the equation is projected onto.
The coefficient matrices are
\begin{align}
    \label{eq:chpt5: Nonlinear Matrix Coefficients A}
    A_q^{nm,p} &= \G \eta_n \int d^dx\ u_n(\bx) u_m(\bx) u_p^*(\bx) u_q(\bx) F_s(\bx) \\
    B_q^{n,m} &= \G(\eta_n + \g D_0) \int d^dx\ u_n(\bx) u_m^*(\bx) u_q(\bx) \vph(\bx) F_s(\bx) \\
    C_q^{nm,} &= \G \eta_n \int d^dx\ u_n(\bx) u_m(\bx) u_q(\bx) \vph^*(\bx) F_s(\bx) \\
    E_q^{,n} &= \g D_0 \G \int d^dx\ u_n^*(\bx) u_q(\bx) \vph^2(\bx) F_s(\bx) \\
    \label{eq:chpt5: Nonlinear Matrix Coefficients G}
    G_q^{n,} &= (\eta_n - \g D_0) \delta_{qn} + \eta_n \G \int d^dx\ u_n(\bx) u_q(\bx) |\vph(\bx)|^2 F_s(\bx)
\end{align}
These have been substantially simplified by a specifice choice for the CF pump function $F_s(\bx)$: the physical pump profile $F(\bx)$ saturated by the given solution $\vph(\bx)$,
\be
\label{eq:chpt5: Fs}
	F_s(\bx) \defn \frac{F(\bx)}{1+\G|\vph(\bx)|^2}.
\ee
The absence of a constant term (independent of the $a$'s) in \cref{eq:chpt5: full nonlinear CF coefficient equation} means that $a_n\equiv0$ $\forall n$ is always a solution, which we also saw in the \scSPA.

As a sanity check, let us compare this to the previously known case of lasing, as described within SALT: a non-vanishing outgoing field even when the incident field is absent. 
To make this comparison, we take $\vph(\bx)\rightarrow 0$, in which case $F_s(\bx) \rightarrow F(\bx)$ and \cref{eq:chpt5: full nonlinear CF coefficient equation} reduces to
\be
	D_0a_n = \sum_{n\pr} \tau_{nn\pr} ({\bf a})  a_{n\pr}
\ee
where the nonlinear map $\tau_{nn\pr}({\bf a})$ is the inverse of the ``lasing map'' of the SALT-algorithm, namely, 
\be
	\tau_{nn\pr}({\bf a}) =  \frac{\eta_n}{\g} \left[ \delta_{nn\pr} + \G \int d^dx\  |\z(\bx)|^2 u_n(\bx) F(\bx) u_{n\pr}(\bx) \right].
\ee
Therefore the \iSPA\ reduces to SPA-SALT~\cite{ge_2010} in the limit of a vanishing incident field, as expected

%It is tempting to think that a suitable method for solving the nonlinear scattering problem \Eq{eq:chpt5: fundamental equation} is to find the roots of \eqref{eq:chpt5: full nonlinear CF coefficient equation}, but this is not so.
%This is because~\eqref{eq:chpt5: full nonlinear CF coefficient equation} was predicated on $\vph(\bx)$ itself being a solution of the nonlinear problem, which is why it lacks a constant term, independent of the $a$'s, and therefore $a_n=0$, $\forall n$, is a valid solution.
%However, a different choice for $\vph$, say $\vph\pr$, which is not a solution of~\eqref{eq: Saturable Scattering: Fundamental Equation}, but instead is given by $(\nabla^2 + \left[ \e_{\rm c} + \g D_0 F \right] \w^2)\vph\pr=-j$, will lead to an equation similar to~\eqref{eq: Saturable Scattering: full nonlinear CF coefficient equation}, but with renormalized coefficients, and an additional constant term.
%Finding the roots of this new polynomial equation could serve as a method for finding the nonlinear solutions with for fixed input.
%However, it was found that in this case the outgoing field $\zeta\pr \equiv \psi\pr - \vph\pr$ was not sufficiently localized in the CF basis, even in the high-\Q limit, so that the requisite truncation of the CF basis induces a large error unless a great many states are included.

%In contrast, Eq.~\eqref{eq: Saturable Scattering: full nonlinear CF coefficient equation} has numerically been found to be consistent with $\zeta$'s that are well-localized in the CF basis.
As with the \scSPA\ in the high-\Q\ limit, $\z(\bx)$ is expected to be well-approximated by a single CF state, $u_0(\bx)$, with the other CF coefficients being negligibly small: $|a_{i\neq0}/a_0| \ll 1$.
For now, let us assume that they are zero (we will return to this assumption in \Cref{sec:chpt5: failure of ispa}), and also that all integrals involving products of odd numbers of a given CF function self-cancel upon integration (for example, $\int d^dx\, u_0 u_1 u_2 u_3^* \sim \int d^dx\, |u_0|^2 u_0 u_1 \sim 0$, but $\int d^dx\, u_0^2 u_1^2 \neq 0$).
Then the array of polynomial equations \cref{eq:chpt5: full nonlinear CF coefficient equation} reduces to a single equation for $a_0=\rho\cis{\theta}$:
\be
    \label{eq:chpt5: SPA nonlinear CF coefficient equation}
    \tilde A \rho^3 + \tilde B \cis[-]{(\theta-\alpha)} \rho^2 + \tilde C \cis{(\theta-\alpha)}\rho^2  + \tilde E \cis[-2]{(\theta-\alpha)} \rho + \tilde G \rho = 0,
\ee
where
\begin{align}
    \label{eq:chpt5: Nonlinear Scalar Coefficients A}
    \tilde A &= \eta_0 \G \int d^dx\ |u_0(\bx)|^2 u_0^2(\bx)  F_s(\bx) \\
    \tilde B &= (\eta_0 + \g D_0) \G \int d^dx\ |u_0(\bx)|^2 u_0(\bx) \sigma(\bx) F_s(\bx) \\
    \tilde C &= \eta_0 \G \int d^dx\ u_0^3(\bx) \sigma(\bx) F_s(\bx) \\
    \tilde E &= \g D_0 \G \int d^dx\ |u_0(\bx)|^2 \sigma^2(\bx) F_s(\bx) \\
    \label{eq:chpt5: Nonlinear Scalar Coefficients G}
    \tilde G &= (\eta_0 - \g D_0) + \eta_0 \G \int d^dx\ u_0^2(\bx) \sigma^2(\bx) F_s(\bx).
\end{align}
We are adopting the notation and high-\Q\ assumptions made earlier in \cref{eq:chpt5: high-Q expression in SPA} when developing the \scSPA\ , namely, $\alpha = \frac{1}{2} \int d^dx\, \vph^2(\bx) F(\bx)$ and $\sigma(\bx) = \cis[-]{\alpha}\vph(\bx)$.\footnote{Strictly this is not $\sigma$, but since $\sigma$ only appears as part of an integrand, we can neglect the traveling wave component $\tau$.}
%In analogy to the single-pole approximation of SALT for lasing (SPA-SALT)~[CITE], we may here consider a single-pole approximation of~\eqref{eq: Saturable Scattering: full nonlinear CF coefficient equation}, in which all but one of the $a_n$'s vanish.
%In the high-\Q limit, the CF states $u$ and the original scattering solution $\vph$ are mostly real inside the cavity, in the sense that 
%\be
%\label{eq: Saturable Scattering: high-\Q expression in SPA}
%	\vph(\bx) = e^{i\alpha}\tau(\bx)(1 + iw(\bx))\qquad u(\bx) = \pm \re{u(x)}(1+z(\bx))
%\ee
%where $\tau(\bx)$ is real and $w(\bx),z(\bx)\ll 1$ everywhere inside the cavity, so that we can neglect them.
%In this case, Eq.~\eqref{eq: Saturable Scattering: Nonlinear Matrix Coeffiecients A}--\eqref{eq: Saturable Scattering: Nonlinear Matrix Coeffiecients G} reduce to scalar coefficients:
%\begin{align}
%\label{eq: Saturable Scattering: Nonlinear Scalar Coefficients A}
%	\tilde A &= \eta \G \int d^dx\ |u(\bx)|^2 u^2(\bx)  F_s(\bx) \\
%	\tilde B &= (\eta + \g D_0) \G \int d^dx\ |u(\bx)|^2 u(\bx) \sigma(\bx) F_s(\bx) \\
%	\tilde C &= \eta \G \int d^dx\ u(\bx)^3 \sigma(\bx) F_s(\bx) \\
%	\tilde E &= \g D_0 \G \int d^dx\ |u(\bx)|^2 \sigma^2(\bx) F_s(\bx) \\
%\label{eq: Saturable Scattering: Nonlinear Scalar Coeffiecients G}
%	\tilde G &= (\eta - \g D_0) + \eta \G \int d^dx\ u^2(\bx) \sigma^2(\bx) F_s(\bx).
%\end{align}
We have introduced the variables with tildes, which are independent of the global phase $\alpha$, for later convenience.
All the integrals in \crefrange{eq:chpt5: Nonlinear Scalar Coefficients A}{eq:chpt5: Nonlinear Scalar Coefficients G} are approximately real in the \iSPA, as is $\eta_0/\g$.
As stated earlier, $a_n\equiv0$ solves \cref{eq:chpt5: full nonlinear CF coefficient equation}, since $\vph(\bx)$ is a solution of the nonlinear scattering problem, and therefore $\rho=0$ is a solution to the \iSPA\ \cref{eq:chpt5: SPA nonlinear CF coefficient equation}.
Any additional solutions must satisfy the complex quadratic
\be
    \label{eq:chpt5: SPA additional solutions}
    \rho^2 + \left( \frac{\tilde B \cis[-]{(\theta-\alpha)}  + \tilde C \cis{(\theta-\alpha)}}{\tilde A} \right) \rho + \left( \frac{\tilde E \cis[-2]{(\theta-\alpha)} + \tilde G}{\tilde A} \right) = 0,
\ee
where $\rho$, $\theta$ are constrained to be real.
Note that under our assumptions $\tilde A$, $\tilde B$,  $\tilde C$,   $\tilde E$, and $\tilde G$ all have the same phase, so that \cref{eq:chpt5: SPA additional solutions} implies the two real conditions
\begin{gather}
    \label{eq:chpt5: rho-theta inverse SPA equations 1}
    \rho^2 + \left( \frac{\tilde B + \tilde C}{\tilde A}\right) \cos(\theta-\alpha) \rho + \left( \frac{\tilde E \cos(2[\theta-\alpha]) + \tilde G}{\tilde A} \right) = 0 \\
    \label{eq:chpt5: rho-theta inverse SPA equations 2}
    \rho \left( \tilde B - \tilde C \right) \sin(\theta-\alpha) + \tilde E \sin(2[\theta-\alpha]) = 0.
\end{gather}

There are two kinds of solutions to \cref{eq:chpt5: rho-theta inverse SPA equations 1,eq:chpt5: rho-theta inverse SPA equations 2}.
The first has $\sin(\alpha-\theta)=0$, which satisfies \cref{eq:chpt5: rho-theta inverse SPA equations 2} for all $\rho$ and all coefficients, so that $(\theta-\alpha) \in \{0,\pi\}$.
This is the same condition we encountered before with the \scSPA\  in the high-\Q\ limit: the outgoing CF amplitude is either in-phase or out-of-phase with the exciting field.
The second class satisfies a different phase relationship, which was not found to be relevant in any of the examples presented in this chapter.
We will discuss it briefly at the end of this section.

Using $(\theta-\alpha) \in \{0,\pi\}$ in \cref{eq:chpt5: rho-theta inverse SPA equations 1} gives
\be
    \label{eq:chpt5: quadratic rho}
    \rho^2 \pm \left( \frac{\tilde B + \tilde C}{\tilde A}\right) \rho + \left( \frac{\tilde E + \tilde G}{\tilde A} \right) = 0. 
\ee
%with the two positive solutions $\rho = | \Sigma \pm \sqrt{\Delta} |/2$  (recall that $\rho=|a_0|$).
The solutions for $\theta$ and $\rho$ can be consolidated into a solution for the one CF coefficient:
\be
    \label{eq:chpt5: inverse SPA solution}
    a_{0,\pm}^{\rm inv\,SPA} = \cis{\alpha}(\Sigma \pm \sqrt{\Delta})/2,
\ee
where $\Sigma$ is the sum of the roots, $\Pi$ the product, $\Delta$ the discriminant, and all are real functionals of $\vph(\bx)$:
\begin{gather}
    \label{eq:chpt5: sigma functional}
    \Sigma \defn (\tilde B + \tilde C)/\tilde A \\
    \label{eq:chpt5: pi functional}
    \Pi \defn (\tilde E + \tilde G)/\tilde A \\
    \label{eq:chpt5: delta discriminant}
    \Delta \defn \Sigma^2 - 4\Pi.
\end{gather}

It follows that in the \iSPA, there are at most three solutions, and that the original solution $\vph(\bx)$ is unique unless $\Delta[\vph(\bx)]>0$.
Whenever $\Delta[\vph(\bx)]$ passes through zero, a pair of solutions to \cref{eq:chpt5: fundamental equation} is created or annihilated in a saddle-node bifurcation~\cite{2015_strogatz_book}, which extends the result from conventional, space-independent saturable bistability theory.
In the high-\Q\ limit, using the approximations in \cref{eq:chpt5: high-Q expression in SPA} gives
\begin{align}
\label{eq:chpt5: Delta functional}
    \Delta[\vph(\bx)] &= \left( 2+\frac{\g D_0}{\eta_0} \right)^2 \left( \frac{\int d^dx\, u_0^3(\bx) \sigma(\bx) F_s(\bx)}{\int d^dx\, u_0^4(\bx) F_s(\bx)}\right)^2 + \\
    & \qquad - 4 \frac{(1 + \g D_0/\eta_0)\int d^dx\, u_0^2(\bx) \sigma^2(\bx) F_s(\bx) + (1-\g D_0/\eta_0)/\G}{\int d^dx\, u_0^4(\bx) F_s(\bx)}. \nonumber
\end{align}
There are several implicit dependencies hidden in this equation:
\begin{enumerate}
    \item The integrals contain $\sigma$, not $\vph$.
    The two are related by a global phase.
    \item $F_s(\bx)$ has implicit $\vph$-dependence [\cref{eq:chpt5: Fs}], scaling like the inverse of the input power for large power.
        Therefore, by CF normalization, $u_0^2$ scales proportionally with the power for large powers.
        In fact, we might expect the proportionality constant to be a little larger than one, since the denominator in the normalization integral for $u_0$ is not $|\sigma(\bx)|^2$, but $|1+\sigma(\bx)|^2$.
    \item $\eta_0$ is the smallest CF eigenvalue of the {\it saturated} problem, i.e. using $F_s(\bx)$ instead of $F(\bx)$.
    	It, too, implicitly depends on $\vph$, scaling as the square of the input power.
\end{enumerate}

With these hidden dependencies in mind, we can analyze \cref{eq:chpt5: Delta functional} for asymptotically low and high input powers.
The low power limit for bistability acts as a sanity check, since we already know that bistability in the absence of any input is actually laser oscillation.
At low powers, $F_s(\bx) \to F(\bx)$, $\sigma(\bx) \to 0$, and the condition for bistability is
\be
    \Delta[\vph(\bx)\rightarrow0] > 0 \implies \frac{\g D_0}{\eta_0} > 1,
\ee
which is just the threshold lasing condition from SALT expressed in terms of the CF eigenvalue, as expected~\cite{ge_2010}, and which we know requires gain.
%For the purely outgoing CF problem with real $\e_c(\bx)$, the CF eigenvalues are generally negative-imaginary, as is $\g(\w)$ near the center of the gain curve.
%Therefore for absorbing two-level media with $D_0<0$, we never expect to find bistability, while for amplifying two-level media with no injected light, bistability is lasing: $\g D_0 > \eta_0$.
The reason that a two-level amplifier shows bistability in this limit, but not a two-level absorber, is because the nonlinearity that supports bistability is not significant unless the intracavity intensity is high, which does not happen for absorbers with small injected intensities, but can happen in lasing above threshold.
%An absorbing cavity in the limit of weak input power has very little saturation, and so there is not bistability in this limit.
%On the other hand, an amplifying cavity will tend to support bistability at much lower input powers, since the intracavity intensity can be high even for weak input; the extreme case of this is lasing.

In the other limit, with large input powers, the $(1-\g D_0/\eta_0)$ term in \cref{eq:chpt5: Delta functional} becomes negligible compared to the rest, while the ratios of each of the integrals approach some positive constant $c$.
In that case, the bistability condition becomes
\be
    \Delta[\vph(\bx) \rightarrow \infty] > 0 \implies 4c(c-1) > 0.
\ee
Since the $u_0$'s are slightly larger than the $\sigma$'s for large input power, as mentioned above, $c$ should be less than unity, and $\Delta[\vph(\bx) \rightarrow \infty]<0$, thus ensuring monostability for very large input intensities.

In summary, we expect the saturating nonlinearity to induce bistability over at most a finite range of input powers, which is consistent with what is found in \cref{fig:chpt5: one-sided fields 1,fig:chpt5: two-sided fields}, and with what is known in the literature for conventional cases without spatial complexity~\cite{1982_Abraham_repprog}.

Recall that the above solutions, which are either in- or out-of-phase with the incident wave, belong to one of two solution classes.
The other class satisfies \cref{eq:chpt5: rho-theta inverse SPA equations 2} with a different phase relation:
\be
    \theta = \alpha + \arccos \left( \frac{\rho}{2}\left[\frac{\tilde C -\tilde B}{\tilde E}\right]\right),
\ee
in which case \cref{eq:chpt5: rho-theta inverse SPA equations 1} implies
\be
    \label{eq:chpt5: alternative bistability solution}
    \rho^2  = \frac{\tilde E - \tilde G}{ \tilde A + \tilde C ( \tilde C - \tilde B )/ \tilde E}.
\ee
This solution is subject to the condition $|\cos(\theta-\alpha)|^2 = \frac{1}{4}\frac{(\tilde E - \tilde G)(\tilde C - \tilde B)^2}{ \tilde A \tilde E^2 + \tilde C \tilde E ( \tilde C - \tilde B )} \le 1$.
In every case analyzed in this chapter, this inequality was grossly violated, and so we ignore this class of solution.
It is an open question whether this is always the case, or whether sometimes the onset of bistability is described by \cref{eq:chpt5: alternative bistability solution} instead of \cref{eq:chpt5: inverse SPA solution}, that is, by a pitchfork bifurcation~\cite{2015_strogatz_book}, instead of the usual saddle-node bifurcation associated with optical bistability.

    \subsection{Failure of the Inverse Single Pole Approximation \label{sec:chpt5: failure of ispa}}

The \iSPA\ successfully shows that the saturable scattering problem supports a single solution for asymptotically small and large input powers (excepting lasing), and can support up to three solutions over some finite range of input powers.
It also successfully shows that the boundaries of the bistable regime are described by saddle-node bifucations, controlled by the zero-crossings of the discriminant functional $\Delta[\vph(\bx)]$.

\begin{figure}[t!]
\centering
    \centerline{ \includegraphics[width=\textwidth]{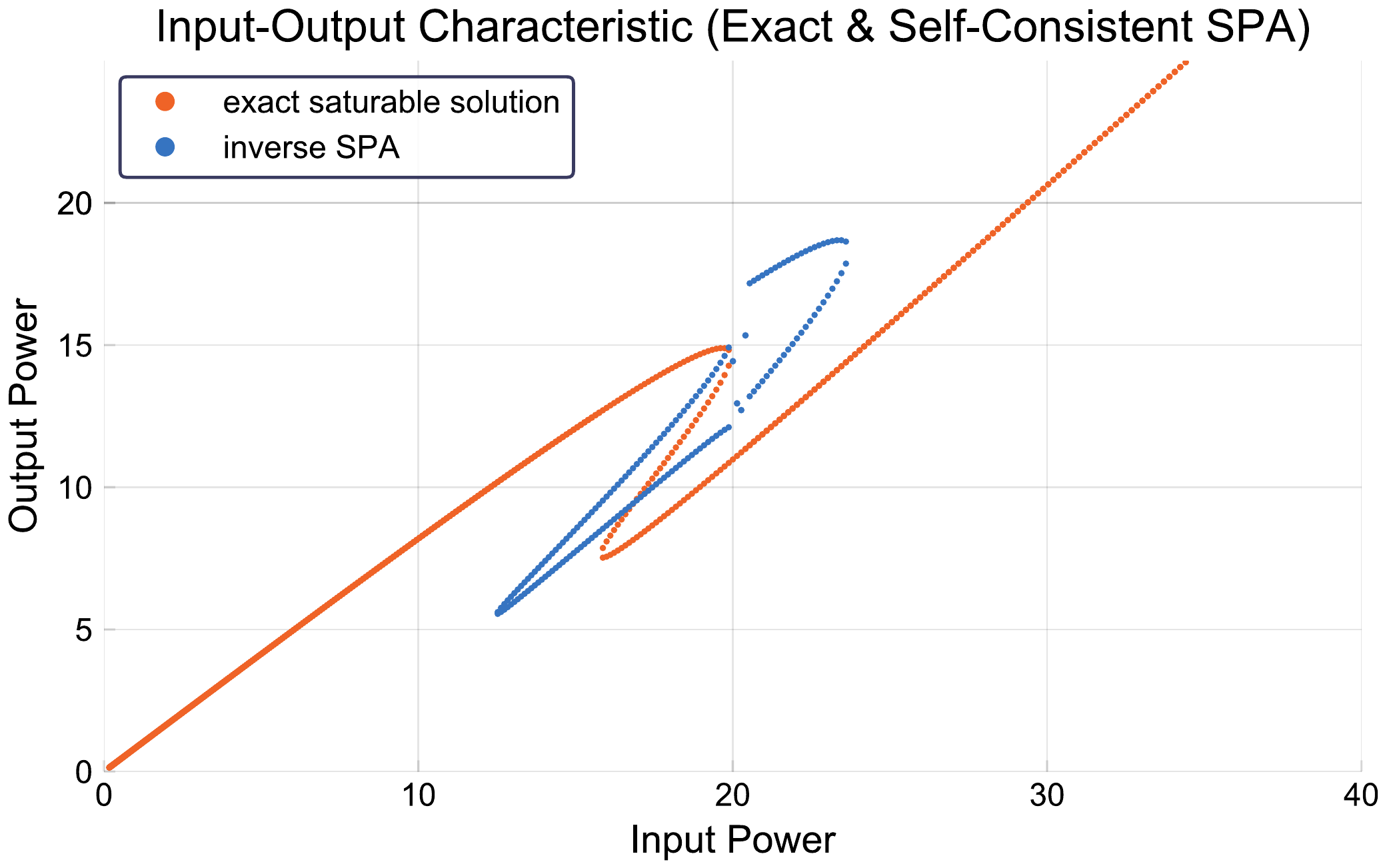} }
    \caption[Comparison of exact bistable saturable scattering with inverse single pole approximation]{Input-output curve for bistable nonlinear scattering.
    The scatterer is the lossy symmetric compound structure described in \vref{fig:chpt5: structure and solutions}a, and is illuminated from the left with varying input power ($x$-axis).
    CPA is not achieved here since the illumination is one-sided.
    Orange is exact solution of \cref{eq:chpt5: fundamental equation} using the iterative method.
    Blue is result from \iSPA, using upper branch as given solution $\vph(\bx)$.
    The \iSPA\ does not agree well with exact solution, and significantly overestimates the size of the bistability region.
    }
    \label{fig:chpt5: one-sided fields}
\end{figure}

Despite these successes, the \iSPA\ gives rather poor estimates, which is surprising given that the \scSPA\  works well, and both are built on the same assumptions.
\Cref{fig:chpt5: one-sided fields} compares the \iSPA\ prediction for the bistable region of the input-output curve with the exact solution for the same compound cavity used in the previous figures.
The \iSPA\ has a relative error in the CF coefficient $a_0$ of $\sim32\%$, while the \scSPA\ estimate for the same coefficient has a relative error of $\le1\%$.
In both cases, the error is measured relative to the relevant CF coefficient of the exact solution.

The reason for this failure is that, in truncating the infinite array of infinite polynomials [\crefrange{eq:chpt5: full nonlinear CF coefficient equation}{eq:chpt5: Nonlinear Matrix Coefficients G}], we have not only assumed that the non-dominant CF coefficients are negligibly small, but that the {\it sum} over such coefficients is small, which is evidently invalid.
A more accurate development of the \iSPA\ would have included an additional contribution from the non-dominant CF coefficients for each CF projection $q$:
\begin{align}
    \label{eq:chpt5: inverse SPA equation with non-dominant terms}
A_q^{00,0}\ &|a_0|^2 a_0 + B_q^{0,0}\ |a_0|^2 + C_q^{00,}\ a_0^2 +  E_q^{,0}\ a_0^*  +  G_q^{0,}\ a_0\ + \nonumber \\
    \bigg\{&\sum_{n\neq0}\ \left[A_q^{00,n}\ a_0^2 a_n^* + (A_q^{0n,0} + A_q^{n0,0})\ |a_0|^2 a_n \right]\ + \nonumber \\
    &\sum_{n,m\neq0} \left[ (A_q^{0n,p} + A_q^{n0,p})\ a_0 a_n a_p^* + A_q^{nm,0}\ a_0^* a_n a_m \right]\   + \nonumber \\
    &\sum_{n,m,p\neq0} A_q^{nm,p}\  a_n a_m a_p^*\  + \nonumber \\
    &\sum_{n\neq0}\ \left[B_q^{0,n}\ a_0 a_n^* + B_q^{n,0}\ a_0^* a_n \right]\ + 
    \sum_{n,m\neq0}\ B_q^{n,m}\ a_n a_m^*\ + \ldots \bigg\} = 0,
\end{align}
where we are explicitly writing out the terms containing factors of $a_0$.
In the naive \iSPA, $B_q^{0,0}$ and $C_q^{00,}$ are the quadratic coefficients, but upon examining the terms in curly brackets, we see that the first $A$-sum also contains terms quadratic in $a_0$, $a_0^*$.
An improved \iSPA\ uses the coefficients $\bar B$ and $\bar C$ which are renormalized from their original values of $B$ and $C$:
\begin{gather}
    \bar B_q^{0,0} = B_q^{0,0} + \sum_{n\neq0}\ (A_q^{0n,0} + A_q^{n0,0})a_n \\
    \bar C_q^{00,} = C_q^{00,} + \sum_{n\neq0}\ A_q^{00,n}a_n^*.
\end{gather}
Similarly, the coefficients $A_q^{nm,0}$ renormalize $E_q^{,0}$, while $A_q^{0n,p} + A_q^{n0,p}$ renormalize $G_q^{0,}$.
Terms arising from the $B$ and $C$ sums further renormalize $G$, $E$.
In each case, the correction will be finite: from \scSPA\ we know that the $|a_m|$ scale approximately with $\eta_m^{-1}$ for large $m$, and we also know that generically $|\sqrt{\eta_m}|\sim|\w_m-\w_0|$ for large $m$~\cite{ge_thesis_2010}, so that $|a_m| \sim 1/\bar s^2 m^2$, and the sum over all such coefficients (weighted by overlap integrals) will be convergent.
The frequency $\w_0$ is the closest resonance, associated with CF eigenvalue $\eta_0$, $\w_m$ is the resonance associated with $\eta_m$, and $\bar s$ is average level spacing.

The non-negligible corrections to the \iSPA\ coefficients cannot be determined without prior knowledge of all the $a_{i\neq0}$, even though each of them can be individually discarded.
The aggregate effect of the non-dominant CF states is not well-controlled in the sense that it cannot be made arbitrarily small, even as the high-\Q\ assumption is better and better satisfied.
It seems unlikely that the \iSPA\ can be systematically modified to account for this.

\begin{figure}[t!]
\centering
    \centerline{ \includegraphics[width=\textwidth]{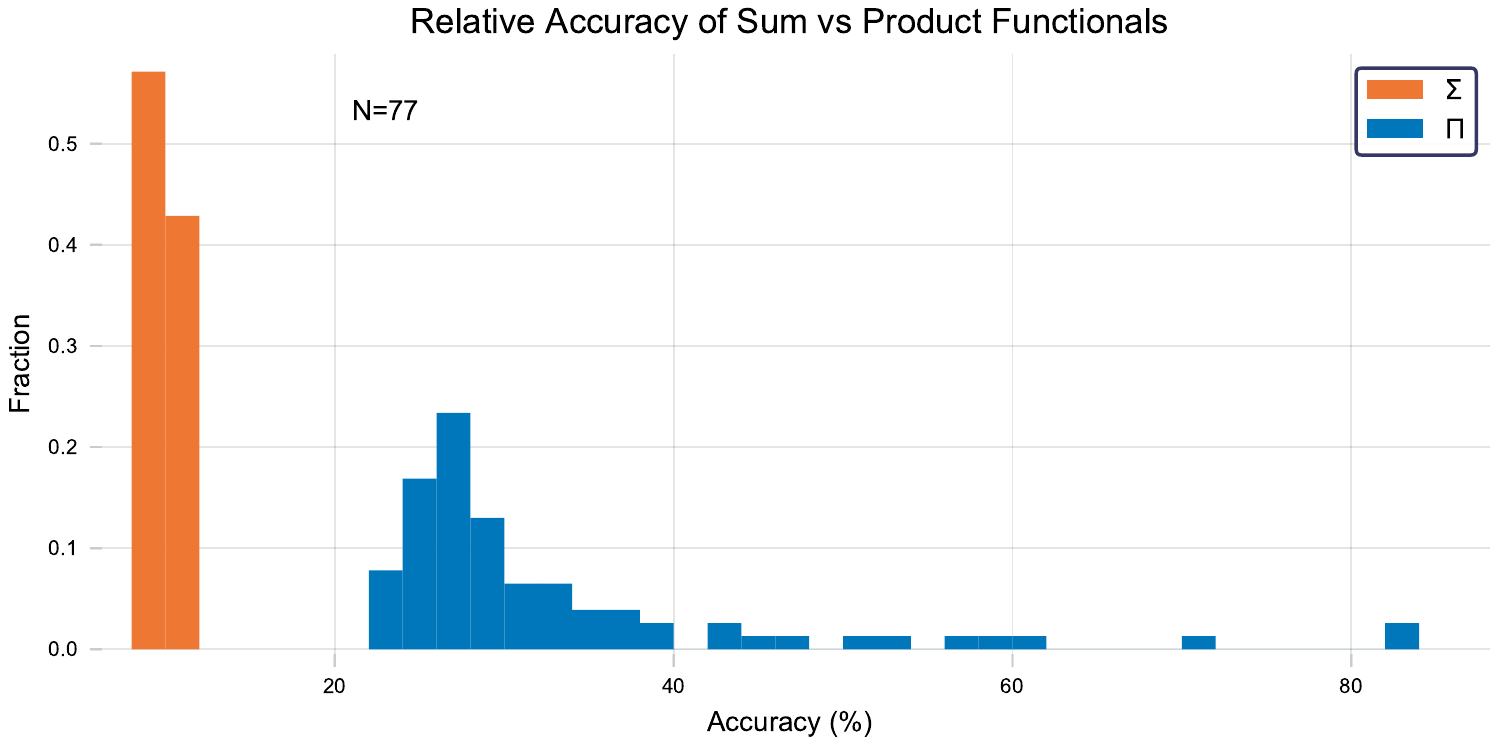} }
    \caption[Comparing the accuracy of the sum-of-roots with the product-of-roots functionals within the inverse single pole approximation]{Comparison of accuracy for the sum of roots ($\Sigma$, orange) and product ($\Pi$, blue) in the \iSPA.
    Shown is a histogram of the accuracy of \iSPA\ estimates over the unstable region of the curves shown in \cref{fig:chpt5: one-sided fields}.
    The mean error in $\Sigma$ is $\sim9\%$, while for $\Pi$ it is $\sim31\%$, consistent with the predictions of \cref{sec:chpt5: failure of ispa}.
    }
    \label{fig:chpt5: sum-v-product}
\end{figure}

Faced with these shortcomings, we must reexamine the validity of the \iSPA\ in extending the results of conventional optical bistability to the case with arbitrary space-dependence.
The form of the predicted bistability curve depended on the quadratic nature of \cref{eq:chpt5: quadratic rho} for $|a_0|$, which itself required that the cubic \cref{eq:chpt5: full nonlinear CF coefficient equation} had a vanishing constant term.
Therefore, if the renormalization of the polynomial coefficients $A$, $B$, etc. either does not generate a new constant term at all, or if that term can be neglected because it is much smaller than all the other renormalized \iSPA\ coefficients, then the qualitative predictions of \iSPA\ remain valid.
This is in fact the case, since in the isolated resonance limit $A_q^{nm,p}$ is negligibly small due to self-cancellation unless all of the indices appear an even number of times: $A_0^{10,1}$ cannot be neglected, but $A_0^{10,0}$ can, in comparison to the others.
When we take this effect into account, the $A$, $B$, and $C$ coefficients for $q=0$ remain unchanged to leading order, while the others become
\begin{gather}
    \bar E_0^{0,} = E_0^{,0} + \sum_{n\neq0} A_0^{nn,0}a_n^2\\
    \bar G_0^{0,} = G_0^{0,} + \sum_{n\neq0} (A_0^{0n,n} + A_0^{n0,n})\ |a_n|^2.
\end{gather}
No new constant term independent of $a_0$ is generated.
It follows that the sum-of-roots functional $\Sigma[\vph(\bx)]$, which depends only on $A$, $B$, and $C$, can be expected to be much more accurate than the product-of-roots $\Pi[\vph(\bx)]$ and the discriminant $\Delta[\vph(\bx)]$, which both depend on $E$ and $G$, as demonstrated in \vref{fig:chpt5: sum-v-product}.

The \iSPA\  fails because it cannot effectively capture the combined effect of the many small $a_{i\neq0}$'s. A possible remedy is to consider a {\it multitude} of CF states, say $n_{\rm CF}>1$ of them, which defines an inverse ``few mode'' approximation.
The infinite array of polynomial equations \crefrange{eq:chpt5: full nonlinear CF coefficient equation}{eq:chpt5: Nonlinear Matrix Coefficients G} will now be truncated to $n_{\rm CF}$ polynomials in the $n_{\rm CF}$ coefficients and their conjugates.
The truncation error will presumably be reduced as $n_{\rm CF}$ is made larger, and the approximation will therefore be more accurate.
One might hope that by including $\mathcal{O}(N)$ states, where $N$ is some computationally feasible number, the inverse few mode approximation could be made substantially more accurate than the poorly performing \iSPA.
Unfortunately, as shown in \vref{tab:chpt5: cf convergence table}, there is no significant improvement in the accuracy of the approximation until $n_{\rm CF}>250$.
To achieve an accuracy better than $1\%$, comparable to the \scSPA, requires $n_{\rm CF}>1000$.
The complexity of constructing and solving these equations for a that many complex CF coefficients is prohibitive.

\begin{table}[t]
\centering
    \begin{tabular}{r c c || r c c}
\toprule
$n_{\rm CF}$ & $|a_0|$ & Error (\%) & $n_{\rm CF}$ & $|a_0|$ & Error (\%) \\
\midrule
1 & 4.10873165 & 32.11
 & 550 & 3.36606946 & 8.23 \\
50 & 4.10656820 & 32.04
 & 600 & 3.33218815 & 7.14 \\
100 & 4.10777713 & 32.08
 & 650 & 3.28121062 & 5.50 \\
150 & 4.10869145 & 32.11
 & 700 & 3.25717018 & 4.73 \\
200 & 4.10902013 & 32.12
 & 750 & 3.23684021 & 4.08 \\
250 & 4.10912925 & 32.13
 & 800 & 3.20222008 & 2.96 \\
300 & 3.78703172 & 21.77
 & 850 & 3.18404535 & 2.38 \\
350 & 3.58326025 & 15.22
 & 900 & 3.17607201 & 2.12 \\
400 & 3.58227253 & 15.19
 & 950 & 3.16374549 & 1.73 \\
450 & 3.51184881 & 12.92
 & 1000 & 3.15344714 & 1.40 \\
500 & 3.42797740 & 10.22
 & 1050 & 3.14431921 & 1.10 \\
\bottomrule
\end{tabular}

    \caption[Error in the inverse ``many pole approximation'' as a function of the number of retained poles]{Relative error in the estimate of the dominant CF coefficient $a_0$, as computed by truncating \cref{eq:chpt5: full nonlinear CF coefficient equation} to the $n_{\rm CF}$ states with the smallest CF eigenvalue (\iSPA\ has $n_{\rm CF}=1$).
    The saturable scattering problem solved for each $n_{\rm CF}$ is the same as in \cref{fig:chpt5: structure and solutions}, for fixed input power equal to the saturable CPA power.
    The inverse ``many-pole'' approximation does not perform any better than the inverse single-pole case until $n_{\rm CF}>250$, and is not comparable to the \scSPA\ until $n_{\rm CF}>1000$.
    }
    \label{tab:chpt5: cf convergence table}
\end{table}

In summary, the implicit assumption built into the \iSPA, that sums over the non-dominant CF amplitudes can be neglected, is false, and dooms its accuracy.
The renormalized equation for the single CF amplitude cannot be constructed without detailed prior knowledge of the non-dominant amplitudes.
Nevertheless, the \iSPA\ shares the same functional form with the correct renormalized equation, and for this reason correctly predicts the number of solutions and the nature of their onset.

    \section{Summary of Algorithm \label{sec:chpt5: summary of algorithm}}

Given a cavity dielectric function $\e_c(\bx)$ and an incident wavefront at frequency $\w$, represented by the equivalent  source $j(\bx)$, we seek all the solutions to the Helmholtz equation with a saturable nonlinearity \cref{eq:chpt5: fundamental equation}:
\be
    \left\{\nabla^2\ + \left[ \e_c({\bx}) +\frac{\g(\w) D_0 F(\bx)}{1+\G(\w)|\vph(\bx,\w)|^2} \right] \w^2 \right\} \vph(\bx) = j({\bx}).
\ee
The end-to-end proposed algorithm for this is as follows:
\begin{enumerate}
    \item \label{it:chpt5: solution step 1} Choose an initial guess $\vph_0(\bx,\w)$.
The results shown in this chapter's figures mostly used $\vph_0(\bx,\w)\equiv0$.
With this choice, the result of the first iteration, $\vph_1(\bx)$, is the solution to the problem with amplification or absorption but no saturation.
If a better guess is at hand, for example if a solution is known for a slightly different input power or with slightly perturbed parameters, then initializing with that solution will reduce the convergence time, and can be used to influence which solution branch is converged to, if more than one exists.
    \item Generate the sequence $\{\vph_n(\bx,\w)\}$ using \cref{eq:chpt5: iterative solution method}, until $\vph_{n+1}(\bx,\w) - \vph_n(\bx,\w)$ is by some measure sufficiently small.
For example, in this chapter the convergence criterion was
\be
    \max_{\bx}(|\vph_{n+1}(\bx,\w)-\vph_n(\bx,\w)|)< 10^{-11}.
\ee
Any $\vph_n(\bx,\w)$ satisfying this condition is a solution.
    \item Extract the solution phase $\alpha$ using \cref{eq:chpt5: extract phase} and construct the overlap integrals in \crefrange{eq:chpt5: Nonlinear Scalar Coefficients A}{eq:chpt5: Nonlinear Scalar Coefficients G}.
This requires solving the CF eigenproblem for the smallest CF eigenvalue $\eta_0$ and its eigenfunction $u_0$, using the saturated $F_s(\bx)$ from \cref{eq:chpt5: Fs}.
Construct the \iSPA\ solutions $a_{0,\pm}^{\rm inv\,SPA}$ from \crefrange{eq:chpt5: inverse SPA solution}{eq:chpt5: delta discriminant}.
    \item The previous step provides poor estimates $a_{0,\pm}^{\rm inv\,SPA}$ for candidate extra solutions.
However, these can be refined by using them as initial guesses for the \scSPA\  [\cref{eq:chpt5: self-consistent SPA theta,eq:chpt5: self-consistent SPA rho}] which gives much better estimates, $a_{0,\pm}$.
    \item If the \scSPA\  fails to converge, it is a sign that likely there are no additional solutions beside what was found in Step~\ref{it:chpt5: solution step 1}, so that the system is in the monostable regime.
On the other hand, if it does converge, then the new solutions $\vph\pr_\pm(\bx,\w)=\vph(\bx,\w) + a_{0,\pm} u_0(\bx,\w)$ are used as initial guesses for the iteration of Step~\ref{it:chpt5: solution step 1}.
The results of this final iteration, together with the original solution $\vph_n(\bx,\w)$ from Step \ref{it:chpt5: solution step 1}, are what we sought: the complete set of solutions to \cref{eq:chpt5: fundamental equation}.
\end{enumerate}

\chapter{Summary and Outlook \label{chp:chpt6: outlook}}
In this thesis we have demonstrated a number of important results for electromagnetic scattering from \nh\ systems, with a special focus on treating structures with arbitrary spatial complexity and quality factor.

We have presented a new class of electromagnetic eigenvalue problems: the $R$-zero and reflectionless scattering mode (RSM) problems.
This novel concept is a broad generalization of impedance matching to systems with arbitrary geometry and numbers of channels.
Unlike the previously studied resonance and $S$-matrix zero problems, RSMs do not necessarily require intrinsic gain or loss to realize, which means that they are relevant even for passive systems.
We analyzed the symmetry properties of the RSMs, and found that they supported a new kind of exceptional point (EP), which can be engineered in both active and passive systems, unlike resonant or scattering EPs.
One robust way to engineer these is in structures with ${\cal PT}$ symmetry, in which case the EP represents the critical point separating the broken and unbroken symmetry phases.
A case of special interest has both ${\cal P}$ and ${\cal T}$ symmetries, which also robustly supports EPs, but is accompanied by a flattening of the transmission lineshape.

We showed that through \nh\ engineering of a cavity, two perfectly absorbed states (CPA modes) can be brought together at an EP, which is a specific example of the RSM EP mentioned above, at which the absorption lineshape can be flattened from quadratic to quartic.
These new CPA EPs share many properties with the conventionally studied resonant EPs, including the chiral nature of the degenerate eigenmode in disk resonators.
We proposed a patterned ring resonator that exploited this chiral behavior to be a nearly perfect absorber for one sense of chirality only, while being a strongly scattering in the other.

In addition to the novel linear problems listed above, we also treated nonlinear two-level media, which exhibit saturation.
First we extended CPA from cavities with a linear dielectric response to include the saturating nonlinearity and dispersion of a two-level absorbing medium.
We augmented the CPA theorem, which relates linear CPA in a lossy cavity to threshold lasing in an amplifying cavity, to account for both saturation and dispersion.
We then used it to show that the SALT algorithm in the single-mode regime can also be used to find the saturable CPA modes through a simple mapping.
We demonstrated that between a lower and upper threshold for loss one can maintain CPA by continuously adjusting the pump strength.
We also clarified the behavior of the $S$-matrix zeros in the bad-cavity limit of dispersive, but linear, CPA, identifying new modes that are hybrids of the cavity and atomic degrees of freedom, with a strongly dispersive response to changes in the pump.

Finally, we proposed a solution algorithm for the general problem of scattering from a saturable amplifying or absorbing two-level medium.
We found, using the inverse single pole approximation (SPA), that the conventional phenomenology of bistability that has mainly been studied in structures with little to no spatial complexity generalizes to arbitrary geometry in the high-\Q\ limit.
We carefully analyzed the validity of the inverse SPA, and found that it is not a well-controlled approximation, even in the high-\Q\ limit, though it makes correct qualitative predictions about bistability.

There are many open questions and directions in which to take this research.
First, the $R$-zero/RSM concept is especially ripe for potential application as a tool in optical design.
That impedance-matched solutions are always accessible with single-parameter tuning has not been appreciated prior to this work, and, in conjunction with hermitian and \nh\ engineering, could be used to design impedance-matched states with targeted outputs under rather general conditions.
One intriguing possibility is already evident in the RSM chaotic cavity waveguide junction shown in \vref{fig:chpt2: octopus}: the radiative gain/loss of many of the waveguides is evidently small, causing relatively little vertical motion of the $R$-zeros in the high-\Q\ case.
This is likely related to the eigenmode statistics of time-reversal-invariant chaotic cavities, which are known to be peaked at zero amplitude~\cite[57]{1991_haake_book}.
Generalizing from this, chaotic cavities should be amenable to the creation of impedance-matched states which go beyond the RSM boundary condition, i.e.,~which do not scatter into all outgoing channels, but only a subset, and different subsets at different frequencies.
Such solutions are not guaranteed by the analytic properties of the $S$-matrix employed in the general theory, but still may be relatively easy to find with a conventional optimization scheme, combined with an RSM code. 
% to the simultaneous engineering of several different RSMs within the same cavity, with relatively independent tuning degrees of freedom. 
Another avenue to be explored is the role of ${\cal P, T}$ symmetry for RSM EPs in lossless cavities.
So much work has been done on the effects of exceptional points in cavities with gain and/or loss; the possibility of an exceptional point in a cavity without loss or gain, where the transmission lineshape is tunable is promising.
The cavity polariton zeros from saturable CPA are also largely unexplored, and a hybrid resonant EP between the electromagnetic and atomic degrees of freedom is an intriguing possibility, as a change in lineshape in the frequency domain is accompanied by a change in the response function in the time-domain.
Thus, a cavity polariton EP could be a potential mechanism for changing the polarization dynamics of the atomic ensemble through macroscopic cavity engineering.
Finally, the work on saturable absorption and amplification in arbitrary geometries is new, and many aspects are unexplored.
The question of dynamical stability in arbitrary geometries is open, though likely the answer mirrors the well-known ring-laser geometries to some extent.
A very important question with immediate potential impact for the applicability of SALT in laser design is the reassessment of the inverse single pole approximation (SPA).
This approximation is known to diverge from the SALT prediction when several modes are lasing, and it has been assumed that this is due to a breakdown of the approximation that the Green function is dominated by a single pole.
This new work suggests that a modification of the SPA-SALT, which is far more efficient than the full SALT algorithm, to use the \scSPA\ instead of the \iSPA, could possibly extend the range over which SPA-SALT is accurate deeper into the multimode regime.

\appendix

\chapter{Determinants of $S(\w)$ and $R_{\rm in}(\w)$ \label{chp:app: linalg} }
In this appendix we derive the determinant relations \cref{eq:chpt1: det(S)} for an $N$-channel $S$-matrix, and \cref{eq:chpt2: det(R)} for an $N_{\rm in}$-channel generalized reflection matrix $R_{\rm in}$, from their respective wave-operator representations \cref{eq:chpt1: Heidelberg,eq:chpt2: Rin_S}.
To that end, we first derive two identities from linear algebra that will be used for both $S$ and $R_{\rm in}$.

    \subsection*{Linear Algebra Identities}

In this section, matrix $A \in {\mathbb C}^{N \times N}$ is invertible, and matrices $B \in {\mathbb C}^{N \times M}$ and $C \in {\mathbb C}^{M \times N}$ are arbitrary.
\paragraph*{First Identity}
The first identity we will need is
\begin{equation}
    \label{eq:app: push-through}
     (A+BC)^{-1} B=A^{-1} B ( I_M + CA^{-1} B)^{-1}.
\end{equation}
It is a generalization of the ``push-through identity''~\cite{Bernstein_Matrix_book}
\begin{equation}
    \label{eq:app: baby push_through}
    (I_N + BD)^{-1} B = B( I_M + DB)^{-1},
\end{equation}
with $D \in {\mathbb C}^{M \times N}$ an arbitrary matrix.
It is named for its action on $B$ relative to the inverse, and follows trivially from noting that 
\begin{equation}
    B( I_M + DB) = ( I_N + BD) B.
\end{equation}
This can be generalized for $A,B,C$ by starting with 
\begin{equation}
    ( A + BC)^{-1} = A^{-1}( I_N + BCA^{-1})^{-1}.
\end{equation}
Applying \cref{eq:app: baby push_through}, with $D = CA^{-1}$, we arrive at
\begin{equation}
    \label{eq:app: genearlized push-through}
    ( A + BC)^{-1} B = A^{-1} B ( I_M + CA^{-1} B)^{-1},
\end{equation}
which proves the first identity.

\paragraph*{Second Identity}

The second identity we will need is
    \begin{equation}
        \label{eq:app: Sylvester}
         \det( I_N- BC) = \det( I_M - CB).
    \end{equation}
It can be derived as a special case of Schur's determinant formula~\cite{Handbook_of_LA}:
\begin{equation}
    \label{eq:app: Schur}
    \det A \det( D - CA^{-1} B) = \det D \det( A - BD^{-1} C),
\end{equation}
where $D \in {\mathbb C}^{M \times M}$ is invertible.
The judicious choice $ A = I_N$, $ D = I_M$ gives
\begin{equation}
    \det ( I_N - BC) = \det ( I_M - CB),
\end{equation}
which is what we wanted to show.

    \subsection*{Evaluating $\det S(\w)$ \label{sec:app: detailed_derivation_S}}

This section refers to the matrices $A\pr_0$, $\Delta$, $\G$, $W_p$ that are defined in \cref{eq:chpt1: Aeff_and_Sigma}, though these definitions are not needed here to derive of the determinant relation \cref{eq:chpt1: det(S)}.
We apply \cref{eq:app: push-through} to push $W_p$ through $( A_0\pr - \Delta + i \G)^{-1}$, recalling that $ \G = \pi W_p W_p^\dagger$, which gives
\begin{equation}
    ( A_0\pr - \Delta + i \G)^{-1} W_p = G_0\prpr W_p ( I_N + i \pi W_p^\dagger G\prpr_0 W_p)^{-1},
\end{equation}
where $ G_0\prpr \defn ( A_0\pr - \Delta)^{-1}$.
Plug this into the expression for the $S$-matrix [\cref{eq:chpt1: Heidelberg,eq:chpt1: Aeff_and_Sigma}], and factor out $( I_N + i \pi W_p^\dagger G_0\prpr W_p)^{-1}$:
\begin{equation}
    S = ( I_N - i \pi W_p^\dagger G_0\prpr W_p)/( I_N + i \pi W_p^\dagger G_0\prpr W_p),
\end{equation}
which is the $K$-matrix representation of \cref{eq:chpt1: k-matrix rep}~\cite{1967_MacDonald_PR, 1982_Newton_book}.
Taking the determinant and applying \cref{eq:app: Sylvester} to the numerator and denominator, we have
\begin{align}
    \det S &= \frac{\det ( I - i \pi G_0\prpr W_p W_p^\dagger)}{\det( I + i \pi G_0\prpr  W_p W_p^\dagger)}\nonumber\\
    & = \frac{\det ( A_0\pr - \Delta - i \G)}{\det( A_0\pr - \Delta + i \G)},
\end{align}
where $I$ is the identity on the (infinite-dimensional) closed-cavity Hilbert space.
In the last step we multiplied the numerator and denominator by $\det G_0^{\prime\prime-1}$.
This proves \cref{eq:chpt1: det(S)}, which is what we wanted to show.

    \subsection*{Evaluating $\det R_{\rm in}(\w)$ \label{sec:app: detailed_derivation_Rin}}

We proceed as we did with the $S$-matrix, but starting with $R_{\rm in}$ from \cref{eq:chpt2: Rin first}: push $W_F$ through $ G_{\rm eff}$, factor out a common inverse, take the determinant and apply \cref{eq:app: Sylvester}.

Writing $ G_{\rm eff}$ in \cref{eq:chpt2: Rin first} in terms of $ W_F$ and $ W_{\fbar}$,
\begin{equation}
    G_{\rm eff}^{-1} = (A\pr_0 - \Delta +  i \G_{\fbar}) +  i W_F W_F^\dagger,
\end{equation}
and using the identity \cref{eq:app: push-through} yields
\begin{equation}
    (\bar A_0+i \G_F)^{-1} W_F = \bar A_0^{-1} W_F( I_{N_{\rm in}} + i \pi W_F^\dagger \bar A_0^{-1} W_F)^{-1},
\end{equation}
where 
\begin{gather}
    \G_F \defn \pi W_F W_F^\dagger, \\
    \bar A_0 \defn A_0\pr - \Delta + i \Gamma_{\fbar}.
\end{gather}

Plugging this into \cref{eq:chpt2: Rin first} and factoring out $( I_{N_{\rm in}} + i \pi W_F^\dagger \bar A_0^{-1} W_F)^{-1}$ gives a $K$-matrix representation for $ R_{\rm in}$:
\begin{equation}
    R_{\rm in} = ( I_{N_{\rm in}} - i \pi W_F^\dagger \bar A_0^{-1} W_F)/( I_{N_{\rm in}} + i \pi W_F^\dagger \bar A_0^{-1} W_F).
\end{equation}

Taking the determinant, using the identity \cref{eq:app: Sylvester}, and multiplying the numerator and denominator by $\det \bar A_0$ results in
\begin{align}
    \det  R_{\rm in} &= \frac{\det ( I - i \pi \bar A_0^{-1} W_F W_F^\dagger)}{\det ( I + i\pi \bar A_0^{-1} W_F W_F^\dagger)}\\
    &=\frac{\det ( A\pr_0 - \Delta+i \G_{\fbar}-i \G_F)}{\det ( A_0\pr - \Delta +i \G_{\fbar}+i \G_F)} \\
    &=\frac{\det ( A\pr_0 - \Delta+i \G_{\fbar}-i \G_F)}{\det ( A_0\pr - \Delta +i \G)}.
\end{align}
In the last step we used the identities for the filters $F$ and $\bar F$ \cref{eq:chpt2: F + Fbar matrix id} to simplify the denominator.
This proves \cref{eq:chpt2: det(R)}.

\chapter{Boundary Matching from $\hat A_{\rm eff}(\w)$ \label{chp:app: boundary matching} }
In this appendix we will derive the general boundary-matching condition given in \cref{eq:chpt1: general boundary matching} from the effective Hamiltonian formalism of~\cref{sec:chpt1: effective hamiltonian}.

This formalism divides space into two regions: a scattering region and an asymptotic region, with the boundary between them being the last scattering surface (LSS).
The wave operator $\hat A$ has three components, $\hat A_0$ in the interior region, $\hat A_c$ in the asymptotic channels, and $\hat V$ at the LSS, with $\hat A(\w) = \hat A_0(\w) + \hat A_c(\w) + \hat V(\w)$.
For simplicity, we will derive the boundary-matching condition in one-dimension for a single asymptotic region $x>0$, with scattering region $x<0$ and LSS $x=0$.
The result we obtain readily generalizes to a two-sided geometry and to higher dimensions.

We follow an approach similar to the Feshbach projection method used in~\cite{2000_Dittes_PR, 2002_Saving_arxiv, 2009_Rotter_JPA, 2017_Rotter_RMP}.
Divide the space by using projectors $\hat P$ and $\hat Q \defn 1- \hat P$ for the scattering and channel regions, respectively ($\hat P \hat Q = \hat Q \hat P \equiv 0$).
In coordinate space, these are
\begin{equation}
    \braket{x | \hat P | x\pr} = \delta(x-x\pr) \theta(-x), \qquad \braket{x | \hat Q | x\pr} = \delta(x-x\pr) \theta(x),
\end{equation}
where $\theta$ is the step-function: $\theta(x) = 1$ for $x>0$, and zero otherwise.
Using projectors $\hat P$ and $\hat Q$ on the Helmholtz operator $\hat A \equiv (\hat P + \hat Q)\, \hat A\, (\hat P + \hat Q)$, we have
\begin{gather}
    \hat A  = \hat A_0 + \hat A_c + (\hat V_{QP} + \hat V_{PQ}), \\
    \hat A_0 \defn \hat P \hat A \hat P,
        \qquad 
    \hat A_c \defn \hat Q \hat A \hat Q \\
    \hat V_{PQ} \defn \hat P \hat A \hat Q, 
        \qquad
    \hat V_{QP} \defn \hat Q \hat A \hat P.
\end{gather}
The Helmholtz equation is satisfied both inside and outside scattering region:
\begin{align}
    \hat P \hat A \ket{\psi} = 0\quad &\implies \quad \hat A_0 \ket{\psi} = -\hat V_{PQ} \ket{\psi}, \\
    \hat Q \hat A \ket{\psi}=0\quad  &\implies  \quad \hat A_c \ket{\psi} = -\hat V_{QP} \ket{\psi} \label{eq:app: a_c vqp}. 
\end{align}
These equations contain terms like $\theta(-x)\delta(x)\psi\pr(x)$.
It is tempting to evaluate these as $\theta(0)\delta(0)\psi\pr(0)$, but this incorrect, since the specific value of $\theta(0)$ is arbitrarily defined and immaterial.
Instead, we evaluate them in a limiting sense:
\begin{equation}
    \theta(\pm x)\delta(x)\psi\pr(x) = \lim_{dx \to 0^+} \delta(x\mp dx)\psi\pr(0 \mp dx).
\end{equation}
The matrix elements of the operators $\hat A_0$, $\hat A_c$, $\hat V_{PQ}$, and $\hat V_{QP}$ are therefore
\begin{gather}
    \braket{x\pr | \hat A_0 | x} = \delta(x-x\pr) \lim_{dx\to0^+}\{ \theta(-x) [\nabla^2 + \e(x)\w^2] + \delta\pr(x+dx) \}, \\
    \braket{x\pr | \hat A_c | x} = \delta(x-x\pr) \lim_{dx\to0^+}\{ \theta(x) [\nabla^2 + \e(x)\w^2] - \delta\pr(x-dx) \}, \label{eq:app: ac matrix element}\\
    \braket{x\pr | \hat V_{PQ} | x} =  - \lim_{dx\to0^+}  \delta\pr(x+dx), \qquad
    \braket{x\pr | \hat V_{QP} | x} =  + \lim_{dx\to0^+}  \delta\pr(x-dx).  \label{eq:app: qp matrix element}
\end{gather}
Plugging these into \cref{eq:app: a_c vqp}, and using $\delta\pr(x)\psi(x) = -\delta(x)\psi\pr(0)$, we have%$\int dx\,\delta\pr(x)f(x) = -f\pr(0) $ for any smooth test function $f$, we have
\begin{equation}
    \label{eq:app: prelim bc}
    \psi(x) = \lim_{dx\to0^+} G_c^R(x,dx) \psi\pr(dx),
\end{equation}
where $\hat G_c^R$ is the retarded inverse of $\hat A_c$, consistent with purely outgoing radiation in the asymptotic regions.
The solution to the original, unpartitioned Helmholtz equation is continuously differentiable at the LSS, so that $\lim_{dx\to0}\psi\pr(dx) = \psi\pr(0)$, and \cref{eq:app: prelim bc}, evaluated at $x=0$, becomes
\begin{equation}
    \label{eq:app: prelim bc 2}
    \psi(0) - G_c^R(0,0) \psi\pr(0) = 0.
\end{equation}

We have not yet specified the boundary condition on $\hat G_c^R$ at the LSS.
In the asymptotic domain, i.e.,~for $x,x\pr>0$, $G_c^R(x,x\pr)$, with $x<x\pr$, must satisfy 
\begin{equation}
     0 = \lim_{dx\to0^+} \{[\nabla^2 + \e(x)\w^2] - \delta\pr(x+dx) \} G_c^R(x,x\pr),
\end{equation}
Upon integrating from $0$ to $2dx$, 
\begin{equation}
    \label{eq:ap: G neumann}
    0 = \lim_{dx\to0^+} \{  \partial_x G_c^R(2dx,x\pr) -  \partial_x G_c^R(0,x\pr) +  \partial_x G_c^R(dx,x\pr) \} = \partial_x  G_c^R(0,x\pr),
\end{equation}
where $ \partial_x$ indicates the derivative with respect to the first argument of $G_c^R$.
\Cref{eq:ap: G neumann} implies a Neumann boundary condition at the LSS for $\hat G_c^R$.
The boundary-matching condition that $\psi$ satisfies is therefore
\begin{equation}
    \psi(0) - G^{R,N}_c(0,0) \psi\pr(0) = 0.
\end{equation}
For a scattering region $\Omega$ and LSS $\partial\Omega$ in any dimension, this generalizes to
\begin{equation}
    \label{eq:app: general boundary matching}
    \left[\ \psi(\bx) - \oint_{\partial\Omega} \ G^{R,N}_c(\bx,\bx\pr) \nabla \psi(\bx\pr) \cdot d{\bf S}\pr\ \right]_{\bx \in \partial\Omega} = 0,
\end{equation}
where $d{\bf S}$ is the outwardly-directed area element, which is \cref{eq:chpt1: general boundary matching}.

\chapter{On the Free-Space Outgoing Boundary Conditions in Two-Dimensions \label{chp:app: free space} }
In \Cref{chp:app: boundary matching} we derived \cref{eq:chpt1: general boundary matching}:
\begin{equation}
    \left[\ \psi(\bx) - \oint_{\partial\Omega} \ G^{R,N}_c(\bx,\bx\pr) \nabla \psi(\bx\pr) \cdot d{\bf S}\pr\ \right]_{\bx \in \partial\Omega} = 0.
\end{equation}
In the main text it was noted that the outgoing boundary condition for two-dimensional free-space \cref{eq:chpt1: free space bc},
\begin{equation}
    \left[ \psi(r,\theta) - \int \left(\frac{1}{2\pi kR}\sum_m\cis{m(\theta-\theta\pr)} \frac{H_m^+(k R)}{H_m^{+\prime}(k R)} \right) \partial_r\psi(r,\theta)\ R\, d\theta\pr \right]_{r=R} = 0,
\end{equation}
is not manifestly of this form.
In this appendix we will show that \cref{eq:chpt1: free space bc} does indeed follow from \cref{eq:chpt1: general boundary matching}.

We will first construct $G^{R,N}$, the retarded Green function in free space satisfying Neumann boundary conditions at the LSS defined by $r=R$.
We will do this by decomposing $G^{R,N}$ into the sum of two terms: the fundamental solution, which satisfies $\hat A \hat G^R = \hat 1$ with purely outgoing boundary conditions at $|\bx|\to\infty$, and an operator $\hat F$ satisfying $\hat A \hat F = \hat 0$, also with purely outgoing conditions, chosen so that $G^R(\bx,\bx\pr) + F(\bx,\bx\pr)$ satisfies the Neumann boundary condition at $r=R$.

The fundamental solution for the Helmholtz operator in two-dimensions is~\cite[811]{1981_morse_book}\footnote{The convention used in~\cite{1981_morse_book} is $\hat A \hat G^R = -4\pi \hat 1$, so that our expression is is different by a factor of $-1/4\pi$.}
\begin{equation}
    G^R(\bx,\bx\pr) = -\frac{i}{4}H_0^+(k|\bx-\bx\pr|),
\end{equation}
where $k=\w/c$.
By Graf's addition theorem~\cite[363]{1972_Abramowitz_book},
\begin{equation}
    H_0^+(k|\bx-\bx\pr|) = \sum_{m=-\infty}^\infty H_m^+(k|\bx\pr|)J_m(k|\bx|)\cis{m(\theta-\theta\pr)},
\end{equation}
which is valid for $|\bx|\le|\bx\pr|$ (we will ultimately we will take both to $R$).

$F(\bx,\bx\pr)$ can be expanded in outgoing Hankel functions:
\begin{equation}
    F(\bx,\bx\pr) = \sum a_m(\bx\pr) H^+(k|\bx|)\cis{m(\theta-\theta\pr)}.
\end{equation}
The coefficients $a_m(\bx\pr)$ are fixed by requiring that $\partial_r\left[G^R(\bx,\bx\pr) + F(\bx,\bx\pr)\right]_{|\bx|=R}=0$, which gives
\begin{equation}
    F(\bx,\bx\pr) = \frac{i}{4}\sum_{m=-\infty}^\infty \frac{J_m\pr(kR)}{H^{+\prime}_m(kR)}H^+_m(k|\bx\pr|)H^+(k|\bx|)\cis{m(\theta-\theta\pr)}.
\end{equation}
Adding $F$ and $G^R$ and evaluating on the LSS, i.e.,~$|\bx|=|\bx\pr|=R$, gives
\begin{equation}
    \label{eq:chpt1: intermediate G 2-dim}
    G^{R,N}(\bx,\bx\pr) \bigg|_{\rm LSS} = -\frac{i}{4}\sum_{m=-\infty}^\infty \left[ J_m(kR) H_m^{+\prime}(kR) - J_m\pr(kR) H_m^+(kR) \right] \frac{H_m^+(kR)}{H_m^{+\prime}(kR)} \cis{m(\theta-\theta\pr)}.
\end{equation}
The term in brackets is the Wronskian of $J_m$ and $H_m^+$, which is~\cite[360]{1972_Abramowitz_book}
\begin{equation}
    W\{J_m,H^+_m\}(z) = i W\{J_m,Y_m\}(z)=\frac{2i}{\pi z}.
\end{equation}
Plugging this into \cref{eq:chpt1: intermediate G 2-dim} yields
\begin{equation}
    G^{R,N}(\bx,\bx\pr) \bigg|_{\rm LSS} = \frac{1}{2\pi kR}\sum_{m=-\infty}^\infty \frac{H_m^+(kR)}{H_m^{+\prime}(kR)} \cis{m(\theta-\theta\pr)}.
\end{equation}
When used in \cref{eq:chpt1: general boundary matching}, this yields the same \cref{eq:chpt1: free space bc} that we derived in the main text from using continuity of the interior solution with the channel states.

\backmatter

\printbibliography[heading=bibintoc]

\end{document}